\documentclass[draft,tightenlines,nofootinbib,preprint,aps,eqsecnum,amsmath,amssymb]{revtex4}

\begin{document}

\newcommand{\beq}{\begin{equation}}
\newcommand{\eeq}{\end{equation}}
\newcommand{\bea}{\begin{eqnarray}}
\newcommand{\eea}{\end{eqnarray}}
\newcommand{\cir}{{\buildrel \circ \over =}}

\newcommand{\sgn}{\mbox{\boldmath $\epsilon$}}

\newcommand{\on}{\stackrel{\circ}{=}}
\baselineskip 20pt

\title{Simultaneity, Radar 4-Coordinates and the 3+1
Point of View about Accelerated Observers in Special Relativity.}

\author{David Alba}

\affiliation {Dipartimento di Fisica\\ Universita' di Firenze\\
Via G. Sansone 1\\ 50019 Sesto Fiorentino (FI), Italy\\ E-mail:
ALBA@FI.INFN.IT}

\author{Luca Lusanna}

\affiliation
 {Sezione INFN di Firenze\\ Via G. Sansone 1\\ 50019
Sesto Fiorentino (FI), Italy\\ E-mail: LUSANNA@FI.INFN.IT}

\begin{abstract}

After a review of the 1+3 point of view on non-inertial observers
and of the problems of rotating reference frames, we underline
that was is lacking in their treatment is a good global notion of
simultaneity due to the restricted validity (coordinate
singularities show up) of the existing 4-coordinates associated to
an accelerated observer (like the Fermi normal ones).

We show that the relativistic Hamiltonian 3+1 point of view, based
on a 3+1 splitting of Minkowski space-time with a foliation whose
space-like leaves are both simultaneity and Cauchy surfaces,
allows to find a solution to such problems, if we take into
account M$\o$ller's definition of allowed 4-coordinate
transformations extended to radar 4-coordinates. Each admissible
choice of simultaneity implies an associated definition of
instantaneous 3-space (and of spatial distance) and of one-way
velocity of light.

Rigidly rotating relativistic reference frames are shown not to
exist. We give explicit foliations, with simultaneity surfaces
(also space-like hyper-planes) non orthogonal to the arbitrary
non-inertial observer world-line, which correspond to a good
notion of simultaneity for suitable (mutually balancing)
translational and rotational accelerations. Viceversa, given one
such admissible foliation, we can determine the modification of
Einstein's convention implied by its associated notion of
simultaneity. This treatment allows:

i) To give the 3+1 description of both the rotating disk (its
3-geometry depends on the choice of simultaneity)) and the Sagnac
effect.

ii) To show how a GPS system of spacecrafts may establish a grid
of admissible radar 4-coordinates, namely an empirical notion of
simultaneity.

iii) How, given an admissible empirical notion of simultaneity
adapted to Earth's rotation, instead of assuming Einstein's
convention plus Sagnac corrections, it is possible to determine
the associated time delay (including the Shapiro delay as a
post-Newtonian effect) between an Earth station and a satellite.
Its comparison with the future measurements of the ACES mission
will allow to synchronize the clocks according to this empirical
simultaneity.

We show that in parametrized Minkowski theories all the admissible
notions of simultaneity are gauge equivalent ({\it conventionality
of simultaneity as a gauge theory}) and, as an example, we
describe Maxwell theory in non-inertial systems with any
admissible notion of simultaneity, like those needed for a correct
treatment of the magnetosphere of pulsars. These considerations
can be extended to canonical metric gravity on globally hyperbolic
space-times, where, however, the admissible notions of
simultaneity are {\it dynamically} determined by the ADM Hamilton
equations, equivalent to Einstein;s equations.

\today

\end{abstract}

\maketitle

\newpage

\section{Introduction}

The increasing importance of special relativity and post-Newtonian
gravity in fields connected with space navigation and experiments
in the Solar System, clock synchronization and, more in general,
with the rotational aspects of relativistic kinematics  in
astrophysics requires a revisitation of the topics connected with
the notion of simultaneity and with the problem of how an
accelerated observer can build a good system of radar
4-coordinates compatible with a given notion of simultaneity. This
paper is devoted to such a revisititation and to an attempt to
find a unified treatment of these problems. Therefore we start
with a review of the open problems and then we state our
viewpoint.
\bigskip

We shall use the signature  $\eta^{\mu\nu} = \sgn\, (+,---)$, with
$\sgn = \pm$ according to whether the particle physics or general
relativity convention is adopted, for the Minkowski 4-metric and
we shall put $c = 1$. Nevertheless, we keep $c$ in various
formulas for the sake of clarity.
\bigskip

Newtonian mechanics in Galilei space-time (fusion of an absolute
space with an absolute time) and special relativity in Minkowski
space-time (an absolute space-time) rely both on a {\it relativity
principle}, according to which the laws of physics are the same in
every {\it inertial system} (an {inertial observer} with its time
axis and a choice of space axes), namely in a special family of
{\it rigid systems of reference} in uniform translational motion
one with respect to the other. That is the laws of nature are
covariant and there is no preferred inertial observer. In any
inertial system, according to the law of inertia, a material
particle, not acted upon by any agent, will continue to move in a
straight line with constant velocity. Special coordinates are
associated to each inertial system: either Cartesian 3-coordinates
plus time or pseudo-Cartesian (Lorentzian) 4-coordinates. The
transition from an inertial system to another one is performed
with a kinematical group of global transformations, either the
Galilei group or the Poincare' group.

\bigskip

Then, the empirical point of view needed to establish a theory of
measurements requires the replacement of abstract ideal concepts
like (either absolute or dynamical) {\it time} and {\it space}
with actual metrological standards like a {\it clock} and a {\it
rod}: physical time intervals and spatial distances are only
relative quantities with respect to the chosen reference standard
units (only ratios of quantities are physically meaningful), which
are constantly upgraded following the developments of both theory
and technology.

\bigskip

In Newtonian mechanics the absolute space may be identified as an
inertial system associated to the fixed stars. In special
relativity, where it is the space-time to be absolute, one makes a
conventional choice of a {\it quasi-inertial (non-rotating)}
reference system \cite{1a,2a} \footnote{ According to the IAU 2000
Resolutions \cite{3a}, it is a coordinate system named  the {\it
Solar System Barycentric Celestial Reference System}, which is
materialized in the {\it Solar System Barycentric Celestial
Reference Frame} by specifying its axes by means of fixed stars
(quasars) in the Hypparcos catalog \cite{4a}. For processes in the
vicinity of the Earth the (non-inertial but non-rotating) {\it
Geocentric Celestial Reference System} is used. In the definition
of these coordinate systems the post-Newtonian approximation to
general relativity is taken into account \cite{5a}. As a
consequence the metrology of general relativity \cite{6a} (see
Ref.\cite{7a} for an older point of view) has to be used to define
measurable quantities (compatible with general covariance) inside
the solar system, since the special relativistic approximation is
no more sufficient to describe phenomena like time delays. In
particular, given the time-like world-line of an observer, the
{\it proper time} of the clock and the {\it proper length} of an
infinitesimal rod, together with associated coordinate-independent
units, carried by the observer, have to be defined. Assuming the
value of the two-ways invariant velocity of light $c$ as a
conventional constant, the unit of length, the {\it proper meter},
is derived from the unit of time, the {\it proper second}. For the
synchronization of distant clocks {\it Einstein's convention is
used}, while for the definition of the distance of events at
finite space-like separation both the world-line of an observer
associated to one of these events and a space-like path joining
them is needed besides the synchronization convention. In
practice, special coordinate systems, like the two previous ones,
are introduced and coordinate-dependent units of time (the TAI
second) and of length are used, with suitable instructions to
connect them to the proper units.}. In both theories the
chrono-geometrical structure is absolute, i.e. non dynamical.
Given some standard of length and time ({\it rods and clocks}), we
can perform any measurement we like on a system independently from
its dynamics. Only in general relativity the chrono-geometrical
structure becomes dynamical \cite{8a,9a}.

\bigskip

The main difference between the two theories lies in the notion of
{\it simultaneity of two events}. Due to the absolute nature of
Newtonian time, the points on a $t = const.$ section of Galilei
space-time are all simultaneous (instantaneous absolute 3-space),
whichever inertial system we are using. As a consequence, the
causal notions of {\it before} and {\it after} a certain event are
absolute. Instead in special relativity there is no absolute
notion of simultaneity. Given an event, all the points outside the
light cone with vertex in that event are not causally connected
with that event (they have space-like separation from it), so that
the notions of {\it before} and {\it after} an event become {\it
observer-dependent}. Once we have chosen to describe physics with
respect to any inertial system ($x^{\mu}$ are the associated
Cartesian 4-coordinates), the events simultaneous for the inertial
observer chosen as origin are usually assumed to be those lying in
the space-like hyper-planes $x^o = c\, t = const.$ in accord with
Einstein's convention for the synchronization of distant clocks.
\medskip

As a consequence, the synchronization of two space-like separated
clocks  has to be defined, not being implied by the
chrono-geometrical structure of Minkowski space-time. Usually this
is done by means of {\it Einstein's convention} \cite{10a} (see
for instance Refs.\cite{11a,12a} ) based on the choice of the rays
of light \footnote{The {\it conformal} structure of Minkowski
space-time is selected by the {\it two independent postulates} of
special relativity that the round-trip velocity of light is the
{\it same} in every inertial system (the round trip postulate) and
{\it isotropic} (the light postulate). Let us remark that only the
{\it round-trip} ({\it two-way}) speed of light has a physical
significance, since the {\it one-way} velocity between two events
$A$ and $B$ (and its being or not isotropic) depends on the
definition of synchronization of the two clocks at those points,
i.e. from the notion of simultaneity used.} as preferred tools to
measure time and length. In a given inertial system the clock $A$,
associated to the time-like world-line $\gamma_A$, emits a light
signal at its time $x^o_{Ai}$, corresponding to an event $Q_i$ on
$\gamma_A$, towards the time-like world-line $\gamma_B$ carrying
the clock $B$ \footnote{The clocks are assumed to be standard
clocks measuring proper time. See Refs.\cite{13a,14a,6a} for their
mathematical characterization in special and general relativity.}.
When the signal arrives at a point $P$ on $\gamma_B$, it is
reflected towards $\gamma_A$, where it is detected at time
$x^o_{Af}$, corresponding to an event $Q_f$ on $\gamma_A$. Then
the clock $B$ at the event $P$ on $\gamma_B$ is synchronized to
the time $x^o_A = {1\over 2}\, (x^o_{Ai} + x^o_{Af})$,
corresponding to an event $Q$ in between $Q_i$ and $Q_f$. It can
be checked that $Q$ and $P$ {\it lie on the same space-like
hyper-plane orthogonal to the world-line} $\gamma_A$, i.e. that
they are simultaneous events for the chosen inertial observer
\footnote{Einstein's convention has been criticized by Reichenbach
and Grundbaum \cite{15a}, who said that any convention $x^o_A =
E\, x^o_{Af} + (E - 1)\, x^o_{Ai} = x^o_{Ai} + E\, (x^o_{Af} -
x^o_{Ai})$ with $0 < E = const. < 1$ can be used ({\it
conventionalism of simultaneity}) without leading to any
contradiction. In general light propagation becomes anisotropic,
i.e. direction dependent, because from $x^o_B = x^o_A$, $x^o_B -
x^o_{Ai} = E\, (x^o_{Af} - x^o_{Ai})$, $x^o_{Af} - x^o_B = (1 -
E)\, (x^o_{Af} - x^o_{Ai})$ we get $c_{AiB} = {c \over 2}\, E$,
$c_{BAf} = {c \over 2}\, (1 - E)$, ${2\over c} = {1\over
{c_{AiB}}} + {1\over {c_{BAf}}}$ with either $c_{AiB} > c$ or
$c_{BAf} > c$ (but not both). See
Refs.\cite{16a,17a,18a,19a,20a,21a,22a} and their rich
bibliography for the various aspects of the debate about the
conventionalist point of view. In particular let us stress the
following points: a) The constant $E$ may be generalized to a
point-dependent function (see for instance Ref.\cite{23a}); b)
Anderson and Stedman \cite{18a} underline how the notions of
spatial distance depend on the choice of the notion of
simultaneity made by the observer (see footnote 1); c) Giulini
\cite{17a} shows that in the relativistic case Malament's
non-conventionalist notion of {\it absolute simultaneity} \cite
{16a} (as an equivalence relation implied by the causal
automorphisms preserving the light-cone structure) has to be
replaced with a notion of {\it relative simultaneity} with respect
to some additional structure on space-time; d) both
Anderson-Stedman \cite{18a} and Minguzzi \cite{22a} propose a {\it
gauge interpretation} of simultaneity and of the one-way velocity
of light; e) in Ref.\cite{19a}, where there is a review of the
various viewpoints on the conventionality of simultaneity, it is
underlined the conventional nature of the statements about the
isotropy or anisotropy of light propagation and of the
measurements of length.}. In general relativity Einstein's
convention has been generalized to non-inertial observers by
Martzke and Wheeler \cite{24a} to define local 4-coordinates,
which can be called {\it radar coordinates} due to the technology
implied to build them. An alternative to the use of light rays is
the synchronization of clocks by their {\it slow transport}: we
shall not deal with it, since in Ref. \cite{25a} it is shown its
equivalence to Einstein's convention notwithstanding claims of the
contrary (see  also Section 2.1 of Ref.\cite{19a}).

\bigskip

In both theories the concept of inertial observers is an {\it
idealization}. Every actual observer is always accelerated and in
practice (for instance in astronomy) one speaks of {\it
quasi-inertial systems} \cite{1a,2a,3a,4a}. In Newtonian mechanics
they are defined as those {\it rigid} systems of reference, in
which the sensibility of the measuring apparatuses does not allow
to detect any inertial force, like the Coriolis one, which modify
Newton's law ($m\, \vec a = \vec F\,\, \mapsto\,\, m\, {\vec
a}^{'} = {\vec F}^{'} + m\, \vec f$) when dynamics is described by
an observer carrying a rigid accelerated system of reference (see
for instance Ref.\cite{26a}, Section 39). In special relativity
the notion of quasi-inertial system is more problematic, because
there is no accepted definition of inertial forces seen by an
accelerated observer when a manifestly Lorentz covariant
description of relativistic mechanics is used \footnote{See
Ref.\cite{27a} for the problems, like the no-interaction-theorem,
associated to the description of the motion of massive
relativistic particles either free or with action-at-a-distance
relativistic interactions. Again the main problem is how to
perform a simultaneous description of the interacting particles.}.
It is only in the Hamiltonian version of relativistic mechanics
that we can re-introduce a notion of non-inertial forces, because
Hamilton equations define a force law.
\medskip

Due to the absence of any statement about non-inertial systems
(replacing the relativity principle), usually special relativity
is seen as an approximation to general relativity, valid locally
near an observer in free fall \footnote{Also general relativity
makes no positive statement about non-inertial systems: the laws
of nature are now {\it generally covariant} (namely they assume
the same form in every 4-coordinate system), but this {\it only}
implies the elimination of rigid inertial systems (only local
inertial systems for an observer in free fall remain). Moreover,
now the chrono-geometrical structure of space-time becomes {\it
dynamical} (it is described by the metric tensor, which is also
the potential for the gravitational field), space-time itself
looses its reality and we need a physical identification of
space-time points as point events ({\it space-time is the
gravitational field itself}). See Ref.\cite{8a,9a} for a full
discussion of these aspects of general relativity. }. The
equivalence principle is invoked to say that an uniform
gravitational field and an uniform acceleration are locally
indistinguishable \footnote{But more realistically (see
Ref.\cite{28a}) this is true only on the geodesic of an observer
in free fall, due to the gravitational tidal forces evidentiated
by the geodesic deviation equation.} and that it is meaningless to
speak of inertial forces in general relativity: but again at the
Hamiltonian level this is possible with respect  to non-rigid
systems of reference as shown in Ref.\cite{8a}. Since the
transition to general relativity adds new problems without solving
the special relativistic ones, let us concentrate our discussion
on special relativity without gravity as an autonomous theory.

\bigskip

\subsection{The Locality Hypothesis.}

Since the actual observers are accelerated, we need some statement
correlating the measurements made by them to those made by
inertial observers, the only ones with a general framework for the
interpretation of their experiments. This statement is usually the
{\it hypothesis of locality} which is expressed in the following
terms according to Mashhoon \cite{29a} (see also
Refs.\cite{30a,31a}): {\it an accelerated observer at each instant
along its world-line is physically equivalent to an otherwise
identical momentarily comoving inertial observer}, namely a
non-inertial observer passes through a continuous infinity of
hypothetical momentarily comoving inertial observers \footnote{For
Einstein's comments on this point see Stachel \cite{32a}.
M$\o$ller (\cite{12a}, p.223) makes the assumption that the length
of the measuring rods are independent of the accelerations
relative to an inertial system. For Klauber \cite{33a} it is the
{\it surrogate frame postulate}.}. While this hypothesis is
verified in Newtonian mechanics and in those relativistic cases in
which a phenomenon can be reduced to point-like coincidences of
classical point particles and light rays (geometrical optic
approximation), its validity is questionable in presence of
electro-magnetic waves. As emphasized by Mashhoon
\cite{29a,30a,31a}, in this case we can trust the locality
hypothesis only when the wave-length $\lambda$ of the
electro-magnetic wave is much shorter of the acceleration length
${\cal L}$ of the observer, describing the degree of variation of
its state \footnote{${\cal L} = {{c^2}\over a}$ for an observer
with translational acceleration $a$; ${\cal L} = {c\over
{\Omega}}$ for an observer rotating with frequency $\Omega$.},
i.e. when $\lambda << {\cal L}$. When $\lambda << {\cal L}$ holds,
so that the period of the wave satisfies ${{\lambda}\over c} <<
{{{\cal L}}\over c}$, the observer state does not change
appreciably on the time scale needed to detect a few oscillations
of the wave and to measure its frequency. Instead in the case of
the electro-magnetic waves radiated by an accelerating charged
particle with acceleration length ${\cal L}$, we have $\lambda
\approx {\cal L}$. In this case it is highly problematic to
consider the particle momentarily equivalent to an identical
comoving inertial particle. This fact is confirmed by the
causality problems (pre-acceleration, runaway solutions) of the
classical Abraham - Lorentz - Dirac equation of motion of the
particle (see for instance Ref.\cite{34a}), which depend on the
time derivative of the acceleration, and by the still going on
discussions \cite{35a} on the energy balance and the back-reaction
in these radiative phenomena, due, besides the problem of the
self-energies, to the absence of a clear notion of simultaneity
for the particle and electro-magnetic degrees of freedom allowing
to define a well posed Cauchy problem \footnote{See Refs.
\cite{36a} for a semi-classical Hamiltonian approach to these
problems by using Grassmann-valued electric charges to regularize
the self-energies.}.
\medskip

Also the measurement of time dilation based on the muon lifetime
can be shown \cite{29a} to give the standard result $\tau_{(\mu )}
= \gamma\, \tau^o_{(\mu )}$ ($\tau^o_{(\mu )}$ is the lifetime in
an inertial system) only modulo negligible corrections of order
$(\lambda / {\cal L})^2$ ($\lambda = \hbar / m\, c$ is the muon
Compton wavelength). The hypothesis of locality is clearly valid
in many Earth-based experiments since $c^2 / g_{Earth} \approx 1\,
lyr$, $c / \Omega_{Earth} \approx 20\, AU$. \medskip

As we shall see, there are simultaneity conventions which satisfy
the locality hypothesis and others in which the associated
observers are not a sequence of comoving inertial ones. Only in a
theory in which all the simultaneity conventions are equivalent in
some sense the locality hypothesis can be fully justified.

\bigskip

\subsection{The 1+3 Point of View.}

Therefore it is far from clear which is the description of
physical phenomena given by a non-inertial observer, especially a
rotating one. The fact that we can describe phenomena only locally
near the observer and that the actual observers are accelerated
leads to the {\it 1+3 point of view} (or {\it threading
splitting}) \cite{37a,38a}.
\medskip

Given the world-line $\gamma$ of the accelerated observer, we
describe it with Lorentzian coordinates $x^{\mu}(\tau )$,
parametrized with an affine parameter $\tau $, with respect to a
given inertial system. Its unit 4-velocity is
$u^{\mu}_{\gamma}(\tau ) = {\dot x}^{\mu}(\tau ) / \sqrt{\sgn\,
{\dot x}^2(\tau )}$ [$\quad {\dot x}^{\mu} = {{d x^{\mu}}\over
{d\tau}}$]. The observer proper time $\tau_{\gamma}(\tau )$ is
defined by $\sgn\, {\dot {\tilde x}}^2(\tau_{\gamma}) = 1$, if we
use the notations $x^{\mu}(\tau ) = {\tilde x}^{\mu}(\tau
_{\gamma}(\tau ))$ and $u^{\mu}(\tau ) = {\tilde
u}^{\mu}(\tau_{\gamma}(\tau )) = d\, {\tilde
x}^{\mu}(\tau_{\gamma}) / d\, \tau_{\gamma}$, and it is indicated
by a {\it comoving standard atomic clock}. At each point of
$\gamma$ with proper time $\tau_{\gamma}(\tau )$, the tangent
space to Minkowski space-time in that point has the 1+3 splitting
in vectors parallel to ${\tilde u}^{\mu}(\tau_{\gamma})$ and
vectors lying in the 3-dimensional (so-called {\it local observer
rest frame}) subspace $R_{\tilde u(\tau_{\gamma})}$ orthogonal to
${\tilde u}^{\mu}(\tau_{\gamma})$ \footnote{Let us remark that
there is {\it no notion of a 3-space simultaneous with a point of
$\gamma$} and whose tangent space at that point is $R_{\tilde
u(\tau_{\gamma})}$: i) at a geometrical level $R_{\tilde
u(\tau_{\gamma})}$ may at best be considered as a local
approximate substitute of it in an infinitesimal neighborhood; ii)
actually, when the locality hypothesis holds, the acceleration
radii determine the effective dimension of the neighborhood and
this fact has led some authors, see for instance Ref.\cite{21a},
to identify  $R_{\tilde u(\tau_{\gamma})}$ with the simultaneity
3-space of the observer.}. By a conventional choice of three
spatial axes $E^{\mu}_{(a)}(\tau ) = {\tilde
E}^{\mu}_{(a)}(\tau_{\gamma}(\tau ))$, $a=1,2,3$, orthogonal to
$u^{\mu}(\tau ) = E^{\mu}_{(o)}(\tau ) = {\tilde
u}^{\mu}(\tau_{\gamma}(\tau )) = {\tilde
E}^{\mu}_{(o)}(\tau_{\gamma}(\tau ))$, the non-inertial observer
is endowed with an ortho-normal tetrad \footnote{This amounts to a
choice of three comoving gyroscopes in addition to the comoving
standard atomic clock. Usually the spatial axes are chosen to be
Fermi-Walker transported as a standard of non-rotation, which
takes into account the Thomas precession \cite{39a}. Let us remark
that this physical reference frame of an observer has not to be
confused with the Celestial Reference Frames (see footnote 1) used
in astronomy and geodesy.} $E^{\mu}_{(\alpha )}(\tau ) = {\tilde
E}^{\mu}_{(\alpha )}(\tau_{\gamma}(\tau ))$, $\alpha = 0,1,2,3$.
\medskip

The matter or field tensors seen by the observer are the
(coordinate-independent) tetradic components of these tensors: for
the electro-magnetic field strength $F_{\mu\nu}(z){|}_{z=\tilde
x(\tau_{\gamma})}$ they are $F_{(\alpha )(\beta )}(\tilde
x(\tau_{\gamma})) = F_{\mu\nu}(\tilde x(\tau_{\gamma}))\, {\tilde
E}^{\mu}_{(\alpha )}(\tau_{\gamma})\, {\tilde E}^{\nu}_{(\beta
)}(\tau_{\gamma})$. In the case of a vector field $X^{\mu}(z)$,
the physical observables \cite{33a} for the observer are the
scalar quantities formed with $X_{(o)}(\tau_{\gamma}) =
X_{\mu}(\tilde x(\tau_{\gamma}))\, {\tilde
E}^{\mu}_{(o)}(\tau_{\gamma}) = X_{\mu}(\tilde x(\tau_{\gamma}))\,
{\tilde u}^{\mu}(\tau_{\gamma})$ and $\sum_{(a)}\,
X_{(a)}^2(\tau_{\gamma})$ with $X_{(a)}(\tau_{\gamma}) =
X_{\mu}(\tilde x(\tau_{\gamma}))\, {\tilde
E}^{\mu}_{(a)}(\tau_{\gamma})$. For instance, if we have two
incident light rays with tangent vectors $k^{\mu}$, $k^{{'}\mu}$ [
$\eta_{\mu\nu}\, k^{\mu}\, k^{\nu} = \eta_{\mu\nu}\, k^{{'}\mu}\,
k^{{'}\nu} = 0$], their observable angle $\phi_{\gamma}$ seen by
the observer is defined by $cos\, \phi_{\gamma} = \sum_{(a)}\,
k_{(a)}\, k^{'}_{(a)} / |k_{(o)}|\, |k^{'}_{(o)}|$.

\bigskip

However the threading point of view says nothing on how to define
sets of events of Minkowski space-time {\it simultaneous} with
each point of the world-line $\gamma$ of the non-inertial observer
and this is a source of problems both for the synchronization of
clocks and for the definition of measurements of length. Let us
see what is known in the literature.

\bigskip

\subsection{4-Coordinates for Accelerated Observers.}

A) A first attempt (both in special and general relativity) to
treat this problem is based on the {\it Fermi normal coordinates}
\cite{40a}. In each point of the world-line $\gamma$ of the
accelerated observer one considers the {\it hyper-plane
orthogonal} to the observer unit 4-velocity $u^{\mu}(\tau )$, i.e.
the local observer rest frame at that point. Then, on each
hyper-plane one considers three space-like geodesics as spatial
coordinate lines. In this way we can coordinatize a world-tube
around the world-line $\gamma$, whose {\it radius} is determined
by the intersection of hyper-planes at different times.
Notwithstanding various efforts to ameliorate the construction
\cite{41a}, in this way we obtain only local coordinates and a
notion of simultaneity valid only inside the world-tube (see also
Section 4.1 of Ref.\cite{19a}). Let us remark that similar
coordinates are employed in the attempts to define the
relativistic center of mass of an extended object \cite{42a}
\footnote{See Refs.\cite{43a} for the study of the problem of the
relativistic center of mass without using this type of
coordinates. }.

\bigskip

B) Mashhoon \cite{29a,31a} has introduced a variant of the
previous coordinates, i.e. the {\it geodesic coordinates for
rotating observers}. Since in this construction the observer has
spatial axes $E^{\mu}_{(a)}(\tau )$ obtained from those of an
inertial observer with a Lorentz boost to a comoving observer
along the axis $2$ (tangent to the circular orbit), the observer
can define an {\it acceleration tensor} ${\cal A}_{(\alpha )(\beta
)}$ by means of the equation ${{d\, E^{\mu}_{(\alpha )}(\tau
)}\over {d\, \tau}} = {\cal A}_{(\alpha )}{}^{(\beta )}(\tau )\,
E^{\mu}_{(\beta )}(\tau )$. The translational acceleration and
rotational frequency of the observer are $a_{(i)} = {\cal
A}_{(o)(i)}$ and $\Omega_{(i)} = {1\over 2}\,
\epsilon_{(i)(j)(k)}\, {\cal A}_{(j)(k)}$, respectively. The two
invariants $I_1 = {\vec \Omega}^2 - {\vec a}^2$ and $I_2 = \vec a
\cdot \vec \Omega$ may be interpreted as the two {\it acceleration
radii} determining the world-tube of validity of these geodesic
coordinates, namely the region where the hyper-planes orthogonal
to the observer world-line can be considered as simultaneity
3-spaces (see footnote 11).

\bigskip

C) A third approach is the one of Pauri and Vallisneri \cite{44a}
\footnote{See this paper,  Ref.\cite{20a} and their bibliography
for the role of the notion of simultaneity in the interpretation
of the twin paradox. In particular in Ref.\cite{20a} it is shown
the independence of the solution of the paradox from the choice of
the notion of simultaneity.} . It is a refinement of the
Martzke-Wheeler method \cite{24a} to build radar coordinates.
Given the observer world-line $x^{\mu}(\tau )$, the simultaneity
surface through $x^{\mu}(0)$ is built as the union of the
intersections of the past light-cone of $x^{\mu}(+\Delta \tau )$
with the future light-cone of $x^{\mu}(- \Delta \tau )$ when
$\Delta \tau$ varies (if the world-line is a straight line one
recovers the hyper-planes of Einstein's convention). In this way
it is possible to build a foliation of Minkowski space-time with
space-like hyper-surfaces (simultaneity surfaces). However, in the
example explicitly worked out by these authors the embedding
$x^{\mu} = z^{\mu}(\tau ,|\vec \sigma |\, \hat n)$ ($|\vec \sigma
|$ is the radial distance from $\gamma$ and $\hat n$ a unit
3-vector) describing these hyper-surfaces in Minkowski space-time
is a periodic function of $|\vec \sigma |$ with an oscillating
limit for $|\vec \sigma |\, \rightarrow\, \infty$. Again this
signals the presence of a coordinate singularity after one spatial
period and a limited range of validity of these coordinates too.
\bigskip

\subsection{The Rotating Disk.}

Finally there is the enormous amount of bibliography, reviewed in
Ref.\cite{44ab}, about the problems of the {\it rotating disk} and
of the {rotating coordinate systems}. Independently from the fact
whether the disk is a material extended object or a geometrical
congruence of time-like world-lines (integral lines of some
time-like unit vector field), the idea followed by many
researchers \cite{11a,12a,45a} is to start from the Cartesian
4-coordinates of a given inertial system, to pass to cylindrical
3-coordinates and then to make a either Galilean (assuming a
non-relativistic behaviour of rotations at the relativistic level)
or Lorentz transformation to comoving rotating 4-coordinates, with
a subsequent evaluation of the 4-metric in the new coordinates. In
other cases \cite{46a} a suitable global 4-coordinate
transformation is postulated, which avoids the so-called {\it
horizon problem} (the points where all the previous 4-metrics have
either vanishing or diverging components, when the rotational
frequency reaches the velocity of light). Various authors (see for
instance Refs.\cite{47a}) do not define a coordinate
transformation but only a rotating 4-metric. Just starting from
M$\o$ller rotating 4-metric \cite{12a}, Nelson \cite{48a} was able
to deduce a 4-coordinate transformation implying it.
\bigskip

The problems arise when one tries to define measurements of
length, in particular that of the circumference of the disk.
Einstein \cite{49a} claims that while the rods along the radius
$R_o$ are unchanged those along the rim of the disk are Lorentz
contracted: as a consequence more of them are needed to measure
the circumference, which turns out to be greater than $2\pi\, R_o$
(non-Euclidean 3-geometry even if Minkowski space-time is 4-flat)
and not smaller. This was his reply to Ehrenfest \cite{50a}, who
had pointed an inconsistency in the accepted special relativistic
description of the disk \footnote{If $R$ and $R_o$ denote the
radius of the disk in the rotating and inertial frame
respectively, then we have $R = R_o$ because the velocity is
orthogonal to the radius. But the circumference of the rim of the
disk is Lorentz contracted so that $2\pi\, R < 2\pi\, R_o$
inconsistently with Euclidean geometry. } in which it is the
circumference to be Lorentz contracted: as a consequence this fact
was named the {\it Eherenfest paradox} (see the historical paper
of Gr$\o$n in Ref.\cite{44ab}).

\medskip

As underlined in Ref. \cite{51a} (see also Refs.\cite{52a,53a})
there is an intertwining of the following problems: i) Does the
rim of the disk in the rotating system define the same geometrical
circumference as the set of the positions of the points of the rim
as seen by a given inertial observer? ii) How is defined the {\it
instantaneous 3-space of the rotating disk} if we take into
account that the associated congruence of time-like world-lines
describing its points in time has not zero vorticity, so that it
is not surface-forming? iii) How to define the 3-geometry of the
rotating disk? iv) How to measure the length of the circumference?
v) Which time and notion of simultaneity has to be used to
evaluate the velocity of (massive or massless) particles in
uniform motion along the circumference? vi) Do standard rods
undergo Lorentz contraction (validity of the locality hypothesis)?
In Refs. \cite{54a,44ab} there is a rich bibliography on the
existing answers to these questions and the remark that the actual
standard rods used for measurements are rods with free ends (not
be confused with arcs of circumference), which a) remain unchanged
when slowly transported; b) are assumed not to be influenced by an
acceleration (the locality hypothesis) or a local gravitational
field. In Section VIB more details on the rotating disk are given.

\bigskip

\subsection{The Sagnac Effect.}

The other important phenomenon connected with the rotating disk is
the {\it Sagnac effect} \cite{55a} (see the recent review in
Ref.\cite{56a} for how many interpretations of it exist), namely
the phase difference generated by the difference in the time
needed for a round-trip by two light rays, emitted in the same
point, one co-rotating and the other counter-rotating with the
disk \footnote{For monochromatic light in vacuum with wavelength
$\lambda$ the fringe shift is $\delta z = 4\, \vec \Omega \cdot
\vec A / \lambda\, c$, where $\vec \Omega $ is the Galilean
velocity of the rotating disk supporting the interferometer and
$\vec A$ is the vector associated to the area $|\vec A|$ enclosed
by the light path. The time difference is $\delta t = \lambda\,
\delta z /c = 4\, \vec \Omega \cdot \vec A/c^2$, which agrees, at
the lowest order, with the proper time difference $\delta \tau =
(4\, A\, \Omega /c^2)\, (1 - \Omega^2\, R^2/c^2)^{-1/2}$, $A =
\pi\, R^2$, evaluated in an inertial system with the standard
rotating disk coordinates. This proper time difference is twice
the time lag due to the {\it synchronization gap} predicted for a
clock on the rim of the rotating disk with a non-time orthogonal
metric. See Refs.\cite{56a,57a,58a} for more details. See also
Ref.\cite{59a} for the corrections included in the GPS protocol to
allow the possibility of making the synchronization of the entire
system of ground-based and orbiting atomic clocks in a reference
local inertial system. Since  usually, also in GPS, the rotating
coordinate system has $t^{'} = t$ ($t$ is the time of an inertial
observer on the axis of the disk) the gap is a consequence of the
impossibility to extend Einstein's convention of the inertial
system also to the non-inertial one rotating with the disk: after
one period two nearby synchronized clocks on the rim are out of
synchrony.}. This effect, which has been tested (see the
bibliography of Refs.\cite{56a,60a}) for light, X rays and matter
waves (Cooper pairs, neutrons, electrons and atoms), has important
technological applications and must be taken into account for the
relativistic corrections to space navigation, has again an
enormous number of theoretical interpretations (both in special
and general relativity) like for the solutions of the Ehrenfest
paradox. Here the lack of a good notion of simultaneity leads to
problems of {\it time discontinuities or desynchronization
effects} when comparing clocks on the rim of the rotating disk.
\bigskip

Moreover, various authors use the Sagnac effect, together with the
Foucault pendulum, as a clear hint that, contrary to translations,
the rotations of the reference frame have an {\it absolute
character} so that non-rotating frames are preferred frames
\cite{33a} \footnote{Let us remark that this is an attitude
opposite to that of the supporters of Mach principle \cite{61a}.}.
Another disturbing aspect of rotating frames for these authors is
that the (coordinate) velocity of light is no more isotropic (see
footnote 4) when the rotating 4-metric is {\it not
time-orthogonal}, i.e. when $g_{oi} \not= 0$. In general
relativity, where the $g_{oi}$'s are the (gauge, i.e. not
determined by Einstein's equations) {\it shift functions}, this
fact implies the addition of {\it gravito-magnetic effects}
(dragging of inertial frames, Lense-Thirring effect)
\cite{62a,63a} to the anisotropy of light propagation (not to
mention new phenomena like the gravitomagnetic clock effects
\cite{64a,65a} and the spin-rotation couplings \cite{66a}; see the
review \cite{67a}). Let us remark that in general relativity
synchronous 4-coordinates, for which the shift functions vanish,
are subject to develop singularities in short times, when one
attempts to do numerical gravity.

\medskip

Another area which is in a not well established form is
electrodynamics in non-inertial systems either in vacuum or in
material media ({\it problem of the non-inertial constitutive
equations}). Its clarification is needed both to derive the Sagnac
effect from Maxwell equations without gauge ambiguities \cite{57a}
and to determine which types of experiments can be explained by
using the locality principle to evaluate the electro-magnetic
fields in the comoving system (see the Wilson experiment
\cite{68a} and the associated controversy \cite{69a} on the
validity of the locality principle) without the need of a more
elaborate treatment like for the radiation of accelerated charges.
It would also help in the tests of the validity of special
relativity (for instance on the possible existence of a preferred
frame) based on Michelson-Morley - type experiments \cite{70a}
(see also Ref.\cite{67a}).

\bigskip

We do not accept the interpretation of rotations as absolute and
refuse the points of view implying deviations from standard
special relativity like the new postulates (no Lorentz contraction
under rotations and preferred nature of non-rotating frames) of
Klauber \cite{33a} or of Selleri \cite{71a}. Instead (see also
Ref.\cite{57a}) we remark that the Sagnac effect and the Foucault
pendulum are {\it experiments which signal the rotational
non-inertiality of the frame}. The same is true for neutron
interferometry \cite{72a}, where different settings of the
apparatus are used to {\it detect either rotational or
translational non-inertiality of the laboratory}. As a consequence
a null result of these experiments can be used to give a
definition of {\it relativistic quasi-inertial system}.

\bigskip

Let us remark that the disturbing aspects of rotations are rooted
in the fact that there is a deep difference between translations
and rotations at every level both in Newtonian mechanics and
special relativity:  the generators of translations satisfy an
Abelian algebra, while the rotational ones a non-Abelian algebra.
As shown in Refs.\cite{42a,73a}, at the Hamiltonian level we have
that the translation generators are the three components of the
momentum, while the generators of rotations are a pair of
canonical variables ($L^3$ and $arctg\, {{L^2}\over {L^1}}$) and
an unpaired variable ($|\vec L|$). As a consequence we can
separate globally the motion of the 3-center of mass of an
isolated system from the relative variables, but we cannot
separate in a global and unique way three Euler angles describing
an overall rotation, because the residual vibrational degrees of
freedom are not uniquely defined.

\medskip
Let us also remark that general relativity has been completely
re-expressed in the 1+3 point of view (starting from the works of
Cattaneo \cite{74a} and ending with Refs.\cite{75a}): the real
open problem of this approach is that no one is able to formulate
a Cauchy problem in this setting.

\bigskip

In conclusion the 1+3 point of view has to face a big group of
problems most of which originated by the absence of a good notion
of simultaneity. They are not academic theoretical problems. The
Global Positioning System \cite{59a} (GPS and its European
counterpart Galileo) and the future mission ACES of the European
Space Agency on the synchonization of clocks \cite{76a}, due to
the level of time keeping accuracy of the order $10^{-15}$ (and
higher) reached by the standard laser cooled atomic clocks
\cite{77a}, requires to take into account relativistic effects
till the order $1/c^3$ \cite{78a,79a,80a}. But this has to be done
after the introduction of a good notion of simultaneity in general
relativity, which becomes a special relativistic notion of
simultaneity in presence of weak gravitational fields compatible
with the (non-inertial non-rotating) Geocentric Celestial
Reference System and the (inertial non-rotating) Solar System
Barycentric Celestial Reference System. This fact, together with
the increasing interest in astronomy for relativistic corrections
to reference frames and light propagation \cite{81a} and in
astrophysics for relativistic rotating stars and black holes
\cite{82a}, points to the necessity of a re-formulation of the
previous problems in a framework allowing a good control of the
notion of simultaneity.

\bigskip

\subsection{The 3+1 Point of View, M$\o$ller Admissible
Coordinates and Parametrized Minkowski Theories.}

The aim of this paper is to try to show that the framework, in
which these problems find a natural co-existence with the standard
treatment of special relativity, requires an inversion of
attitude. Let us consider the 3+1 splittings of Minkowski
space-time associated to its foliations with arbitrary space-like
hyper-surfaces and not only with space-like hyper-planes. Each of
these hyper-surfaces is both {\it a simultaneity surface and a
Cauchy surface} for the equations of motion of the relativistic
systems of interest. After the choice of a foliation, i.e. of a
notion of simultaneity, we can determine, as we shall see, which
are the non-inertial observers compatible with that notion of
simultaneity.  Having given a {\it notion of simultaneity}, there
will be associated notions of {\it one-way velocity of light}, of
{\it synchronization of distant clocks}, of {\it instantaneous
3-space} and of {\it spatial length}.

\medskip

Moreover in this framework we will show that it is possible

a) to give an operational method, generalizing Einstein's
convention, for building the radar coordinates adapted to an
arbitrary foliation \footnote{See Refs.\cite{83a} for
epistemological and mathematical supports to the notion of {\it
radar coordinates}.} (see Subsection A of Section VI).

b) to solve the following inverse problem: given a single
non-inertial observer or a congruence of non-inertial observers
(like in the case of the rotating disk) find which are the
foliations, i.e. the notions of simultaneity, compatible with them
(see Sections III and V, respectively).

\bigskip

The 3+1 point of view is less physical (it is impossible to
control the initial data on a non-compact space-like Cauchy
surface), but it is the only known way to establish a well posed
Cauchy problem for the dynamics, so to be able to use the
mathematical theorems on the existence and uniqueness of the
solutions of field equations for identifying the predictability of
the theory. A posteriori, a non-inertial observer can try to
separate the part of the dynamics, implied by these solutions,
which is determined at each instant from the (assumed known)
information coming from its causal past (see Ref.\cite{84a} for an
attempt to re-phrase the instant form of dynamics in a form
employing only data from the causal past of an observer) from the
part coming from the rest of the universe.

\bigskip

As emphasized by Havas \cite{23a}, to implement this program we
have to come back to M$\o$ller's formalization \cite{12a} (Chapter
VIII, Section 88) of the notion of simultaneity, based on previous
work by Hilbert \cite{84ab}. Given an inertial system with
Cartesian 4-coordinates $x^{\mu}$ in Minkowski space-time and with
the $x^o = const.$ simultaneity hyper-planes, M$\o$ller defines
the {\it admissible coordinates transformations} $x^{\mu}\,
\mapsto\, y^{\mu} = f^{\mu}(x)$ [with inverse transformation
$y^{\mu}\, \mapsto\, x^{\mu} = h^{\mu}(y)$] as those
transformations whose associated metric tensor $g_{\mu\nu}(y) =
{{\partial h^{\alpha}(y)}\over {\partial y^{\mu}}}\, {{\partial
h^{\beta}(y)}\over {\partial y^{\nu}}}\, \eta_{\alpha\beta}$
satisfies the following conditions

\bea
 && \sgn\, g_{oo}(y) > 0,\nonumber \\
 &&{}\nonumber \\
 && \sgn\, g_{ii}(y) < 0,\qquad \begin{array}{|ll|} g_{ii}(y)
 & g_{ij}(y) \\ g_{ji}(y) & g_{jj}(y) \end{array}\, > 0, \qquad
 \sgn\, det\, [g_{ij}(y)]\, < 0,\nonumber \\
 &&{}\nonumber \\
 &&\Rightarrow det\, [g_{\mu\nu}(y)]\, < 0.
 \label{I1}
 \eea

These are the necessary and sufficient conditions for having
${{\partial h^{\mu}(y)}\over {\partial y^o}}$ behaving as the
velocity field of a relativistic fluid, whose integral curves, the
fluid flux lines, are the world-lines of time-like observers.
Eqs.(\ref{I1}) say:

i) the observers are time-like because $\sgn g_{oo} > 0$;

ii) that the hyper-surfaces $y^o = f^{o}(x) = const.$ are good
space-like simultaneity surfaces.

\medskip

Moreover we must ask that $g_{\mu\nu}(y)$ tends to a finite limit
at spatial infinity on each of the hyper-surfaces $y^o = f^{o}(x)
= const.$ If, like in the ADM canonical formulation of metric
gravity \cite{85a}, we write $g_{oo} = \sgn\, (N^2 - g_{ij}\,
N^i\, N^j)$, $g_{oi} = g_{ij}\, N^j$ introducing the lapse ($N$)
and shift ($N^i$) functions, this requirement says that the lapse
function (i.e. the proper time interval between two nearby
simultaneity surfaces) and the shift functions (i.e. the
information about which points on two nearby simultaneity surfaces
are connected by the so-called {\it evolution vector field}
${{\partial h^{\mu}(y)}\over {\partial y^o}}$) do not diverge at
spatial infinity. This implies that at spatial infinity on each
simultaneity surface there is {\it no asymptotic either
translational or rotational acceleration} \footnote{They would
contribute \cite{86a} to the asymptotic line element with the
diverging terms $[A_i(y^o)\, y^i + B_{ij}(y^o)\, y^i\, y^j]\,
(dy^o)^2$ and $\epsilon_{ijk}\, \omega^j(y^o)\, y^k\, dy^o\,
dy^i$, respectively. } and the asymptotic line element is $ds^2 =
g_{\mu\nu}(y)\, dy^{\mu}\, dy^{\nu}\, \rightarrow_{spatial\,
infinity}\, \sgn\, \Big( F^2(y^o)\, (dy^o)^2 + 2\, G_i(y^o)\,
dy^o\, dy^i - {d \vec y}^2\Big)$. But this would break manifest
covariance unless $F(y^o) = 1$ and $G_i(y^o) = 0$. As a
consequence, {\it the simultaneity surfaces must tend to
space-like hyper-planes at spatial infinity}.

\bigskip

In this way all the admissible notions of simultaneity of special
relativity are formalized as 3+1 splittings of Minkowski
space-time by means of foliations whose leaves are space-like
hyper-surfaces tending to hyper-planes at spatial infinity. Let us
remark that admissible coordinate transformations $x^{\mu} \mapsto
y^{\mu} = f^{\mu}(x)$ constitute the most general extension of the
Poincare' transformations $x^{\mu} \mapsto y^{\mu} = a^{\mu} +
\Lambda^{\mu}{}_{\nu}\, x^{\nu}$ compatible with special
relativity. A special family of admissible transformations are the
sub-group of the {\it frame-preserving} ones: $x^o\, \mapsto\, y^o
= f^o(x^o, \vec x)$, $\vec x\, \mapsto\, \vec y = \vec f(\vec x)$.

\bigskip

It is then convenient to describe \cite{87a,27a,86a} the
simultaneity surfaces of an admissible foliation (3+1 splitting of
Minkowski space-time) with {\it naturally adapted Lorentz scalar
admissible coordinates} $x^{\mu}\, \mapsto \sigma^A = (\tau ,\vec
\sigma ) = f^A(x)$ [with inverse $\sigma^A\, \mapsto\, x^{\mu} =
z^{\mu}(\sigma ) = z^{\mu}(\tau ,\vec \sigma )$] such that:

i) the scalar time coordinate $\tau$ labels the leaves
$\Sigma_{\tau}$ of the foliation ($\Sigma_{\tau} \approx R^3$);

ii) the scalar curvilinear 3-coordinates $\vec \sigma = \{
\sigma^r \}$ on each $\Sigma_{\tau}$ are defined with respect to
an arbitrary time-like centroid $x^{\mu}(\tau )$ chosen as their
origin;

iii) if $y^{\mu} = f^{\mu}(x)$ is any admissible coordinate
transformation describing the same foliation, i.e. if the leaves
$\Sigma_{\tau}$ are also described by $y^o = f^o(x) = const.$,
then, modulo reparametrizations, we must have $y^{\mu} =
f^{\mu}(z(\tau ,\vec \sigma )) = {\tilde f}^{\mu}(\tau ,\vec
\sigma ) = A^{\mu}{}_A\, \sigma^A$ with $A^o{}_{\tau} = const.$,
$A^o{}_r = 0$, so that we get $y^o = const.\, \tau$, $y^i =
A^i{}_A(\tau ,\vec \sigma )\, \sigma^A$. Therefore, modulo
reparametrizations, the $\tau$ and $\vec \sigma$ adapted
admissible coordinates are {\it intrinsic coordinates}, which are
mathematically allowed as charts in the atlas for Minkowski
space-time. They are called {\it radar-like 4-coordinates} (see
Subsection A of Section VI for the justification of this name)
and, probably, they were introduced for the first time by Bondi
\cite{88a}. The use of adapted Lorentz-scalar radar coordinates
allows to avoid all the (often ambiguous) technicalities connected
with the extrapolations of effects like Lorentz-contraction or
time dilation from inertial to non-inertial frames.

\medskip

The use of these Lorentz-scalar adapted coordinates allows to make
statements depending only on the foliation but not on the
4-coordinates $y^{\mu}$ used for Minkowski space-time.

\bigskip

The simultaneity hyper-surfaces $\Sigma_{\tau}$ are described by
their embedding $x^{\mu} = z^{\mu}(\tau ,\vec \sigma )$ in
Minkowski space-time [$(\tau ,\vec \sigma ) \mapsto z^{\mu}(\tau
,\vec \sigma )$, $R^3\, \mapsto \, \Sigma_{\tau} \subset M^4$] and
the induced metric is $g_{AB}(\tau ,\vec \sigma ) = z^{\mu}_A(\tau
,\vec \sigma )\, z^{\nu}_B(\tau ,\vec \sigma )\, \eta_{\mu\nu}$
with $z^{\mu}_A = \partial z^{\mu} / \partial \sigma^A$ (they are
flat tetrad fields over Minkowski space-time). Since the vector
fields $z^{\mu}_r(\tau ,\vec \sigma )$ are tangent to the surfaces
$\Sigma_{\tau}$, the time-like vector field of normals
$l^{\mu}(\tau ,\vec \sigma )$ is proportional to
$\epsilon^{\mu}{}_{\alpha\beta\gamma}\, z^{\alpha}_1(\tau ,\vec
\sigma )\, z^{\beta}_2(\tau ,\vec \sigma )\, z^{\gamma}_3(\tau
,\vec \sigma )$. Instead the time-like evolution vector field is
$z^{\mu}_{\tau}(\tau ,\vec \sigma ) = N(\tau ,\vec \sigma )\,
l^{\mu}(\tau ,\vec \sigma ) + N^r(\tau ,\vec \sigma )\,
z^{\mu}_r(\tau ,\vec \sigma )$, so that we have $dz^{\mu}(\tau
,\vec \sigma ) = z^{\mu}_{\tau}(\tau ,\vec \sigma )\, d\tau +
z^{\mu}_r(\tau ,\vec \sigma )\, d\sigma^r = N(\tau ,\vec \sigma
)\, d\tau\, l^{\mu}(\tau ,\vec \sigma ) + (N^r(\tau ,\vec \sigma
)\, d\tau + d\sigma^r)\, z^{\mu}_r(\tau ,\vec \sigma )$.

Since the 3-surfaces $\Sigma_{\tau}$ are {\it equal time} 3-spaces
with all the clocks synchronized, the spatial distance between two
equal-time events will be $dl_{12} = \int^2_1 dl\,
\sqrt{{}^3g_{rs}(\tau ,\vec \sigma (l))\, {{d\sigma^r(l)}\over
{dl}}\, {{d\sigma^s(l)}\over {dl}}}\,\,$ [$\vec \sigma (l)$ is a
parametrization of the 3-geodesic $\gamma_{12}$ joining the two
events on $\Sigma_{\tau}$]. Moreover, by using test rays of light
we can define the {\it one-way} velocity of light between events
on different $\Sigma_{\tau}$'s.

\medskip

When we have a Lagrangian description of an isolated system on
arbitrary space-like hyper-surfaces (the {\it parametrized
Minkowski theories} of Refs.\cite{87a,27a} and of the Appendix of
Ref.\cite{86a}), {\it the physical results about the system do not
depend on the choice of the notion of simultaneity}. In this
approach, besides the configuration variables of the isolated
system, there are the embeddings $z^{\mu}(\tau ,\vec \sigma )$ as
extra {\it gauge configuration} variables in a suitable Lagrangian
determined in the following way. Given the Lagrangian of the
isolated system in the Cartesian 4-coordinates of an inertial
system, one makes the coupling to an external gravitational field
and then replaces the external 4-metric with $g_{AB} = z^{\mu}_A\,
\eta_{\mu\nu}\, z^{\nu}_B$. Therefore the resulting Lagrangian
depends on the embedding through the associated metric $g_{AB}$.
It can be shown that, due to the presence of {\it a special-
relativistic type of general covariance} (reparametrization
invariances of the action under frame-preserving diffeomorphisms),
the transition from a foliation to another one (i.e. a change of
the notion of simultaneity) is a gauge transformation of the
theory. Therefore, in parametrized Minkowski theories the {\it
conventionalism of simultaneity} is rephrased as a {\it gauge
problem} (in a way different from Refs.\cite{18a,22a}), i.e. as
the {\it arbitrary choice of a gauge fixing selecting a well
defined notion of simultaneity among those allowed by the gauge
freedom}. Moreover, for every isolated system there is a preferred
notion of simultaneity, namely the one associated with the 3+1
splitting whose leaves are the {\it Wigner hyper-planes}
orthogonal to the conserved 4-momentum of the system: {\it this
preferred simultaneity, intrinsically selected by the isolated
system, leads to the Wigner-covariant rest-frame  instant form of
dynamics}.
\bigskip

The main property of {\it each foliation with simultaneity
surfaces} associated to an admissible 4-coordinate transformation
is that the embedding of the leaves of the foliation automatically
determine two time-like vector fields and therefore {\it two
congruences of (in general) non-inertial time-like observers}:

i) The time-like vector field $l^{\mu}(\tau ,\vec \sigma )$ of the
normals to the simultaneity surfaces $\Sigma_{\tau}$ (by
construction surface-forming, i.e. irrotational), whose flux lines
are the world-lines of the so-called (in general non-inertial)
Eulerian observers. The simultaneity surfaces $\Sigma_{\tau}$ are
(in general non-flat) Riemannian 3-spaces in which the physical
system is visualized and in each point the tangent space to
$\Sigma_{\tau}$ is the local observer rest frame $R_{\tilde
l(\tau_{\gamma})}$ of the Eulerian observer through that point.
This 3+1 viewpoint is called {\it hyper-surface 3+1 splitting}.

ii) The time-like evolution vector field $z^{\mu}_{\tau}(\tau
,\vec \sigma ) / \sqrt{\sgn\, g_{\tau\tau}(\tau ,\vec \sigma ) }$,
which in general is not surface-forming (i.e. it has non-zero
vorticity like in the case of the rotating disk). The observers
associated to its flux lines have the local observer rest frames
$R_{\tilde u(\tau_{\gamma})}$ not tangent to $\Sigma_{\tau}$:
there is no notion of 3-space for these observers (1+3 point of
view or {\it threading splitting}) and no visualization of the
physical system in large. However these observers can use the
notion of simultaneity associated to the embedding $z^{\mu}(\tau
,\vec \sigma )$, which determines their 4-velocity. This 3+1
viewpoint is called {\it slicing 3+1 splitting}. In the case of
the uniformly rotating disk all the existing rotating
4-coordiinate systems have a {\it coordinate singularity} (the
horizon problem: $g_{oo}(y^o, \vec y) = 0$) where $\omega\, r =
c$: there the time-like observers of the congruence would become
null observers like on the horizon of a Schartzschild black hole
and this is not acceptable in absence of a horizon.

\bigskip

As a consequence the plan of the paper is the following one.

\medskip

In Section II we give some preliminaries on reference frames and
on the notions of simultaneity in the 3+1 and 1+3 points of view
with the associated method of synchronization of clocks and of
definition of spatial distances.

In Section III we shall define the class of foliations
implementing the idea behind the locality hypothesis that a
non-inertial observer is equivalent to a continuous family of
comoving inertial observers (as shown by Havas \cite{23a}, there
are other admissible foliations not in this class). Eqs.(\ref{I1})
will put restrictions on the comoving observers. The main
byproduct of these restrictions will be that there exist
admissible 4-coordinate transformations interpretable as {\it
rigid systems of reference with arbitrary translational
acceleration}. However there is {\it no admissible 4-coordinate
transformation corresponding to a rigid system of reference with
rotational motion}. When rotations are present, the admissible
4-coordinate transformations give rise to a continuum of local
systems of reference like it happens in general relativity ({\it
differential rotations}). Moreover our parametrization of this
class of foliations uses an arbitrary centroid $x^{\mu}(\tau )$
with a time-like world-line as origin of the 3-coordinates on the
simultaneity surfaces $\Sigma_{\tau}$: {\it therefore these
foliations describe possible notions of simultaneity for the
arbitrary non-inertial observer $x^{\mu}(\tau )$}.

\medskip

In Section IV we describe the simplest foliations of the previous
class, whose simultaneity surfaces are space-like hyper-planes
with differentially rotating coordinates.
\medskip

In Section V we solve the following inverse problem: given a
time-like unit vector field, i.e. a (in general not irrotational)
congruence of non-inertial observers like that associated with a
rotating disk, find an admissible foliation with simultaneity
surfaces such that $z^{\mu}_{\tau}(\tau ,\vec \sigma )$ is
proportional to the given vector field.
\medskip

Section VI contains some applications of our results. In
Subsection A an operational method, generalizing Einstein's
convention to arbitrary simultaneity surfaces, is proposed to
build a grid of radar 4-coordinates to be used by a set of
satellites of the GPS type. In Subsection B we give the 3+1 point
of view on the rotating disk, while in Subsection C we give its
description in the foliation of Section IV and we evaluate the
Sagnac effect. In Subsection D the foliation of Section IV is used
to describe Earth rotation (instead of assuming Einstein's
convention plus Sagnac corrections as it happens in GPS) as an
empirical admissible notion of simultaneity (see Subsection A) and
it is applied to the determination of the one-way time transfer
(including the Shapiro delay as a post-Newtonian effect) for the
propagation of light from an Earth station to a satellite. Its
comparison with the future measurements of the ACES mission will
allow to synchronize the clocks according to this empirical
simultaneity. In Subsection E we describe electro-magnetism as a
parametrized Minkowski theory and we arrive at Maxwell equations
in non-inertial frames as a result of gauge fixings determining an
arbitrary admissible notion of simultaneity.

\medskip

In the Conclusions we give a general overview of the results
obtained and we discuss the {\it dynamical} nature of the
admissible  notions of simultaneity in general relativity.

\medskip

In Appendix A there is a sketch of the derivation of the Sagnac
effect from the non-inertial Maxwell equations.

In Appendix B there is the study of a family of admissible
embeddings, whose leaves are parallel hyper-planes, closed under
the action of the Lorentz group.

\vfill\eject

\section{More about 3+1 versus 1+3 Notions of Simultaneity.}

In this Section we shall collect some differential geometry tools
needed to compare the 1+3 approximate notion of simultaneity with
the exact 3+1 admissible ones. Due to the non-uniformity in the
nomenclature used in the literature, let us first introduce some
definitions following Ref.\cite{39a}.

An {\it inertial observer} in Minkowski space-time $M^4$ is a
time-like future-oriented straight line $\gamma$. Any point $P$ on
$\gamma$ together with the unit time-like tangent vector
$e^{\mu}_{(o)}$ to $\gamma$ at $P$ is an {\it instantaneous
inertial observer}. Let us choose a point $P$ on $\gamma$ as the
origin of an {\it inertial system} $I_P$ having $\gamma$ as time
axis and three orthogonal space-like straight lines orthogonal to
$\gamma$ in $P$, with unit tangent vectors $e^{\mu}_{(r)}$,
$r=1,2,3$ as space axes. Let $x^{\mu}$ be a Cartesian 4-coordinate
system referred to these axes, in which the line element has the
form $ds^2 = \eta_{\mu\nu}\, dx^{\mu}\, dx^{\nu}$ with
$\eta_{\mu\nu} = \sgn\, (+---)$, $\sgn = \pm 1$. Associated to
these coordinates there is a {\it reference frame} (or {\it system
of reference} or {\it platform} \cite{2a} \footnote{According to
Ref.\cite{2a} a reference frame is a platform with a tetrad field
defining a tetrad for each observer.} ) given by the congruence of
time-like straight lines parallel to $\gamma$, namely a unit
vector field $u^{\mu}(x)$. Each of the integral lines of the
vector field is identified by a fixed value of the three spatial
coordinates $x^i$ and represent an observer: this is a {\it
reference point} according to M$\o$ller \cite{12a}.

A reference frame $l$, i.e. a time-like vector field $l^{\mu}(x)\,
{{\partial}\over {\partial x^{\mu}}}$ with its  congruence of
time-like world-lines and its associated 1+3 splitting of $TM^4$,
admits the decomposition \footnote{$P_{\mu\nu}(x) = \eta_{\mu\nu}
- \sgn\, l_{\mu}(x)\, l_{\nu}(x)$ is the 3-metric in the
rest-frame in the point $x^{\mu}$,  i.e. in the tangent 3-plane
orthogonal to $l^{\mu}(x)$; $D^{(\eta )}$ is the Levi-Civita
covariant derivative on Minkowski space-time.}

\bea
 D^{(\eta )}_{\mu}\, l_{\nu} &=&  l_{\mu}\, a_{\nu} + {1\over
3}\, \Theta \, P_{\mu\nu} + \sigma_{\mu\nu} +
\omega_{\mu\nu},\nonumber \\
 &&a^{\mu} = l^{\nu}\,  D^{(\eta )}_{\nu}\, l^{\mu},
 \nonumber \\
 &&\Theta =  D^{(\eta )}_{\mu}\, l^{\mu},\nonumber \\
 &&\sigma_{\mu\nu} = {1\over 2}\, (a_{\mu}\, l_{\nu} + a_{\nu}\,
 l_{\mu}) + {1\over 2}\, ( D^{(\eta )}_{\mu}\,
l_{\nu} +  D^{(\eta )}_{\nu}\, l_{\mu}) - {1\over 3}\, \Theta\,
P_{\mu\nu},\nonumber \\
 && with\,\, magnitude\,\, \sigma^2={1\over 2} \sigma_{\mu\nu}\sigma^{\mu\nu},\nonumber \\
 &&\omega_{\mu\nu} = - \omega_{\nu\mu} = \epsilon_{\mu\nu\alpha\beta}\,
 \omega^{\alpha}\, l^{\beta} = {1\over 2}\, (a_{\mu}\, l_{\nu} - a_{\nu}\,
 l_{\mu}) + {1\over 2}\, ( D^{(\eta )}_{\mu}\, l_{\nu} -  D^{(\eta )}_{\nu}\,
  l_{\mu}),\nonumber \\
 && \omega^{\mu} = {1\over 2}\, \epsilon^{\mu\alpha\beta\gamma}\, \omega_{\alpha\beta}\,
l_{\gamma},
 \label{II1}
 \eea

\noindent where $a^{\mu}$ is the 4-acceleration, $\Theta$ the {\it
expansion} \footnote{It measures the average expansion of the
infinitesimally nearby world-lines surrounding a given world-line
in the congruence.}, $\sigma _{\mu\nu}$ the {\it shear}
\footnote{It measures how an initial sphere in the tangent space
to the given world-line, which is Lie transported along $l^{\mu}$
(i.e. it has zero Lie derivative with respect to $l^{\mu}\,
\partial_{\mu} $), is distorted towards an ellipsoid with principal axes
given by the eigenvectors of $\sigma^{\mu}{}_{\nu}$, with rate
given by the eigenvalues of $\sigma^{\mu}{}_{\nu}$.} and
$\omega_{\mu\nu}$ the {\it twist or vorticity or rotation}
\footnote{It measures the rotation of the nearby world-lines
infinitesimally surrounding the given one.}; $\sigma_{\mu\nu}$ and
$\omega _{\mu\nu}$ are purely spatial ($\sigma_{\mu\nu} l^{\nu} =
\omega_{\mu\nu} l^{\nu} = 0$). Due to the Frobenius theorem, the
congruence is (locally) {\it hyper-surface orthogonal} if and only
if $\omega_{\mu\nu}=0$. The equation ${1\over l}\, l^{\mu}\,
\partial_{\mu}\, l = {1\over 3}\, \Theta$ defines a representative
length $l$ along the world-line of $l^{\mu}$, describing the
volume expansion (or contraction) behaviour of the congruence.
\bigskip

Another important characterization of a reference frame $l$ is its
{\it synchronizability}. If we define the 1-form $\alpha_l =
\eta_{\mu\nu}\, l^{\nu}(x)\, dx^{\mu} = l_{\mu}(x)\, dx^{\mu}$,
then the reference frame $l$ is said to be \cite{39a}:

i) {\it locally synchronizable} iff $\alpha_l \wedge d\alpha_l =
0$, i.e. $d\alpha_l = \beta \wedge \alpha_l$;

ii) {\it locally proper time synchronizable} iff $d\alpha_l = 0$,
i.e. $\alpha_l = d \beta_l$;

iii) {\it synchronizable} iff there are $C^{\infty}$ functions
$h$, $t$ from $M^4$ to $R$ such that $\alpha_l = h\, dt$ and
$\sgn\, h > 0$;

iv) {\it proper time synchronizable} iff $\alpha_l = dt$.
\medskip

Since we have $\alpha_l \wedge d\alpha_l = 0\,\,\,
\Leftrightarrow\,\,\, \omega_l = \omega_{\mu\nu}(x)\, dx^{\mu}
\wedge dx^{\nu} = 0$, rotating reference frames, i.e. with nonzero
vorticity, are {\it not} locally synchronizable, i.e. these
reference frames are not surface-forming (or in other words the
1+3 splitting does not imply a 3+1 splitting with simultaneity
surfaces {\it orthogonal} to the integral lines in the
congruence). Therefore to give a reference frame (a congruence) in
general does not define a notion of simultaneity and an allowed
coordinate transformation. In Section V, given  a
non-synchronizable reference frame $u^{\mu}(x)$, we will show how
it is possible to find embeddings $z^{\mu}(\tau ,\vec \sigma )$
whose leaves $\Sigma_{\tau}$ [{\it not orthogonal} to the integral
lines of $u^{\mu}(x)$] are possible  notions of simultaneity
related to the non-synchronizable reference frame.

\bigskip

An {\it inertial reference frame} is defined as a covariantly
constant  vector field $I$, i.e. $D^{(\eta )}\, I = 0$
\footnote{In general relativity there is no frame $I$ satisfying
$D^{(g)}\, I = 0$, so that inertial reference frames do not exist.
As shown in Ref.\cite{89a} in general relativity we can define:
\medskip
 i) {\it pseudo-inertial reference frames} $Q$, which have
all the integral lines in free fall, are non-rotating and at least
locally synchronizable [i.e. $D_Q\, Q = 0$ and $\alpha_Q \wedge
d\alpha_Q = 0$, where $\alpha_Q = {}^4g_{\mu\nu}(x)\, Q^{\mu}(x)\,
dx^{\mu}$ if $Q = Q^{\mu}(x)\, {{\partial}\over {\partial
x^{\mu}}}$]; in naturally adapted 4-coordinates we have $Q =
{{\partial}\over {\partial y^o}}$ ;
 \medskip
ii) {\it local Lorentz reference frames} $L$ associated to a
geodetic line $\gamma$ with associated {\it local Lorentz (or
Riemann normal)} 4-coordinates, for which only $\gamma$ is in free
fall, while outside $\gamma$ we have $D_L\, L \not= 0$ and
$\alpha_L \wedge d\alpha_L \not= 0$ (not synchronizable); along
the geodesic  $\gamma$ we can build a {\it Fermi-Walker
transported inertial moving frame} as a standard of no-rotation. }
, and is identified as $(I_P, e^{\mu}_{(A)}, x^{\mu})$,
$A=0,1,2,3$ and for any point $Q$ we have $Q - P = x_{\mu}\,
e^{\mu}_{(A)} \in T_PM^4$ \footnote{The vector fields
$e^{\mu}_{(A)}(x)$ form an orthonormal basis of the tangent spaces
$T_PM^4$ and are called an {\it orthonormal moving frame}.} .
Other inertial systems with origin $P$ are obtained from $I_P$
with global rigid Lorentz transformations, while Poincare'
transformations describe also a change of origin.
\medskip

To each time-like observer with world-line $x^{\mu}(\tau )$ we can
associate an orthonormal tetrad whose time axis is
$E^{\mu}_{\tau}(\tau ) = {{{\dot x}^{\mu}(\tau )}\over
{\sqrt{\sgn\, {\dot x}^2(\tau ) }}}$. The spatial axes are
realized with the rotation axes of three mutually orthogonal
gyroscopes (see footnote 12). The orientation of the gyroscopes is
a matter of convention. Two widely used conventions are:

$\alpha$) the three axes always point the fixed stars (like in
Gravity Probe B) and are not Fermi-Walker transported;

$\beta$) the three axes point the fixed stars at $\tau = 0$ and
then are Fermi-Walker transported along $x^{\mu}(\tau )$ (standard
of non-rotation).

About gyroscopes see Ref.\cite{90a} and its bibliography.

\bigskip

As said in the Introduction, according to M$\o$ller \cite{12a} the
more general {\it admissible coordinate transformations} in
special relativity are those transformations $x^{\mu} \mapsto
y^{\mu} = f^{\mu}(x)$ [with inverse $x^{\mu} = h^{\mu}(y)$] such
that the associated 4-metric $g_{\mu\nu}(y) = {{\partial
h^{\alpha}(y)}\over {\partial y^{\mu}}}\, {{\partial
h^{\beta}(y)}\over {\partial y^{\nu}}}\, \eta_{\alpha\beta} $
satisfies the conditions (\ref{I1}) and the asymptotic conditions
at spatial infinity described after Eqs.(\ref{I1}). In this way we
get a new reference frame: the time-like lines of the new
congruence are identified by $y^i = f^i(x) = const.$ Therefore
{\it a reference-frame-preserving coordinate transformation} must
be of the type $y^o = f^o(x^{\mu})$, $y^i = f^i(x^k)$. In the new
admissible coordinate system the simultaneity surfaces are
determined by $y^o = f^o(x) = const.$: they are space-like
hyper-surfaces, with the one $y^o = f^o(x) = 0$ passing through
$P$.
\medskip

As emphasized by Havas \cite{23a} the transition from a global
inertial system, with its standard clock (showing the same
proper-time rate when placed at rest at the same place and with
the proper time equal to the Cartesian coordinate time since
$\eta_{oo} = \sgn$ and with Einstein's synchronization of all the
clocks with $x^o = ct = const.$), to arbitrary allowed
4-coordinates only means to use arbitrary curvilinear spatial
3-coordinates, non-standard clocks (with the proper time -
coordinate time connection $s_{12} = c\, {\cal T}_{12} =
\int_1^2\, \sqrt{\sgn\, g_{oo}(y)}\, dy^o$), a non-standard
definition of simultaneity ($dy^o = 0$ so that the space-like
simultaneity surfaces are defined by $y^o = f^o(x) = const.$) and
({\it non-rigid}) arbitrary reference frames (defined by the
time-like vector field $l^{\mu}(y)\, {{\partial}\over {\partial
y^{\mu}}}$ describing the field of unit normals to the
simultaneity surfaces) as in general relativity as emphasized also
by Fock \cite{91a}.

Since the line element in the allowed 4-coordinates is $ds^2 =
g_{\mu\nu}(y)\, dy^{\mu}\, dy^{\nu}$, in general with $g_{oi}(y)
\not= 0$ ({\it non-time-orthogonal metric}), the determination of
{\it spatial length} and the synchronization of clocks can be done
in two different ways like in general relativity:
\medskip

A) {\it globally}, but in a coordinate-dependent way, by using the
$dy^o = 0$ exact notion of (coordinate) simultaneity, defined
using Einstein's convention, associated to the chosen allowed
4-coordinates (the 3+1 point of view) and the related notion of
instantaneous 3-space;

B) {\it locally}, but in a coordinate-independent, by means of
light signals by using Einstein's convention to define local
simultaneity (the 1+3 point of view), as done for instance in
Landau-Lifschitz \cite{11a}, but with the lacking notion of
intantaneous 3-space replaced by the local rest frame of some
observer.
\bigskip

Let us compare these two notions of simultaneity and their
implications for the synchronization of clocks and for the
definition of spatial distance between different points in
Minkowski space-time \footnote{ Let us remember that in an
inertial system with Cartesian 4-coordinates, all the clocks on
the simultaneity surface $x^o = ct = const.$ are synchronized with
Einstein's convention using light signals or, equivalently, by
slow transport of clocks. This is equivalent to the standard
statement of relativity books that the absolute chrono-geometrical
structure of Minkowski space-time is realized by putting a clock
and rods in each point with the clocks on the hyper-planes $x^o =
const.$ synchronized.}.
\medskip

A) All the events on the space-like hyper-surface $\Sigma_{y^o}$,
$y^o = f^o(x) = const.$, are simultaneous, namely simultaneity is
realized with the condition $dy^o = 0$ ($c\, d{\cal T} =
\sqrt{\sgn\, g_{oo}(y)}\, dy^o = 0$) on the coordinate time.
Therefore, there must be a generalization of Einstein's convention
using light rays implying the possibility to realize the
synchronization of all the clocks lying on the instantaneous
3-space $y^o(x) = const.$ . We will see in Section VI, Subsection
A, how to define such a generalization. Two simultaneous nearby
events $A$ and $B$ will have 4-coordinates $(y^o; y_A^i)$ and
$(y^o; y_B^i = y_A^i + dy^i)$ respectively and their (coordinate)
spatial distance will be

\beq
 dl_{AB} = \sqrt{-\sgn\, g_{ij}(y_A)\, dy^i\, dy^j}.
 \label{II2}
 \eeq

\medskip

Let us remark that in each event the coordinate time increment is
parallel to the normal $l^{\mu}(y)$ to the simultaneity surface
through that event. \bigskip

If the simultaneous events $A$ and $B$ are at finite distance on
$\sigma_{y^o}$, to each space-like path ${\cal P}$ joining them is
associated a distance $L^{({\cal P})}_{AB} = \int_{\cal P}\,
dl_{AB}$, as said in footnote 1, which is extremized by choosing
the 3-geodesics joining them on $\Sigma_{y^o}$.

\bigskip

B) The usual strategy to define local simultaneity in general
relativity (where it works only locally for nearby pairs of events
due to the gravitational field \footnote{But for globally
hyperbolic space-times the global simultaneity A) of the 3+1 point
of view can be defined and is used \cite{86a,92a} in canonical ADM
\cite{85a} metric gravity.}), as exemplified in Ref.\cite{11a}, is
the Martzke-Wheeler extension \cite{24a}, adapted to accelerated
observers, of Einstein's convention using test light rays (see for
instance Ref.\cite{93a}).

Given an event $A$ with 4-coordinates $y^{\mu} = f^{\mu}(x)$ and
proper time

\beq
 {\cal T}_A = {1\over c}\, \int^A_o\, \sqrt{\sgn\, g_{oo}(y)}\,
dy^o,
 \label{II3}
 \eeq

\noindent a nearby (locally) simultaneous event $B$ will have
4-coordinates $y_B^{\mu} = y^{\mu} + \Delta\, y^{\mu}$ with
$\Delta y^o \not= 0$ determined by Einstein's convention in the
following way.

If $A$ receives a light signal from a nearby event $B_{-}$ of
4-coordinates $y^{\mu}_{-} = y^{\mu} + \delta\, y^{\mu}_{-} = (y^o
+ \delta\, y^o_{-}; y^i + \Delta y^i)\,\,\,$ with $\delta y^o_{-}
< 0$ , then $\delta y^o_{-}$ is determined \footnote{The two
solutions of $ds^2 = 0$ are $\hat \delta y^o_{(\pm)} = -
{{g_{oi}(y)}\over { g_{oo}(y)}}\, \Delta y^i\, \pm {1\over {
g_{oo}(y)}}\, \sqrt{\, [g_{oi}(y)\, g_{oj}(y) - g_{oo}(y)\,
g_{ij}(y)]\, \Delta y^i\, \Delta y^j }$. With our conventions on
the signature of the metric ($\sgn\, g_{oo}(y) > 0$, $-\sgn\,
g_{ij}(y) > 0$), for $\sgn = +$ we get $\hat \delta y^o_{(+)} > 0$
and $\hat \delta y^o_{(-)} < 0$, while for $\sgn = -$ we get $\hat
\delta y^o_{(-)} > 0$ and $\hat \delta y^o_{(+)} < 0$. This
implies Eq.(\ref{II4}).} by the condition $ds^2 = 0$ with the
result

\bea
 \delta\, y^o_{-} &=& {1\over { g_{oo}(y)}}\Big( -
g_{oi}(y)\, \Delta y^i - \sgn\, \sqrt{\, [g_{oi}(y)\, g_{oj}(y) -
 g_{oo}(y)\, g_{ij}(y)]\, \Delta y^i\, \Delta y^j }\Big) =\nonumber \\
&&\nonumber\\
 &=& - {{g_{oi}(y)}\over { g_{oo}(y)}}\, \Delta y^i -
 {{\sgn}\over {\sqrt{\sgn\, g_{oo}(y)}}}\, \sqrt{-\sgn\, \Big(
 g_{ij}(y) - {{g_{oi}(y)\, g_{oj}(y)}\over {g_{oo}(y)}}\Big)\,
 \Delta y^i\, \Delta y^j}\, < 0.
 \label{II4}
 \eea

\noindent Then $A$ re-transmits the signal to a nearby event
$B_{+}$, where the light signal is absorbed,  of 4-coordinates
$y^{\mu}_{+} = y^{\mu} + \delta\, y^{\mu}_{+} = (y^o + \delta\,
y^o_{+}; y^i + \Delta\, y^i)$ with $\delta y^o_{+} > 0$. Now $ds^2
= 0$ gives the following expression for $\delta y^o_{+}$

\bea
 \delta\, y^o_{+} &=& {1\over {g_{oo}(y)}}\Big( - g_{oi}(y)\,
\Delta y^i + \sgn\, \sqrt{\, [g_{oi}(y)\, g_{oj}(y) - g_{oo}(y)\,
g_{ij}(y)]\, \Delta y^i\, \Delta y^j }\Big)  =\nonumber \\
&&\nonumber\\
&=& - {{g_{oi}(y)}\over {g_{oo}(y)}}\, \Delta y^i +
 {{\sgn}\over {\sqrt{\sgn\, g_{oo}(y)}}}\, \sqrt{-\sgn\, \Big(
 g_{ij}(y) - {{g_{oi}(y)\, g_{oj}(y)}\over {g_{oo}(y)}}\Big)\,
 \Delta y^i\, \Delta y^j}\, > 0.
 \label{II5}
 \eea

\noindent Then the coordinate time $y^o_B = y^o + \Delta\, y^o$ of
the {\it (locally) simultaneous event} $B$ (in between events
$B_{-}$ and $B_{+}$) is determined by the {\it Einstein
convention}

\beq
 \Delta\, y^o\, {\buildrel {def}\over =}\,
 {1\over 2}\, (\delta\, y^o_{+} + \delta\, y^o_{-}) = -
 {{g_{oi}(y)}\over {g_{oo}(y)}}\, \Delta\, y^i.
 \label{II6}
 \eeq

This corresponds to replace the exact global simultaneity
definition $dy^o = 0$ (or $d{\cal T} = {1\over c}\,
\sqrt{\sgn\, g_{oo}(y)}\, dy^o = 0$) given in A) with the
local approximate one implied by the vanishing of the so-called
{\it Einstein synchronized proper-time pseudo-interval}

\beq
 c\, \Delta\, {\widetilde {\cal T}} =
\sqrt{\sgn\, g_{oo}(y)}\, \Delta\, y^o + {{\sgn\, g_{oi}(y)}\over
{\sqrt{\sgn\, g_{oo}(y) }}}\, \Delta\, y^i\, {\buildrel {def}\over
=}\, \sqrt{\sgn\, g_{oo}(y)}\, \Delta\, y^o + \Delta_{y^o} = 0,
 \label{II7}
 \eeq

\noindent which is not proportional to an exact 1-form (it is not
an interval) like it happens for $c\, d\, {\cal T} = 0$.
Therefore, with this definition of simultaneity, the statement
that two nearby events $A$ and $B$ are (locally) simultaneous
requires the use of two events $B_{-}$ and $B_{+}$ (in the past
and in the future of $B$ respectively) with the following
difference of coordinate time

\bea
 \delta\, y^o &=& \delta\, y^o_{+} - \delta\,
y^o_{-} = {{2\, \sgn}\over { g_{oo}(y)}}\, \sqrt{ [g_{oi}(y)\,
g_{oj}(y) - g_{oo}(y)\, g_{ij}(y)]\, \Delta y^i\, \Delta y^j }
 =\nonumber \\
&&\nonumber\\
 &=& {2\over {\sqrt{\sgn\, g_{oo}(y)}}}\, \sqrt{- \sgn\,
 \Big(g_{ij}(y) - {{g_{oi}(y)\, g_{oj}(y)}\over {g_{oo}(y)}}\Big)\,
 \Delta y^i\, \Delta y^j}.
 \label{II8}
 \eea

\medskip

Since we can re-write the line element $ds^2 = g_{\mu\nu}(y)\,
dy^{\mu}\, dy^{\nu}$ as

\beq
 ds^2 = \sgn\, \Big(  \sqrt{\sgn\, g_{oo}(y)}\,
dy^o + {{\sgn\, g_{oi}(y)}\over {\sqrt{\sgn\, g_{oo}(y)}}}\,
dy^i\Big)^2 +  \Big( g_{ij}(y) - {{g_{oi}(y)\, g_{oj}(y)}\over
{g_{oo}(y)}}\Big)\, dy^i\, dy^j,
 \label{II9}
 \eeq

\noindent the {\it (pseudo-proper) spatial distance} between the
(locally) simultaneous events $A$ and $B$ implied by the local
simultaneity notion $c\, \Delta\, {\widetilde {\cal T}} = 0$ is

\beq
 \Delta l_{AB} = \sqrt{-\sgn\, \Big( g_{ij}(y) -
{{g_{oi}(y)\, g_{oj}(y)}\over {g_{oo}(y)}}\Big)\, \Delta y^i\,
\Delta y^j }\, {\buildrel {def}\over =}\,
\sqrt{{}^3\gamma_{ij}(y)\, \Delta y^i\, \Delta y^j}.
 \label{II10}
 \eeq

\noindent This justifies the use of the spatial 3-metric
${}^3\gamma_{ij}(y) = -\sgn\, [g_{ij}(y) - {{g_{oi}(y)\,
g_{oj}(y)}\over {g_{oo}(y)}}]$ of signature $(+++)$,  which
satisfies the analogue of the spatial conditions of Eq.(\ref{I1})
\cite{11a}.
\medskip

In each point, remembering $y^{\mu} = f^{\mu}(x)$ with inverse
$x^{\mu} = h^{\mu}(y)$, we have

\bea
 c\, \Delta\, {\widetilde {\cal T}} &=& {{\sgn}\over
{\sqrt{\sgn\, g_{oo}(y)}}}\, {{\partial h^{\alpha}(y)}\over
{\partial y^o}}\, \eta_{\alpha\beta}\, \Big[ {{\partial
h^{\beta}(y)}\over {\partial y^o}}\, \Delta y^o + {{\partial
 h^{\beta}(y)}\over {\partial y^i}}\, \Delta y^i\Big] =\nonumber \\
 &=& {{\sgn}\over {\sqrt{\sgn\, g_{oo}(y)}}}\, {{\partial
h_{\alpha}(y)}\over {\partial y^o}}\, \Delta x^{\alpha} \,
{\buildrel {def}\over =}\, \sgn\, u_{\mu}(y)\, \Delta y^{\mu}.
 \label{II11}
 \eea

The time-like vector field $u^{\mu}(y) = {1\over {\sqrt{\sgn\,
g_{oo}(y)}}}\, {{\partial h^{\mu}(y)}\over {\partial y^o}}$
defines a (in general non-surface-forming) reference frame (1+3
point of view), which in each point is orthogonal to the
space-like vector

\beq
 \Delta l^{\mu}_{AB\perp} = \Big( {{\partial h^{\mu}(y)}\over
{\partial y^i}} - {{g_{oi}(y)}\over {g_{oo}(y)}}\, {{\partial
h^{\mu}(y)}\over {\partial y^o}}\Big)\, \Delta y^i =
[\delta^{\mu}_{\nu} - \sgn\, u^{\mu}(y)\, u_{\nu}(y)]\, \Delta
x^{\nu},
 \label{II12}
 \eeq

\noindent lying in the rest frame tangent plane at $A$ in $T_AM^4$
(the local rest frame $R_u$). The spatial distance can be
re-written as \footnote{This equation give us a expression of the
metric ${}^3\gamma_{ij}(y)$ in terms of $\delta y^o_+$, $\delta
y^o_-$. Instead from Eqs.(\ref{II5}),(\ref{II6}) we get the
following expression \cite{28a} (Chapter III, Section 40) for
$dl_{AB}$ of Eq.(\ref{II2})
\[
(dl_{AB})^2 = - \sgn\, g_{ij}(y)\, \Delta y^i\, \Delta y^j =
- \sgn\, g_{oo}(y)\, \delta y^o_+\cdot\delta y^o_-.
\]} $(\Delta l_{AB})^2 = -\sgn\, \Delta l_{AB\perp}^{\mu}\,
\eta_{\mu\nu}\, \Delta l_{AB\perp}^{\nu} = {{\sgn\,
g_{oo}(y)}\over 4}\, (\delta y^o)^2 =  {{\sgn\, g_{oo}(y)}\over
4}\, [\delta y^o_{+} - \delta y^o_{-}]^2$. This shows that we can
define  the line pseudo-element with the coordinate-independent
time-orthogonal decomposition

\beq (\Delta s_{AB})^2 = \sgn\, \Big[c^2\, (\Delta {\tilde {\cal
T}}_{AB})^2 - (\Delta l_{AB})^2\Big].
 \label{II13}
 \eeq
\bigskip

However, since the local observer rest frame $R_u$ is not tangent
to an instantaneous 3-space, we cannot define the spatial distance
of non-nearby simultaneous events. In phenomenological
calculations $R_u$ is identified with the instantaneous 3-space of
the observer (see footnote 11) and the finite spatial distance of
a point P from the observer is defined by considering the
3-geodesic joining P to the observer.

\bigskip

As said in the Introduction, instead of the 4-coordinates
$y^{\mu}$, it is convenient to use the Lorentz scalar (radar)
4-coordinates $\sigma^A = (\tau ,\vec \sigma ) = \sigma^A(x)$ and
the embedding (inverse coordinate transformation) $x^{\mu} =
z^{\mu}(\sigma ) = z^{\mu}(\tau ,\vec \sigma )$, naturally adapted
to the simultaneity leaves $\Sigma_{\tau}$ of the 3+1 splitting.
In these 4-coordinates the previous two cases A) and B) are
re-formulated in the following way
\medskip

A) We use the synchronization $l_{\mu}(\sigma )\, dz^{\mu}(\sigma
) = N\, d\tau = 0$ (synchronizzability of the reference frame of
normals to the simultaneity surfaces), whose associated spatial
distance of simultaneous events is

\beq
 dl_{AB} = \sqrt{-\sgn\, g_{rs}(\tau ,\vec \sigma )\,
d\sigma^r\, d\sigma^s}.
 \label{II14}
 \eeq

\medskip

B) We use the time-like 4-vector

\beq
 u^{\mu}(\tau ,\vec \sigma ) = {{z^{\mu}_{\tau}(\tau ,\vec \sigma
)}\over {\sqrt{\sgn\, g_{\tau\tau}(\tau ,\vec \sigma ) }}},
 \label{II15}
 \eeq

\noindent which defines a reference frame, in general
non-synchronizable, and we use the  local notion of simultaneity
($\Delta \tau_E$ is an Einstein synchronized non-proper-time
pseudo-interval)

\bea
 c\, \Delta {\widetilde {\cal T}}
&=& \sgn\, u_{\mu}(\sigma )\, \Delta z^{\mu}(\sigma )\,
{\buildrel {def}\over =}\, \sqrt{\sgn\, g_{\tau\tau}(\sigma ) }\,
\Delta \tau + \Delta_{\tau}\,  =\nonumber \\
&&\nonumber\\
 &=& {{\sgn\, z_{\tau \mu}(\tau ,\vec \sigma )}\over {\sqrt{\sgn\,
 g_{\tau\tau}(\tau ,\vec \sigma )}}}\, \Big[z^{\mu}_{\tau}(\tau ,\vec \sigma )\,
 \Delta \tau +  z^{\mu}_r(\tau ,\vec \sigma )\, \Delta \sigma^r\Big] =
\nonumber \\
&&\nonumber\\
 &=& \sqrt{\sgn\, g_{\tau\tau}(\tau ,\vec \sigma )}\, \Delta \tau +
 {{g_{\tau r}(\tau ,\vec \sigma )}\over {\sqrt{\sgn\, g_{\tau\tau}(\tau
 ,\vec \sigma )}}}\, \Delta \sigma^r\, {\buildrel {def}\over =}\,
 \sqrt{\sgn\, g_{\tau\tau}(\sigma ) }\, \Delta \tau_E = 0.
 \label{II16}
 \eea

\noindent Now  we get the solutions $\delta \tau_{\pm}  = -
{{g_{\tau r}}\over {g_{\tau\tau}}}\, \Delta \sigma^r \pm
{{\sgn}\over {\sqrt{\sgn\, g_{\tau\tau}}}}\, \sqrt{- \epsilon\,
\Big(g_{rs} - {{g_{\tau r}\, g_{\tau s}}\over
{g_{\tau\tau}}}\Big)\, \Delta \sigma^r\, \Delta \sigma^s } $
($\delta \tau_{+} > 0$, $\delta \tau_{-} < 0$) of $ds^2 = 0$, so
that the Einstein convention becomes

\beq
 \Delta \tau\, {\buildrel {def}\over =}\,
 {1\over 2}\, (\delta\, \tau_{+} + \delta\, \tau_{-})
 = - {{ g_{\tau r}(\tau ,\vec \sigma )}\over
{ g_{\tau\tau}(\tau ,\vec \sigma ) }}\, \Delta \sigma^r .
 \label{II17}
  \eeq

\noindent By defining

\beq
 \Delta l^{\mu}_{AB\perp} = \Big( z^{\mu}_r - {{g_{\tau r}}\over
{g_{\tau\tau}}}\, z^{\mu}_{\tau}\Big)\, \Delta \sigma^r,
 \label{II18}
 \eeq

\noindent we arrive at the following definition of  {\it pseudo
spatial distance} ( $\delta \tau = \delta\, \tau_{+} - \delta\,
\tau_{-}$) \footnote{Moreover we have $(dl_{AB})^2 = - \sgn\,
g_{rs}(\tau ,\vec \sigma )\, d\sigma^r\, d\sigma^s = - \sgn\,
g_{\tau\tau}(\tau ,\vec \sigma )\, \delta \tau_{+}\, \delta
\tau_{-}$.}

\bea
 \Delta l_{AB\perp}^2 &=& - \sgn\, \Big(g_{rs}(\tau ,\vec
\sigma ) - {{g_{\tau r}(\tau ,\vec \sigma )\, g_{\tau s}(\tau
,\vec \sigma ) }\over {g_{\tau\tau}(\tau ,\vec \sigma )}}\Big)\,
\Delta \sigma^r\, \Delta \sigma^s =\nonumber \\
&&\nonumber\\
 &=& {}^3\gamma_{rs}(\tau ,\vec \sigma )\, \Delta\, \sigma^r\,
\Delta\, \sigma^s = {{\sgn\, g_{\tau\tau}(\tau ,\vec \sigma
)}\over 4}\, (\delta \tau )^2,
 \label{II19}
 \eea

\noindent with the 3-metric ${}^3\gamma_{rs}$ of signature
$(+++)$.

\bigskip

Let us define the {\it one-way velocity of a ray of light}
according to the two points of view  A) and B) on simultaneity. A
ray of light has a null geodesic $z^{\mu}(\lambda ) = z^{\mu}(\tau
(\lambda ), \vec \sigma (\lambda ))$ [$\lambda$ affine parameter
along the null geodesic] of Minkowski space-time as world-line,
defined by the condition [use Eq.(\ref{II9})]
\medskip

\bea
 &&ds^2 = \nonumber\\
&&\nonumber\\
&=&\left[ g_{\tau\tau}(\sigma (\lambda ))\, \left({{d\tau
 (\lambda )}\over {d\lambda^2}}\right)^2 + 2\, g_{\tau r}(\sigma
 (\lambda ))\, {{d\tau (\lambda )}\over {d\lambda}}\,
 {{d\sigma^r(\lambda )}\over {d\lambda}} + g_{rs}(\sigma (\lambda
 ))\, {{d\sigma^r(\lambda )}\over {d\lambda}}\,
 {{d\sigma^s(\lambda )}\over {d\lambda}}\right]\, (d\lambda )^2
 =\nonumber \\
&&\nonumber\\
 &=& \left[ \sgn\, \left(\sqrt{\sgn\,
 g_{\tau\tau}(\tau ,\vec \eta (\tau ))}\, + {{\sgn\, g_{\tau r}(\tau
 ,\vec \eta (\tau ))}\over {\sqrt{\sgn\, g_{\tau\tau}(\tau
 ,\vec \eta (\tau ))}}}\, {{d\eta^r(\tau )}\over {d\tau}}\right)^2+\right.\nonumber\\
&&\nonumber\\
 &&- \sgn\, \left.\, {}^3\gamma_{rs}(\tau ,\vec \eta (\tau ))\,
 {{d\eta^r(\tau )}\over {d\tau}}\, {{d\eta^s(\tau )}\over
 {d\tau}}\right]\, \left({{d\tau (\lambda )}\over
 {d\lambda}}\right)^2\, (d\lambda )^2 = 0,
 \label{II20}
 \eea
\medskip

\noindent where we have introduced the 3-coordinates $\vec \eta
(\tau )$ for the light ray by means of $z^{\mu}(\tau (\lambda ),
\vec \sigma (\lambda )) = z^{\mu}(\tau (\lambda ), \vec \eta (\tau
(\lambda )))$. This condition implies
\medskip

\beq
 g_{rs}(\tau ,\vec \eta (\tau ))\, {{d\eta^r(\tau )}\over
 {d\tau}}\, {{d\eta^s(\tau )}\over {d\tau}} + 2\, g_{\tau r}(\tau
 ,\vec \eta (\tau ))\, {{d\eta^r(\tau )}\over {d\tau}} +
 g_{\tau\tau}(\tau ,\vec \eta (\tau )) = 0
\label{II21}
\eeq

\noindent or

 \beq
 {}^3\gamma_{rs}(\tau ,\vec \eta (\tau
 ))\, {{d\eta^r(\tau )}\over {d\tau}}\, {{d\eta^s(\tau )}\over
 {d\tau}} - \left(\sqrt{\sgn\,
 g_{\tau\tau}(\tau ,\vec \eta (\tau ))}\, + {{\sgn\, g_{\tau r}(\tau
 ,\vec \eta (\tau ))}\over {\sqrt{\sgn\, g_{\tau\tau}(\tau
 ,\vec \eta (\tau ))}}}\, {{d\eta^r(\tau )}\over {d\tau}}\right)^2=
 0.
 \label{II22}
 \eeq
\medskip

The solution of this equation defines a different {\it one-way
direction-dependent coordinate velocity of light} according to the
points of view A) and B): ${{d\eta^r(\tau )}\over {d\tau}} = {{d
\eta (\tau )}\over {d\tau}}\, {\hat n}^r$ with the 3-direction of
propagation $\hat n$ normalized to the
unity according to the chosen A) or B) notion of spatial distance.
\medskip

A) We use the definition of one-way velocity $v_A = {{d
l_{AB}}\over {d\tau}} $ implied by Eq.(\ref{II14}). We put $-
\sgn\, g_{rs}(\tau ,\vec \eta (\tau ))\, {\hat n}_A^r\, {\hat
n}_A^s = 1$. Eq.(\ref{II21}) and its (future-pointing) solution
are
\medskip

\bea
 &&\left( {{d\eta_A(\tau )}\over {d\tau}}\right)^2 - 2\, \sgn\,
 g_{\tau r}(\tau ,\vec \eta (\tau ))\, {\hat n}_A^r\,
 {{d\eta_A(\tau )}\over {d\tau}} - \sgn\, g_{\tau\tau}(\tau
 ,\vec \eta (\tau )) = 0,\nonumber \\
 &&\nonumber \\
 v^r_A(\tau ) &=& {{d\eta_A(\tau )}\over {d\tau}}\, {\hat n}^r_A =
 \left[\sgn\, g_{\tau s}(\tau ,\vec \eta (\tau ))\, {\hat n}^s_A
 + \sqrt{[\,g_{\tau s}(\tau ,\vec \eta (\tau ))\, {\hat n}^s_A\,]^2
 +\sgn\, g_{\tau\tau}(\tau ,\vec \eta (\tau ))}\,
 \right]\, {\hat n}^r_A =\nonumber \\
&&\nonumber\\
 &=& \left[ {{\sgn\, g_{\tau\tau}(\tau ,{\vec \eta}(\tau ))}\over
 {- \sgn\, g_{\tau s}(\tau ,\vec \eta (\tau ))\, {\hat n}^s_A
 + \sqrt{[g_{\tau s}(\tau ,\vec \eta (\tau ))\, {\hat n}^s_A]^2
 +\sgn\, g_{\tau\tau}(\tau ,\vec \eta (\tau ))}}}\right]\, {\hat n}_A^r,\nonumber \\
 &&\nonumber \\
 {\tilde v}_A(\tau ) &=& {c\over {\sqrt{\sgn\, g_{\tau\tau}(\tau
 ,\vec \eta (\tau ))}}}\, {{d\eta_A(\tau )}\over {d\tau}} = \nonumber\\
&&\nonumber\\
&=&c\,
 {{ \sgn\, g_{\tau s}(\tau ,\vec \eta (\tau ))\, {\hat n}^s_A
 + \sqrt{(g_{\tau s}(\tau ,\vec \eta (\tau ))\, {\hat n}^s_A)^2
 +\sgn\, g_{\tau\tau}(\tau ,\vec \eta (\tau ))}}\over
 {\sqrt{\sgn\, g_{\tau\tau}(\tau ,\vec \eta (\tau ))}}}\nonumber \\
 &&{}\nonumber \\
 &\rightarrow& \, c \qquad\mbox{ if } g_{\tau r} = 0.
 \label{II23}
 \eea

\medskip
B) We use the definition of one-way velocity $v_B = {{\Delta
l_{AB} }\over {d\tau}} $ implied by Eq.(\ref{II19}). We put $
{}^3\gamma_{rs}(\tau ,\vec \eta (\tau ))\, {\hat n}_B^r\, {\hat
n}_B^s = 1$. Eq.(\ref{II22}) and its (future-pointing) solution
are (see Refs. \cite{12a,23a})

\bea
 &&\Big({{d\eta_B(\tau )}\over
 {d\tau}}\Big)^2 - \left(\sqrt{\sgn\,
 g_{\tau\tau}(\tau ,\vec \eta (\tau ))}\, + {{\sgn\, g_{\tau r}(\tau
 ,\vec \eta (\tau ))}\over {\sqrt{\sgn\, g_{\tau\tau}(\tau
 ,\vec \eta (\tau ))}}}\,\hat{n}_B\, {{d\eta_B(\tau )}\over {d\tau}}\right)^2=
 0,
 \nonumber \\
 &&{}\nonumber \\
 v^r_B(\tau ) &=& {{d\eta_B(\tau )}\over {d\tau}}\, {\hat n}^r_B =
 {{g_{\tau\tau}(\tau ,\vec \eta (\tau ))}\over {\sqrt{\sgn\,
 g_{\tau\tau}(\tau ,\vec \eta (\tau ))} - \sgn\, g_{\tau s}(\tau ,\vec \eta (\tau ))
 \, {\hat n}^s_B }}\,\, {\hat n}_B^r,\nonumber \\
 &&{}\nonumber \\
 {\tilde v}_B(\tau )& = &{c\over {\sqrt{\sgn\, g_{\tau\tau}(\tau
 ,\vec \eta (\tau ))}}}\, {{d\eta_B(\tau )}\over {d\tau}} = c\,
 {\sqrt{{\sgn\, g_{\tau\tau}(\tau ,\vec \eta (\tau ))}}\over {\sqrt{\sgn\,
 g_{\tau\tau}(\tau ,\vec \eta (\tau ))} - \sgn\, g_{\tau s}(\tau ,\vec \eta (\tau ))
 \, {\hat n}^s_B}}\nonumber \\
 &&{}\nonumber \\
 &\rightarrow&\, c \qquad\mbox{ if } g_{\tau r} = 0.
 \label{II24}
 \eea

\medskip

Havas \cite{23a} showed that, if we use admissible coordinate
transformations $y^{\mu} = f^{\mu}(x)$ instead of adapted
coordinates $\sigma^A$, then the analogue of Eq.(\ref{II24})
corresponds to a generalization of Einstein's convention of the
type $x^o_A = x^o_{B_{-}} + E\, (x^o_{B_{+}} - x^o_{B_{-}})$, $0 <
E < 1$ with a position- and direction- dependent $E =
{{x^o_{B_{-}} - x^o_A}\over {x^o_{B_{+}} - x^o_{B_{-}} }}$ in
place of the constant $E$ of footnote 4. Instead in adapted
coordinates we get $\tau_A = \tau_{B_{-}} + {\cal E}\,
(\tau_{B_{+}} - \tau_{B_{-}})$, $0 < {\cal E} < 1$. See also
Ref.\cite{21a} for the study of this type of non-standard
synchrony  and of the associated 3-metric to be used for measuring
spatial distances.

\bigskip

Finally  the local notion B) of simultaneity  and the definition
(\ref{II19}) of spatial distance imply the following null
pseudo-interval [see Eq.(\ref{II13})]

\beq
 (\Delta s_{AB})^2
= \sgn\, [c^2\, (\Delta {\widetilde {\cal T}}_{AB})^2 - (\Delta
l_{AB})^2] = 0,
 \label{II25}
 \eeq

\noindent and allow to define an invariant isotropic one-way
pseudo-velocity

\beq
 v_{AB} = {{\Delta l_{AB}}\over {\Delta \widetilde{\cal T}_{AB}}} = c,
 \label{II26}
 \eeq

\noindent even for non-time-orthogonal metrics. This is a formal
answer to Klauber's criticism \cite{33a}, but, since $(\Delta
{\tilde {\cal T}}_{AB})$ is neither the proper time of $A$ nor
that of $B$, the interpretation of this pseudo-velocity is not
clear.

\vfill\eject

\section{Admissible 4-Coordinates and the Locality Hypothesis:
Non-Existence of Rigid Rotating Reference Frames.}

In Ref.\cite{23a} Havas proposed the following two examples (the
second one is a time-dependent transformation) of simultaneity
foliations associated with admissible 4-coordinates, i.e. whose
4-metric satisfies Eq.(\ref{I1}):

\bea
 1)&&{}\nonumber \\
 &&{}\nonumber \\
 &&y^o =x^o + {1\over {c^2}}\, f(x^i),\qquad y^i = x^i,\qquad
 \mbox{with inverse } x^o = y^o - {1\over {c^2}}\, f(y^i),\quad
  x^i = y^i,\nonumber \\
  &&{}\nonumber \\
 &&\mbox{and with associated 4-metric } \nonumber\\
&&\nonumber\\
&& g_{oo}(y) = - c^2,\quad g_{oi}(y) = {{\partial f(y)}\over
{\partial y^i}}\,\, (\mbox{ if }  g_{oi} = const.,\mbox{ then } (g_{oi})^2 <
c^2),\nonumber \\
&&\nonumber\\
 &&g_{ij}(y) =\delta_{ij} - {1\over {c^2}}\,
{{\partial f}\over {\partial y^i}}\, {{\partial f}\over {\partial
y^j}};
 \label{III1}
 \eea
\bigskip

\medskip
 \bea
 2)&&{}\nonumber \\
 &&{}\nonumber \\
 &&y^o = x^o\, f^{-1}(x^i),\qquad y^i = x^i,\qquad
 \mbox{with inverse } x^o = y^o\,
f(y^i),\quad x^i = y^i,\qquad \nonumber\\
&&{}\nonumber \\
&&\mbox{and with associated
4-metric }\nonumber\\
&&\nonumber\\
&&
 g_{oo}(y) = - c^2\, f^2,\qquad g_{oi}(y) = - y^o\, c^2\,
  f\, {{\partial f}\over {\partial y^i}},\qquad
 g_{ij}(y) = \delta_{ij} - (y^o)^2\, c^2\, {{\partial f}\over
{\partial y^i}}\, {{\partial f}\over {\partial y^j}}.
 \label{III2}
 \eea
\medskip

Both of them are examples of admissible 4-coordinate systems not
interpretable in terms of comoving observers as required by the
locality hypothesis.

\bigskip

Let us now consider a class of 4-coordinate transformations which
implements the idea of accelerated observers as sequences of
comoving observers  (the locality hypothesis) and let us determine
the conditions on the transformations to get a set of admissible
4-coordinates. From now on we shall use Lorentz-scalar radar-like
4-coordinates $\sigma^A = (\tau ; \vec \sigma )$ adapted to the
foliation, whose simultaneity leaves are denoted $\Sigma_{\tau}$.
\medskip

As we have said, the admissible embeddings $x^{\mu} = z^{\mu}(\tau
,\vec \sigma )$ [inverse transformations of $x^{\mu} \mapsto
\sigma^A(x)$], defined with respect to a given inertial system,
must tend to parallel space-like hyper-planes at spatial infinity.
If $l^{\mu} = l^{\mu}_{(\infty )} = \epsilon^{\mu}_{\tau}$,
$l^2_{(\infty )} = \sgn$, is the asymptotic normal, let us define
the asymptotic orthonormal tetrad $\epsilon^{\mu}_A$, $A=\tau
,1,2,3$, by using the standard Wigner boost for time-like
Poincare' orbits $L^{\mu}{}_{\nu}(l_{(\infty )}, {\buildrel \circ
\over l}_{(\infty )})$, ${\buildrel \circ \over l}_{(\infty )} =
(1; \vec 0)$: $\epsilon^{\mu}_A\, {\buildrel {def}\over =}\,
L^{\mu}{}_A(l_{(\infty )}, {\buildrel \circ \over l}_{(\infty
)})$, $\eta_{AB}=\epsilon^{\mu}_A\, \eta_{\mu\nu}\,
\epsilon^{\nu}_B$.
\medskip

Then a parametrization of the asymptotic hyper-planes is $z^{\mu}
= x^{\mu}_o + \epsilon^{\mu}_A\, \sigma^A = x^{\mu}(\tau ) +
\epsilon^{\mu}_r\, \sigma^r$ with $x^{\mu}(\tau ) = x^{\mu}_o +
\epsilon^{\mu}_{\tau}\, \tau$ a time-like straight-line (an
asymptotic inertial observer). Let us define a family of 3+1
splittings of Minkowski space-time by means of the following
embeddings
\medskip

\bea
 z^{\mu}(\tau ,\vec \sigma ) &=& x^{\mu}_o +
\Lambda^{\mu}{}_{\nu}(\tau ,\vec \sigma )\, \epsilon^{\nu}_A\,
\sigma^A\, = {\tilde x}^{\mu}(\tau ) + F^{\mu}(\tau ,\vec \sigma
),\nonumber \\
 &&{}\nonumber \\
 &&{\tilde x}^{\mu}(\tau ) = x^{\mu}_o + \Lambda^{\mu}{}_{\nu}(\tau ,\vec
 0)\, \epsilon^{\nu}_{\tau}\, \tau,\nonumber \\
 && F^{\mu}(\tau ,\vec
 \sigma ) = [\Lambda^{\mu}{}_{\nu}(\tau ,\vec \sigma ) -
 \Lambda^{\mu}{}_{\nu}(\tau ,\vec 0)]\, \epsilon^{\nu}_{\tau}\,
 \tau + \Lambda^{\mu}{}_{\nu}(\tau ,\vec \sigma )\,
 \epsilon^{\nu}_r\, \sigma^r,\nonumber \\
 &&{}\nonumber \\
 &&\Lambda^{\mu}{}_{\nu}(\tau ,\vec \sigma )\, {\rightarrow}_{|\vec
\sigma | \rightarrow \infty}\, \delta^{\mu}_{\nu},\quad
\Rightarrow\,\, z^{\mu}(\tau ,\vec \sigma )\, {\rightarrow}_{|\vec
\sigma | \rightarrow \infty}\,\, x^{\mu}_o + \epsilon^{\mu}_A\,
 \sigma^A = x^{\mu}(\tau ) + \epsilon^{\mu}_r\, \sigma^r,\nonumber \\
 &&{}
 \label{III3}
 \eea

\medskip

\noindent where $\Lambda^{\mu}{}_{\nu}(\tau ,\vec \sigma )$ are
Lorentz transformations ($\Lambda^{\mu}{}_{\alpha}\,
\eta_{\mu\nu}\, \Lambda^{\nu}{}_{\beta} = \eta_{\alpha\beta}$)
belonging to the component connected with the identity of
$SO(3,1)$. While the functions $F^{\mu}(\tau ,\vec \sigma )$
determine the form of the simultaneity surfaces $\Sigma_{\tau}$,
the centroid ${\tilde x}^{\mu}(\tau )$, corresponding to an
arbitrary time-like observer chosen as origin of the 3-coordinates
on each $\Sigma_{\tau}$, determines how these surfaces are packed
in the foliation.

\medskip
A variant are the embeddings
\medskip

\beq
 z^{\mu}(\tau ,\vec \sigma ) = x_o^{\mu}(\tau ) +
\Lambda^{\mu}{}_{\nu}(\tau ,\vec \sigma )\, \epsilon^{\nu}_A\,
\sigma^A \, {\rightarrow}_{|\vec \sigma | \rightarrow \infty}\,
x^{\mu}_o(\tau ) + \epsilon^{\mu}_{\tau}\, \tau +
\epsilon^{\mu}_r\, \sigma^r = x^{\mu}(\tau ) + \epsilon^{\mu}_r\,
\sigma^r,
 \label{III4}
 \eeq
\medskip

\noindent with $x^{\mu}(\tau ) = x^{\mu}_o(\tau ) +
\epsilon^{\mu}_{\tau}\, \tau = \epsilon^{\mu}_{\tau}\, [\tau +
f(\tau )]$ an arbitrary time-like straight-line (an inertial
observer) not parametrized in terms of the proper time. The only
difference now is that the asymptotic hyper-planes are no more
uniformly spaced like in the case $x^{\mu}(\tau ) = x^{\mu}_o +
\epsilon^{\mu}_{\tau}\, \tau$ ($z^{\mu}_{\tau} =
\epsilon^{\mu}_{\tau}\, \mapsto\, z^{\mu}_{\tau}(\tau ) =
\epsilon^{\mu}_{\tau}\, [1 + \dot f(\tau )]$).

\bigskip

Since the asymptotic foliation with parallel hyper-planes, having
a constant vector field $l^{\mu} = \epsilon^{\mu}_{\tau}$ of
normals, defines an inertial reference frame, we see that the
foliation (\ref{III3}) with its associated non-inertial reference
frame is obtained from the asymptotic inertial frame by means of
{\it point-dependent Lorentz transformations}. As a consequence,
the integral lines, i.e. the non-inertial observers  and
(non-rigid) non-inertial reference frames associated to this
special family of simultaneity notions, are parametrized as a
continuum of comoving inertial observers as required by the
locality hypothesis \footnote{The second non-inertial and
non-surface-forming reference frame (the skew one) associated with
these embeddings, with vector field $z^{\mu}_{\tau}(\tau ,\vec
\sigma )/\sqrt{\sgn\, g_{\tau\tau}(\tau ,\vec \sigma )}$,
asymptotically tends to the same asymptotic inertial reference
frame because $z^{\mu}_{\tau}(\tau ,\vec \sigma )\,
{\rightarrow}_{|\vec \sigma | \rightarrow \infty}\,
\epsilon^{\mu}_{\tau} = l^{\mu}_{(\infty )}$ if $\partial_{\tau}\,
\Lambda^{\mu}{}_{\nu}(\tau ,\vec \sigma )\, {\rightarrow}_{|\vec
\sigma | \rightarrow \infty}\, 0$. }.

\medskip

Let us remark that when an arbitrary isolated system is described
by a Minkowski parametrized theory, in which the embeddings
$z^{\mu}(\tau ,\vec \sigma )$ are {\it gauge} configuration
variables, the transition from the description of dynamics in one
of these non-inertial reference frames compatible with the
locality hypothesis to another arbitrary allowed reference frame,
like the one of Eqs.(\ref{III1}), is a {\it gauge transformation}:
{\it therefore in this case the locality hypothesis can always be
assumed valid modulo gauge transformations}.
\medskip

An equivalent parametrization of the embeddings of this family of
reference frames is
\medskip

\bea
 z^{\mu}(\tau ,\vec \sigma ) &=& x^{\mu}_o + \epsilon^{\mu}_B\,
\Lambda^B{}_A(\tau ,\vec \sigma )\, \sigma^A = x^{\mu}_o +
U^{\mu}_A(\tau ,\vec \sigma )\, \sigma^A
 = {\tilde x}^{\mu}(\tau ) + F^{\mu}(\tau ,\vec \sigma ),\nonumber \\
 &&{}\nonumber \\
 &&{\tilde x}^{\mu}(\tau ) = x^{\mu}_o +
U^{\mu}_{\tau}(\tau ,\vec 0)\, \tau ,\nonumber \\
 &&F^{\mu}(\tau ,\vec \sigma ) =
 [U^{\mu}_{\tau}(\tau ,\vec \sigma ) -
 U^{\mu}_{\tau}(\tau ,\vec 0)]\, \tau + U^{\mu}_r(\tau ,\vec
 \sigma )\, \sigma^r,
\label{III5}
\eea

\noindent where we have defined:

 \bea
 \Lambda^B{}_A(\tau ,\vec \sigma ) &=& \epsilon^B_{\mu}\,
  \Lambda^{\mu}{}_{\nu}(\tau ,\vec \sigma )\,
\epsilon^{\nu}_A,\qquad  U^{\mu}_A(\tau ,\vec \sigma )\,
 \eta_{\mu\nu}\, U^{\nu}_B(\tau ,\vec \sigma ) =
 \epsilon^{\mu}_A\, \eta_{\mu\nu}\,
 \epsilon^{\nu}_B = \eta_{AB}, \nonumber \\
 &&{}\nonumber \\
 U^{\mu}_A(\tau ,\vec \sigma ) &=& \epsilon^{\mu}_B\,
 \Lambda^B{}_A(\tau ,\vec \sigma )\, {\rightarrow}_{|\vec \sigma |
  \rightarrow \infty}\, \epsilon^{\mu}_A,
 \label{III6}
 \eea

\noindent where $\epsilon^B_{\mu} = \eta_{\mu\nu}\, \eta^{BA}\,
\epsilon^{\nu}_A$ are the inverse tetrads.

\bigskip

A slight generalization of these embeddings allows to find
Nelson's \cite{48a} 4-coordinate transformation [but extended from
$\vec \sigma$-independent Lorentz transformations
$\Lambda^{\mu}{}_{\nu} = \Lambda^{\mu}{}_{\nu}(\tau )$ to $\vec
\sigma$-dependent ones!] implying M$\o$ller rotating 4-metric
\footnote{$g_{oo} = \sgn ([(1+ {{\vec a \cdot \vec x}\over
{c^2}})^2 - {{(\omega \times \vec x)^2}\over {c^2}})$, $g_{oi} =
-\sgn\, {1\over c}\, (\vec \omega \times \vec x)^i$, $g_{ij} =
-\sgn\, \delta_{ij}$, where $\vec a$ is the time-dependent
acceleration of the observer's frame of reference relative to the
comoving inertial frame and $\vec \omega$ is the time-dependent
angular velocity of the observer's spatial rotation with respect
to the comoving frame; $\vec x$ is the position vector of a
spatial point with respect to the origin of the observer's
accelerated frame.}

\bea
 z^{\mu}(\tau ,\vec \sigma ) &=& x^{\mu}_o + \epsilon^{\mu}_A\,
\Big[ \Lambda^A{}_B(\tau ,\vec \sigma )\, \sigma^B + V^A(\tau
,\vec \sigma )\Big],\nonumber \\
 &&{}\nonumber \\
 &&V^{\tau}(\tau ,\vec \sigma ) = \int^{\tau}_o d\tau_1\,
\Lambda^{\tau}{}_{\tau}(\tau_1 ,\vec \sigma ) -
\Lambda^{\tau}{}_{\tau}(\tau ,\vec \sigma )\, \tau,\nonumber \\
&&\nonumber\\
&&V^r(\tau ,\vec \sigma ) = \int^{\tau}_o d\tau_1\,
\Lambda^r{}_{\tau}(\tau_1, \vec \sigma ) - \Lambda^r{}_{\tau}(\tau
,\vec \sigma )\, \tau .
 \label{III7}
 \eea
\bigskip

Let us study the conditions imposed by Eqs.(\ref{I1})  on the
foliations of the type (\ref{III5}) (for the others it is similar)
to find which ones correspond to allowed 4-coordinate
transformations. We shall represent each Lorentz matrix $\Lambda$
as the product of a Lorentz boost $B$ and a rotation matrix ${\cal
R}$ {\it to separate the translational from the rotational
effects} ($\vec \beta = \vec v/c$ are the boost parameters,
$\gamma (\vec \beta ) = 1/\sqrt{1- {\vec \beta}^2}$, ${\vec
\beta}^2 = (\gamma^2 - 1)/\gamma^2$, $B^{-1}(\vec \beta ) =
B(-\vec \beta )$; $\alpha$, $\beta$, $\gamma$ are three Euler
angles and $R^{-1} = R^T$)

\bea
 &&\Lambda (\tau ,\vec \sigma ) = B(\vec \beta (\tau ,\vec \sigma
 ))\, {\cal R}(\alpha (\tau ,\vec \sigma ), \beta (\tau ,\vec
 \sigma ), \gamma (\tau ,\vec \sigma )),\nonumber \\
 &&{}\nonumber \\
 &&B^A{}_B(\vec \beta ) = \left( \begin{array}{cc}
\gamma (\vec \beta )& \gamma (\vec \beta )\, \beta^s\\ \gamma
(\vec \beta )\, \beta^r& \delta^{rs} + {{\gamma^2(\vec \beta )\,
\beta^r\, \beta^s}\over {\gamma (\vec \beta ) + 1}}
\end{array} \right),\qquad
 {\cal R}^A{}_B(\alpha ,\beta ,\gamma ) = \left(
 \begin{array}{cc} 1 & 0\\ 0& R^r{}_s(\alpha ,\beta ,\gamma )
 \end{array} \right),\nonumber \\
 &&{}\nonumber \\
 &&R(\alpha ,\beta ,\gamma ) =\label{III8}\\
&&\nonumber\\
 &&= \left( \begin{array}{ccc}
\cos \alpha \cos \beta \cos \gamma -\sin \alpha \sin \gamma  &
\sin \alpha \cos \beta \cos \gamma +\cos \alpha \sin \gamma  &
-\sin \beta \cos \gamma
\\ -\cos \alpha \cos \beta \sin \gamma -\sin \alpha \cos \gamma  &
-\sin \alpha \cos \beta \sin \gamma +\cos \alpha \cos \gamma  &
\sin \beta \sin \gamma \\ \cos \alpha \sin \beta  & \sin \alpha
\sin \beta  & \cos \beta \end{array} \right).\nonumber
 \eea

Then we get

\bea
 z^{\mu}_{\tau}(\tau ,\vec \sigma ) &=& U^{\mu}_{\tau}(\tau ,\vec
 \sigma ) + \partial_{\tau}\, U^{\mu}_A(\tau ,\vec \sigma )\,
 \sigma^A =\nonumber \\
&&\nonumber\\
 &=& U^{\mu}_{\tau}(\tau ,\vec \sigma ) + U^{\mu}_B(\tau ,\vec \sigma
 )\, \Omega^B{}_A(\tau ,\vec \sigma )\, \sigma^A,\nonumber \\
 &&{}\nonumber \\
 z^{\mu}_r(\tau ,\vec \sigma ) &=& U^{\mu}_r(\tau ,\vec \sigma ) +
 \partial_r\, U^{\mu}_A(\tau ,\vec \sigma )\, \sigma^A =\nonumber \\
&&\nonumber\\
 &=& U^{\mu}_r(\tau ,\vec \sigma ) + U^{\mu}_B(\tau ,\vec \sigma
 )\, \Omega^B_{(r) A}(\tau ,\vec \sigma )\, \sigma^A,\nonumber \\
 &&{}\nonumber \\
 l^{\mu}(\tau ,\vec \sigma ) &=&
 {1\over {\sqrt{|\det\, g_{rs}(\tau ,\vec \sigma )|}}}\,
 \epsilon^{\mu}{}_{\alpha\beta\gamma}\, [ z^{\alpha}_1\,
 z^{\beta}_2\, z^{\gamma}_3](\tau ,\vec \sigma ),\nonumber \\
 &&{}\nonumber \\
  &&\mbox{ (normal to the simultaneity surfaces\, )},
 \label{III9}
 \eea

\noindent where we have introduced the following matrices

\begin{eqnarray*}
 &&\Omega^A{}_B = (\Lambda^{-1}\, \partial_{\tau}\, \Lambda
 )^A{}_B = ( {\cal R}^{-1}\, \partial_{\tau}\, {\cal R} + {\cal
 R}^{-1}\, B^{-1}\, \partial_{\tau}\, B\, {\cal R})^A{}_B
 =( \Omega_{\cal R} + {\cal R}^{-1}\, \Omega_B\, {\cal
 R})^A{}_B,\nonumber \\
 &&{}\nonumber \\
 &&\qquad \Omega_{\cal R} = {\cal R}^{-1}\, \partial_{\tau}\, {\cal R} =
 \left( \begin{array}{cc} 0& 0\\ 0& \Omega_R =
 R^{-1}\, \partial_{\tau}\, R \end{array} \right),
\end{eqnarray*}

\begin{eqnarray*}
 &&\qquad \Omega_B = B^{-1}(\vec \beta )\, \partial_{\tau}\, B(\vec \beta
 ) = - \partial_{\tau}\, B(-\vec \beta )\, B^{-1}(-\vec \beta ) =
 \nonumber \\
&&\nonumber\\
 &&\qquad = \left( \begin{array}{cc} 0& \gamma\, (\partial_{\tau}\beta^s +
 {{\gamma^2\, \vec \beta \cdot \partial_{\tau}\vec \beta\,
 \beta^s}\over {\gamma + 1}})\\ \gamma\, (\partial_{\tau}\beta^r +
 {{\gamma^2\, \vec \beta \cdot \partial_{\tau}\vec \beta\,
 \beta^r}\over {\gamma + 1}})& - {{\gamma^2}\over {\gamma + 1}}\,
 (\beta^r\, \partial_{\tau}\beta^s - \partial_{\tau}\beta^r\,
 \beta^s) \end{array} \right),\nonumber \\
 &&{}\nonumber \\
 &&{}\nonumber \\
 && \qquad \Omega = \left( \begin{array}{cc} 0&
\gamma\, (\partial_{\tau}\beta^u +
 {{\gamma^2\, \vec \beta \cdot \partial_{\tau}\vec \beta\,
 \beta^u}\over {\gamma + 1}})\, R^u{}_s\\
 R^{T r}{}_u\, \gamma\, (\partial_{\tau}\beta^u +
 {{\gamma^2\, \vec \beta \cdot \partial_{\tau}\vec \beta\,
 \beta^u}\over {\gamma + 1}})& \Omega^r_{R\, s} - {{\gamma^2}\over
 {\gamma + 1}}\, R^{T\, r}{}_u\, (\beta^u\, \partial_{\tau}\beta^v
 - \partial_{\tau}\beta^u\, \beta^v)\, R^v{}_s \end{array} \right),
 \end{eqnarray*}

\bea
 &&\Omega^A_{(r) B} = (\Lambda^{-1}\, \partial_r\, \Lambda )^A{}_B
  = ( {\cal R}^{-1}\, \partial_r\, {\cal R} + {\cal
 R}^{-1}\, B^{-1}\, \partial_r\, B\, {\cal R})^A{}_B
 =( \Omega_{{\cal R}\, (r)} + {\cal R}^{-1}\, \Omega_{B\, (r)}\, {\cal
 R})^A{}_B = \nonumber \\
&&\nonumber\\
 &&= \left( \begin{array}{cc} 0&
\gamma\, (\partial_r\beta^u +
 {{\gamma^2\, \vec \beta \cdot \partial_r\vec \beta\,
 \beta^u}\over {\gamma + 1}})\, R^u{}_s\\
 R^{T w}{}_u\, \gamma\, (\partial_r\beta^u +
 {{\gamma^2\, \vec \beta \cdot \partial_r\vec \beta\,
 \beta^u}\over {\gamma + 1}})& \Omega^w_{R\, s} - {{\gamma^2}\over
 {\gamma + 1}}\, R^{T\, w}{}_u\, (\beta^u\, \partial_r\beta^v
 - \partial_r\beta^u\, \beta^v)\, R^v{}_s \end{array}\right),
 \label{III10}
 \eea
\medskip

\noindent assumed to vanish at spatial infinity, $\Omega^A{}_B
(\tau ,\vec \sigma ), \Omega^A_{(r)\, B}(\tau ,\vec \sigma )\,
{\rightarrow}_{|\vec \sigma | \rightarrow \infty}\, 0$. {\it The
matrix $\Omega_B$  describes the translational velocity ($\vec
\beta$) and acceleration ($\partial_{\tau}\vec \beta$), while the
matrix $\Omega_R$ the rotational angular velocity}.

The $z^{\mu}_A$'s and the associated 4-metric are

\bea
 z^{\mu}_{\tau}(\tau ,\vec \sigma ) &=& \Big([1 + \Omega^{\tau}{}_r\,
 \sigma^r]\, U^{\mu}_{\tau} + \Omega^r{}_A\, \sigma^A\, U^{\mu}_r
 \Big)(\tau ,\vec \sigma ),\nonumber \\
&&\nonumber\\ z^{\mu}_r(\tau ,\vec \sigma ) &=&
\Big(\Omega^{\tau}_{(r)\, s}\, \sigma^s\,
 U^{\mu}_{\tau} + [\delta_r^s + \Omega^s_{(r)\, A}\, \sigma^A]\,
 U^{\mu}_s\Big)(\tau ,\vec \sigma ),
\label{III11}
\eea

\noindent and

 \begin{eqnarray*}
 g_{\tau\tau}(\tau ,\vec \sigma ) &=& \Big(z^{\mu}_{\tau}\,
 \eta_{\mu\nu}\, z^{\nu}_{\tau}\Big)(\tau ,\vec \sigma
 )= \sgn\, \Big([ 1+ \Omega^{\tau}{}_r\, \sigma^r]^2 -
  \sum_r\, [\Omega^r{}_A\, \sigma^A]^2\Big)(\tau ,\vec \sigma ),\nonumber \\
&&\nonumber\\
 g_{r\tau}(\tau ,\vec \sigma ) &=& \Big( z^{\mu}_r\,
 \eta_{\mu\nu}\, z^{\nu}_{\tau}\Big)(\tau ,\vec \sigma )
 = \sgn\, \Big(\Omega^{\tau}_{(r)\, s}\, \sigma^s\, [1 + \Omega^{\tau}{}_u\, \sigma^u]
 -\nonumber \\
&&\nonumber\\
 &-& \sum_s\, \Omega^s{}_A\, \sigma^A\, [\delta^s_r + \Omega^s_{(r)\, A}\, \sigma^A]
 \Big)(\tau ,\vec \sigma ),
 \end{eqnarray*}

\bea
 g_{rs}(\tau ,\vec \sigma ) &=& \Big( z^{\mu}_r\, \eta_{\mu\nu}\,
 z^{\nu}_s\Big)(\tau ,\vec \sigma ) =
  \sgn\, \Big(- \delta_{rs} - [\Omega^r_{(s)\, A} + \Omega^s_{(r)\, A}]\,
 \sigma^A +\nonumber \\
&&\nonumber\\
 &+& \Omega^{\tau}_{(r)\, u}\, \Omega^{\tau}_{(s)\, v}\, \sigma^u\, \sigma^v -
 \sum_u\, \Omega^u_{(r)\, A}\, \Omega^u_{(s)\, A}\, \sigma^A\, \sigma^B
 \Big)(\tau ,\vec \sigma ).
 \label{III12}
 \eea

\medskip

Eqs.(\ref{I1}) are complicated restrictions on the parameters
$\vec \beta (\tau ,\vec \sigma )$, $\alpha (\tau ,\vec \sigma )$,
$\beta (\tau ,\vec \sigma )$, $\gamma (\tau ,\vec \sigma )$ of the
Lorentz transformations, which say that {\it translational
accelerations and rotational frequencies are not independent} but
must {\it balance each other} if Eqs.(\ref{III5}) describe the
inverse of an allowed 4-coordinate transformation.
\medskip

Let us consider two extreme cases.

\bigskip

A)  {\it Rigid non-inertial reference frames with translational
acceleration exist}. An example are the following embeddings,
which are compatible with the locality hypothesis only for $f(\tau
) = \tau$ (this corresponds to $\Lambda = B(\vec 0)\, {\cal
R}(0,0,0)$, i.e. to an inertial reference frame)

\medskip

\bea
 z^{\mu}(\tau ,\vec \sigma ) &=& x^{\mu}_o +
\epsilon^{\mu}_{\tau}\, f(\tau ) + \epsilon^{\mu}_r\,
\sigma^r,\nonumber \\
 &&{}\nonumber \\
 &&g_{\tau\tau}(\tau ,\vec \sigma ) = \sgn\,
 \Big({{d f(\tau )}\over {d\tau}}\Big)^2,\quad g_{\tau r}(\tau ,\vec \sigma )
 =0,\quad g_{rs}(\tau ,\vec \sigma ) = -\sgn\, \delta_{rs}.
 \label{III13}
 \eea

\medskip

This is a foliation with parallel hyper-planes with respect to a
centroid $x^{\mu}(\tau ) = x^{\mu}_o + \epsilon^{\mu}_{\tau}\,
f(\tau )$ (origin of 3-coordinates). The hyper-planes have
translational acceleration ${\ddot x}^{\mu}(\tau ) =
\epsilon^{\mu}_{\tau}\, \ddot f(\tau )$, so that they are not
uniformly distributed like in the inertial case $f(\tau ) = \tau$.

\bigskip

B) On the other hand  {\it rigid rotating reference frames do not
exist}. Let us consider the embedding (compatible with the
locality hypothesis) with $\Lambda = B(\vec 0)\, {\cal R}(\alpha
(\tau ),\beta (\tau ),\gamma (\tau ))$ and $x^{\mu}(\tau ) =
x^{\mu}_o + \epsilon^{\mu}_{\tau}\, \tau$
\medskip

\begin{eqnarray*}
 z^{\mu}(\tau ,\vec \sigma ) &=& x^{\mu}(\tau ) + \epsilon^{\mu}_r\,
R^r{}_s(\tau )\, \sigma^s,\nonumber \\
 &&{}\nonumber \\
 z^{\mu}_{\tau}(\tau ,\vec \sigma ) &=& {\dot x}^{\mu}(\tau ) +
 \epsilon^{\mu}_r\, {\dot R}^r{}_s(\tau )\, \sigma^s,\qquad
 z^{\mu}_r(\tau ) = \epsilon^{\mu}_s\, R^s{}_r(\tau ),\nonumber \\
 &&{}\nonumber \\
 g_{\tau\tau}(\tau ,\vec \sigma ) &=& \sgn\,
 \Big({\dot x}^2(\tau ) + 2\, {\dot x}_{\mu}(\tau )\,
 \epsilon^{\mu}_r\, {\dot R}^r{}_s(\tau )\, \sigma^s -
  \sgn \, {\dot R}^r{}_u(\tau )\, {\dot R}^r{}_v(\tau )\,
 \sigma^u\, \sigma^v\Big),
 \end{eqnarray*}

\bea
 g_{\tau r}(\tau ,\vec \sigma ) &=& \sgn\, \Big({\dot x}_{\mu}(\tau )\,
 \epsilon^{\mu}_s\, R^s{}_r(\tau ) - \epsilon\, {\dot R}^v{}_u(\tau )\,
 R^v{}_r(\tau )\, \sigma^u \Big),\nonumber\\
&&\nonumber\\
 g_{rs}(\tau ,\vec \sigma )
 &=& - \sgn\, R^u{}_r(\tau )\, R^u{}_s(\tau ),
 \label{III14}
 \eea
\medskip

\noindent which corresponds to a foliation with parallel
space-like hyper-planes with normal $l^{\mu} =
\epsilon^{\mu}_{\tau}$. It can be verified that it is not the
inverse of an allowed 4-coordinate transformation, because the
associated $g_{\tau\tau}(\tau ,\vec \sigma )$ has a zero at
\footnote{We use the notations $\vec \sigma = \sigma\, \hat
\sigma$, $\sigma = |\vec \sigma |$, $\vec \Omega = \Omega\, \hat
\Omega$, ${\hat \sigma}^2 = {\hat \Omega}^2 = 1$, $\Omega^u = -
{1\over 2}\, \epsilon^{urs}\, (\dot R\, R^{-1})^r{}_s$,
$b^{\mu}_r(\tau ) = \epsilon^{\mu}_s\, R_r{}^s(\tau )$.}
\medskip

\beq
 \sigma = \sigma_R = {1\over {\Omega (\tau )}}\, \Big[- {\dot x}
_{\mu}(\tau )\, b^{\mu}_r(\tau )\, (\hat \sigma \times \hat
\Omega(\tau ))^r + \sqrt{{\dot x}^2(\tau ) + [{\dot x}_{\mu}(\tau
)\, b^{\mu}_r(\tau )\, (\hat \sigma \times \hat \Omega (\tau
))^2}]^2 \,\, \Big],
 \label{III15}
 \eeq

\noindent with $\quad \sigma_R \rightarrow \infty$ for $\Omega
\rightarrow 0$. At $\sigma = \sigma_R$ the time-like vector
$z^{\mu}_{\tau}(\tau ,\vec \sigma )$ becomes light-like (the {\it
horizon problem}), while for an admissible foliation with
space-like leaves it must always remain time-like.

\medskip

This pathology (the so-called horizon problem) is common to most
of the rotating coordinate systems quoted in Subsection D of the
Introduction. Let us remark that an analogous pathology happens on
the event horizon of the Schwarzschild black hole. Also in this
case we have a coordinate singularity where the time-like Killing
vector of the static space-time becomes light-like. For the
rotating Kerr black hole the same coordinate singularity happens
already at the boundary of the ergosphere \cite{94a}. Also in the
existing theory of rotating relativistic stars \cite{95a}, where
differential rotations are replacing the rigid ones in model
building, it is assumed that in certain rotation regimes an
ergosphere may form \cite{96a}: again, if one uses 4-coordinates
adapted to the Killing vectors, one gets a similar coordinate
singularity.

\bigskip

In the next Section we shall consider the minimal modification of
Eq.(\ref{III14}) so to obtain the inverse of an allowed
4-coordinate transformation.

\vfill\eject

\section{The Simplest Notion of Simultaneity when Rotations are
Present.}

Let us look for the simplest embedding $x^{\mu} = z^{\mu}(\tau
,\vec \sigma )$, inverse of an admissible 4-coordinate
transformation $x^{\mu} \mapsto \sigma^A$ compatible with the
locality hypothesis, which contains a rotating reference frame,
with also translational acceleration, of the type of
Eq.(\ref{III12}). The minimal modification of Eq.(\ref{III12}) is
to replace the rotation matrix $R(\tau )$ with $R(\tau ,|\vec
\sigma|)$, namely the rotation varies as a function of some radial
distance $|\vec \sigma |$ ({\it differential rotation}) from the
arbitrary time-like world-line $x^{\mu}(\tau )$, origin of the
3-coordinates on the simultaneity surfaces. Since the
3-coordinates $\sigma^r$ are Lorentz scalar we shall use the
radial distance $\sigma = |\vec \sigma| = \sqrt{\delta_{rs}\,
\sigma^r\, \sigma^s }$, so that $\sigma^r = \sigma \, {\hat
\sigma}^r$ with $\delta_{rs}\, {\hat \sigma}^r\, {\hat \sigma}^s =
1$. Therefore let us replace Eq.(\ref{III12}) with the following
embedding

\medskip

\bea
 &&z^{\mu}(\tau ,\vec \sigma ) = x^{\mu}(\tau ) + \epsilon^{\mu}_r\,
R^r{}_s(\tau , \sigma )\, \sigma^s\,
 {\buildrel {def}\over =}\, x^{\mu}(\tau ) + b^{\mu}_r(\tau
 ,\sigma )\, \sigma^r,\nonumber \\
 &&{}\nonumber \\
 &&R^r{}_s(\tau ,\sigma ) {\rightarrow}_{\sigma \rightarrow
 \infty} \delta^r_s,\qquad \partial_A\, R^r{}_s(\tau
 ,\sigma )\, {\rightarrow}_{\sigma \rightarrow
 \infty}\, 0,\nonumber \\
 &&{}\nonumber \\
 &&b^{\mu}_s(\tau ,\sigma ) = \epsilon^{\mu}_r\, R^r{}_s(\tau
 ,\sigma )\, {\rightarrow}_{\sigma \rightarrow
 \infty}\, \epsilon^{\mu}_s,\quad [b^{\mu}_r\, \eta_{\mu\nu}\, b^{\nu}_s](\tau ,\sigma )
 = - \sgn\, \delta_{rs}.
 \label{IV1}
 \eea
\medskip

Since $z^{\mu}_r(\tau ,\vec \sigma ) = \epsilon^{\mu}_s\,
\partial_r\, [R^s{}_u(\tau ,\sigma )\, \sigma^u]$, it follows that
the normal to the simultaneity surfaces is $l^{\mu} =
\epsilon^{\mu}_{\tau}$, namely the hyper-surfaces are {\it
parallel space-like hyper-planes}. These hyper-planes have
translational acceleration ${\ddot x}^{\mu}(\tau )$ and a rotating
3-coordinate system with rotational frequency

\begin{eqnarray*}
 \Omega^r(\tau ,\sigma ) &=& - {1\over 2}\, \epsilon^{ruv}\, \Big[
 R^{-1}(\tau ,\sigma )\, {{\partial R(\tau ,\sigma )}\over
 {\partial \tau}}\Big]^{uv}\, {\rightarrow}_{\sigma \rightarrow
 \infty}\, 0,\nonumber \\
 &&{}\nonumber \\
 &&\Downarrow \nonumber \\
 &&{}\nonumber \\
 {{\partial b^{\mu}_s(\tau ,\sigma )}\over {\partial \tau}} &=&
 \epsilon^{\mu}_r\, {{\partial R^r{}_s(\tau ,\sigma )}\over
 {\partial \tau}} = - \epsilon^{suv}\, \Omega^u(\tau ,\sigma )\,
 b^{\mu}_v(\tau ,\sigma ),
 \end{eqnarray*}

\bea
 \Omega^1(\tau ,\sigma ) &=&\Big[ \partial_{\tau} \beta\, \sin\,
 \gamma - \partial_{\tau} \alpha\, \sin\, \beta\, \cos\,
 \gamma\Big](\tau ,\sigma ),\nonumber \\
&&\nonumber\\
 \Omega^2(\tau ,\sigma ) &=& \Big[\partial_{\tau} \beta\, \cos\,
 \gamma + \partial_{\tau} \alpha\, \sin\, \beta\, \sin\,
 \gamma\Big](\tau ,\sigma ),\nonumber \\
&&\nonumber\\
 \Omega^3(\tau ,\sigma ) &=& \Big[ \partial_{\tau} \gamma +
 \partial_{\tau} \alpha\, \cos\, \beta \Big](\tau ,\sigma ).
  \label{IV2}
  \eea

\noindent In the last three lines we used Eqs.(\ref{III8}) to find
the angular velocities. Moreover we can define

 \bea
&&\Omega_{(r)}(\tau ,\sigma ) = \Big[ R^{-1}\, \partial_r\,
 R\Big](\tau ,\sigma ) = 2{\hat \sigma}^r\, \Big[R^{-1}\,
 {{\partial R}\over {\partial \sigma }}\Big](\tau ,\sigma )\,
 {\rightarrow}_{\sigma \rightarrow \infty}\, 0,\nonumber \\
&&\nonumber\\
 &&\Omega^u_{(r) v}(\tau ,\sigma )\, \sigma^v = \Phi_{uv}(\tau
 ,\sigma )\, {{\sigma^r\, \sigma^v}\over {\sigma}},\qquad
 \Phi_{uv} = - \Phi_{vu},
\label{IV3}
\eea

\medskip

As a consequence we have

\bea
 {\dot x}^{\mu}(\tau ) &=& \sgn\, \Big([{\dot x}_{\nu}(\tau
 )\, l^{\nu}]\, l^{\mu} - \sum_r\, [{\dot x}_{\nu}(\tau )\,
 \epsilon^{\nu}_r]\, \epsilon^{\mu}_r \Big),\nonumber \\
 &&{}\nonumber \\
 z^{\mu}_{\tau}(\tau ,\vec \sigma ) &=& N(\tau ,\vec \sigma )\,
 l^{\mu} + N^r(\tau ,\vec \sigma )\, z^{\mu}_r(\tau ,\vec \sigma )
 =\nonumber \\
&&\nonumber\\
 &=& {\dot x}^{\mu}(\tau ) - \epsilon^{suv}\, \Omega^u(\tau
 ,\sigma )\, b^{\mu}_v(\tau ,\sigma )\, \sigma^s =\nonumber \\
&&\nonumber\\
 &=& {\dot x}^{\mu}(\tau ) - (\vec \sigma \times \vec \Omega (\tau ,\sigma ))^r\,
 b^{\mu}_r(\tau ,\sigma )\, {\rightarrow}_{\sigma \rightarrow
 \infty}\, {\dot x}^{\mu}(\tau ),\nonumber \\
&&{}\nonumber \\
 z^{\mu}_r(\tau ,\vec \sigma ) &=& \epsilon^{\mu}_s\, \Big[
 R^s{}_r(\tau ,\sigma ) + \partial_r\, R^s{}_u(\tau ,\sigma )\,
 \sigma^u\Big] =\nonumber \\
&&\nonumber\\
 &=& b^{\mu}_s(\tau ,\sigma )\, \Big[ \delta^s_r + \Omega^s_{(r)
 u}(\tau ,\sigma )\, \sigma^u\Big]\, {\rightarrow}_{\sigma \rightarrow
 \infty}\, \epsilon^{\mu}_r,
\label{IV4}
\eea

\noindent and then we obtain

 \bea
 g_{\tau\tau}(\tau ,\vec \sigma ) &=& {\dot x}^2(\tau ) - 2\,
 {\dot x}_{\mu}(\tau )\, b^{\mu}_r(\tau ,\sigma )\, (\vec \sigma
 \times \vec \Omega (\tau ,\sigma ))^r - \sgn\,
(\vec{\sigma}\times\vec{\Omega})^2
=\nonumber \\
&&\nonumber\\
 &=& \Big[ N^2 - g_{rs}\, N^r\, N^s\Big](\tau ,\vec
 \sigma ),\nonumber \\
&&\nonumber\\
 g_{\tau r}(\tau ,\vec \sigma ) &=& \Big[ g_{rs}\, N^s\Big](\tau ,\vec \sigma )
 = \nonumber\\
&&\nonumber\\
&=&{\dot x}_{\mu}(\tau )\,
 b^{\mu}_r(\tau ,\sigma )\, \Big[ \delta^v_r + \Omega^v_{(r)
 u}(\tau ,\sigma )\, \sigma ^u\Big] + \sgn\,
 [\vec \sigma \times \vec \Omega (\tau ,\sigma
 )]^s\, \Big[ \delta^s_r + \Omega^s_{(r) u}(\tau ,\sigma )\,
 \sigma^u\Big],\nonumber \\
&&\nonumber\\
 - \sgn\,g_{rs}(\tau ,\vec \sigma ) &=&  \delta_{rs} +
 \Big(\Omega^r_{(s) u}(\tau ,\sigma ) + \Omega^s_{(r) u}(\tau
 ,\sigma )\Big)\, \sigma^u
 + \sum_w\, \Omega^w_{(r) u}(\tau ,\sigma )\, \Omega^w_{(s)
 v}(\tau ,\sigma )\, \sigma^u\, \sigma^v.\nonumber \\
 &&{}
 \label{IV5}
 \eea

 The requirement that $g_{\tau\tau}(\tau ,\vec \sigma )$ and
 $g_{\tau r}(\tau ,\vec \sigma )$ tend to finite limits at spatial
 infinity puts the restrictions

 \begin{eqnarray*}
 | \vec \Omega (\tau ,\sigma )|,\,&& |\Omega^u_{(r) v}(\tau
 ,\sigma )|\,\, {\rightarrow}_{\sigma \rightarrow \infty}\,
 O(\sigma^{-(1+\eta )}),\quad \eta > 0,\nonumber \\
 &&{}\nonumber \\
 &&\Downarrow \nonumber \\
 &&{}\nonumber \\
 \partial_A\, R^r{}_s(\tau ,\sigma ) &{\rightarrow}_{\sigma \rightarrow
 \infty}& O(\sigma^{-(1+\eta )}),\, \Rightarrow\, R^r{}_s(\tau
 ,\sigma )\, {\rightarrow}_{\sigma \rightarrow \infty}\,
 O(\sigma^{-(1+\eta )}),\nonumber \\
 &&{}\nonumber \\
 z^{\mu}_{\tau}(\tau ,\vec \sigma ) &{\rightarrow}_{\sigma \rightarrow
 \infty}& {\dot x}^{\mu}(\tau ) + O(\sigma^{-\eta}),\nonumber \\
&&\nonumber\\
 b^{\mu}_r(\tau ,\sigma ) &{\rightarrow}_{\sigma \rightarrow
 \infty}& \epsilon^{\mu}_r + O(\sigma^{-(1+\eta )}),\qquad
 z^{\mu}_r(\tau ,\vec \sigma )\, {\rightarrow}_{\sigma \rightarrow \infty}\,
 \epsilon^{\mu}_r + O(\sigma^{-(1+\eta )}),\nonumber \\
&&\nonumber\\
 N^r(\tau ,\vec \sigma )\, z^{\mu}_r(\tau ,\vec \sigma )\,
 &{\rightarrow}_{\sigma \rightarrow \infty}&\, - \sgn\, {\dot
 x}_{\nu}(\tau )\, \epsilon^{\nu}_r\, \epsilon^{\mu}_r +
 O(\sigma^{-(1+2\eta )}),\nonumber \\
&&\nonumber\\
 N^r(\tau ,\vec \sigma ) &{\rightarrow}_{\sigma \rightarrow
 \infty}& -\sgn\, \delta^{rs}\, {\dot x}_{\nu}(\tau )\,
 \epsilon^{\nu}_s + O(\sigma^{-\eta}),
 \end{eqnarray*}

\bea
 N(\tau ,\vec \sigma )&& l^{\mu} = [z^{\mu}_{\tau} - N^r\,
 z^{\mu}_r](\tau ,\vec \sigma )\, {\rightarrow}_{\sigma
 \rightarrow \infty}\, \sgn\, [{\dot x}_{\nu}(\tau )\,
 l^{\nu}]\, l^{\mu} + O(\sigma^{-\eta}),\nonumber \\
 &&{}\nonumber \\
 g_{\tau\tau}(\tau ,\vec \sigma ) &{\rightarrow}_{\sigma \rightarrow
 \infty}& {\dot x}^2(\tau ) + O(\sigma^{-2\eta}),\nonumber \\
&&\nonumber\\
 g_{\tau r}(\tau ,\vec \sigma ) &{\rightarrow}_{\sigma \rightarrow
 \infty}& {\dot x}_{\mu}(\tau )\, \epsilon^{\mu}_r +
 O(\sigma^{-\eta}),\nonumber \\
&&\nonumber\\
 g_{rs}(\tau ,\vec \sigma ) &{\rightarrow}_{\sigma \rightarrow
 \infty}& - \sgn\, \delta_{rs} + O(\sigma^{-\eta}).
  \label{IV6}
  \eea

\bigskip

Let us look for a family of rotation matrices $R^r{}_s(\tau
,\sigma )$ satisfying the condition $\sgn\, g_{\tau\tau}(\tau
,\vec \sigma ) > 0$ of Eqs.(\ref{I1}).
\medskip

Let us make the ansatz that the Euler angles of $R(\alpha ,\beta
,\gamma )$ have the following factorized dependence on $\tau$ and
$\sigma$

\beq
 \alpha (\tau ,\sigma) =F(\sigma )\, \tilde \alpha (\tau
 ),\qquad
 \beta (\tau ,\sigma ) = F(\sigma )\, \tilde \beta (\tau
 ),\qquad
 \gamma (\tau ,\sigma )=F(\sigma )\, \tilde \gamma (\tau
 ),
 \label{IV7}
 \eeq

\noindent with

 \beq
  F(\sigma ) > 0,\quad {{d F(\sigma )}\over {d
\sigma }}\not= 0,\qquad F(\sigma )\, {\rightarrow}_{\sigma
\rightarrow \infty}\,
 O(\sigma^{-(1+\eta )}).
 \label{IV8}
 \eeq

We get

 \bea
  \Omega^1(\tau ,\sigma ) &=& F(\sigma )\, \Big( {\dot {\tilde
 \beta}}(\tau )\, \sin\, [F(\sigma )\, \tilde \gamma (\tau )] -
 {\dot {\tilde \alpha}}(\tau )\, \sin\, [F(\sigma )\, \tilde \beta
 (\tau )]\, \cos\, [F(\sigma )\, \tilde \gamma (\tau
 )]\Big),\nonumber \\
&&\nonumber\\
 \Omega^2(\tau ,\sigma ) &=& F(\sigma )\, \Big( {\dot {\tilde
 \beta}}(\tau )\, \cos\, [F(\sigma )\, \tilde \gamma (\tau )] +
 {\dot {\tilde \alpha}}(\tau )\, \sin\, [F(\sigma )\, \tilde \beta
 (\tau )]\, \sin\, [F(\sigma )\, \tilde \gamma (\tau
 )]\Big),\nonumber \\
&&\nonumber\\
 \Omega^3(\tau ,\sigma ) &=& F(\sigma )\, \Big( {\dot {\tilde
 \gamma}}(\tau ) + {\dot {\tilde \alpha}}(\tau )\, \cos\, [F(\sigma
 )\, \tilde \beta (\tau )]\Big),\nonumber \\
 &&{}\nonumber \\
 &&\Downarrow \nonumber \\
 &&{}\nonumber \\
 \Omega^r(\tau ,\sigma ) &=& F(\sigma )\, \tilde \Omega (\tau
 ,\sigma )\, {\hat n}^r(\tau ,\sigma ),\qquad {\hat n}^2(\tau
 ,\sigma ) = 1,\nonumber \\
 &&{}\nonumber \\
 0 &<& \tilde \Omega (\tau ,\sigma )\, \leq\, 2\, \max\, \Big( {\dot
 {\tilde \alpha}}(\tau ), {\dot {\tilde \beta}}(\tau ), {\dot
 {\tilde \gamma}}(\tau ) \Big) = 2\, M_1.
 \label{IV9}
 \eea
\medskip

Since $l^{\mu} = \epsilon^{\mu}_{\tau}\, {\buildrel {def}\over
=}\, b^{\mu}_{\tau}$ and $b^{\mu}_r(\tau ,\sigma )$ form an
orthonormal tetrad [$b^{\mu}_A(\tau ,\sigma )\, \eta_{\mu\nu}\,
b^{\nu}_B(\tau ,\sigma ) = \eta_{AB}$], let us decompose the
future time-like 4-velocity ${\dot x}^{\mu}(\tau )$ on it
($v_l(\tau )$ is the asymptotic lapse function)

\bea
 {\dot x}^{\mu}(\tau ) &=& v_l(\tau )\, l^{\mu} -
 \sum_r\, v_r(\tau ,\sigma )\, b^{\mu}_r(\tau ,\sigma
 )\nonumber \\
 &&{}\nonumber \\
 v_l(\tau ) &=& \sgn\, {\dot x}_{\mu}(\tau )\, l^{\mu} > 0,\qquad
 v_r(\tau ,\sigma ) = \sgn\, {\dot x}_{\mu}(\tau )\, b^{\mu}_r(\tau
 ,\sigma ),\nonumber \\
 &&{}\nonumber \\
 \sgn\, {\dot x}^2(\tau ) &=& v^2_l(\tau ) - \sum_r\,
 v^2_r(\tau ,\sigma ) > 0,\, \Rightarrow \sum_r\, v^2_r(\tau
 ,\sigma ) = {\vec v}^2(\tau ,\sigma ) \equiv {\vec v}^2(\tau ) <
 v^2_l(\tau ),
\label{IV10}
\eea

We add the condition

 \beq
  |\vec v(\tau )| \leq {{v_l(\tau )}\over
K},\qquad K > 1.
 \label{IV11}
 \eeq

This condition is slightly stronger than the last of
Eqs.(\ref{IV10}), which  does not exclude the possibility that the
observer in $\vec{\sigma}=0$ has a time-like 4-velocity ${\dot
x}^{\mu}(\tau )$ which, however, becomes light-like at $\tau = \pm
\infty$ \footnote{This is the case of a (non-time-like)  Rindler
observer with uniform 4-acceleration, see Ref.\cite{97a}}. The
condition (\ref{IV11}) excludes this possibility. In other words
the condition (\ref{IV11}) tell us that the observer is without
event-horizon, namely he can explore all the Minkowski space-time
by light-signal.

\medskip

Then the condition $\sgn\, g_{\tau\tau}(\tau ,\vec \sigma ) > 0$
becomes

\bea
 && \sgn\, g_{\tau\tau}(\tau ,\vec \sigma ) =\nonumber\\
&&\nonumber\\ &=& \sgn\, {\dot
 x}^2(\tau ) - 2\,  \sigma\, F(\sigma )\, \tilde \Omega
 (\tau ,\sigma )\, \sum_r\, v_r(\tau ,\sigma )\, \Big[\hat \sigma
 \times \hat n(\tau ,\sigma )\Big]^r
-\sigma^2\, {\tilde \Omega}^2(\tau ,\sigma )\, F^2(\sigma )\,
 \Big[\hat \sigma \times \hat n(\tau ,\sigma )\Big]^2 =\nonumber\\
&&\nonumber\\
 &=& c^2(\tau ) - 2\, b(\tau ,\vec \sigma )\, X(\tau ,\sigma ) -
 a^2(\tau ,\vec \sigma )\, X^2(\tau ,\sigma ) > 0
\label{IV12}
\eea

\noindent where we have defined

 \bea
 &&c^2(\tau ) =  \sgn\, {\dot x}^2(\tau ) = v^2_l(\tau ) -
 {\vec v}^2(\tau ) > 0,\qquad c^2(\tau ) \geq {{K^2-1}\over
 {K^2}}\, v^2_l(\tau ),\nonumber \\
 &&{}\nonumber \\
 &&b(\tau ,\vec \sigma ) = \sum_r\, v_r(\tau ,\sigma
 )\, \Big[ \hat \sigma \times \hat n(\tau ,\sigma
 )\Big]^r,\nonumber \\
&&\nonumber\\
  && |b(\tau ,\vec \sigma )| \leq |\vec v(\tau )| < v_l(\tau
  ),\,\mbox{ or }\, |b(\tau ,\vec \sigma )| \leq {{v_l(\tau )}\over
  K},\,\,  K > 1,\nonumber \\
  &&{}\nonumber \\
  &&a^2(\tau ,\vec \sigma ) =\Big[\hat \sigma \times \hat n(\tau
  ,\sigma )\Big]^2 > 0,\quad a^2(\tau ,\vec \sigma ) \leq 1,\quad
  b^2(\tau ,\vec \sigma ) + a^2(\tau ,\sigma )\, c^2(\tau ) >
  0,\nonumber \\
  &&{}\nonumber \\
  &&X(\tau ,\sigma ) = \sigma\, F(\sigma )\, \tilde \Omega (\tau
  ,\sigma ).
 \label{IV13}
 \eea

\medskip

The study of the equation $a^2\, X^2 + 2\, b\, X - c^2 = A^2\, (X
- X_{+})\, (X - X_{-}) = 0$, with solutions $X_{\pm} = {1\over
{a^2}}\, (- b \pm \sqrt{b^2 + a^2\, c^2})$, shows that $\sgn\,
g_{\tau\tau} > 0$ implies $X_{-} < X < X_{+}$; being $X_-<0$ and
$X>0$ [see Eq.(\ref{IV9})], we have that a half of the conditions
($X_-<X$) is always satisfied. We have only to discuss the
condition $X<X_+$.
\medskip

Since $- v_{l}/K \leq b \leq v_l/K$, when $b$ increases in this
interval $X_{+}$ decrease with $b$. This implies
\[
X_{+} > {1\over {a^2}}\, \Big(- {{v_l}\over K} +
\sqrt{{{v^2_l}\over {K^2}} + a^2\, c^2 }\Big),
\]
so that $c^2 \geq {{K^2 - 1}\over {K^2}}\, v^2_l$ implies that we
will have $g_{\tau\tau} > 0$ if $0< X < {{v_l}\over {K\, a^2}}\,
(\sqrt{1 + (K^2 - 1)\, a^2} - 1)$, namely if the function
$F(\sigma )$ satisfies the condition

\[
 | F(\sigma )| <
{{v_l(\tau )}\over {K\, \sigma\, a^2(\tau ,\vec \sigma )\,
 \tilde \Omega (\tau ,\sigma )}}\,
\Big(\sqrt{1 + (K^2 - 1)\, a^2(\tau ,\vec \sigma )} - 1\Big)=
{{v_l(\tau )}\over {K}\,\tilde \Omega (\tau ,\sigma )}\,g(a^2).
\]

\medskip
Since $a^2\leq 1$ and $g(x)=(1/x)(\sqrt{1+(K^2-1)x}-1)$ is
decreasing for $x$ increasing in the interval $0<x<1$ ($K>1$), we
get $g(a^2)>g(1)=K-1$ and the stronger condition
\[
| F(\sigma )| < {{v_l(\tau )}\over {K\,\tilde \Omega (\tau ,\sigma
)}}\,(K-1).
\]

\medskip

The condition (\ref{IV9}) on the Euler angles and the fact that
 Eq.(\ref{IV11}) implies $\min v_l(\tau )=m>0$  lead to the
 following final condition on $F(\sigma)$

\bea
 &&0< F(\sigma ) <
 {m\over {2\, K\, M_1\, \sigma}}\,(K-1)=\frac{1}{M\,\sigma},\qquad
 {{d F(\sigma )}\over {d \sigma }} \not= 0,\nonumber \\
 &&{}\nonumber \\
 \mbox{ or }\qquad&&| \partial_{\tau} \alpha (\tau ,\sigma )|,
  | \partial_{\tau} \beta (\tau ,\sigma )|,
   | \partial_{\tau} \gamma (\tau ,\sigma )| <
   {{m}\over {2\, K\, \sigma}}\,(K-1),\nonumber \\
   &&{}\nonumber \\
 \mbox{ or }\qquad&&| \Omega^r(\tau ,\sigma ) | < {{m}\over { K\,
   \sigma}}\,(K-1).
 \label{IV14}
 \eea
\bigskip

This means that, while the linear velocities ${\dot x}^{\mu}(\tau
)$ and the translational accelerations ${\ddot x}^{\mu}(\tau )$
are arbitrary, the allowed rotations $R(\alpha ,\beta ,\gamma )$
on the leaves of the foliation {\it have the rotational
frequencies}, namely the angular velocities $\Omega^r(\tau ,\sigma
)$, {\it limited by an upper bound proportional to the minimum of
the linear velocity $v_l(\tau ) = {\dot x}_{\mu}(\tau )\, l^{\mu}$
orthogonal to the parallel hyper-planes}.

\bigskip

Instead of checking the conditions (\ref{I1}) on $g_{rs}(\tau
,\vec \sigma )$, let us write

\bea
 z^{\mu}(\tau ,\vec \sigma ) &=& \xi_l(\tau ,\vec
 \sigma )\, l^{\mu} - \sum_r\, \xi_r(\tau ,\vec \sigma )\,
 \epsilon^{\mu}_r,\nonumber \\
 &&{}\nonumber \\
 \xi_l(\tau ,\vec \sigma ) &=& \sgn\, z_{\mu}(\tau ,\vec \sigma )\,
 l^{\mu} = \sgn\, x_{\mu}(\tau )\, l^{\mu} = x_l(\tau ),\nonumber \\
&&\nonumber\\
 \xi_r(\tau ,\vec \sigma ) &=& \sgn\, z_{\mu}(\tau ,\vec \sigma )\,
 \epsilon^{\mu}_r = \sgn\, x_{\mu}(\tau )\, \epsilon^{\mu}_r +
 R^r{}_s(\tau ,\sigma )\, \sigma^s = x_{\epsilon\, r}(\tau ) + R^r{}_s(\tau
 ,\sigma )\, \sigma^s,
\label{IV15}
\eea

\noindent so that we get

 \bea
 \partial_{\tau}\, \xi_l(\tau ,\vec \sigma ) &=& {\dot x}_l(\tau )
 = v_l(\tau ),\qquad \partial_r\, \xi_l(\tau ,\vec \sigma ) = 0,\nonumber \\
 &&{}\nonumber \\
 \partial_u\, \xi_r(\tau ,\vec \sigma ) &=& R^r{}_u(\tau ,\sigma )
 + \partial_uR^r{}_s(\tau ,\sigma )\, \sigma^s =\nonumber\\
&&\nonumber\\
&=& R^r{}_v(\tau ,\sigma
 )\, \Big[ \delta^v_u + \omega^v_{(u) w}(\tau ,\sigma )\,
 \sigma^w\Big] =\nonumber \\
\nonumber\\
&=& R^r{}_v(\tau ,\sigma )\, \Big[ \delta^v_u +
 \Phi_{uv}(\tau ,\sigma )\, {{\sigma^u\, \sigma^w}\over
 {\sigma}}\Big] \, {\buildrel {def}\over =}\, \Big(R(\tau
 ,\sigma )\, M(\tau ,\vec \sigma )\Big)_{ru},
 \label{IV16}
 \eea
\medskip

\noindent and let us show that $\sigma^A = (\tau ,\vec \sigma )
\mapsto (\xi_l(\tau ,\vec \sigma ), \xi_r(\tau ,\vec \sigma )\, )$
is a coordinate transformation with positive Jacobian. This will
ensure that these foliations with parallel hyper-planes are
defined by embeddings such that $\sigma^A \mapsto x^{\mu} =
z^{\mu}(\tau ,\vec \sigma )$ is the inverse of an admissible
4-coordinate transformation $x^{\mu} \mapsto \sigma^A$.
\medskip

Therefore we have to study the Jacobian

\bea
 J(\tau ,\vec \sigma ) &=& \left( \begin{array}{cc} {{\partial\,
 \xi_l(\tau ,\vec \sigma )}\over {\partial \tau}} & {{\partial\,
 \xi_s(\tau ,\vec \sigma )}\over {\partial \tau}}\\
 {{\partial\, \xi_l(\tau ,\vec \sigma )}\over {\partial \sigma^r}}
 & {{\partial\, \xi_s(\tau ,\vec \sigma )}\over {\partial \sigma^r}}
\end{array} \right) = \left( \begin{array}{cc} v_l(\tau )& {{\partial\,
 \xi_s(\tau ,\vec \sigma )}\over {\partial \tau}}\\
 0_r&  \Big(R(\tau ,\sigma )\, M(\tau ,\vec \sigma )\Big)_{rs}
\end{array} \right),\nonumber \\
 &&{}\nonumber \\
  &&{}\nonumber \\
   \det\, J(\tau ,\vec \sigma ) &=&
v_l(\tau )\, \det\, R(\tau ,\sigma )\, \det\, M(\tau ,\vec \sigma )
= v_l(\tau )\, \det\, M(\tau ,\vec \sigma ).
 \label{IV17}
 \eea
\medskip

To show that $\det\, M(\tau ,\vec \sigma ) \not= 0$, let us look
for the null eigenvectors $W_r(\tau ,\vec \sigma )$ of the matrix
$M(\tau ,\vec \sigma )$, $M_{rs}(\tau ,\vec \sigma )\, W_s(\tau
,\vec \sigma ) = 0$ or $W_r(\tau ,\vec \sigma ) - \Phi_{uv}(\tau
,\sigma ) \, {{\sigma^u}\over {\sigma}}\, \sigma^s\, W_s(\tau
,\vec \sigma ) = 0$ [see Eq.(\ref{IV3})]. Due to $\Phi_{uv} = -
\Phi_{vu}$, we get $\sigma^s\, W_s(\tau ,\vec \sigma ) = 0$ and
this implies $W_r(\tau ,\vec \sigma ) = 0$, i.e. the absence of
null eigenvalues. Therefore $\det\, M(\tau ,\vec \sigma ) \not= 0$
and an explicit calculation shows that $\det\, M(\tau ,\vec \sigma
) = 1$. As a consequence, we get $\det\, J(\tau ,\vec \sigma ) =
v_l(\tau ) > 0$. Therefore, $x^{\mu} \mapsto \sigma^A$ is an
admissible 4-coordinate transformation.

\bigskip

Since in Eq.(\ref{IV1}) $x^{\mu}(\tau )$ is interpretable as the
world-line of an arbitrary non-inertial time-like accelerated
observer, these allowed foliations with parallel space-like
hyper-planes ({\it not orthogonal to the world-line}) define  good
notions of simultaneity, replacing the attempts based on Fermi
coordinates, for an accelerated observer with arbitrary time-like
world-line $x^{\mu}(\tau )$.

\bigskip

Let us remark that the congruence of time-like world-lines
associated to the constant normal $l^{\mu}$ defines an inertial
reference frame: each inertial observer is naturally endowed with
the orthonormal tetrad $b^{\mu}_A = (l^{\mu}; \epsilon^{\mu}_r)$.

\bigskip

Let us consider the second skew congruence, whose observer
world-lines are $x^{\mu}_{\vec \sigma}(\tau ) = z^{\mu}(\tau ,\vec
\sigma )$, and let us look for an orthonormal tetrad
$V^{\mu}_A(\tau ,\vec \sigma ) = (z^{\mu}_{\tau}(\tau ,\vec \sigma
)/\sqrt{\sgn\, g_{\tau\tau}(\tau ,\vec \sigma)}; V^{\mu}_r(\tau
,\vec \sigma ))$ to be associated to each of its time-like
observers. Due to the orthonormality we have $V^{\mu}_A(\tau ,\vec
\sigma ) = \Lambda^{\mu}{}_{\nu = A}(\tau ,\vec \sigma )$ with
$\Lambda (\tau ,\vec \sigma )$ a Lorentz matrix. Therefore we can
identify them with $SO(3,1)$ matrices parametrized as the product
of a pure boost with a pure rotation as in Eqs. (III8). If we
introduce

\begin{eqnarray}
&&\vec{E}_r(\tau,\vec{\sigma})=\{E^k_r(\tau,\vec{\sigma})\}=R^{s=k}_r(\alpha_m
(\tau ,\sigma ),\beta_m (\tau ,\sigma ),\gamma_m (\tau ,\sigma ))
\nonumber\\
 &&{}\nonumber \\
 && \Rightarrow\,\, {{\partial {\vec E}_r(\tau ,\vec \sigma
 )}\over {\partial \tau}} = \stackrel{def}{=}
 \vec{\omega}_{m}
 (\tau) \times\vec{E}_r(\tau,\vec{\sigma}),\nonumber \\
 &&\nonumber\\
 &&B^{jk}(\vec{\beta}_m(\tau,\vec{\sigma}))=\delta^{ij}+
\frac{\gamma^2(\vec{\beta}_m(\tau ,\sigma
))}{\gamma(\vec{\beta}_m(\tau ,\sigma ))+1}\,\beta^i_m(\tau
,\sigma )\, \beta^j_m(\tau ,\sigma ),
 \label{IV18}
\end{eqnarray}

\noindent  we can write

\begin{equation}
V^\mu_{A}(\tau,\vec{\sigma})=\Lambda^\mu_{\nu=A}
(\tau,\vec{\sigma})= \left(
\begin{array}{cc}
\frac{1}{\sqrt{1-\vec{\beta}_m^2(\tau,\vec{\sigma})}}&
\frac{\vec{\beta}_m(\tau,\vec{\sigma})\cdot\vec{E}_{
r}(\tau,\vec{\sigma})}{\sqrt{1-\vec{\beta}_m^2(\tau,\vec{\sigma})}}\\
\frac{\beta^j_m(\tau,\vec{\sigma})}{\sqrt{1-\vec{\beta}_m^2(\tau,\vec{\sigma})}}&\,
B^{jk}({\vec \beta}_m (\tau,\vec{\sigma}))\,
E^k_{r}(\tau,\vec{\sigma})
\end{array}
\right).
 \label{IV19}
\end{equation}

We stress that for every observer $x^\mu_{\vec{\sigma}}(\tau)$ the
choice of the $V^\mu_r(\tau,\vec{\sigma})$'s, and therefore also
of the $\vec{E}_r(\tau,\vec{\sigma})$'s, is arbitrary. As a
consequence the angular velocity $\vec{\omega}_m(\tau)$ {\em
defined} by the second of the Eqs.(\ref{IV18}) is in general not
related with the angular velocity (\ref{IV2}) defined by the
embedding. On the contrary, the parameter
$\vec{\beta}_m(\tau,\vec{\sigma})$ is related to the embedding by
the relation $\beta^i_m(\tau,\vec{\sigma}) =
z^i_\tau(\tau,\vec{\sigma}) /z^o_\tau(\tau,\vec{\sigma})$.

\bigskip

For every observer $x^{\mu}_{\vec \sigma}(\tau )$ of the
congruence, endowed with the orthonormal tetrad $E^{\mu}_{\vec
\sigma\, A}(\tau ) = V^{\mu}_A(\tau ,\vec \sigma )$, we get

\bea
 &&
 {{d E^{\mu}_{\vec \sigma\, A}(\tau )}\over {d\tau}} =
 {{\cal A}_{\vec \sigma\, A}}^B(\tau )\, V^{\mu}_{\vec \sigma\, B}(\tau ),
 \nonumber \\
 &&{}\nonumber \\
 &\Rightarrow & {\cal A}_{\vec \sigma \, AB}(\tau ) = - {\cal
 A}_{\vec \sigma\, BA}(\tau ) = {{d E^{\mu}_{\vec \sigma\, A}(\tau
 )}\over {d\tau}}\,\eta_{\mu\nu}
 E^\nu_{\vec \sigma\, B}(\tau ),
 \label{IV20}
  \eea

Using the (\ref{IV19}) we obtain [$\gamma(\tau,\vec{\sigma}) = 1/
\sqrt{1 - {\vec \beta}_m^2(\tau,\vec{\sigma})}$,
$\dot{\vec{\beta}}_m(\tau,\vec{\sigma})=
d{\vec{\beta}}_m(\tau,\vec{\sigma})/d\tau$]

 \bea
 a_{\vec \sigma\, r}(\tau ) &=& {\cal A}_{\vec \sigma\, \tau
 r}(\tau ) =\left[
 -\gamma \,(\dot{\vec{\beta}}_m\cdot\vec{E}_r)-
 \frac{\gamma^3}{\gamma+1}\,
 (\dot{\vec{\beta}}_m\cdot\vec{\beta}_m)
 (\vec{\beta}_m\cdot\vec{E}_r)\right](\tau,\vec{\sigma})
 \nonumber \\
 &&\nonumber\\
 \Omega_{\vec \sigma\, r}(\tau ) &=& {1\over 2}\, \epsilon_{ruv}\,
 {\cal A}_{\vec \sigma\, uv}(\tau ) =\nonumber\\
 &&\nonumber\\
 &=&
 \left[
 -\vec{\omega}_m\cdot\vec{E}_r-
 \frac{\gamma^2}{\gamma+1}
 \epsilon^{rsu}(\vec{\beta}_m\cdot\vec{E}_s)
 (\dot{\vec{\beta}}_m\cdot\vec{E}_u)
 \right](\tau,\vec{\sigma})
 \label{IV21}
 \eea

Therefore the acceleration radii (see Subsection C of the
Introduction) of these observers are

\bea
 I_1 &=&{\vec \Omega}_{\vec \sigma}^2 - {\vec a}_{\vec \sigma}^2 =
 \left[
 \vec{\omega}^2_m+
 2\frac{\gamma^2}{\gamma+1}
 \vec{\omega}_m\cdot(\dot{\vec{\beta}}_m\times\vec{\beta}_m)+
 \gamma^2(\gamma-2)\,\dot{\vec{\beta}}_m^2-
 \frac{\gamma^6}{\gamma+1}\,
 (\dot{\vec{\beta}}_m\cdot\vec{\beta}_m)^2
 \right](\tau,\vec{\sigma})
 \nonumber \\
 &&{}\nonumber \\
 I_2 &=& {\vec a}_{\vec \sigma} \cdot {\vec \Omega}_{\vec \sigma} = \left[
 \gamma\,(\dot{\vec{\beta}}_m\cdot\vec{\omega}_m)+
 \frac{\gamma^3}{\gamma+1}\,
 (\dot{\vec{\beta}}_m\cdot\vec{\beta}_m)
 (\vec{\beta}_m\cdot\vec{\omega}_m)
 \right](\tau,\vec{\sigma})
 \label{IV22}
 \eea

 \bigskip

The non-relativistic limit of the embedding (\ref{IV1}) can be
obtained by choosing $\epsilon^{\mu}_r = (0; e^i_r)$. We obtain a
generalization of  the standard translating and rotating
3-coordinate systems on the hyper-planes of constant absolute
Newtonian time

\bea
 t^{'}(\tau ) &=& t(\tau ),\nonumber \\
 z^i(\tau ,\vec \sigma ) &=& x^i(\tau ) + e^i_r\, R^r{}_s(\tau
 ,\sigma )\, \sigma^s,
 \label{IV23}
 \eea

\noindent without any restriction on rotations, namely with $R =
R(\tau )$ allowed.

\vfill\eject

\section{Notion of Simultaneity Associated to Rotating Reference
Frames.}

In this Section we consider the inverse problem of finding a
foliation of Minkowski space-time with simultaneity surfaces
associated to a given  arbitrary reference frame with non-zero
vorticity, namely to a time-like vector field whose expression in
Cartesian 4-coordinates in an inertial system is ${\tilde
u}^{\mu}(x)$ with ${\tilde u}^2(x) = \sgn$. In other words we are
looking for embeddings $z^{\mu}(\tau ,\vec \sigma )$, inverse of
an admissible 4-coordinate transformation, such that we have
${\tilde u}^{\mu}(z(\tau,\vec{\sigma})) = u^{\mu}(\tau ,\vec
\sigma ) = z^{\mu}_{\tau}(\tau ,\vec \sigma )/\sqrt{\sgn\,
g_{\tau\tau}(\tau ,\vec \sigma )}$. Let us remark that if the
vorticity is zero, the vector field ${\tilde u}^{\mu}(x)$ is
surface-forming, there is  a foliation whose surfaces have the
normal field proportional to $u^\mu(\tau,\vec{\sigma})$ and these
surfaces automatically give an admissible foliation with
space-like hyper-surfaces of Minkowski space-time.

\bigskip

Let us first show that, given an arbitrary time-like vector field
${\tilde u}^{\mu}(x)$, the looked for foliation exists. Let us
consider the equation

\beq
 {\tilde u}^{\mu}(x) \, {{\partial s(x)}\over {\partial x^{\mu}}}
 = 0,
 \label{V1}
 \eeq

\noindent where $s(x)$ is a scalar field. This equation means that
$s(x)$ is constant along the integral lines $x^{\mu}(s)$ [$d
x^{\mu}(s)/ {ds} = {\tilde u}^{\mu}(x(s))$] of the vector field,
i.e. it is a comoving quantity, since

\beq
 {{d s(x(s))}\over {ds}} = {\tilde u}^{\mu}(x(s))\, {{\partial
s}\over {\partial x^{\mu}}}(x(s)) = 0.
 \label{V2}
  \eeq

Since Eq.(\ref{V1}) has three independent solutions $s^{(r)}(x)$ ,
$r=1,2,3$, they can be used to identify three coordinates $
\sigma^r(x) = s^{(r)}(x)$. Moreover the three 4-vectors
${{\partial \sigma^r(x)}\over {\partial x^{\mu}}}$ are space-like
by construction.
\bigskip

Since Minkowski space-time is globally hyperbolic, there exist
{\it time-functions} $\tau (x)$ such that i) $\tau (x) = const.$
defines space-like hyper-surfaces; ii) ${{\partial \tau (x)}\over
{\partial x^{\mu}}}$ is a time-like 4-vector.
\medskip

As a consequence we can build an invertible 4-coordinate
transformation $x^{\mu} \mapsto \sigma^A(x) = (\tau (x),
\sigma^r(x))$, with inverse $\sigma^A = (\tau , \sigma^r) \mapsto
x^{\mu} = z^{\mu}(\tau ,\vec \sigma )$ for every choice of $\tau
(x)$. It can be shown that we get always a non-vanishing Jacobian
\footnote{Let us show that the equations
\[
\alpha\,\frac{\partial\tau(x)}{\partial x^\mu}+\beta_r\,
\frac{\partial\sigma^r(x)}{\partial x^\mu}=0
\]
implies $\alpha=\beta_r=0$. If we multiply for $\tilde{u}^\mu(x)$,
we get $\alpha\,\tilde{u}^\mu(x)\,\frac{\partial\tau(x)}{\partial
x^\mu}=0$. But $\frac{\partial\tau(x)}{\partial x^\mu}$ and
$\tilde{u}^\mu(x)$ are both time-like with
$\tilde{u}^\mu(x)\,\frac{\partial\tau(x)}{\partial x^\mu}\neq 0$,
so that we get $\alpha=0$. We remain with the equations $\beta_r\,
\frac{\partial\sigma^r(x)}{\partial x^\mu}=0$, which imply
 $\beta_r=0$ since the
$\frac{\partial\sigma^r(x)}{\partial x^\mu}$ are independent by
construction.}

\beq
J = \det\, \Big(
{{\partial \tau (x)}\over {\partial x^{\mu}}}, {{\partial
\sigma^r(x)}\over {\partial x^{\mu}}}\Big) \not= 0.
\label{V3}
\eeq

By using

\beq {{\partial \sigma^A(x)}\over {\partial x^{\nu}}}\, {{\partial
x^{\mu}}\over {\partial \sigma^A}}(\sigma (x)) = \eta^{\mu}_{\nu},
 \label{V4}
 \eeq

\noindent and Eq.(\ref{V1}) we get the desired result

\beq
 {\tilde u}^{\mu}(x) = {\tilde u}^{\nu}(x)\, {{\partial
\sigma^A(x)}\over {\partial x^{\nu}}}\, {{\partial x^{\mu}}\over
{\partial \sigma^A}}(\sigma (x)) = \Big( {\tilde u}^{\nu}(x)\,
{{\partial \tau (x)}\over {\partial x^{\nu}}}\Big)\, {{\partial
z^{\mu}(\tau ,\vec \sigma )}\over {\partial \tau}} =
{{z^{\mu}_{\tau}(\tau ,\vec \sigma )}\over {\sqrt{\sgn \,
g_{\tau\tau}(\tau ,\vec \sigma )}}}.
 \label{V55}
  \eeq

\bigskip

Given a unit time-like vector field ${\tilde u}^{\mu}(x) =
u^{\mu}(\tau ,\vec \sigma )$ such that $u^{\mu}(\tau ,\vec \sigma
)\, {\rightarrow}_{|\vec \sigma | \rightarrow \infty}\,
n^{\mu}(\tau )$ and ${{\partial u^{\mu}(\tau ,\vec \sigma )}\over
{\partial \sigma^r}}\,   {\rightarrow}_{|\vec \sigma | \rightarrow
\infty}\, 0$, to find the embeddings $z^{\mu}(\tau ,\vec \sigma )$
we must integrate the equation

\beq
 {{\partial z^{\mu}(\tau ,\vec \sigma )}\over {\partial
\tau}} = f(\tau ,\vec \sigma )\, u^{\mu}(\tau ,\vec \sigma
),\qquad u^2(\tau ,\vec \sigma ) = \sgn,
 \label{V6}
 \eeq

\noindent where $f(\tau ,\vec \sigma )$ is an integrating factor.
\medskip

Since Eq.(\ref{V6}) implies $\sgn \, g_{\tau\tau}(\tau ,\vec
\sigma ) = f^2(\tau ,\vec \sigma ) > 0$, the only restrictions on
the integrating factor are:

i) it must never vanish;

ii) $f(\tau ,\vec \sigma )\,  {\rightarrow}_{|\vec \sigma |
\rightarrow \infty}\, f(\tau )$ finite.
\bigskip

The integration of Eq.(\ref{V2}) gives

\begin{eqnarray*}
 z^{\mu}(\tau ,\vec \sigma ) &=& g^{\mu}(\vec \sigma ) +
 \int_o^{\tau} d\tau_1\, f(\tau_1, \vec \sigma )\, u^{\mu}(\tau_1,
 \vec \sigma ),\nonumber \\
 &&{}\nonumber \\
 &&\Downarrow\nonumber \\
 &&{}\nonumber \\
 z^{\mu}_r(\tau ,\vec \sigma )&=& \partial_r\, g(\vec \sigma ) +
 \int_o^{\tau} d\tau_1\, \partial_r\, [f(\tau_1, \vec \sigma )\,
 u^{\mu}(\tau ,\vec \sigma )],
 \end{eqnarray*}

\bea
 g_{\tau r}(\tau ,\vec \sigma ) &=&f(\tau ,\vec \sigma )\,
 u_{\mu}(\tau ,\vec \sigma )\, \Big[ \partial_r\, g(\vec \sigma ) +
 \int_o^{\tau} d\tau_1\, \partial_r\, [f(\tau_1, \vec \sigma )\,
 u^{\mu}(\tau ,\vec \sigma )]\Big]\nonumber \\
&&\nonumber\\
 && {\rightarrow}_{|\vec \sigma | \rightarrow \infty}\,
 f(\tau )\, n_{\mu}(\tau )\, \Big[ \lim_{|\vec \sigma | \rightarrow \infty}\,
\partial_r\, g(\vec \sigma )\Big],
 \label{V7}
 \eea

\noindent where $g(\vec \sigma )$ is arbitrary and we have assumed
that the integrating factor satisfies $\partial_r\, f(\tau ,\vec
\sigma )\, {\rightarrow}_{|\vec \sigma | \rightarrow \infty}\, 0$.
\bigskip

For the sake of simplicity let us choose $g(\vec \sigma ) =
\epsilon^{\mu}_r\, \sigma^r$ with the constant 4-vectors
$\epsilon^{\mu}_r$ belonging to an orthonormal tetrad
$\epsilon^{\mu}_A$. Then $g_{\tau r}(\tau ,\vec \sigma )$ has the
finite limit $f(\tau )\, n_{\mu}(\tau )\, \epsilon^{\mu}_r$.

With this choice for $g(\vec \sigma )$ we get

\bea
 z^{\mu}_r(\tau ,\vec \sigma ) &=& [\delta_{rs} + \alpha_{rs}(\tau
 ,\vec \sigma )]\, \epsilon^{\mu}_s + \beta_r(\tau ,\vec \sigma
 )\, \epsilon^{\mu}_{\tau},\nonumber \\
 &&{}\nonumber \\
 &&\quad \alpha_{rs}(\tau ,\vec \sigma ) =
 \int_o^{\tau} d\tau_1\, \partial_r\, [f(\tau_1 ,\vec \sigma )\,
 \epsilon_{s \mu}\, u^{\mu}(\tau_1, \vec \sigma )],\nonumber \\
&&\nonumber\\
 &&\quad \beta_r(\tau ,\vec \sigma ) =
 \int_o^{\tau} d\tau_1\, \partial_r\, [f(\tau_1 ,\vec \sigma )\,
 \epsilon_{\tau \mu}\, u^{\mu}(\tau_1, \vec \sigma )].
 \label{V8}
 \eea
 \medskip

 Since $u^{\mu}(\tau ,\vec \sigma )$ and $\epsilon^{\mu}_{\tau}$
 are future time-like [$\sgn\, u^o(\tau ,\vec \sigma ) > 0$,
  $\sgn\, \epsilon^o_{\tau} > 0$], we have $u^{\mu}(\tau ,\vec \sigma ) =
 \sgn\, a(\tau ,\vec \sigma )\, \epsilon^{\mu}_{\tau} +
 b_r(\tau ,\vec \sigma )\, \epsilon^{\mu}_r$ with $a(\tau ,\vec
 \sigma ) > 0$ and without zeroes.
\bigskip

 Let us determine the integrating factor $f(\tau ,\vec \sigma )$
 by requiring $\beta_r(\tau ,\vec \sigma ) = 0$ as a consequence
 of the equation

 \bea
 0 &=& \sgn\, \partial_r\, [f(\tau ,\vec \sigma )\, \epsilon_{\tau \mu}\,
 u^{\mu}(\tau ,\vec \sigma )] = f(\tau ,\vec \sigma )\,
 \partial_r\, a(\tau ,\vec \sigma ) + \partial_r\, f(\tau ,\vec
 \sigma )\, a(\tau ,\vec \sigma ),\nonumber \\
 &&{}\nonumber \\
 &&\Downarrow\nonumber \\
 &&{}\nonumber \\
 f(\tau ,\vec \sigma ) &=& e^{c(\tau )}\, a(\tau ,\vec \sigma
 ),\nonumber \\
 &&{}\nonumber \\
 z^{\mu}_r(\tau ,\vec \sigma ) &=& [\delta_{rs} + \alpha_{rs}(\tau
 ,\vec \sigma )]\, \epsilon^{\mu}_s,\nonumber \\
&&\nonumber\\
 \alpha_{rs}(\tau ,\vec \sigma ) &=& \int_o^{\tau} d\tau_1\,
 e^{c(\tau_1)}\, \partial_r\, [a(\tau_1,\vec \sigma )\,
 \epsilon_{s \mu}\, u^{\mu}(\tau_1, \vec \sigma )],\nonumber \\
 &&{}\nonumber \\
 g_{rs}(\tau ,\vec \sigma ) &=& -\sgn\, \Big(\delta_{rs} +
 \alpha_{rs}(\tau ,\vec \sigma ) +  \alpha_{sr}(\tau ,\vec \sigma )
 + \sum_u\,  \alpha_{ru}(\tau ,\vec \sigma )\,  \alpha_{su}(\tau ,\vec \sigma )
\Big).
 \label{V9}
 \eea
\medskip

 Let us choose the arbitrary function $C(\tau ) = e^{c(\tau )}$ so
 small that $|  \alpha_{rs}(\tau ,\vec \sigma )| << 1$ for every
 $r$, $s$, $\tau$, $\vec \sigma$, so that all the conditions on
 $g_{rs}(\tau ,\vec \sigma )$ from Eqs.(\ref{I1}) are satisfied.
 \bigskip

 In conclusion given an arbitrary congruence of time-like
 world-lines, described by a vector field ${\tilde u}^{\mu}(x)$,
 an embedding defining a good notion of simultaneity is
 [$x^{\mu}(\tau ) \, {\buildrel {def}\over =}\, z^{\mu}(\tau ,\vec 0)$]

 \bea
  z^{\mu}(\tau ,\vec \sigma ) &=& \epsilon^{\mu}_r\, \sigma^r +
  \int_o^{\tau} d\tau_1\, C(\tau_1)\, \epsilon_{\tau \nu}\,
  u^{\nu}(\tau_1, \vec \sigma )\, u^{\mu}(\tau_1, \vec \sigma )
  =\nonumber \\
  &=& x^{\mu}(\tau ) +  \epsilon^{\mu}_r\, \sigma^r +
  \int_o^{\tau} d\tau_1\, C(\tau_1)\, \epsilon_{\tau \nu}\,
  \Big[u^{\nu}(\tau_1, \vec \sigma )\, u^{\mu}(\tau_1, \vec \sigma )
  - u^{\nu}(\tau ,\vec 0)\, u^{\mu}(\tau ,\vec 0)\Big],\nonumber  \\
 &&{}
  \label{V10}
  \eea

\noindent for sufficiently small $C(\tau )$. Here
$\epsilon^{\mu}_A$ is an arbitrary orthonormal tetrad.

\vfill\eject

\section{Applications.}

In this Section we shall apply the 3+1 point of view to the
description of GPS, to the problem of the rotating disk and of the
Sagnac effect, to the determination of the time delay for light
propagation between an Earth station and a satellite and finally
to Maxwell theory.

\subsection{The Global Positioning System and the Determination
of a Set of Radar Coordinates.}

In Eqs.(\ref{IV1}) we gave a family of embeddings $x^{\mu} =
z^{\mu}(\tau ,\vec \sigma )$ defining possible notions of
simultaneity, i.e. admissible 3+1 splittings of Minkowski
space-time with foliations with space-like hyper-planes
$\Sigma_{\tau}$ as leaves, to be associated to the world-line
$x^{\mu}(\tau )$ of an arbitrary time-like observer $\gamma$,
chosen as origin of the 3-coordinates on each simultaneity leaf
$\Sigma_{\tau}$, i.e. $x^{\mu}(\tau ) = z^{\mu}(\tau ,\vec 0)$.
The space-like hyper-planes $\Sigma_{\tau}$ are not orthogonal to
$\gamma$: if $l^{\mu} = \epsilon^{\mu}_{\tau}$ is the normal to
$\Sigma_{\tau}$ we have $l_{\mu}\, {{{\dot x}^{\mu}(\tau )}\over
{\sqrt{\sgn\, {\dot x}^2(\tau )}}} \not= \sgn$ except in the
limiting case of an inertial observer with 4-velocity proportional
to $l^{\mu}$.
\medskip

If $\tau$ is the scalar coordinate-time of the foliation, the
proper time of the standard atomic clock $C$ of $\gamma$ will be
defined by $d{\cal T}_{\gamma} = \sqrt{\sgn\, g_{\tau\tau}(\tau ,
\vec 0)}\, d\tau$ [$x^{\mu}(\tau ) = {\tilde x}^{\mu}({\cal
T}_{\gamma})$]. This defines ${\cal T}_{\gamma} = {\cal
F}_{\gamma}(\tau )$ as a monotonic function of $\tau$, whose
inverse will be denoted $\tau = {\cal G}({\cal T}_{\gamma})$.
Moreover, we make a conventional choice of a tetrad ${}_{(\gamma
)}E^{\mu}_A(\tau )$ associated to $\gamma$ with ${}_{(\gamma
)}E^{\mu}_{\tau}(\tau ) = {{{\dot x}^{\mu}(\tau ) }\over
{\sqrt{\sgn\, {\dot x}^2(\tau )}}}$.

\bigskip

Let us consider a set of $N$ arbitrary time-like world-lines
$x^{\mu}_i(\tau )$, $i=1,..,N$, associated to observers
$\gamma_i$, so that $\gamma$ and the $\gamma_i$'s can be imagined
to be the world-lines of $N + 1$ spacecrafts (like in GPS
\cite{59a}) with $\gamma$ chosen as a reference world-line. Each
of the  world-lines $\gamma_i$ will have an associated standard
atomic clock $C_i$ and a conventional tetrad
${}_{(\gamma_i)}E^{\mu}_A(\tau )$.

\medskip

To compare the distant clocks $C_i$ with $C$ in the chosen notion
of simultaneity, we define the 3-coordinates ${\vec \eta}_i(\tau
)$ of the $\gamma_i$

\beq
 x^{\mu}_i(\tau ) {\buildrel {def}\over =} z^{\mu}(\tau ,{\vec \eta}_i(\tau )).
 \label{VI1}
 \eeq

\noindent Then the proper times ${\cal T}_{\gamma_i}$ of the
clocks $C_i$ will be expressed in terms of the scalar coordinate
time $\tau$ of the chosen simultaneity as

\beq
 d{\cal T}_{\gamma_i} = \sqrt{\sgn\, \Big[ g_{\tau\tau}(\tau
,{\vec \eta}_i(\tau ))+ 2g_{\tau r}(\tau ,{\vec \eta}_i(\tau
))\,\dot{\eta}^r(\tau)+ g_{rs}(\tau ,{\vec \eta}_i(\tau
))\,\dot{\eta}^r(\tau)\,\dot{\eta}^s(\tau) \Big]}\, d\tau,
 \label{VI2}
  \eeq

\noindent so that with this notion of simultaneity the proper
times ${\cal T}_{\gamma_i}$ are connected to the proper time
${\cal T}_{\gamma}$ by the following relations

\beq
 d{\cal T}_{\gamma_i} = \left.\sqrt{{{g_{\tau\tau}(\tau ,{\vec \eta}_i(\tau ))+
2g_{\tau r}(\tau ,{\vec \eta}_i(\tau ))\,\dot{\eta}_i^r(\tau)+
g_{rs}(\tau ,{\vec \eta}_i(\tau
))\,\dot{\eta}_i^r(\tau)\,\dot{\eta}_i^s(\tau) } \over
{g_{\tau\tau}(\tau ,\vec 0)}}}\, \right|_{\tau = {\cal G}({\cal
 T}_{\gamma})}\, d{\cal T}_{\gamma}.
 \label{VI3}
 \eeq
\medskip

This determines the synchronization of the $N + 1$ clocks once we
have expressed the 3-coordinates ${\vec \eta}_i(\tau )$ in terms
of the given world-lines $x^{\mu}(\tau )$, $x^{\mu}_i(\tau )$ and
of the embedding (\ref{IV1}). From the definition

\beq
 x_i^{\mu}(\tau ) =
 z^{\mu}(\tau ,{\vec \eta}_i(\tau )) = x^{\mu}(\tau ) +
  \epsilon^{\mu}_r\, R^r{}_s(\tau ,|{\vec \eta}_i(\tau )|)\, \eta^s_i(\tau ),
 \label{VI4}
 \eeq

\noindent   we get [$|{\vec \eta}_i(\tau )| \, {\buildrel
{def}\over =}\, \sqrt{\delta_{rs}\, \eta^r_i(\tau )\,
\eta^s_i(\tau )}$, $\eta^r_i(\tau ) =  |{\vec \eta}_i(\tau )|\,
{\hat  n}_i^r(\tau )$, $\delta_{rs}\, {\hat n}^r_i(\tau )\, {\hat
n}^s_i(\tau ) = 1$]
\medskip

\beq
  \eta^u_i(\tau ) = - \sum_w\, [R^{-1}(\tau , |{\vec \eta}_i(\tau
  )|)]^u{}_w\, \epsilon^{\nu}_w\, [x_{i\nu}(\tau ) - x_{\nu}(\tau
  )].
 \label{VI5}
 \eeq

Then, if we put the solution

\beq
 |{\vec \eta}_i(\tau )| =
  F_i\Big[\epsilon^{\mu}_r\, \Big(x_{i\mu}(\tau ) - x_{\mu}(\tau
  )\Big)\Big],
 \label{VI6}
 \eeq

\noindent of the equations

\bea
 |{\vec \eta}_i(\tau )|^2 &=& \delta_{rs}\,
 \sum_{mn}\,  [R^{-1}(\tau ,|{\vec \eta}_i(\tau
  )|)]^r{}_m\, [R^{-1}(\tau ,|{\vec \eta}_i(\tau
  )|)]^s{}_n\nonumber \\
  &&\epsilon^{\mu}_m\, [x_{i\mu}(\tau ) - x_{\mu}(\tau )]\,
  \epsilon^{\nu}_n\, [x_{i\nu}(\tau ) - x_{\nu}(\tau )],
 \label{VI7}
 \eea

\noindent into Eqs.(\ref{VI5}), we obtain the looked for
expression of the 3-coordinates ${\vec \eta}_i(\tau )$

  \beq
\eta^u_i(\tau ) = - \sum_m\, \Big[ R^{-1}(\tau ,
  F_i[\epsilon^{\alpha}_w\, (x_{i\alpha}(\tau ) - x_{\alpha}(\tau
  ))])\Big]^u{}_m\, \epsilon^{\nu}_m\, [x_{i\nu}(\tau ) -
  x_{\nu}(\tau )].
 \label{VI8}
 \eeq

  \bigskip

We will now define {\it an operational method to build a grid of
radar 4-coordinates} associated with the arbitrarily given
time-like world-line $x^{\mu}(\tau )$ of the spacecraft $\gamma$
and with an admissible embedding $z^{\mu} (\tau ,\vec \sigma )$
(we use Eq.(\ref{IV1}) as an example), by using light signals
emitted by $\gamma$ and reflected towards $\gamma$ from the other
spacecrafts $\gamma_i$. This will justify the name radar
4-coordinates and will show how the simultaneity convention
(\ref{IV1}) deviates from Einstein's convention.

\bigskip

To this end, given an embedding $x^{\mu} = z^{\mu} (\tau ,\vec
\sigma )$ of the family (\ref{IV1}) and one of its simultaneity
leaves $\Sigma_{\tau}$ with the point $Q$ of 4-coordinates
$x^{\mu}(\tau )$ on $\gamma$ as origin of the 3-coordinates $\vec
\sigma$, let us consider a point $P$ on $\Sigma_{\tau}$ with
coordinates $z^{\mu}(\tau ,\vec \sigma )$ (for $\vec \sigma =
{\vec \eta}_i(\tau )$ it corresponds to the spacecraft
$\gamma_i$). We want to express the adapted 4-coordinates $\tau =
\tau (P)$, $\vec \sigma = \vec \sigma (P)$ of $P$ in terms of data
on the world-line $\gamma$ corresponding to the emission of a
light signal in $Q_{-}$ at $\tau_{-} < \tau$ and to its reception
in $Q_{+}$ at $\tau_{+} > \tau$ after reflection at $P$.
\medskip

Let $x^{\mu}(\tau_{-})$ be the intersection of the world-line
$\gamma$ with the past light-cone through $P$ and
$x^{\mu}(\tau_{+})$ the intersection with the future light-cone
through $P$. To find $\tau_{\pm}$ we have to solve the equations
$\Delta^2_{\pm} = [x^{\mu}(\tau_{\pm}) - z^{\mu}(\tau ,\vec \sigma
)]^2 = 0$ with $\Delta^{\mu}_{\pm} = x^{\mu}(\tau_{\pm}) -
z^{\mu}(\tau ,\vec \sigma )$. We are interested in the solutions
$\Delta^o_{+} = |{\vec \Delta}_{+}|$ and $\Delta^o_{-} = - |{\vec
\Delta}_{-}|$. Let us remark that on the simultaneity surfaces
$\Sigma_{\tau}$ we have $x^o(\tau ) \not= z^o(\tau ,\vec \sigma )$
for the Cartesian coordinate times.

\bigskip

Let us show that the adapted coordinates $\tau$ and $\vec \sigma$
of the event $P$ with Cartesian 4-coordinates $z^{\mu}(\tau ,\vec
\sigma )$ in an inertial system can be determined in terms of the
{\it emission scalar time} $\tau_{-}$ of the light signal, the
{\it emission unit 3-direction} ${\hat n}_{(\tau_{-})}(
\theta_{(\tau_{-})} ,\phi_{(\tau_{-})} )$ [so that
$\triangle^{\mu}_{-} = |{\vec \triangle}_{-}|\, (-\sgn ; {\hat
n}_{(\tau_{-})})$ ] and the {\it reception scalar time} $\tau_{+}$
registered by the observer $\gamma$ with world-line $x^{\mu}(\tau
)$. These data are usually given in terms of the proper time
${\cal T}(\tau )$ of the observer $\gamma$ by using $d{\cal T} =
\sqrt{\sgn\, g_{\tau\tau}(\tau ,\vec 0) }\, d\tau$.
\bigskip

Let us introduce the following parametrization by using
Eqs.(\ref{IV1})

\begin{eqnarray*}
 z^{\mu}(\tau ,\vec \sigma ) &=& x^{\mu}(\tau ) +
 \epsilon^{\mu}_r\, R^r{}_s(\tau ,\sigma )\, \sigma^s\, {\buildrel
 {def}\over =}\nonumber \\
 &&{}\nonumber \\
 &{\buildrel {def}\over =}& \Big[ \xi_l(\tau ,\vec
 \sigma )\, l^{\mu} + \xi^r(\tau ,\vec \sigma )\,
 \epsilon^{\mu}_r\Big] =\nonumber \\
&&\nonumber\\
 &=&  \Big[ x_l(\tau )\, l^{\mu} + \sum_r\, [x_{\epsilon}^r(\tau ) + \zeta^r(\tau
 ,\vec \sigma )]\, \epsilon^{\mu}_r\Big],\nonumber \\
 &&{}\nonumber \\
 \xi_l(\tau ,\vec \sigma ) &=& \sgn\,z_{\mu}(\tau ,\vec \sigma )\, l^{\mu} =
 \sgn\, x_{\mu}(\tau )\, l^{\mu} = x_l(\tau ),
 \end{eqnarray*}

\bea
  \xi^r(\tau ,\vec \sigma ) &=& \sgn\,z_{\mu}(\tau ,\vec \sigma )\,
  \epsilon^{\mu}_r = x^r_{\epsilon}(\tau ) + \zeta^r(\tau ,\vec
  \sigma ),\nonumber \\
&&\nonumber\\
  x_{\epsilon}^r(\tau ) &=&\sgn\,
 x_{\mu}(\tau )\, \epsilon^{\mu}_r,\qquad \zeta^r(\tau ,\vec
 \sigma ) = R^r{}_s(\tau ,\sigma )\, \sigma^s\,
 {\rightarrow}_{\sigma \rightarrow \infty}\, \sigma^r.
 \label{VI9}
 \eea

\medskip

Then the two equations $\triangle^2_{\pm} = [x^{\mu}(\tau_{\pm}) -
z^{\mu}(\tau ,\vec \sigma )]^2 = \sgn\, ([x_l(\tau_{\pm}) -
x_l(\tau )]^2 - [{\vec x}_{\epsilon}(\tau_{\pm}) - {\vec
x}_{\epsilon}(\tau ) - \vec \zeta (\tau ,\vec \sigma )]^2) = 0$
can be rewritten in the form

 \bea
  x_l(\tau_{+}) &=& x_l(\tau ) + |{\vec \triangle}_{+}|
  = x_l(\tau ) + | {\vec x}_{\epsilon}(\tau_{+}) - \vec \xi
  (\tau ,\vec \sigma )|,\nonumber \\
   &&\qquad |{\vec \Delta}_{+}| = |{\vec x}_{\epsilon}(\tau_{+}) -
  \vec \xi (\tau ,\vec \sigma )|\, {\rightarrow}_{\sigma \rightarrow \infty}\,
  |{\vec x}_{\epsilon}(\tau_{+}) - {\vec x}_{\epsilon}(\tau ) - \vec \sigma |,\nonumber \\
  &&{}\nonumber \\
  x_l(\tau_{-}) &=& x_l(\tau ) - |{\vec \triangle}_{-}|
  = x_l(\tau ) - | {\vec x}_{\epsilon}(\tau_{-}) - \vec \xi
  (\tau ,\vec \sigma )|,\nonumber \\
   &&\qquad |{\vec \Delta}_{-}| = |{\vec x}_{\epsilon}(\tau_{-}) -
  \vec \xi (\tau ,\vec \sigma )|\, {\rightarrow}_{\sigma \rightarrow \infty}\,
  |{\vec x}_{\epsilon}(\tau_{-}) - {\vec x}_{\epsilon}(\tau ) - \vec \sigma |.
  \label{VI10}
  \eea

It can be shown \cite{98a} that, if no observer is allowed to
become a Rindler observer \cite{97a}, then each equation admits a
unique \footnote{ If we introduce the function
\[
g_{\pm}(y)=x_l(y)-x_l(\tau ) \pm|{\vec x}_{\epsilon}(y)- \vec\xi
  (\tau ,\vec \sigma )|,
\]
Eqs. (\ref{VI10}) are equivalent to $g_{\pm}(y)=0$. The solution
is unique because the functions $g_{\pm}(y)$ are decreasing in
$y$, since we have
\[
\frac{dg_{\pm}(y)}{dy}=
-v_l(y)\pm\sum_r\,v_r(y)\,\frac{x^r_{\epsilon}(y)- \xi^r
  (\tau ,\vec \sigma )}{|{\vec x}_{\epsilon}(y)- \vec\xi
  (\tau ,\vec \sigma )|}.
\]
Using  Eq.(\ref{IV11}) in the form
\[
\sum_r\,v_r(y)\,\frac{x^r_{\epsilon}(y)- \xi^r
  (\tau ,\vec \sigma )}{|{\vec x}_{\epsilon}(y)- \vec\xi
  (\tau ,\vec \sigma )|}
\leq |\vec{v}(y)|<v_l(y),
\]
we get $ \frac{dg_{\pm}(y)}{dy}<0$, since $v_l(y)>0$. } solution
$\tau_{\pm} = T_{\pm}(\tau ,\vec \sigma )$.

Therefore the following four data measured by the observer
$\gamma$

  \begin{eqnarray*}
  \tau_{\pm} &=& T_{\pm}(\tau ,\vec \sigma ),\nonumber \\
&&{}\nonumber \\
 \end{eqnarray*}

\bea
  {\hat n}_{(\tau_{-})}(\theta_{(\tau_{-})}, \phi_{(\tau_{-})})
  &=& \Big( \sin\, \theta_{(\tau_{-})}\, \sin\, \phi_{
  (\tau_{-})}, \sin\, \theta_{(\tau_{-})}\, \cos\, \phi_{(\tau_{-})},
  \cos\, \theta_{(\tau_{-})}\Big) =\nonumber \\
&&\nonumber\\
  &=& {{{\vec \triangle}_{-}}\over {|{\vec \triangle}_{-}|}}
  = \left.{{{\vec x}_{\epsilon}(\tau_{-}) - {\vec x}_{\epsilon}(\tau )
  - \vec \zeta (\tau ,\vec  \sigma )}\over {|{\vec x}
  _{\epsilon}(\tau_{-}) - {\vec x}_{\epsilon}(\tau ) - \vec \zeta (\tau ,\vec
  \sigma )|}}\right|_{\tau_{-} = T_{-}(\tau ,\vec \sigma )} = \hat m(\tau ,\vec \sigma ),
  \label{VI11}
  \eea

\noindent can be inverted to get the adapted coordinates $\tau
(P)$, $\vec \sigma (P)$ of the event $P$ with 4-coordinates
$z^{\mu}(\tau ,\vec \sigma )$ in terms of the data (Einstein's
convention for the radar time would be ${\cal E} = {1\over 2}$)

\bea
 \tau (P) &=& \tau (\tau_{-}, {\hat n}_{(\tau_{-})}, \tau_{+})
{\buildrel {def}\over =} \tau_{-} + {\cal E}(\tau_{-}, {\hat
n}_{(\tau_{-})}, \tau_{+})\, [\tau_{+} - \tau_{-}],\nonumber \\
 &&{}\nonumber \\
 \vec \sigma (P)&=& {\vec {\cal G}}(\tau_{-}, {\hat n}_{(\tau_{-})},
 \tau_{+})\, {\rightarrow}_{\tau_{+} \rightarrow \tau_{-}}\, 0.
 \label{VI12}
 \eea

\bigskip

Let us remark that

i) for $x^{\mu}(\tau ) = \tau\, l^{\mu}$ (inertial observer with
world-line {\it orthogonal} to $\Sigma_{\tau}$; ${\vec
x}_{\epsilon}(\tau ) = 0$) we get the Einstein's convention for
the radar time, because we have
\medskip

\begin{eqnarray*}
 \tau_{\pm} &=& \tau \pm |\vec \zeta (\tau ,\vec \sigma )|,\qquad \tau =
{1\over 2}\, (\tau_{+} + \tau_{-}),\qquad \sigma = |\vec \zeta
(\tau ,\vec \sigma )| = {1\over 2}\, (\tau_{+} -
\tau_{-}),\nonumber \\
  {\cal E} &=& {1\over 2},\qquad \zeta^r(\tau
,\vec \sigma ) = - {1\over 2}\, (\tau_{+} - \tau_{-})\, {\hat
n}^r_{(\tau_{-})},\nonumber \\
 \sigma^r &=& {\cal G}^r = {1\over
2}\, (\tau_{+} - \tau_{-})\, (R^{-1})^r{}_s({{\tau_{+} +
\tau_{-}}\over 2}, {{\tau_{+} - \tau_{-}}\over 2})\, {\hat
n}^s_{(\tau_{-})};
 \end{eqnarray*}

ii) for $x^{\mu}(\tau ) = \tau\, [l^{\mu} + \epsilon^{\mu}_r\,
a^r]$ (inertial observer with world-line {\it non-orthogonal} to
$\Sigma_{\tau}$; ${\vec x}_{\epsilon}(\tau ) = \tau\, \vec a$),
after some straightforward calculations, we get

\begin{eqnarray*}
 \tau_{\pm} &=& \tau + {1\over {1 - {\vec a}^2}}\, \Big[- \vec a \cdot
\vec \zeta (\tau ,\vec \sigma ) \pm \sqrt{(\vec a \cdot \vec \zeta
 (\tau ,\vec \sigma ) )^2 + (1 - {\vec a}^2)\, \sigma^2}\Big],\nonumber \\
 \tau &=& {1\over 2}\, \Big[\tau_{+} + \tau_{-} + {{\tau_{+} -
\tau_{-}}\over {1 - {\vec a}^2}}\, \sqrt{{{{\vec a}^2 + \vec a
\cdot {\hat n}_{(\tau_{-})}}\over {1 + {\vec a}^2 - {\vec a}^4 +
(3 - 2\, {\vec a}^2)\, \vec a \cdot {\hat n}_{(\tau_{-}})
}}}\Big],\nonumber \\
 {\cal E} &=& {1\over 2}\, \Big[ 1 + {1\over {1 - {\vec a}^2}}\,
\sqrt{{{{\vec a}^2 + \vec a \cdot {\hat n}_{(\tau_{-})}}\over {1 +
{\vec a}^2 - {\vec a}^4 + (3 - 2\, {\vec a}^2)\, \vec a \cdot
{\hat n}_{(\tau_{-}}) }}}\Big],
 \end{eqnarray*}

\begin{eqnarray*}
 \sigma &=& |\vec \zeta (\tau ,\vec \sigma )| = {1\over 2}\,
(\tau_{+} - \tau_{-})\,  \sqrt{{{1 + {\vec a}^2 + 2\, \vec a \cdot
{\hat n}_{(\tau_{-})}}\over {1 + {\vec a}^2 - {\vec a}^4 + (3 -
 2\, {\vec a}^2)\, \vec a \cdot {\hat n}_{(\tau_{-}}) }}},\nonumber \\
 {{\zeta^r(\tau ,\vec \sigma )}\over {|\vec \zeta (\tau ,\vec
\sigma )|}} &=& - {{  \sqrt{{\vec a}^2 + \vec a \cdot {\hat
n}_{(\tau_{-})}} + \sqrt{1 + {\vec a}^2 - {\vec a}^4 + (3 - 2\,
{\vec a}^2)\, \vec a \cdot {\hat n}_{(\tau_{-}}) } }\over {\sqrt{1
+ {\vec a}^2 + 2\, \vec a \cdot {\hat n}_{(\tau_{-})} }}}\, {{a^r
+ {\hat n}^r_{(\tau_{-})} }\over {1 - {\vec a}^2}},\nonumber \\
 \sigma^r &=& {\cal G}^r = - {1\over 2}\, (\tau_{+} - \tau_{-})\, \Big(
1 + \sqrt{ {{{\vec a}^2 + \vec a \cdot {\hat n}_{(\tau_{-})}}\over
{1 + {\vec a}^2 - {\vec a}^4 + (3 - 2\, {\vec a}^2)\, \vec a \cdot
{\hat n}_{(\tau_{-}}) }} }\Big)\nonumber \\
 &&(R^{-1})^r{}_s\Big({1\over 2}\,
\Big[\tau_{+} + \tau_{-} + {{\tau_{+} - \tau_{-}}\over {1 - {\vec
a}^2}}\, \sqrt{{{{\vec a}^2 + \vec a \cdot {\hat
n}_{(\tau_{-})}}\over {1 + {\vec a}^2 - {\vec a}^4 + (3 - 2\,
{\vec a}^2)\, \vec a \cdot {\hat n}_{(\tau_{-}})
}}}\Big],\nonumber \\
 &&{1\over 2}\, (\tau_{+} - \tau_{-})\,  \sqrt{{{1 + {\vec a}^2 + 2\,
\vec a \cdot {\hat n}_{(\tau_{-})}}\over {1 + {\vec a}^2 - {\vec
a}^4 + (3 - 2\, {\vec a}^2)\, \vec a \cdot {\hat n}_{(\tau_{-}})
}}} \Big)\quad {{a^s + {\hat n}^s_{(\tau_{-})}}\over {1 - {\vec
a}^2}};
\end{eqnarray*}

iii) for non-inertial trajectories $x^{\mu}(\tau ) = f(\tau )\,
l^{\mu} + \epsilon^{\mu}_r\, g^r(\tau )$ [$\sgn\, [{\dot f}^2(\tau
) - \sum_r\, {\dot g}^r(\tau )\, {\dot g}^r(\tau )] > 0$] the
evaluation of ${\cal E}$ and ${\vec {\cal G}}$ cannot be done
analytically, but only numerically.

\bigskip

Let us now consider an infinitesimal displacement $\delta z^{\mu}
= z^{\mu}(\tau + \delta \tau , \vec \sigma + \delta \vec \sigma )
- z^{\mu}(\tau ,\vec \sigma )$ of $P$ on $\Sigma_{\tau}$ to
$P^{'}$ on $\Sigma_{\tau + \delta \tau}$. The event $P^{'}$ will
receive light signals from the event $Q(\tau_{-} + \delta
\tau_{-})$ on $\gamma$ and will reflect them towards the event
$Q(\tau_{+} + \delta \tau_{+})$ on $\gamma$. Now, using
$\Delta_{\pm}^2 = 0$, we have $\Delta^{'\, \mu}_{\pm} =
\Delta^{\mu}_{\pm} + {\dot x}^{\mu}(\tau_{\pm})\, \delta
\tau_{\pm} - \delta z^{\mu}$ and $\Delta^{'\, 2}_{\pm} = 2\,
\Delta^{\mu}_{\pm}\, [{\dot x}_{\mu}(\tau_{\pm})\, \delta
\tau_{\pm} - \delta z_{\mu}] + (higher\, order\, terms)$. As a
consequence we get (see Ref.\cite{44a})

\bea
 {{\partial \tau_{\pm}}\over {\partial z^{\mu}}} &=& {{\Delta
 _{\pm\, \mu}}\over {\Delta_{\pm} \cdot {\dot x}(\tau_{\pm})}},\nonumber \\
 &&{}\nonumber \\
 with&& \sgn\, \Delta_{+} \cdot \Delta_{-} < 0,\quad
 \sgn\, \dot x(\tau_{+}) \cdot \Delta_{+} > 0,\quad
 \sgn\, \dot x(\tau_{-}) \cdot \Delta_{-} < 0.
 \label{VI13}
 \eea

\bigskip

Since ${{\partial \tau (P)}\over {\partial z^{\mu}}}$ is a
time-like 4-vector orthogonal to $\Sigma_{\tau}$, it must be
proportional to the normal $l^{\mu}$ to the space-like
hyper-planes of the foliation (\ref{IV1}) till now considered. For
a general admissible foliation we have (from $\triangle^2_{-} = 0$
we get $\triangle_{-} \cdot {{\partial \triangle_{-}}\over
{\partial z^{\mu}}} = 0$ and then $\triangle^{\mu}_{-}\,
{{\partial {\hat n}_{\tau_{-}}}\over {\partial z^{\mu}}} = 0$;
instead in general $\triangle^{\mu}_{+}\, {{\partial {\hat
n}_{\tau_{-}}}\over {\partial z^{\mu}}} \not= 0$)

\begin{eqnarray*}
 {{\partial \tau (P)}\over {\partial z^{\mu}}} &=& \Big[ {\cal E}
 + (\tau_{+} - \tau_{-})\, {{\partial {\cal E}}\over {\partial
 \tau_{+}}}\Big]\, {{\partial \tau_{+}}\over {\partial z^{\mu}}} +\nonumber \\
 &+& \Big[ 1 - {\cal E} + (\tau_{+} - \tau_{-})\,  {{\partial
 {\cal E}}\over {\partial \tau_{-}}}  \Big]\, {{\partial \tau_{-}}\over {\partial
 z^{\mu}}} +\nonumber \\
 &+& (\tau_{+} - \tau_{-})\, {{\partial {\cal E}}\over {\partial\,
 {\hat n}_{(\tau_{-})} }}\, {{\partial {\hat n}_{(\tau_{-})}}\over
 {\partial z^{\mu}}}  =\nonumber \\
 &=& \Big[ {\cal E}
 + (\tau_{+} - \tau_{-})\, {{\partial {\cal E}}\over {\partial
 \tau_{+}}}\Big]\,  {{\Delta
 _{+\, \mu}}\over {\Delta_{+} \cdot {\dot x}(\tau_{+})}}
 +\nonumber \\
 &+& \Big[ 1 - {\cal E} + (\tau_{+} - \tau_{-})\, \Big( {{\partial
 {\cal E}}\over {\partial \tau_{-}}} + {{\partial {\cal E}}\over {\partial\,
 {\hat n}_{(\tau_{-})}}}\, {{\partial {\hat n}_{(\tau_{-})}}\over
 {\partial \tau_{-}}}\Big) \Big]\,  {{\Delta
 _{-\, \mu}}\over {\Delta_{-} \cdot {\dot
 x}(\tau_{-})}} +\nonumber \\
 &+& (\tau_{+} - \tau_{-})\, {{\partial {\cal E}}\over {\partial\,
 {\hat n}_{(\tau_{-})} }}\, {{\partial {\hat n}_{(\tau_{-})}}\over
 {\partial z^{\mu}}} ,
 \end{eqnarray*}

 \bea
 \sgn\, \Big({{\partial \tau (P)}\over {\partial z^{\mu}}}\Big)^2
 &=& \sgn\, {{\Delta_{+} \cdot \Delta_{-}}\over {\Delta_{+} \cdot {\dot x}(\tau_{+})
 \, \Delta_{-} \cdot {\dot x}(\tau_{-})}}\, \Big[ {\cal E}
 + (\tau_{+} - \tau_{-})\, {{\partial {\cal E}}\over {\partial
 \tau_{+}}}\Big]\nonumber \\
 &&\Big[ 1 - {\cal E} + (\tau_{+} - \tau_{-})\,  {{\partial
 {\cal E}}\over {\partial \tau_{-}}} \Big] +
 (\tau_{+} - \tau_{-})^2\, \Big({{\partial {\cal E}}\over {\partial\,
 {\hat n}_{(\tau_{-})} }}\Big)^2\, \Big({{\partial {\hat n}_{(\tau_{-})}}\over
 {\partial z^{\mu}}}\Big)^2 +\nonumber \\
 &+& 2\, (\tau_{+} - \tau_{-})\, \Big[ {\cal E}
 + (\tau_{+} - \tau_{-})\, {{\partial {\cal E}}\over {\partial
 \tau_{+}}}\Big]\, {{\partial {\cal E}}\over {\partial\,
 {\hat n}_{(\tau_{-})} }}\,
{{\triangle_{+} \cdot {{\partial {\hat n}_{(\tau_{-})}}\over
 {\partial z^{\mu}}}}\over {\triangle_{+} \cdot {\dot x}(\tau_{+})}}
 > 0,\nonumber \\
 &&{}\nonumber \\
 &&\nonumber\\
  &&\quad for\, every\,\,\, \tau_{-}, \theta_{(\tau_{-})}, \phi_{(\tau_{-})},
 \tau_{+}.
 \label{VI14}
 \eea

\noindent This is the condition on the function ${\cal
E}(\tau_{-}, {\hat n}_{(\tau_{-})}, \tau_{+})$ to have an
admissible foliation.
\medskip

Since $ \sgn\, {{\Delta_{+} \cdot \Delta_{-}}\over
{\Delta_{+} \cdot {\dot x}(\tau_{+})
 \, \Delta_{-} \cdot {\dot x}(\tau_{-})}} > 0$, in the special case
 ${{\partial {\cal E}}\over {\partial {\hat n}_{(\tau_{-})}}} = 0$ it must be

 \bea
 &&\Big[ {\cal E}
 + (\tau_{+} - \tau_{-})\, {{\partial {\cal E}}\over {\partial
 \tau_{+}}}\Big]\,
 \Big[ 1 - {\cal E} + (\tau_{+} - \tau_{-})\,  {{\partial
 {\cal E}}\over {\partial \tau_{-}}} \Big] > 0,\nonumber \\
 &&{}\nonumber \\
 &&\Downarrow\nonumber \\
 &&{}\nonumber \\
 && {\cal E} + (\tau_{+} - \tau_{-})\, {{\partial {\cal E}}\over {\partial
 \tau_{+}}} \not= 0,\qquad
 1 - {\cal E} + (\tau_{+} - \tau_{-})\,  {{\partial
 {\cal E}}\over {\partial \tau_{-}}}  \not= 0.
 \label{VI15}
 \eea

\medskip

Finally the functions ${\cal E}$ and ${\vec {\cal G}}$ must have a
finite limit for $\tau_{\pm}\, \rightarrow\, \pm \infty$, i.e. at
spatial infinity on $\Sigma_{\tau}$.

\bigskip

Given the world-line $x^{\mu}(\tau )$ of an observer $\gamma$ and
four functions $0 < {\cal E}(\tau_{-}, {\hat n}_{(\tau_{-})},
\tau_{+}) <1$ and ${\vec {\cal G}}(\tau_{-}, {\hat
n}_{(\tau_{-})}, \tau_{+})\, {\rightarrow}_{\tau_{+} \rightarrow
\tau_{-}}\, 0$, with ${\cal E}$ satisfying Eq.(\ref{VI14}), we can
build the admissible adapted 4-coordinates $\tau$, $\vec \sigma $
of a $\gamma$-dependent notion of simultaneity, because
Eqs.(\ref{VI14}) and (\ref{VI15}) ensure that the surfaces
$\Sigma_{\tau}$ are space-like since their normal ${{\partial \tau
(P)}\over {\partial z^{\mu}}}$ is everywhere time-like.

\bigskip

To reconstruct the embedding associated to this notion of
simultaneity we must invert the Jacobian matrix $b^A_{\mu} =
{{\partial \sigma^A(P)}\over {\partial z^{\mu}}}$ and find the
matrix  $b_A^{\mu} = {{\partial z^{\mu}(P)}\over {\partial
\sigma^A}}$ satisfying the conditions $b^A_{\mu}\, b^{\mu}_B =
\delta^A_B$, $b^A_{\mu}\, b^{\nu}_A = \delta^{\nu}_{\mu}$. Then by
integrating ${{\partial z^{\mu}(P)}\over {\partial \sigma^A}}$ we
get the associated embedding $x^{\mu} = z^{\mu}(\tau ,\vec \sigma
)$. Let us remark that for $\sigma \rightarrow \infty$ $b^{\mu}_A$
must tend in a direction-independent way to the asymptotic tetrad
$\epsilon^{\mu}_A$ associated to the asymptotic space-like
hyper-planes of any admissible foliation. As a consequence
$b^A_{\mu} = {{\partial \sigma^A(P)}\over {\partial z^{\mu}}}$
must tend to the asymptotic cotetrad $\epsilon^A_{\mu}$. This is a
condition on the admissible functions ${\cal E}$ and ${\vec {\cal
G}}$.

\bigskip

If we call $|{\vec \Delta}_{-}|$ the {\it light distance of
$Q_{-}$ on $\gamma$ to $P$} and $|{\vec \Delta}_{+}|$ the {\it
light distance of $P$ to $Q_{+}$ on $\gamma$ } (see Section 4 of
Ref.\cite{21a}) we get the following two one-way velocities of
light (with $c = 1$) in coordinates adapted to the given notion of
simultaneity

\bea
 c_{-} &=& {{|{\vec \Delta}_{-}|}\over {\tau - \tau_{-}}} =
{{|{\vec \Delta}_{-}|}\over {{\cal E}\, (\tau_{+} - \tau_{-})}}
= {{2\, \eta\, |\vec \Delta | }\over {{\cal E}\, (\tau_{+} -
\tau_{-})}},
 \;\;\qquad from\, Q_{-}\, to\, P,\nonumber \\
 &&{}\nonumber \\
 c_{+} &=& {{|{\vec \Delta}_{+}|}\over {\tau_{+} - \tau}} =
{{|{\vec \Delta}_{+}|}\over {(1 - {\cal E})\, (\tau_{+} -
\tau_{-})}} = {{2\, (1 - \eta )\, |\vec \Delta | }\over {(1 -
{\cal E})\, (\tau_{+} - \tau_{-})}},\;\;\qquad from\, P\, to\, Q_{+},\nonumber \\
 &&{}\nonumber \\
 && |\vec \Delta|\, {\buildrel {def}\over =}\, {1\over 2}\,
  (|{\vec \Delta}_{+}| + |{\vec \Delta}_{-}|),\qquad \eta\,
  {\buildrel {def}\over =}\, {{|{\vec \Delta}_{-}|}\over
  {|{\vec \Delta}_{+}| + |{\vec \Delta}_{-}|}}.
  \label{VI16}
  \eea

\noindent If $c_{\tau} = {{2\, |\vec \Delta|}\over {\tau_{+} -
\tau_{-}}}$ is the isotropic average round-trip $\tau$-coordinate
velocity of light, we get $c_{+} = {{1 - \eta}\over {1- {\cal
E}}}\, c_{\tau}$, $c_{-} = {{\eta}\over {{\cal E}}}\, c_{\tau}$.
\medskip

If $x^{\mu}(\tau )$ is a straight-line (inertial observer) we can
adopt Einstein's convention ${\cal E} = {1\over 2}$, i.e. $\tau
(P) = {1\over 2}\, (\tau_{+} + \tau_{-})$ and $|\vec \sigma | =
|{\vec {\cal G}}| = {1\over 2}\, (\tau_{+} - \tau_{-})$
(hyper-planes orthogonal to the observer world-line). This implies
$|{\vec \Delta}_{+}| = |{\vec \Delta}_{-}|$ and $\eta =
{1\over 2}$.

Instead, if we ask $c_{\tau} = c_{+} = c_{-}$, i.e. {\it isotropy}
of light propagation, we get ${\cal E} = \eta$. This shows that
once we have made a convention on {\it two} of the quantities {\it
spatial distance}, {\it one-way speed of light} and {\it
simultaneity}, the third one is automatically determined
\cite{21a}.

\bigskip

In general relativity on globally hyperbolic space-times, we can
define in a similar way the allowed global notions of simultaneity
and the allowed one-way velocities of test light. Then the
knowledge of the factor ${\cal E}$ associated to an allowed notion
of simultaneity will allow an operational determination of the
4-coordinates $(\tau ,\vec \sigma )$ adapted to the chosen notion
of simultaneity with simultaneity surfaces $\tau = const.$ as
radar coordinates. This is a step towards implementing the
operational definition of space-time proposed in Refs.
\cite{8a,9a}. The lacking ingredient is an operational
confrontation of the tetrads ${}_{(\gamma_i)}E^{\mu}_A(\tau )$
with the tetrad ${}_{(\gamma)}E^{\mu}_A(\tau )$ of the reference
world-line: this would allow a determination of the 4-metric in
the built radar 4-coordinates and a reconstruction of a finite
region of space-time around the $N + 1$ spacecrafts of the GPS
type, whose trajectories are supposed known (for instance
determined with the standard techniques of space navigation
\cite{99a} controlled by a station on the Earth). See
Refs.\cite{100a} for other approaches to GPS type coordinates.
\medskip

However, as we shall comment in the Conclusions, in general
relativity the admissible notions of simultaneity are {\it
dynamically} determined by the Hamilton equations, equivalent to
Einstein's equations, of the ADM canonical formulation of metric
gravity.

\newpage

\subsection{The 3+1 Point of View on the Rotating Disk.}

Let us now consider the 3+1 point of view about the problem of the
{\it rotating disk}, which is still under debate after nearly a
century of proposals for the resolution of the {\it Ehrenfest
paradox} (see Ref.\cite{44ab,54a} for a rich bibliography on the
most relevant points of view on the rotating disk). A basic
ambiguity in the formulation of the problem comes from the
non-relativistic notion of a rigid (either geometrical or
material) disk put in global rigid motion (this is possible due to
the existence of arbitrary non-relativistic rigid rotating
reference frames). At the relativistic level we have:
\medskip

1) Rigid bodies do not exist. At best we can speak of {\it Born
rigid motions} \cite{101a} and {\it Born reference frames}
\footnote{A reference frame or platform is {\it Born-rigid}
\cite{2a} if in Eq.(\ref{II1}) the expansion $\Theta$ and the
shear $\sigma_{\mu\nu}$ vanish, i.e. if the spatial distance
between neighboring world-lines remains constant. }. However
Gr$\o$n \cite{102a} has shown that the acceleration phase of a
material disk is not compatible with Born rigid motions. Moreover,
we do not have a well formulated and accepted relativistic
framework to discuss a relativistic elastic material disk (see
Ref.\cite{103a} for a review), so that many statements in the
literature cannot be checked with actual calculations. As a
consequence most of the authors treating the rotating disk (either
explicitly or implicitly)  consider it as a geometrical entity, to
be identified with a congruence of time-like world-lines (helices,
see Ref.\cite{104a}) with non-zero vorticity \footnote{I.e.
non-surface-forming and therefore non-synchronizable. Therefore
the observers associated to this congruence have neither a notion
of global simultaneity nor a notion of instantaneous 3-space
(since we do not have a preferred observer, we cannot use its
local rest frame as 3-space like it was discussed in footnote 11:
each observer will have a different local rest frame). As shown in
Ref.\cite{54a} the only meaningful concept which can be defined is
an {\it abstract relative 3-space}, i.e. the space whose points
are the time-like world-lines of the congruence. Another problem
is the definition of the {\it rods and clocks} of an observer of
the congruence. As shown in Ref.\cite{54a} the two existing
notions are: a) an {\it optical congruence} with the light null
4-geodesics approximated by 3-geodesics (used in Ref.\cite{54a});
b) a {\it congruence of Sevres meters} (i.e. a measurement of
spatial distances with slowly transported rigid rods by definition
not changing their length under acceleration; it was used by
Einstein \cite{10a}, M$\o$ller \cite{12a}, Landau-Lifschitz
\cite{11a}) with {\it free ends} (instead in Ref.\cite{52a} the
rod was identified with a piece of the rim of the disk). }. But
this corresponds to a model of material disk in which it is
composed of a relativistic perfect fluid with zero pressure, i.e.
to a relativistic dust contained in a cylindrical world-tube in
Minkowski space-time (in the Cartesian 4-coordinates of an
inertial system  the restriction is $r = \sqrt{\sum_r\, (x^r)^2}
\leq R$).
\medskip

2) As we have shown in Section III, Eq.(\ref{III2}), relativistic
rigid rotating reference frames do not exist. Therefore all the
rotating reference frames appearing in the literature are only
locally defined due to the horizon problem, so that the vector
fields defining the relativistic frames are only defined on a
sub-manifold Minkowski space-time containing the disk.
\bigskip

The 3+1 point of view looks at these problems in a different way
and suggests the following re-formulation  of the rotating disk.
Let the disk be a relativistic isolated system (either a
relativistic material body or a relativistic fluid or a
relativistic dust as a limit case \footnote{As an example of a
congruence simulating a geometrical rotating disk we can consider
the relativistic dust described by generalized Eulerian
coordinates of Ref.\cite{105a} after the gauge fixing to a family
of differentially rotating parallel hyper-planes.}) with compact
support always contained in a finite world-tube $W$, which in the
Cartesian 4-coordinates of an inertial system is a time-like
cylinder of radius R. At the initial time the disk support
$D_{\tau = 0}$ is just the circle $W_o$ of radius R in the chosen
inertial system; at subsequent times the support could be
different according to the internal dynamics of the isolated
system. Let us consider a parametrized Minkowski theory
\cite{87a,27a,86a}, namely a Lagrangian whose configuration
variables are those of the isolated system plus the embeddings
$z^{\mu}(\tau ,\vec \sigma )$ describing the allowed 3+1
splittings of Minkowski space-time with the associated notions of
Cauchy and simultaneity surfaces as said in the Introduction.
Since the embeddings $z^{\mu}(\tau ,\vec \sigma )$ are {\it gauge
variables} all the allowed notions of simultaneity are gauge
equivalent. The simultaneity surfaces $\Sigma_{\tau}$ (in general,
but not necessarily, curved Riemannian 3-manifolds embedded in the
flat Minkowski space-time) intersect the world-tube $W$ with
3-dimensional sub-manifolds $D_{\tau} \subset \Sigma_{\tau}$
describing the instantaneous 3-space of the disk at $\tau$
according to this notion of simultaneity.

\medskip

Let us choose a particular embedding $z^{\mu}(\tau ,\vec \sigma
)$, i.e. a well defined notion of simultaneity. The congruence of
time-like (in general non-inertial) observers whose world-lines
are  the integral curves of the vector filed $l^{\mu}(\tau ,\vec
\sigma )$ of unit normals to $\Sigma_{\tau}$ is used to define
{\it rods and clocks} for this notion of simultaneity by slow
transport of those pertaining to the asymptotic inertial observers
at spatial infinity (fixed stars), which are the standard rods and
clocks of inertial systems. Alternatively, we can define the radar
4-coordinates $(\tau ,\vec \sigma )$ with the method of Subsection
A. Therefore on $\Sigma_{\tau}$ we can measure spatial distances
with the 3-metric $g_{rs}$, synchronize distant clocks and define
one-way velocity of light between two simultaneity surfaces as
discussed in Section II.

The second congruence associated to the chosen notion of
simultaneity, whose time-like observers have the integral curves
of the vector field $z^{\mu}_{\tau}(\tau ,\vec \sigma )/
\sqrt{\sgn\, g_{\tau\tau}(\tau ,\vec \sigma )}$ as world-lines
\footnote{They are the lines $\vec \sigma = {\vec \sigma}_o =
const.$, i.e. the generalized helices $x^{\mu}_{{\vec
\sigma}_o}(\tau ) = z^{\mu}(\tau ,{\vec \sigma}_o)$ with ${\dot
x}^{\mu}_{{\vec \sigma}_o}(\tau ) / \sqrt{\sgn\, {\dot x}^2_{{\vec
\sigma}_o}(\tau )} = z^{\mu}_{\tau}(\tau ,{\vec \sigma}_o )/
\sqrt{\sgn\, g_{\tau\tau}(\tau ,{\vec \sigma}_o )}$.}, is used to
define a (in general non-surface-forming, non-synchronizable)
reference frame with translational and rotational accelerations,
whose restriction $u^{\mu}_D(\tau ,\vec \sigma )$ to the
world-tube $W$ \footnote{In general for given a world-tube $W$
there will be a preferred family of adapted embeddings such that
the associated vector fields $u^{\mu}_D(\tau ,\vec \sigma )$ have
the property that their integral lines are contained completely in
the world-tube $W$. In the next Subsection we will study a simple
embedding of the type (\ref{IV1}) with this property. } is the 3+1
counterpart of the local rigid rotating reference frame used in
the treatments of the rotating disk (see for instance
Ref.\cite{106a}).

\medskip

Every notion of simultaneity has associated a different notion of
spatial length, and therefore a different radius and circumference
length will appear at each non-inertial observers, namely the disk
3-geometry will be simultaneity-dependent. But this is natural
because in special relativity the notions of simultaneity and
simultaneous spatial distance are reference-frame-dependent i.e.
observer-dependent. Even if all of them are gauge-equivalent in
parametrized Minkowski theories, there is no useful notion of {\it
gauge equivalence class} (see Ref. \cite{8a} for the analogous
problem in general relativity), because an extended physical
laboratory corresponds to a completely fixed gauge and not to an
equivalence class:  its definition requires a definite choice of
the notion of simultaneity and of a reference observer, endowed
with a tetrad, as origin of the coordinates.

\newpage

\subsection{The Simplest Embedding for a Rotating Disk and the Sagnac Effect.}

Let us now consider the following admissible embedding of the type
(\ref{IV1}), corresponding to a foliation with flat parallel
space-like hyper-planes with normal $l^{\mu}$ (defining inertial
Eulerian observers)

\beq
 z^{\mu}(\tau ,\vec \sigma ) = l^{\mu}\, \tau +
 \epsilon^{\mu}_r\, R^r_{(3)\, s}(\tau ,\sigma )\,
 \sigma^s,
 \label{VI17}
 \eeq

\noindent where

 \bea
 &&R^r_{(3)\, s}(\tau ,\sigma ) = \left( \begin{array}{ccc} \cos\,
 \theta (\tau ,\sigma ) & - \sin\,  \theta (\tau ,\sigma )& 0\\
 \sin\,  \theta (\tau ,\sigma )& \cos\,  \theta (\tau ,\sigma )& 0\\
 0& 0& 1 \end{array} \right),\nonumber \\
 &&{}\nonumber \\
 &&\theta (\tau ,\sigma ) = F(\sigma )\, \omega\, \tau,\quad
 F(\sigma ) < {c\over {\omega\, \sigma}},\nonumber \\
 &&{}\nonumber \\
 &&\Omega^r{}_s(\tau ,\sigma ) = \left(R^{-1}_{(3)}\, {{d R_{(3)}}\over
 {d\tau}}\right){}^r{}_s(\tau ,\sigma ) = \omega\, F(\sigma )\,
 \left( \begin{array}{ccc} 0& -1& 0\\ 1& 0& 0\\ 0& 0& 0
 \end{array} \right),\nonumber\\
 &&\nonumber\\
 &&\Omega (\tau ,\sigma ) = \Omega (\sigma ) = \omega\,
 F(\sigma ).
 \label{VI18}
 \eea

\bigskip

A simple choice for the function $F(\sigma )$,  compatible with
the conditions (\ref{IV14}), is $F(\sigma ) = {1\over {1 +
{{\omega^2\, \sigma^2}\over {c^2}}}}$ \footnote{We have introduced
the explicit $c$ dependence. In the rest of the Section we put $c
= 1$.}, so that at spatial infinity we get $\Omega (\tau ,\sigma )
= {{\omega}\over {1 + {{\omega^2\, \sigma^2}\over {c^2}}}}
\rightarrow_{\sigma \rightarrow \infty}\, 0$. Let us remark that
nearly rigid rotating systems can be obtained by using a function
$F(\sigma )$ approximating the step function $\Theta (R - r)$.

\bigskip

By introducing cylindrical 3-coordinates $r$, $\varphi$, $h$ we
get the following form of the embedding

\bea
 z^{\mu}(\tau ,\vec \sigma ) &=& l^{\mu}\, \tau +
 \epsilon^{\mu}_1\, [\cos\, \theta (\tau ,\sigma )\, \sigma^1 -
 \sin\, \theta (\tau ,\sigma )\, \sigma^2] +\nonumber \\
 &&\nonumber\\
 &+& \epsilon^{\mu}_2\, [\sin\, \theta (\tau ,\sigma )\, \sigma^1 +
 \cos\, \theta (\tau ,\sigma )\, \sigma^2] + \epsilon^{\mu}_3\,
 \sigma^3 =\nonumber \\
&&\nonumber\\
 &=& l^{\mu}\, \tau + \epsilon^{\mu}_1\, r\, \cos\, [\theta (\tau
 ,\sigma ) + \varphi ] + \epsilon^{\mu}_2\, r\, \sin\, [\theta
 (\tau ,\sigma ) + \varphi ] + \epsilon^{\mu}_3\, h,\nonumber \\
 &&{}\nonumber \\
  && \sigma^1 = r\, \cos\, \varphi,\quad \sigma^2 = r\, \sin\,
 \varphi,\quad \sigma^3 = h,\quad \sigma = \sqrt{r^2 + h^2}.
 \label{VI19}
 \eea

\medskip

Then we get

 \bea
  \frac{\partial
z^\mu(\tau,\vec{\sigma})}{\partial\tau}&=&
 z^{\mu}_{\tau}(\tau ,\vec \sigma ) = l^{\mu} - \omega\, r\,
 F(\sigma )\, \Big( \epsilon^{\mu}_1\, \sin\, [\theta
 (\tau ,\sigma ) + \varphi ] - \epsilon^{\mu}_2\, \cos\, [\theta
 (\tau ,\sigma ) + \varphi ]\Big),\nonumber \\
&&\nonumber\\
\frac{\partial z^\mu(\tau,\vec{\sigma})}{\partial\varphi}&=&
 z^\mu_\varphi(\tau,\vec{\sigma})=
-\epsilon^\mu_1\,r\,\sin\,[\theta(\tau,\sigma )+\varphi]
+\epsilon^\mu_2\,r\,\cos\,[\theta(\tau,\sigma )+\varphi]
\nonumber\\
&&\nonumber\\
\frac{\partial z^\mu(\tau,\vec{\sigma})}{\partial r}&=&
 z^\mu_{(r)}(\tau,\vec{\sigma})=
-\epsilon^\mu_1\,\left((\cos\,[\theta(\tau,\sigma )+\varphi]
-\frac{r^2\omega\tau}{\sqrt{r^2+h^2}}\,\frac{dF(\sigma)}{d\sigma}
\,\sin\,[\theta(\tau,\sigma )+\varphi]\right)+\nonumber\\
&&\nonumber\\
&+&\epsilon^\mu_2\,\left(\sin\,[\theta(\tau,\sigma )+\varphi]
+\frac{r^2\omega\tau}{\sqrt{r^2+h^2}}\,\cos\,[\theta(\tau,\sigma )+\varphi]\right)
\nonumber\\
&&\nonumber\\
\frac{\partial z^\mu(\tau,\vec{\sigma})}{\partial h}&=&
 z^\mu_h(\tau,\vec{\sigma})=\epsilon^\mu_3
-\epsilon^\mu_1\,\left(\frac{rh\omega\tau}{\sqrt{r^2+h^2}}\,\frac{dF(\sigma)}{d\sigma}
\,\sin\,[\theta(\tau,\sigma )+\varphi]\right)+\nonumber\\
&&\nonumber\\
&+&\epsilon^\mu_2\,\left(\frac{rh\omega\tau}{\sqrt{r^2+h^2}}\,\frac{dF(\sigma)}{d\sigma}
\,\cos\,[\theta(\tau,\sigma )+\varphi]\right),
 \label{VI20}
 \eea

\noindent where we have used the notation $(r)$ to avoid confusion
with the index $r$  used as 3-vector index (for example in
$\sigma^r$).
\bigskip

Then in cylindrical 4-coordinates $\tau$, $r$, $\varphi$ and $h$
the 4-metric is

 \bea
 \sgn\,g_{\tau\tau}(\tau ,\vec \sigma ) &=& 1 - \omega^2\, r^2\,
 F^2(\sigma ),\nonumber \\
 &&\nonumber\\
 \sgn\,g_{\tau\varphi}(\tau ,\vec \sigma ) &=& -\omega\,r^2\,F(\sigma),\nonumber\\
&&\nonumber\\
 \sgn\,g_{\tau (r)}(\tau ,\vec \sigma ) &=& -\frac{\omega^2\,r^3\,\tau}{\sqrt{r^2+h^2}}
\,F(\sigma)\,\frac{dF(\sigma)}{d\sigma},\nonumber\\
 &&\nonumber\\
 \sgn\,g_{\tau h}(\tau ,\vec \sigma ) &=& -\frac{\omega^2\,r^2\,h\,\tau}{\sqrt{r^2+h^2}}
\,F(\sigma)\,\frac{dF(\sigma)}{d\sigma},\nonumber\\
 &&\nonumber\\
 \sgn\,g_{\varphi\varphi}(\tau ,\vec \sigma ) &=&-r^2,\nonumber\\
 &&\nonumber\\
 \sgn\,g_{(r)(r)}(\tau ,\vec \sigma ) &=&-1-\frac{r^4\,\omega^2\,\tau^2}{r^2+h^2}
\left(\frac{dF(\sigma)}{d\sigma}\right)^2,\nonumber\\
 &&\nonumber\\
 \sgn\,g_{hh}(\tau ,\vec \sigma ) &=&-1-\frac{r^2\,h^2\,\omega^2\,\tau^2}{r^2+h^2}
\left(\frac{dF(\sigma)}{d\sigma}\right)^2,\nonumber\\
 &&\nonumber\\
 \sgn\,g_{(r)\varphi}(\tau ,\vec \sigma ) &=&-\frac{\omega\,r^3\,\tau}{\sqrt{r^2+h^2}}
\,\frac{dF(\sigma)}{d\sigma},\nonumber\\
 &&\nonumber\\
 \sgn\,g_{h\varphi}(\tau ,\vec \sigma ) &=&-\frac{\omega^2\,r^2\,h\,\tau}{\sqrt{r^2+h^2}}
\,\frac{dF(\sigma)}{d\sigma},\nonumber\\
 &&\nonumber\\
 \sgn\,g_{h(r)}(\tau ,\vec \sigma ) &=& - \frac{r^3\,h\,\omega^2\,\tau^2}{r^2+h^2}
\left(\frac{dF(\sigma)}{d\sigma}\right)^2,\nonumber \\
 &&{}\nonumber \\
 &&with\, inverse\nonumber \\
 &&{}\nonumber \\
 \sgn\, g^{\tau\tau}(\tau ,\vec \sigma ) &=& 1,
 \qquad \sgn\, g^{\tau\varphi}(\tau ,\vec \sigma ) = - \omega\,
 F(\sigma ),\nonumber \\
 &&{}\nonumber \\
 \sgn\, g^{\tau (r)}(\tau ,\vec \sigma ) &=& \sgn\, g^{\tau h}(\tau ,\vec \sigma )
 = 0,\qquad \sgn\, g^{(r)(r)}(\tau ,\vec \sigma ) = \sgn\, g^{hh}(\tau ,\vec \sigma )
 = - 1,\nonumber \\
 &&{}\nonumber \\
 \sgn\, g^{\varphi\varphi}(\tau ,\vec \sigma ) &=& - {{1 + \omega^2\, r^2\, [\tau^2\,
 ({{dF(\sigma )}\over {d\sigma}})^2 - F^2(\sigma )}\over
 {r^2}},\nonumber \\
 &&{}\nonumber \\
 \sgn\, g^{\varphi (r)}(\tau ,\vec \sigma ) &=& {{\omega\, r\,
 \tau}\over {\sqrt{r^2 + h^2}}}\, {{dF(\sigma )}\over
 {d\sigma}},\qquad \sgn\, g^{\varphi h}(\tau ,\vec \sigma ) = {{\omega\, h\,
 \tau}\over {\sqrt{r^2 + h^2}}}\, {{dF(\sigma )}\over
 {d\sigma}}.
 \label{VI21}
 \eea
\bigskip

It is easy to observe that the congruence of (non inertial)
observers defined by the 4-velocity field

 \beq
  {{z^{\mu}_\tau(\tau
,\vec \sigma )}\over {\sqrt{\sgn\, g_{\tau\tau}(\tau ,\vec \sigma
)}}} = {{l^{\mu} - \omega\, r\, F(\sigma )\, \Big(
\epsilon^{\mu}_1\, \sin\, [\theta
 (\tau ,\sigma ) + \varphi ] - \epsilon^{\mu}_2\, \cos\, [\theta
 (\tau ,\sigma ) + \varphi ]\Big)}\over
{1 - \omega^2\, r^2\, F^2(\sigma )}},
 \label{VI22}
  \eeq

\noindent has the observers moving along the world-lines

 \bea
 &&x^{\mu}_{{\vec\sigma}_o}(\tau ) = z^{\mu}(\tau ,{\vec \sigma}_o)
= \nonumber\\ &&\nonumber\\ &=&l^{\mu}\, \tau + r_o\, \Big(
\epsilon^{\mu}_1\, \cos\, [\omega\, \tau\, F(\sigma_o) +
\varphi_o] + \epsilon^{\mu}_2\, \sin\, [\omega\, \tau\,
F(\sigma_o) + \varphi_o]\Big) + \epsilon^{\mu}_3\, h_o.
 \label{VI23}
  \eea

The world-lines (\ref{VI23}) are labeled by their initial value
$\vec{\sigma}={\vec \sigma}_o=(\varphi_o,r_o,h_o)$ at $\tau = 0$.
\bigskip

In particular for $h_o=0$ and $r_o=R$ these world-lines are
helices on the {\em  cylinder} in the Minkowski space

 \bea
 &&\epsilon_3^\mu\,z_\mu=0,\qquad
\left(\epsilon_1^\mu\,z_\mu\right)^2+\left(\epsilon_2^\mu\,z_\mu\right)^2=R^2,
 \nonumber \\
 &&{}\nonumber \\
 && or \qquad r=R,\qquad h=0.
   \label{VI24}
    \eea

These helices are defined the equations $\varphi=\varphi_o,\,
r=R,\,h=0$ if expressed in the embedding adapted coordinates
$\varphi,r,h$. Then the congruence of observers (\ref{VI22}),
defined by the foliation (\ref{VI17}), defines on the cylinder
(\ref{VI24}) the {\em rotating observers} usually assigned to the
rim of a rotating disk, namely  observes running along the helices
$ x^{\mu}_{{\vec\sigma}_o}(\tau ) = l^{\mu}\, \tau + R\, \Big(
\epsilon^{\mu}_1\, \cos\, [\Omega(R)\, \tau\, + \varphi_o] +
\epsilon^{\mu}_2\, \sin\, [\Omega(R)\, \tau\,  + \varphi_o]\Big)$
after having put $\Omega(R)\equiv\omega\,F(R)$.
\bigskip

On the cylinder (\ref{VI24}) the line element is obtained from the
line  element $ds^2$ for the metric (\ref{VI21}) by putting
$dh=dr=0$ and $r=R$, $h=0$. Therefore the cylinder line element is

 \beq
  \sgn\,(ds_{cyl})^2 = \Big[1 - \omega^2\, R^2\,
F^2(R)\Big]\, (d\tau )^2 - 2\, \omega\, R^2\, F(R)\, d\tau
d\varphi - R^2\, (d\varphi )^2.
 \label{VI25}
  \eeq
\medskip

We can define the light rays on the cylinder, i.e. the null curves
on it, by solving the equation

\begin{equation}
 \sgn\,(ds_{cyl})^2=(1-R^2\,\Omega^2(R))\,d\tau^2-2\,R^2\,\Omega(R)\,d\tau\,d\varphi-
R^2\,d\varphi^2=0,
 \label{VI26}
 \end{equation}

\noindent which implies

\begin{equation}
 R^2\,\left(\,\frac{d\varphi(\tau)}{d\tau}\,\right)^2+
2\,R^2\,\Omega(R)\,\left(\,\frac{d\varphi(\tau)}{d\tau}\,\right)-
(1-R^2\,\Omega(R))=0.
 \label{VI27}
 \end{equation}

\bigskip

The two solutions

\begin{equation}
\frac{d\varphi(\tau)}{d\tau}=\pm\,\frac{1}{R}-\Omega(R),
 \label{VI28}
 \end{equation}

\noindent define the world-lines on the cylinder for {\em
clockwise or anti-clockwise} rays of light.

\begin{equation}
 \begin{array}{l}
\Gamma_1:\qquad \varphi (\tau ) - \varphi_o =
\left(+\frac{1}{R}-\Omega(R)\right)\,\tau ,\\ \\
 \Gamma_2:\qquad
\varphi (\tau ) - \varphi_o =
\left(-\frac{1}{R}-\Omega(R)\right)\,\tau
\end{array}.
 \label{VI29}
 \end{equation}

\bigskip

This is the {\em geometric origin} of the {\em Sagnac Effect}.
Since $\Gamma_1$ describes the world-line of the ray of light
emitted at $\tau=0$ by the rotating observer $\varphi=\varphi_o$
in the increasing sense of $\varphi$ (anti-clockwise), while
$\Gamma_2$ describes that of the ray of light emitted at $\tau=0$
by the same observer in the decreasing sense of $\varphi$
(clockwise), then the two rays of light will be re-absorbed by the
same observer at {\it different $\tau$-times} \footnote{Sometimes
the {\em proper time of the rotating observer} is used: $d{\cal
T}_o=d\tau\sqrt{1-\Omega^2(R)\,R^2}$.} $\tau_{(\pm\, 2\pi ) }$,
whose value, determined by the two conditions $\varphi
(\tau_{(\pm\, 2\pi )}) - \varphi_o= \pm\, 2\pi$, is

\begin{equation}
 \begin{array}{l}
\Gamma_1:\qquad \tau_{(+2\pi )} =
\frac{2\pi\,R}{1-\Omega(R)\,R},\\
\\
\Gamma_2:\qquad \tau_{(-2\pi )}=\frac{2\pi\,R}{1+\Omega(R)\,R}.
\end{array}
 \label{VI30}
 \end{equation}

The {\it time difference} between the re-absorption of the two
light rays is

\begin{equation}
\Delta\tau = \tau_{(+2\pi )} - \tau_{(-2\pi)} =
\frac{4\pi\,R^2\,\Omega(R)}{1-\Omega^2(R)\,R^2} =
\frac{4\pi\,R^2\,\omega\, F(R)}{1-\omega^2\, F^2(R)\,R^2},
 \label{VI31}
 \end{equation}

\noindent and it corresponds to the phase difference named the
{\em Sagnac effect}

\begin{equation}
 \Delta\Phi=\Omega\,\Delta\tau .
 \label{VI32}
 \end{equation}
\medskip

We see that we recover the standard result if we take a function
$F(\sigma )$ such that $F(R) = 1$. In the non-relativistic
applications, where $F(\sigma ) \rightarrow 1$, the correction
implied by admissible relativistic coordinates is totally
irrelevant.

\bigskip

Till now we have described the Sagnac effect by using the $\tau$
time coordinate associated to the notion of simultaneity
(\ref{IV1}). Let us now compare it with the notions of
synchronization for the rotating observers based on the use of the
world-lines (\ref{VI29}) for the two light rays. This will be done
by using the notions of synchronizations of type A) and B)
introduced in Section II. Then we will study the associated
notions of spatial distance of type A) or B) by evaluating the
radius and the circumference of the rotating disk in the two
cases. \bigskip

Let us consider a reference observer $(\varphi_o = const., \tau )$
and another one $(\varphi = const. \not= \varphi_o, \tau )$. If
$\varphi > \varphi_o$ we use the notation $(\varphi_R, \tau )$,
while for $\varphi < \varphi_o$ the notation $(\varphi_L, \tau )$
with $\varphi_R - \varphi_o = - (\varphi_L - \varphi_o)$.

Let us consider the two rays of light $\Gamma_{R\, -}$ and
$\Gamma_{L\, -}$, emitted in the right and left directions at the
event $(\varphi_o,\tau_{-})$ on the rim of the disk and received
at $\tau$ at the events $(\varphi_R,\tau)$ and $(\varphi_L,\tau )$
respectively. Both of them are  reflected towards the reference
observer, so that we have two rays of light $\Gamma_{R\, +}$ and
$\Gamma_{L\, +}$ which will be absorbed at the event
$(\varphi_o,\tau_{+})$.

By using Eq.(\ref{VI30}) for the light propagation, we get

\bea
 &&\begin{array}{l} \Gamma_{R\, -}:\qquad (\varphi  -
\varphi_o) = \frac{1-R\Omega(R)}{R}\,(\tau - \tau_{-}),\\
\\
\Gamma_{R\, +}:\qquad (\varphi  - \varphi_o) =
\frac{1+R\Omega(R)}{R}\,(\tau_{+} - \tau),
\end{array}\nonumber \\
 &&{}\nonumber \\
 &&\begin{array}{l}
\Gamma_{L\, -}:\qquad (\varphi  - \varphi_o) =
-\frac{1+R\Omega(R)}{R}\,(\tau - \tau_{-}),\\
\\
\Gamma_{L\, +}:\qquad (\varphi  - \varphi_o) =
-\frac{1-R\Omega(R)}{R}\,(\tau_{+} - \tau).
\end{array}
 \label{VI33}
 \eea

\medskip

Eqs.(\ref{II17}),(\ref{II18}) define the following local
synchronization of type B) in a neighborhood of the observer
$(\varphi_o ,\tau )$ [$(\varphi ,\tau )$ is an observer in the
neighborhood]

\begin{equation}
 c\, \Delta\,\widetilde{\cal T}= \sqrt{1-R^2\,\Omega^2(R)}\,\Delta
\tau_E= \sqrt{1-R^2\,\Omega^2(R)}\,\Delta \tau-
\frac{R^2\,\Omega^2(R)}{\sqrt{1-R^2\Omega^2(R)}} \,\Delta \varphi.
 \label{VI34}
 \end{equation}
\medskip

Let us see what happens if we try to extend this local
synchronization of type B) to a global one for  two distant
observers $(\varphi_o, \tau )$ and $(\varphi ,\tau )$ in the form
of an Einstein convention for $\tau_E$ (the result is the same
both for $\varphi = \varphi_R$ and $\varphi = \varphi_L$)

\begin{equation}
\tau_E=\frac{1}{2}\,(\tau_{+} + \tau_{-}) = \tau-
\frac{R^2\Omega(R)}{1-R^2\,\Omega^2(R)}(\varphi  - \varphi_o).
 \label{VI35}
 \end{equation}

This is contradictory because the curves defined by
$\tau_E=constant$ are {\it not closed}, since they are helices
that assign the {\it same time} $\tau_E$ to different events on
the world-line of an observer $\varphi_o=constant$. For example
\[
(\varphi_o,\tau)\;\;\;
\mbox{ and }\;\;\;
\left(\varphi_o,\tau+2\pi
\,\frac{R^2\Omega(R)}{1-R^2\,\Omega^2(R)}\right)
\]
are on the same helix $\tau_E=constant$.
\medskip

This {\em desynchronization effect or synchronization gap} is only
the expression of the fact that the observers of the rotating disk
congruence with non-zero vorticity  are not globally
synchronizable, i.e. that the B) synchronization holds only
locally in the form (\ref{VI34}) \footnote{See Ref.\cite{58a} for
a derivation of the Sagnac effect in an inertial system by using
Einstein's synchronization in the locally comoving inertial frames
on the rim of the disk and by asking for the equality of the
one-way velocities in opposite directions.}. As a consequence {\it
usually a discontinuity in the synchronization of clocks is
accepted and taken into account} (see Ref.\cite{59a} for the GPS).
{\it Instead, with an admissible notion of simultaneity, all the
clocks on the rim of the rotating disk lying on a hyper-surface
$\Sigma_{\tau}$ are automatically synchronized}.
\bigskip

The synchronization of the type A) is defined by the condition
$\tau=const.$ and can be built with a generalized operative
procedure as discussed in  Subsection A. In fact by
Eqs.(\ref{VI33}) we can calculate $\tau$ and $\varphi$ as function
on $\tau_{\pm}$ e $n = \pm$ [$n = +$ for $\varphi = \varphi_R$, $n
= -$ for $\varphi = \varphi_L$; $n$ replace ${\hat
n}_{(\tau_{-})}(\theta_{(\tau_{-})}, \phi_{(\tau_{-})})$ of
Eq.(\ref{VI11})] and obtain the following modification of
Einstein's convention for radar time

\bea
 \tau(\tau_{-}, n, \tau_{+}) &=&
\frac{1}{2}(\tau_{+} + \tau_{-}) - \frac{n\,R\,\Omega(R)}{2}\,
(\tau_{+} - \tau_{-})\, {\buildrel {def}\over =}\, \tau_{-} +
 {\cal E}(\tau_{-}, n, \tau_{+})\, (\tau_{+} - \tau_{-}),\nonumber \\
 &&{}\nonumber \\
 with &&\qquad {\cal E}(\tau_{-}, n, \tau_{+}) = {{1 - n\, R\,
 \Omega (R)}\over 2},\qquad \Omega (R) = \omega\, F(R).
 \label{VI36}
 \eea
\bigskip

Let us now consider the evaluation of the radius and the
circumference of the rotating disk with the A) and B) notions of
spatial distance.

\bigskip

With the convention A)  on the hyper-surfaces (\ref{VI17}) we use
the 3-metric $- \sgn\, g_{su}\, (\tau ,u=(r),\varphi,h)$ given by
Eq.(\ref{VI21}). With this metric the length of the circumference

\beq
 \epsilon^\mu_3\,z_\mu=0,\quad
(\epsilon^\mu_1\,z_\mu)^2+(\epsilon^\mu_2\,z_\mu)^2=R,\qquad
 or \qquad h=0,\quad r=R,
 \label{VI37}
 \eeq

\noindent is

\begin{equation}
C=\int_0^{2\pi}\,d\varphi\,\sqrt{-\sgn\,g_{\varphi\varphi}}=2\pi\,R,
 \label{VI38}
 \end{equation}

\bigskip

The curve

\begin{equation}
 h=0,\qquad \varphi=\varphi_o=constant,
 \label{VI39}
 \end{equation}

\noindent has the length

\begin{equation}
\widetilde{R}=\int_0^R\,dr\,\sqrt{-\sgn\,g_{(r)(r)}}=\int_0^R\,dr\,
\sqrt{1-r^2\,\omega^2\,\tau^2 \left(\frac{dF(r)}{dr}\right)^2},
 \label{VI40}
 \end{equation}

\noindent which is equal to $R$ only at $\tau = 0$.

\medskip

However the curve (\ref{VI39}) does  not decribe a ray of the
circumference (\ref{VI37}). The curve (\ref{VI39}) has the
parametric representation with parameter $r$

\beq \epsilon^\mu_3\,z_\mu = 0,\qquad
 \epsilon^\mu_1\,z_\mu =
 r\cos[\varphi_o+\omega\,\tau\,F(r)],\qquad
\epsilon^\mu_2\,z_\mu = r\sin[\varphi_o+\omega\,\tau\,F(r)],
 \label{VI41}
 \eeq

\noindent while a ray of the circumference (\ref{VI37}) has the
parametric representation

\bea
 \epsilon^\mu_3\,z_\mu &=& 0,\nonumber \\
 &&{}\nonumber \\
\epsilon^\mu_1\,z_\mu &=& r\cos\varphi_o=
r\cos[\varphi(\tau,r)+\omega\,\tau\,F(r)],\nonumber \\
 &&{}\nonumber \\
\epsilon^\mu_2\,z_\mu &=& r\sin\varphi_o = r\sin[\varphi(\tau,r) +
\omega\,\tau\,F(r)], \nonumber \\
 &&{}\nonumber \\
 or && h=0,\qquad r=\lambda ,\qquad \varphi (\tau ) =
 \varphi_o - \omega\,\tau\,F(\lambda).
 \label{VI42}
 \eea

The length of the true ray (\ref{VI42}) is [remarkably this result
is independent of the function $F(r)$]

\begin{equation}
\int_0^R\,d\lambda\sqrt{- \sgn\,\left[
g_{(r)(r)}\left(\frac{dr}{d\lambda}\right)^2+
2g_{(r)\varphi}\left(\frac{dr}{d\lambda}\right)\left(\frac{d\varphi}{d\lambda}\right)+
g_{\varphi\varphi}\left(\frac{d\varphi}{d\lambda}\right)^2\right]}=R.
 \label{VI43}
 \end{equation}

\bigskip

Moreover let us observe that the 3-dimensional tensor of curvature
obtained by the 3-metric $g_{rs}$ is null, ${}^3R_{suvw}=0$, since
the rotating coordinates $r,\varphi,h$ can be obtained by a
$\tau$-dependent coordinate transformation by the Cartesian
3-coordinate on the (flat) hyper-plane (\ref{VI17}):
$\zeta^r=\epsilon^r_\mu\,z^\mu$. Therefore, with the admissible
notion of simultaneity (\ref{VI17}) the 3-geometry of every slice
of the rotating disk contained in the simultaneity surfaces
$\Sigma_{\tau}$ is Euclidean.

\bigskip

On the contrary with the convention B), used by most authors but
without the function $F(r )$ ensuring an admissible foliation, we
have to use the 3-metric

\begin{equation}
{}^3\gamma_{uv} = - \sgn\, \Big(g_{uv} - \frac{g_{\tau u}\,g_{\tau
v}}{g_{\tau\tau}}\Big).
 \label{VI44}
 \end{equation}

Since on the plane $h=0$ we get (note the $\varphi$-independence
and also the $\tau$-independence of
${}^3\gamma_{\varphi\varphi}\Big|_{h=0}$)

\begin{eqnarray}
{}^3\gamma_{\varphi\varphi}\Big|_{h=0}&=&\sgn\,\left[
-g_{\varphi\varphi}+\frac{g_{\tau \varphi}\,g_{\tau
\varphi}}{g_{\tau\tau}}
\right]_{h=0}=\frac{r^2}{1-r^2\Omega^2(r)},\nonumber\\
 &&\nonumber\\
  {}^3\gamma_{(r)(r)}\Big|_{h=0}&=&\sgn\,\left[
-g_{(r)(r)}+\frac{g_{\tau (r)}\,g_{\tau (r)}}{g_{\tau\tau}}
\right]_{h=0}=1+\frac{r^2\omega^2\tau^2}{1-r^2\Omega^2(r)}
\left(\frac{dF(r)}{dr}\right)^2,\nonumber\\
 &&\nonumber\\
{}^3\gamma_{\varphi(r)}\Big|_{h=0}&=&\sgn\,\left[
-g_{\varphi(r)}+\frac{g_{\tau \varphi}\,g_{\tau
(r)}}{g_{\tau\tau}} \right]_{h=0}=\frac{\omega \tau
r^2}{1-r^2\Omega^2(r)}\frac{dF(r)}{dr},
 \label{VI45}
 \end{eqnarray}

\medskip

\noindent we obtain the following length for the circumference
(\ref{VI37})

\begin{equation}
C'=\int_0^{2\pi}\,d\varphi\,\sqrt{{}^3\gamma_{\varphi\varphi}}
=\frac{2\pi\,R}{\sqrt{1-R^2\Omega^2(R)}},
 \label{VI46}
 \end{equation}

\noindent while the length of the ray (\ref{VI42}) is [this result
is independent from $F(r)$]

\begin{equation}
\int_0^R\,d\lambda\sqrt{
{}^3\gamma_{(r)(r)}\left(\frac{dr}{d\lambda}\right)^2+
2{}^3\gamma_{(r)\varphi}\left(\frac{dr}{d\lambda}\right)
\left(\frac{d\varphi}{d\lambda}\right)+
{}^3\gamma_{\varphi\varphi}\left(\frac{d\varphi}{d\lambda}\right)^2}=R.
 \label{VI47}
 \end{equation}

\noindent The metric ${}^3\gamma_{uv}$ defines a curvature tensor
${}_\gamma R_{suvw}\neq 0$ (see Ref.\cite{54a}). Therefore, a
non-Euclidean 3-geometry for the rotating disk is obtained if we
approximate the instantaneous 3-space of the disk with anyone of
the local rest frames of the observers of the congruence with
non-zero vorticity (the abstract relative space of Ref.\cite{54a})
on the rim of the disk (Eq.(\ref{VI45}) is $\varphi$-independent).

\newpage

\subsection{Relativistic Theory for Time and Frequency Transfer.}

As a further application of the admissible foliations let us
consider the problem of the evaluation of the time and frequency
transfers \cite{79a} from an Earth station $B$ and a satellite
$A$, because it is relevant for the ACES ESA project on
synchronization of clocks \cite{76a}, which needs corrections of
order $c^{-3}$ due to the level of accuracy in time keeping ($5
\cdot 10^{-17}$ in fractional frequency or $5\, ps$ in time
transfer) reached with laser cooled atomic clocks.
\medskip

As we shall see the ACES mission can be re-interpreted as a
determination of the function ${\cal E}$ of Eq.(\ref{VI12}),
measuring the deviation from Einstein's convention (${\cal E} =
{1\over 2}$), which is associated to a choice of the notion of
simultaneity compatible with admissible differentially rotating
4-coordinates taking into account the rotation of the Earth.
Since, as we have seen, such a choice is one of the conventions
defining an enlarged laboratory in special relativity, it has to
be done {\it a priori} and in the most convenient way.

\bigskip

The existing calculation of these quantities \cite{79a,59a} has
been done in the non-inertial (non-rotating) Geocentric Celestial
Reference Frame \cite{3a} (see footnote 1) considered as an
inertial frame in free fall in post-Newtonian gravity
\footnote{The line element is modified to take into account
post-Newtonian gravitational effects in a suitable harmonic
4-coordinate system {\it identified} with an inertial geocentric
Cartesian (non-rotating) coordinate system. Post-Newtonian gravity
is needed for the evaluation \cite{79a} of photon world-lines and
Shapiro time delay. Strictly speaking, given the post-Newtonian
4-metric, Eistein's convention is not compatible with it and one
should replace the inertial system with an admissible
(non-Cartesian) radar 4-coordinate system generating it, in
analogy with Eq.(\ref{III7}) for M$\o$ller rotating 4-metric. This
radar 4-coordinate system, and its notion of simultaneity, should
then be modified to take into account Earth rotation along the
lines presented in this Subsection.} and uses the hyper-planes of
constant geocentric coordinate time as notion of simultaneity. For
attempts to re-formulate the problem in non-inertial frames,
taking into account the rotation of the Earth, see
Refs.\cite{80a}, especially the first one where for the first time
there is an application of the PPN formalism to the time transfer
problem with estimates of the effects of Earth multipoles to the
ACES project.

\medskip

Let us first review the approach of Ref.\cite{79a}. If $x^{\mu}$
are Cartesian inertial Geocentric 4-coordinates, with the notion
of simultaneity based on the hyper-planes $x^o = c\, t = const.$
(Einstein's convention), the world-line of the Earth station $B$
is parametrized as $x^{\mu}_B(t) = (x^o_B = c\, t; {\vec x}_B(t)\,
)$, while the world-line of the satellite $A$ is $x^{\mu}_A(t) = (
x^o_A = c\, t; {\vec x}_A(t)\, )$. The basic quantity to be
evaluated with the simultaneity $x^o = const.$ is the {\it one-way
time transfer}. If at $t = t_A$ the satellite $A$ emits an
electro-magnetic signal, its reception at the Earth station $B$
will happen at time $t_B > t_A$ such that $\triangle^2_{AB} = (x_A
- x_B)^2 = \sgn\, [c^2\, (t_A - t_B)^2 - {\vec \triangle}^2_{AB}]
= 0$ with ${\vec \triangle}_{AB} = {\vec x}_A(t_A) - {\vec
x}_B(t_B)\, {\buildrel {def}\over =}\, R_{AB}\, {\hat N}_{AB}$,
${\hat N}_{AB}^2 = 1$ (we use a notation like in Ref.\cite{79a}
for comparison). Then in the flat Minkowski space-time limit we
get

\beq
 T_{AB} = t_B - t_A = {1\over c}\, R_{AB}.
 \label{VI48}
 \eeq

Since in real experiments the position ${\vec x}_B(t_A)$ of the
Earth station at the emission time is better known than the
position ${\vec x}_B(t_B)$ at the reception time, the quantity
$R_{AB}$ has to be re-expressed in terms of the {\it
instantaneous} (in the sense of the simultaneity $x^o = const.$)
distance ${\vec D}_{AB} = {\vec x}_A(t_A) - {\vec x}_B(t_A)$,
$D_{AB} = |{\vec D}_{AB}|$. To order $c^{-3}$ we get \cite{79a}

\bea
 R_{AB} &=& | {\vec x}_A(t_A) - {\vec x}_B(t_B)| =
 \sqrt{\Big[{\vec D}_{AB} + {\vec v}_B(t_A)\, R_{AB} + {1\over 2}\,
 {\vec a}_B(t_A)\, R^2_{AB} +O(R^3_{AB}) \Big]^2},\nonumber \\
 &&{}\nonumber \\
 &&\Downarrow\nonumber \\
 &&{}\nonumber \\
 R_{AB} &=& D_{AB} + {1\over c}\, {\vec D}_{AB} \cdot {\vec
 v}_B(t_A) + \nonumber \\
 &+& {1\over {c^2}}\, D_{AB}\, \Big[ {\vec v}_B^2(t_A) +
 {{({\vec D}_{AB} \cdot {\vec v}_B(t_A)}\over {D^2_{AB}}} +
 {\vec D}_{AB} \cdot {\vec a}_B(t_A)\Big]+
  O({1\over {c^3}}),\nonumber \\
 &&{}\nonumber \\
 && {\vec v}_B(t) = {{d {\vec x}_B(t)}\over {dt}},\qquad {\vec
 a}_B(t) = {{d^2 {\vec x}_B(t)}\over {dt^2}}.
 \label{VI49}
 \eea

\bigskip

Finally post-Newtonian gravity contributes with the Shapiro time
delay \cite{79a}, so that the final result is ($M$ is the Earth
mass)

\bea
 T_{AB}(t_A) &=& {1\over c}\, R_{AB} + {{2\, G\, M}\over {c^3}}\, ln\,
 {{|{\vec x}_A(t_A)| + |{\vec x}_B(t_B)| + R_{AB}}\over
 {|{\vec x}_A(t_A)| + |{\vec x}_B(t_B)| - R_{AB}}} =\nonumber \\
 &=&{1\over c}\, D_{AB} + {1\over {c^2}}\, {\vec D}_{AB} \cdot {\vec
 v}_B(t_A) + {1\over {c^3}}\, D_{AB}\, \Big[ {\vec v}_B^2(t_A) +
 {{({\vec D}_{AB} \cdot {\vec v}_B(t_A))^2}\over {D^2_{AB}}} +
 {\vec D}_{AB} \cdot {\vec a}_B(t_A)\Big] +\nonumber \\
 &+& {{2\, G\, M}\over {c^3}}\, ln\,
 {{|{\vec x}_A(t_A)| + |{\vec x}_B(t_A)| + D_{AB}}\over
 {|{\vec x}_A(t_A)| + |{\vec x}_B(t_A)| - D_{AB}}} + O({1\over
 {c^4}})
 \label{VI50}
 \eea

\medskip

The two terms in $T_{AB}$ beyond $D_{AB}/c$ are usually referred
to as the {\it Sagnac terms} of first ($1/c^2$) and second
($1/c^3$) order due the rotations of the Earth and the satellite
(see Ref.\cite{78a} for their derivation by using a standard
non-admissible rotating frame). In the inertial system Earth
rotation is simulated with a term $\omega^2_E\, |{\vec x}_B(t_A)|$
in the acceleration ${\vec a}_B(t_A)$.

\bigskip

In Ref.\cite{79a}, after stating that the experimental uncertainty
in time of ACES will be at the level of $5\, ps$, there is an
estimate, at low elevation of a satellite at $400\, Km$ of
altitude, of $200\, ns$ for the first order Sagnac term, of $11\,
ps$ for the Shapiro time delay and of $5\, ps$ for the second
order Sagnac term.

\bigskip

If we consider a signal emitted at $t_{B^{'}}$ by the Earth
station, reflected at $t_A$ from the satellite and re-absorbed at
$t_B$ by the Earth station and if $T_{AB} = t_B - t_A$ and
$T_{B^{'}A} = t_A - t_{B^{'}}$ are the two one-way time transfers,
then for the two-way process we get

\beq
 t_A = t_{B^{'}} + E\, (t_B - t_{B^{'}}),\quad with\quad E =
 {1\over 2}\, \Big( 1 + T_{B^{'}A} - T_{AB}\Big).
 \label{VI51}
 \eeq

\noindent By measuring $t_{B^{'}}$ and $t_B$ with the atomic clock
of the Earth station and by using a theoretical determination of
the two one-way transfers with the simultaneity $x^o = const.$ it
should be possible to check whether in post-Newtonian gravity
Einstein's convention ($E = {1\over 2}$ and $T_{B^{'}} = T_{AB}$)
holds or is modified. However {\it a priori these calculations
depend on the chosen notion of simultaneity} and may change going
to a non-inertial frame taking into account Earth's rotation.
\medskip

For the determination of the one- and two-way frequency transfer
see Ref.\cite{79a}.

\bigskip
\bigskip

Let us now see what happens if we consider a good notion of
simultaneity, of the type (\ref{IV1}), adapted to the rotation of
the Earth, i.e. admissible transformations $x^{\mu} \mapsto
\sigma^A$ from the Cartesian geocentric inertial coordinates
$x^{\mu}$ to intrinsic (radar-type) coordinates such that the
inverse transformation $\sigma^A \mapsto x^{\mu}$ defines the
embedding

\bea
 x^{\mu} &=& z^{\mu}(\tau ,\vec \sigma ) = x^{\mu}_o + l^{\mu}\,
 \tau + \epsilon^{\mu}_r\, \zeta^r(\tau ,\vec \sigma ),\nonumber \\
 &&{}\nonumber \\
 &&\zeta^r(\tau ,\vec \sigma ) = R_E^r{}_s(\tau ,\sigma )\,
 \sigma^s,
 \label{VI52}
 \eea

\noindent where $x^{\mu}(\tau ) = x^{\mu}_o + l^{\mu}\, \tau$ is
the world-line of the center of mass of the Earth (origin of the
3-coordinates $\vec \sigma$) and $\epsilon^{\mu}_A = (
\epsilon^{\mu}_{\tau } = l^{\mu}; \epsilon^{\mu}_r)$ is an
asymptotic tetrad determined by the fixed stars. Let us remark
that with a suitable $x^{\mu}(\tau ) = x^{\mu}_o + l^{\mu} f(\tau
) + \epsilon^{\mu}_r\, g^r(\tau ) $, with ${\dot f}^2(\tau ) >
\sum_r\, {\dot g}^r(\tau )\, {\dot g}^r(\tau )$, we could describe
the (non-inertial) motion of the center of the Earth with respect
to the (quasi) inertial Solar System Barycentric Celestial
Reference System (see footnote 1): in this case $x^{\mu} \mapsto
\sigma^A$ would be an admissible coordinate transformation from
such an inertial system to an intrinsic coordinated system adapted
to both the linear acceleration and the rotational motions of the
Earth.
\medskip

Since the intrinsic  coordinates are adapted to the motions of the
Earth, the Earth station $B$ has fixed intrinsic 3-coordinates

\bea
 \eta^r_B &=& R_B\, {\hat \eta}^r_B = const.,\qquad {\hat {\vec
 \eta}}_B^2 = 1,\nonumber \\
 &&{}\nonumber \\
 x^{\mu}_B(\tau ) &=& z^{\mu}(\tau ,{\vec \eta}_B) = x^{\mu}(\tau
 ) + \epsilon^{\mu}_r\, \zeta^r(\tau ,{\vec \eta}_B),\nonumber \\
 &&{}\nonumber \\
 \zeta^r(\tau ,{\vec \eta}_B) &=& R^r_{E\, s}(\tau ,R_B)\, R_B\,
 {\hat \eta}_B^s.
 \label{VI53}
 \eea

The matrix $R_E(\tau ,\sigma )$ is a rotation matrix such that
$R_E(\tau ,R_B)$ takes into account the effects of the rotation,
precession and nutation of the Earth at the position $B$ of the
Earth station through its three Euler angles. By ignoring
precession and rotation, the effect of the rotation of the Earth
is parametrized by means of the matrix (corresponding to a
rotation around the third axis)

\bea
 R_{EB}(\tau ) &{\buildrel {def}\over =}& R_E(\tau ,R_B) =
 \left( \begin{array}{ccc}
\cos\Omega_B\,\tau&-\sin\Omega_B\,\tau& 0\\
\sin\Omega_B\,\tau&\cos\Omega_B\,\tau& 0\\ 0&0&1
\end{array} \right),\nonumber \\
 &&{}\nonumber \\
 \Omega_B &=& \omega_E\, F(R_B) = const.,\qquad \theta_B(\tau ) =
 \Omega_B\, \tau .
 \label{VI54}
 \eea

\noindent If we normalize the gauge function $F(\sigma )$ so that
$F(R_B) = 1$, we get $\Omega_B = \omega_E$, where $\omega_E$ is
the angular velocity of the Earth assumed constant. A possible
choice for $F(\sigma )$, respecting Eqs.(\ref{IV14}) and such that
$F(R_B) = 1$, is
\medskip

\beq
 F(\sigma ) = {{1 + \omega^2_E\, R^2_B}\over {1 + \omega^2_E\, \sigma^2}}\,\,
 <\,\, {2\over {1 + \omega^2_E\, \sigma^2}}\,\, <\,\, {1\over
 {\omega_E\, \sigma}},
 \label{VI55}
 \eeq

\noindent since $\omega_E\, R_B < 1$ ($c = 1$).

\bigskip

Since they are needed later on, we give also the velocity and the
acceleration of the Earth station

\bea
 {\dot x}^{\mu}_B(\tau ) &=& {\dot x}^{\mu}(\tau ) + \epsilon^{\mu}_r\,
 {\dot \zeta}^r(\tau ,{\vec \eta}_B) = {\dot x}^{\mu}(\tau ) +
 \epsilon^{\mu}_r\, {\dot R}^r_{EB\, s}(\tau )\, R_B\, {\hat
 \eta}^s_B,\nonumber \\
 &&{}\nonumber \\
 {\ddot x}^{\mu}_B(\tau ) &=& {\ddot x}^{\mu}(\tau ) + \epsilon^{\mu}_r\,
 {\ddot \zeta}^r(\tau ,{\vec \eta}_B) = {\ddot x}^{\mu}(\tau ) +
 \epsilon^{\mu}_r\, {\ddot R}^r_{EB\, s}(\tau )\, R_B\, {\hat
 \eta}^s_B,\nonumber \\
 &&{}\nonumber \\
 {\dot R}_{EB}(\tau ) &=& \Omega_B\,
  \left( \begin{array}{ccc}
- \sin\Omega_B\,\tau&-\cos\Omega_B\,\tau& 0\\ \cos\Omega_B\,\tau&-
\sin\Omega_B\,\tau& 0\\ 0&0&0
\end{array} \right),\quad
 {\ddot R}_{EB}(\tau ) = \Omega^2_B\,
  \left( \begin{array}{ccc}
- \cos\Omega_B\,\tau&\sin\Omega_B\,\tau& 0\\ -
\sin\Omega_B\,\tau&- \cos\Omega_B\,\tau& 0\\ 0&0&0
\end{array} \right).\nonumber \\
 &&{}\nonumber \\
 &&{}
 \label{VI56}
 \eea

For inertial motion of the Earth, $x^{\mu}(\tau ) = x^{\mu}_o +
l^{\mu}\, \tau$, we have ${\dot x}^{\mu}(\tau ) = l^{\mu}$ and
${\ddot x}^{\mu}(\tau ) = 0$.

\bigskip

The adapted intrinsic 3-coordinates of the satellite $A$ are
${\vec \eta}_A(\tau ) = R_A(\tau )\, {\hat \eta}_A(\tau )$, ${\hat
\eta}^2_A(\tau ) = 1$. They are deduced from the assumed known
satellite world-line parametrized with $\tau$, i.e.
$x^{\mu}_A(\tau ) = (c\, t(\tau ); {\vec x}_A(t(\tau ))\, ) =
z^{\mu}(\tau ,{\vec \eta}_A(\tau ))$, by using Eq.(\ref{VI8}). Now
we have

\bea
 x^{\mu}_A(\tau ) &=& z^{\mu}(\tau ,{\vec \eta}_A(\tau )) =
 x^{\mu}(\tau ) + \epsilon^{\mu}_r\, \zeta^r(\tau ,{\vec
 \eta}_A(\tau )),\nonumber \\
 &&{}\nonumber \\
 \zeta^r(\tau , {\vec \eta}_A(\tau )) &=& R^r_{EA}{}_s(\tau )\,
 R_A(\tau )\, {\hat \eta}^s_A(\tau ),\nonumber \\
 &&{}\nonumber \\
 R_{EA}(\tau ) &=& R_E(\tau ,R_A(\tau )) = \left(
\begin{array}{ccc}
\cos\Omega_A(\tau )\,\tau&-\sin\Omega_A(\tau )\,\tau&0\\
\sin\Omega_A(\tau )\,\tau&\cos\Omega_A(\tau )\,\tau&0\\ 0&0&1
\end{array} \right),\nonumber \\
 &&{}\nonumber \\
 \Omega_A(\tau ) &=& \omega_E\, F(R_A(\tau )),\qquad \theta_A(\tau
 ) = \Omega_A(\tau )\, \tau .
 \label{VI57}
 \eea

\medskip

If we put $F(R_A(\tau )) = 1 + G(R_A(\tau ))$, we get for the
angular velocity of the satellite

\beq
  \Omega_A(\tau ) = \omega_E + \omega_A(\tau ),\qquad \omega_A(\tau
 ) = \omega_E\, G(R_A(\tau )),
 \label{VI58}
 \eeq

\noindent and Eq.(\ref{VI55}) implies $\omega_A(\tau ) =
\omega_E\, {{\omega_E^2\, [R^2_B - R^2_A(\tau )]}\over {1 +
\omega_E^2\, R^2_A(\tau )}}$.

\bigskip

With this notion of simultaneity we have $\triangle^{\mu}_{AB} =
x^{\mu}(\tau_B) - x^{\mu}(\tau_A) + \epsilon^{\mu}_r\,
[\zeta^r(\tau_B, {\vec \eta}_B) - \zeta^r(\tau_A, {\vec
\eta}_A(\tau_A)]$ when $x^{\mu}(\tau_B) - x^{\mu}(\tau_A) =
l^{\mu}\, (\tau_B - \tau_A)$ and $\triangle_{AB} = 0$ implies

\bea
 {\cal T}_{AB} &=& {1\over c}\, (\tau_B - \tau_A) = {1\over c}\,
 {\cal R}_{AB},\nonumber \\
 &&{}\nonumber \\
 {\cal R}_{AB} &=& {\cal R}_{AB}(\tau_A, \tau_B) = |\vec \zeta (\tau_B,
 {\vec \eta}_B) - \vec \zeta (\tau_A, {\vec \eta}_A(\tau_A)|.
 \label{VI59}
 \eea
 \medskip

If the Earth follows an non-inertial world-line $x^{\mu}(\tau )$
in the Solar System Barycentric Celestial Reference Frame,
$\triangle_{AB}^2 = 0$ implies $f(\tau_B) - f(\tau_A) = |\vec
g(\tau_B) - \vec g(\tau_A) + \vec \zeta (\tau_B, {\vec \eta}_B) -
\vec \zeta (\tau_A, {\vec \eta}_A(\tau_A)|$ and the discussion is
much more complicated and in general can be done only numerically.

\bigskip

To find the analogue of Eq.(\ref{VI49}), we introduce the
instantaneous distance ${\vec {\cal D}}_{AB} = {\vec {\cal
D}}_{AB}(\tau_A) = \vec \zeta (\tau_A, {\vec \eta}_B) - \vec \zeta
(\tau_A, {\vec \eta}_A(\tau_A)$ with ${\cal D}_{AB} = |{\vec {\cal
D}}_{AB} | = |\vec \zeta (\tau_A, {\vec \eta}_B) - \vec \zeta
(\tau_A, {\vec \eta}_A(\tau_A)|$ and we make a Taylor expansion

\bea
 \vec \zeta (\tau_B, {\vec \eta}_B) &=& \vec \zeta (\tau_A, {\vec
 \zeta}_B) + {\dot {\vec \zeta}}(\tau_A, {\vec \eta}_B)\, c\,
 {\cal T}_{AB} + {1\over 2}\, {\ddot {\vec \zeta}}(\tau_A ,{\vec
 \eta}_B)\, c^2\, {\cal T}_{AB}^2 + O(c^3\, {\cal T}^3_{AB}),\nonumber \\
 &&{}\nonumber \\
 &&{\dot \zeta}^r(\tau_A, {\vec \eta}_B)\, {\buildrel
 {def}\over =}\, {{\partial \vec \zeta (\tau ,{\vec \eta}_B)}\over
 {\partial \tau}}{|}_{\tau = \tau_A} = {{\partial R^r_{E s}(\tau ,R_B)}
 \over {\partial \tau}}{}_{\tau = \tau_A}\,\, R_B\, {\hat \eta}^s_B,\nonumber \\
 &&{\ddot \zeta}^r(\tau_A, {\vec \eta}_B)\, {\buildrel
 {def}\over =}\, {{\partial^2 \vec \zeta (\tau ,{\vec \eta}_B)}\over
 {\partial \tau^2}}{|}_{\tau = \tau_A} = {{\partial^2 R^r_{E s}(\tau ,R_B)}
 \over {{\partial \tau}^2}}{}_{\tau = \tau_A}\,\, R_B\, {\hat \eta}^s_B,\nonumber \\
 &&{}\nonumber \\
 {\cal R}_{AB} &=& \sqrt{[{\vec {\cal D}}_{AB} + {\dot {\vec \zeta}}(\tau_A
 ,{\vec \eta}_B)\, {\cal R}_{AB} + {1\over 2}\, {\ddot {\vec \zeta}}(\tau_A,
 {\vec \eta}_B)\, {\cal R}^2_{AB} + O({\cal R}^3_{AB})]^2}.
 \label{VI60}
 \eea

Therefore we get ($\tau = c\, t$)

\bea
 {\cal T}_{AB}(\tau_A) &=& {1\over c}\,
 {\cal R}_{AB} = {1\over c}\, {\cal D}_{AB} + {1\over {c}}\, {\vec {\cal D}}_{AB} \cdot
 {\dot {\vec \zeta}}(\tau_A, {\vec \eta}_B) +\nonumber \\
 &+& {1\over {2\, c}}\, {\cal D}_{AB}\, \Big[{\dot {\vec \zeta}}^2(\tau_A,
 {\vec \eta}_B) + {{({\vec {\cal D}}_{AB} \cdot {\dot {\vec
 \zeta}}(\tau_A, {\vec \eta}_B))^2}\over {{\cal D}^2_{AB}}} +
 {\vec {\cal D}}_{AB} \cdot {\ddot {\vec \zeta}}(\tau_A, {\vec
 \eta}_B)\Big] +\nonumber \\
 &+& {{2\, G\, M}\over {c^3}}\, ln\, {{R_B + R_A(\tau_A) + {\cal D}_{AB}}
 \over {R_B + R_A(\tau_A) - {\cal D}_{AB}}}
 + O(c^{-4}).
 \label{VI61}
 \eea
\medskip

For $\omega_E\, \rightarrow\, 0$, Eq.(\ref{VI61}) becomes
Eq.(\ref{VI50}).

\medskip

To evaluate explicitly this expression, let us introduce the
matrix

\bea
 {\cal R}_{EAB}(\tau_A,\tau_B)&=&R^t_{EA}(\tau_A)R_{EB}(\tau_B)=\left(
\begin{array}{ccc}
\cos[\Omega_B\,\tau_B-\Omega_A(\tau_A)\, \tau_A
]&-\sin[\Omega_B\,\tau_B-\Omega_A(\tau_A)\, \tau_A] &0\\
\sin[\Omega_B\,\tau_B-\Omega_A(\tau_A)\, \tau_A
]&\cos[\Omega_B\,\tau_B-\Omega_A(\tau_A)\, \tau_A] &0\\ 0&0&1
\end{array}
\right) ,\nonumber \\ &&{}
 \label{VI62}
\eea

\noindent which allows to get the following result

\begin{equation}
{\cal R}^2_{AB} = R_A^2(\tau_A) + R_B^2 - 2\, \eta^r_A(\tau_A)\,
\sigma^s_B\, [{\cal R}_{EAB}(\tau_B,\tau_A)]_{rs}.
 \label{VI63}
 \end{equation}

If we introduce  the cylindrical rotating coordinates

\begin{eqnarray}
\sigma^1_B&=&\,r_B\,\cos\varphi_B,\qquad
 \sigma^2_B=\,r_B\,\sin\varphi_B,\qquad
 \sigma^3_B=h_B,\nonumber\\
  &&\nonumber\\
 \eta^1_A(\tau)&=&\,r_A(\tau)\,\cos\varphi_A(\tau),\qquad
\eta^2_A(\tau)=\,r_A(\tau)\,\sin\varphi_A(\tau),\qquad
 \eta^3_A(\tau)=h_A(\tau),\nonumber\\
  &&\nonumber\\
 \Rightarrow&& R_B=\sqrt{r_B^2+h_B^2},\qquad
 R_A(\tau)=\sqrt{r_A^2(\tau)+h_A^2(\tau)},
  \label{VI64}
\end{eqnarray}

\noindent then Eq.(\ref{VI59}) implies

\begin{eqnarray}
{\cal R}^2_{AB}&=&r_A^2(\tau_A)+r_B^2-2\,r_A(\tau_A)\,r_B\,
\cos[\varphi_B+\Omega_B\tau_B-\varphi_A(\tau_A)-\Omega_A(\tau_A)\,
\tau_A ] + \nonumber\\
 &&\nonumber\\
  &+&(h_A(\tau_A)-h_B)^2.
   \label{VI65}
\end{eqnarray}

\medskip

If we put this expression in Eq.(\ref{VI60}), then with a
straightforward calculation we obtain the following form of
Eq.(\ref{VI61}) [for $\omega_E\, \rightarrow 0$ it gives
Eq.(\ref{VI50}) in cylindrical coordinates]

\bea
 {\cal T}_{AB}(\tau_A) &=& {1\over c}\, {\cal R}_{AB} =\nonumber \\
 &&{}\nonumber \\
 &=& {1\over c}\, \sqrt{r^2_A(\tau_A) + r^2_B + [h_A(\tau_A) - h_B]^2
 + 2\, r_A(\tau_A)\, r_B\, cos\, [(\Omega_B - \Omega_A)\, \tau_A +
 \varphi_B - \varphi_A(\tau_A)]} -\nonumber \\
 &&{}\nonumber \\
 &-& {1\over {c^2}}\, r_A(\tau_A)\, r_B\, \Omega_B\, sin\,
 [(\Omega_B - \Omega_A)\, \tau_A + \varphi_B - \varphi_A(\tau_A)]
 +\nonumber \\
 &&{}\nonumber \\
 &+& {1\over {2\, c^3}}\, \Big( -
  \sqrt{r^2_A(\tau_A) + r^2_B + [h_A(\tau_A) - h_B]^2
 + 2\, r_A(\tau_A)\, r_B\, cos\, [(\Omega_B - \Omega_A)\, \tau_A +
 \varphi_B - \varphi_A(\tau_A)]}\nonumber \\
 &&\Big[ r_A(\tau_A)\, r_B\, \Omega^2_B\, cos\,
 [(\Omega_B - \Omega_A)\, \tau_A + \varphi_B - \varphi_A(\tau_A)]
 \Big] +\nonumber \\
 &+&{{3\,  r_A(\tau_A)\, r_B\, \Omega^2_B\, sin^2\,
  [(\Omega_B - \Omega_A)\, \tau_A + \varphi_B - \varphi_A(\tau_A)]}
  \over {\sqrt{r^2_A(\tau_A) + r^2_B + [h_A(\tau_A) - h_B]^2
 + 2\, r_A(\tau_A)\, r_B\, cos\, [(\Omega_B - \Omega_A)\, \tau_A +
 \varphi_B - \varphi_A(\tau_A)]}}} \Big) -\nonumber \\
 & -& {{2\, G\, M}\over {c^3}}\,
  ln \,{{R_B + R_A(\tau_A) + K }\over
 {R_B + R_A(\tau_A) - K }} + O({1\over {c^4}}). \nonumber \\
 &&{}\nonumber \\
 K &=& \sqrt{r^2_A(\tau_A) + r^2_B + [h_A(\tau_A) - h_B]^2
 + 2\, r_A(\tau_A)\, r_B\, cos\, [(\Omega_B - \Omega_A)\, \tau_A +
 \varphi_B - \varphi_A(\tau_A)]}.\nonumber \\
 &&{}
 \label{VI66}
 \eea
\medskip

The admissibility of the notion of simultaneity introduces an
explicit dependence on the function $F(\sigma ) = 1 +
G(R_A(\tau_A))$ of Eq.(\ref{IV1})in the difference of the angular
velocities of the Earth station and of the satellite

\bea
 &&\Omega_B - \Omega_A(\tau_A) = - \omega_A(\tau_A) = - \omega_E\,
 G(R_A(\tau_A)),\nonumber \\
 &&{}\nonumber \\
 &&\Downarrow\quad if\quad F(\sigma ) = {{1 + {{\omega_E^2}\over {c^2}}\, R^2_B}
 \over {1 + {{\omega_E^2}\over {c^2}}\, \sigma^2}}\, {\rightarrow}_{c \rightarrow \infty}\,
 1 + {{\omega^2_E\, (R^2_B - \sigma^2)}\over {c^2}} + O({1\over
 {c^4}}),\nonumber \\
 &&{}\nonumber \\
 &&\Omega_B - \Omega_A(\tau_A) =  \omega_E\, {{{{\omega_E^2}\over {c^2}}\,
 [R^2_A(\tau_A) - R^2_B]}\over {1 + {{\omega_E^2}\over {c^2}}\,
 R^2_A(\tau_A)}}\,
 = {{\omega^3_E}\over {c^2}}\, [R^2_A(\tau_A) - R^2_B] +O({1\over {c^4}}).
 \label{VI67}
 \eea
\medskip

As a consequence with this notion of simultaneity, for $t_A =
{{\tau_A}\over c} < {{\varphi_B - \varphi_A(\tau_A)}\over {c\,
[\Omega_B - \Omega_A(\tau_A)]}} = {{c\, [\varphi_B -
\varphi_A(\tau_A)]}\over {\omega_E^3\, [R^2_A(\tau_A) - R^2_B]}} +
O({1\over {c^3}})$ (it is an implicit restriction on $\tau_A$) we
get

\bea
  {\cal T}_{AB}(\tau_A) &=& {1\over c}\, {\cal R}_{AB} =\nonumber \\
  &&{}\nonumber \\
 &=& {1\over c}\, \sqrt{r^2_A(\tau_A) + r^2_B + [h_A(\tau_A) - h_B]^2
 + 2\, r_A(\tau_A)\, r_B\, cos\, [
 \varphi_B - \varphi_A(\tau_A)]} -\nonumber \\
 &&{}\nonumber \\
 &-& {1\over {c^2}}\, r_A(\tau_A)\, r_B\, \Omega_B\, sin\,
 [ \varphi_B - \varphi_A(\tau_A)]
 +\nonumber \\
 &&{}\nonumber \\
 &+& {1\over {2\, c^3}}\, \Big( -
  \sqrt{r^2_A(\tau_A) + r^2_B + [h_A(\tau_A) - h_B]^2
 + 2\, r_A(\tau_A)\, r_B\, cos\, [
 \varphi_B - \varphi_A(\tau_A)]}\nonumber \\
 &&\Big[ r_A(\tau_A)\, r_B\, \Omega^2_B\, cos\,
 [ \varphi_B - \varphi_A(\tau_A)] \Big] +\nonumber \\
 &+&{{3\,  r^2_A(\tau_A)\, r^2_B\, \Omega^2_B\, sin^2\,
  [ \varphi_B - \varphi_A(\tau_A)]}
  \over {\sqrt{r^2_A(\tau_A) + r^2_B + [h_A(\tau_A) - h_B]^2
 + 2\, r_A(\tau_A)\, r_B\, cos\, [
 \varphi_B - \varphi_A(\tau_A)]}}} -\nonumber \\
 &-& {{\omega^3_E\, r_A(\tau_A)\, r_B\, [r^2_A(\tau_A)
  - r^2_B + h^2_A(\tau_A) - h^2_B]\, sin\,
  [\varphi_B - \varphi_A(\tau_A)]}\over
 {\sqrt{r^2_A(\tau_A) + r^2_B + [h_A(\tau_A) - h_B]^2
 + 2\, r_A(\tau_A)\, r_B\, cos\, [
 \varphi_B - \varphi_A(\tau_A)]}}}\,\, \tau_A \Big) -\nonumber \\
 & -& {{2\, G\, M}\over {c^3}}\,
  ln\, {{R_B + R_A(\tau_A) + \tilde K }\over
 {R_B + R_A(\tau_A) - \tilde K }} + O({1\over {c^4}}), \nonumber \\
 &&{}\nonumber \\
 \tilde K &=& \sqrt{r^2_A(\tau_A) + r^2_B + [h_A(\tau_A) - h_B]^2
 + 2\, r_A(\tau_A)\, r_B\, cos\, [ \varphi_B - \varphi_A(\tau_A)]}.
 \label{VI68}
 \eea

\bigskip

As already said the effect of the Earth rotation is contained in
the second order Sagnac term. With the choice (\ref{VI67}) of the
admissible notion of simultaneity there is an effect of order
$\tau_A\, \omega^3_E / c^3$.
\medskip

If we put the values $h_A = h_B =0$, $r_B = 6.4\, 10^3\, Km$, $r_A
- r_B \approx 400 Km$, $r_A(\tau_A) \approx const. = r_A = 4.1\,
10^5\, Km$, $\omega_E/c = 7.3 \times 10^{-5}\, radian/s$ [see
Ref.\cite{79a} and the last of Refs.\cite{80a}], so that $\Omega_B
- \Omega_A(\tau_A) = {{\omega_E^3}\over {c^2}}\, (r^2_A - r^2_B) +
O({1\over {c^4}}) = const. {\buildrel {def}\over =}\, \omega_{AB}$
, we get

\bea
 for&& \tau_A < {{\varphi_B - \varphi_A(\tau_A)}\over
 {\omega_{AB}}},\nonumber \\
 &&{}\nonumber \\
 &&{}\nonumber \\
 {\cal T}_{AB}(\tau_A) &\approx& {{r_A}\over c}\, \Big[ 1 -
 {{\omega_E\, r_B\, sin\, \alpha (\tau_A)}\over c} -\nonumber \\
 &&{}\nonumber \\
 &-& {{\omega_E^2\, r_A\, r_B}\over {2c^2}}\, \Big([cos\, \alpha
 (\tau_A) + \omega_E\, \tau_A\, sin\, \alpha (\tau_A)] - 3\,
 {{r_B}\over {r_A}}\, sin^2 \alpha(\tau_A)\Big) \Big]
 -\nonumber \\
 &&{}\nonumber \\
 &-& {{2G\, M}\over {c^3}}\, ln\, {{1 + \sqrt{1 + 2\,
 {{r_B}\over {r_A}}\, cos\, \alpha (\tau_A) + ({{r_B}\over {r_A}})^2}}
 \over {1 - \sqrt{1 + 2\,
 {{r_B}\over {r_A}}\, cos\, \alpha (\tau_A) + ({{r_B}\over {r_A}})^2}}}
 + O({1\over {c^4}}),\nonumber \\
 &&{}\nonumber \\
 &&\alpha (\tau_A) = \varphi_B - \varphi_A(\tau_A),\nonumber \\
 &&{}\nonumber \\
 &&{}\nonumber \\
 for&&  \tau_A \geq {{\varphi_B - \varphi_A(\tau_A)}\over
 {\omega_{AB}}},\nonumber \\
 &&{}\nonumber \\
 &&{}\nonumber \\
  {\cal T}_{AB}(\tau_A) &=& {{K}\over c} - {{\omega_E}\over
  {c^2}}\, r_A\, r_B\, sin\, (\omega_{AB}\, \tau_A + \varphi_B -
  \varphi_A(\tau_A)) +\nonumber \\
  &+& {{r_A\, r_B\, \omega_E^2}\over {c^3}}\, \Big[ - K\,
  cos\, (\omega_{AB}\, \tau_A + \varphi_B -
  \varphi_A(\tau_A)) + {3\over K}\,  sin^2
  (\omega_{AB}\, \tau_A + \varphi_B -
  \varphi_A(\tau_A))\Big] -\nonumber \\
  &-&{{2G\, M}\over {c^3}}\, ln\, {{r_B + r_A + K}\over {r_B + r_A
  - K}} + O({ 1\over {c^4}}),\nonumber \\
 &&{}\nonumber \\
 K &=& \sqrt{r^2_A + r^2_B + 2\, r_A\, r_B\,
 cos\, (\omega_{AB}\, \tau_A + \varphi_B -
  \varphi_A(\tau_A))}.
 \label{VI69}
 \eea
\medskip

The second order Sagnac term varies from $-{{\omega^2_E\, r^2_A\,
r_B}\over {2c^3}}$ for $\alpha (\tau_A) = 0$ (where the first
order Sagnac term is very small) to $-{{\omega^2_E\, r^2_A\,
r_B}\over {2c^3}}\, \omega_E\, \tau_A$ (with a linear dependence
on $\tau_A$) for $\alpha (\tau_A) = {{\pi}\over 2}$; for $tg\,
\alpha (\tau_A) = 1/ \omega_E\, \tau_A$ it is reduced by a factor
$r_B/r_A$.

\bigskip

Let us remark that the contribution of Earth rotation changes with
$F(\sigma )$, i.e. with the choice of the notion of simultaneity
in the admissible family (\ref{IV1}) (or in more general
admissible embeddings). As a consequence, to apply consistently
the formalism one should first of all to establish a grid of
admissible radar 4-coordinates around the Earth with the method of
Subsection A and then estimate the implied effect of the Earth
rotation.

\medskip

Therefore given an admissible notion of simultaneity and
admissible rotating coordinates with a fixed $F(\sigma )$, by
measuring $\tau_B$ and $\tau_{B^{'}}$ and by using $\tau_A =
\tau_{B^{'}} + {\cal E}\, (\tau_B - \tau_{B^{'}})$ we have to find
the resulting [$F(\sigma )$-dependent] ${\cal E}$ from the
equation

\bea
 {\cal E} &{\buildrel {def}\over =}& {1\over 2}\, \Big[1 + c\,
 {\cal T}_{B^{'}A}(\tau_A) - c\, {\cal T}_{AB}(\tau_A)\Big]
 =\nonumber \\
 &&{}\nonumber \\
 &=& {1\over 2}\, \Big[ 1 + c\, {\cal T}_{B^{'}A}(\tau_{B^{'}} +
 {\cal E}\, [\tau_B - \tau_{B^{'}}]) - c\, {\cal T}_{AB}(\tau_{B^{'}} +
 {\cal E}\, [\tau_B - \tau_{B^{'}}]) \Big].
 \label{VI70}
 \eea

\medskip
In conclusion we have first to make a convenient convention on the
notion of simultaneity and a choice of $F(\sigma )$ for the
rotating coordinates, then evaluate the one-way time transfer with
it and finally use the ACES mission to check if the measured
deviation ${\cal E} \not= {1\over 2}$ from Einstein's convention
is just the one implied by the chosen $F(\sigma )$.

\bigskip

Admissible notions of simultaneity like that of Eq.(\ref{IV1})
should be useful also for the description of the optical one-way
time transfer among the three spacecrafts of LISA project
\cite{107a} for the detection of gravitational waves. Since the
spacecrafts follow heliocentric orbits (forming an approximate
equilater triangle), in Eq.(\ref{IV1}) $x^{\mu}(\tau )$ should be
the straight world-line of the Sun in the Solar System Barycentric
Celestial Reference Frame. Since the main problem of the LISA time
delay interferometry is the elimination of the laser phase noise,
introduced by the Doppler tracking scheme used to track the
spacecrafts with laser beams, and since the actual rotating and
flexing configuration produced by the spacecraft orbits makes this
task difficult, it is worthwhile to investigate whether the
presence of the arbitrary function $F(\sigma )$ of Eq.(\ref{IV1})
in the one-way time transfers could help in the reduction of
noise. If there would be a reduction for special forms of
$F(\sigma )$, i.e. for special admissible notions of simultaneity,
then one could try to implement such notions by establishing a
grid of suitable radar 4-coordinates like in Subsection A.

\bigskip

Finally Eq.(\ref{IV1}) should also be instrumental for very long
baseline interferometric (VLBI) experiments, reviewed in
Ref.\cite{108a}.

\newpage

\subsection{Maxwell Equations in Non-Inertial Reference Frames.}

The description of the electro-magnetic field as a parametrized
Minkowski theory  has been given in Ref.\cite{87a} (see also the
Appendix of Ref.\cite{86a}). The configuration variables are the
admissible embeddings $z^{\mu}(\tau ,\vec \sigma )$ (tending to
space-like hyper-planes at spatial infinity) and the
Lorentz-scalar electro-magnetic potential $A_A(\tau ,\vec \sigma )
= z^{\mu}_A(\tau ,\vec \sigma )\, {\tilde A}_{\mu}(z(\tau ,\vec
\sigma ))$ (these potentials know the simultaneity surfaces
$\Sigma_{\tau}$), whose associated field strength is $F_{AB}(\tau
,\vec \sigma ) = \partial_A\, A_B(\tau ,\vec \sigma ) -
\partial_B\, A_A(\tau ,\vec \sigma ) = z^{\mu}_A(\tau ,\vec \sigma
)\, z^{\nu}_B(\tau ,\vec \sigma )\, {\tilde F}_{\mu\nu}(z(\tau
,\vec \sigma ))$.
\medskip

The Lagrangian density ($g = |det\, g_{AB}|$, $g_{AB} =
z^{\mu}_A\, \eta_{\mu\nu}\, z^{\nu}_B$)
\medskip

\bea
 {\cal L}(\tau ,\vec \sigma )
 &=& - {1\over 4}\, \sqrt {g(\tau ,\vec \sigma )}\,
g^{{A}{C}}(\tau ,\vec \sigma )\, g^{{B}{D}} (\tau ,\vec \sigma )\,
F_{{A}{B}}(\tau ,\vec \sigma )\, F_{{C}{D}}(\tau ,\vec \sigma ),
\nonumber \\
 &&{}\nonumber \\
 &&\Rightarrow S = \int d\tau d^3\sigma\, {\cal L}(\tau ,\vec
 \sigma ) = \int d\tau\, L(\tau ),
 \label{VI71}
 \eea

\noindent leads to the canonical momenta \footnote{$\gamma =
|det\, g_{rs}|$, $E_r = F_{r\tau}\, (= -\sgn\, E^r\, on\,
hyper-planes)$, $B_r = {1\over 2}\, \epsilon_{ruv}\, F_{uv}$.}

\begin{eqnarray*}
 \pi^{\tau}(\tau ,\vec \sigma )&=&{ {\partial
 {\cal L}(\tau ,\vec \sigma )}\over {\partial \partial_{\tau}
A_{\tau}(\tau ,\vec \sigma )} }=0,\nonumber \\
 &&{}\nonumber \\
 \pi^{r}(\tau
,\vec \sigma )&=&{ {\partial {\cal L}(\tau ,\vec \sigma )}\over
{\partial \partial _{\tau}A_{r}(\tau ,\vec \sigma )} }=-{ {\gamma
(\tau ,\vec \sigma )} \over {\sqrt {g(\tau ,\vec \sigma )}}
}\gamma^{{r}{s}}(\tau , \vec \sigma )\, (F_{\tau {\ s}} - g_{\tau
{v}}\, \gamma^{{v} {u}}\, F_{{u}{s}})(\tau ,\vec \sigma )
=\nonumber \\
 &=&{ {\gamma (\tau ,\vec
\sigma )}\over {\sqrt {g(\tau ,\vec \sigma )}} }\, \gamma^{{
r}{s}}(\tau ,\vec \sigma )\, (E_{s}(\tau ,\vec \sigma ) + g_{\tau
{v}}(\tau ,\vec \sigma )\, \gamma^{{v}{u}} (\tau ,\vec \sigma )\,
\epsilon_{{u}{s}{t}}\, B_{t} (\tau ,\vec \sigma )),\nonumber \\
 &&{}\nonumber \\
 \rho_{\mu}(\tau ,\vec \sigma )&=&-{ {\partial {\cal L}(\tau ,\vec \sigma )}
\over {\partial z^{\mu}_{\tau}(\tau ,\vec \sigma )} }=\nonumber \\
 &&{}\nonumber \\
 &=& { {\sqrt {g(\tau ,\vec \sigma )}}\over 4}\, [(g^{\tau \tau}\, z_{\tau \mu} +
 g^{\tau {r}}\, z_{r\mu})(\tau ,\vec \sigma )\,
 g^{{A}{C}} (\tau ,\vec \sigma )\, g^{{
B}{D}}(\tau ,\vec \sigma )\, F_{{A} {B}}(\tau ,\vec \sigma )\,
F_{{C}{D}}(\tau ,\vec \sigma ) -\nonumber \\
 &&{}\nonumber \\
 &-&2\, [z_{\tau \mu}(\tau ,\vec \sigma )\, (g^{A\tau}\, g^{\tau
C}\,  g^{{B}{D}} + g^{{A}{C}}\, g^{B\tau}\, g^{\tau  D})(\tau
,\vec \sigma ) +\nonumber \\
 &&{}\nonumber \\
 &+&z_{\check r\mu}(\tau ,\vec \sigma )\, (g^{{
A}{r}}\, g^{\tau {C}} + g^{{A}\tau}\, g^{{r} {C}})(\tau ,\vec
\sigma )\, g^{{B}{D}} (\tau ,\vec \sigma )]\, F_{{A}{ B}}(\tau
,\vec \sigma )\, F_{{C}{D}}(\tau ,\vec \sigma )] =
 \end{eqnarray*}

 \bea
 &=&[(\rho_{\nu}\, l^{\nu})\, l_{\mu} + (\rho_{\nu}\, z^{\nu}_{
 r})\, \gamma^{{r} {s}}\, z_{{s}\mu}](\tau ,\vec \sigma ),\nonumber \\
 &&{}\nonumber \\
 &&\{ z^{\mu}(\tau ,\vec \sigma ), \rho_{\nu}(\tau ,{\vec \sigma}_1)
 \} = - \delta^{\mu}_{\nu}\, \delta^3(\vec \sigma - {\vec
 \sigma}_1),\nonumber \\
 &&\{ A_A(\tau ,\vec \sigma ), \pi^B(\tau ,{\vec \sigma}_1)\} =
 \delta^B_A\, \delta^3(\vec \sigma - {\vec \sigma}_1),
 \label{VI72}
 \eea

\medskip

\noindent to the canonical Hamiltonian $H_c = - \int d^3\sigma\,
A_{\tau}(\tau ,\vec \sigma )\, \Gamma (\tau ,\vec \sigma )$ with
$\Gamma (\tau ,\vec \sigma ) = \partial_r\, \pi^r(\tau ,\vec
\sigma )$ and to the following generators of the Poincare' group
[the suffix "s" is for surface]

\bea
 P_s^{\mu}&=&\int d^3\sigma \, \rho^{\mu}(\tau ,\vec \sigma ),\nonumber \\
J_s^{\mu\nu}&=&\int d^3\sigma \, (z^{\mu}(\tau ,\vec \sigma
)\rho^{\nu} (\tau ,\vec \sigma )-z^{\nu}(\tau ,\vec \sigma
)\rho^{\mu}(\tau ,\vec \sigma )).
 \label{VI73}
 \eea

\bigskip

 There are five primary and one secondary first class constraints

\bea
 \pi^{\tau}(\tau ,\vec \sigma ) &\approx& 0,\nonumber \\
 \Gamma (\tau ,\vec \sigma ) &=& \partial_r\, \pi^r(\tau ,\vec
 \sigma ) \approx 0,\nonumber \\
 &&{}\nonumber \\
 {\cal H}_{\mu}(\tau ,\vec \sigma )&=& \rho_{\mu}(\tau ,\vec \sigma
)- l_{\mu}(\tau ,\vec \sigma )\, T_{\tau\tau}(\tau ,\vec \sigma )
- z_{{r}\mu}(\tau ,\vec \sigma )\, \gamma^{{r}{s}} (\tau ,\vec
\sigma )\, T_{\tau s}(\tau ,\vec \sigma )\, {\buildrel {def}\over
=}\nonumber \\
 &{\buildrel {def}\over =}& \rho_{\mu}(\tau ,\vec \sigma ) -
 {\cal G}_{\mu}[z^\mu_r(\tau,\vec{\sigma});\,A_r(\tau,\vec{\sigma}),
 \pi^s(\tau,\vec{\sigma})] \approx 0,\nonumber \\
 &&{}\nonumber \\
 &&T_{\tau \tau}(\tau ,\vec \sigma )
= - {1\over 2}\, ({1\over {\sqrt {\gamma}} }\pi^{r}g_{{ r}{s}}
\pi^{s} - { {\sqrt {\gamma}}\over 2}\, \gamma^{{r}{s}}\,
\gamma^{{u}{v}}\, F_{{r}{u}}\, F_{{s}{v}}) (\tau ,\vec \sigma
),\nonumber \\
 &&{}\nonumber \\
 &&T_{\tau {s}}(\tau ,\vec \sigma
)= F_{{s}{t}}(\tau ,\vec \sigma )\, \pi^{t}(\tau ,\vec \sigma ) =
\epsilon_{{s}{t} {u}}\, \pi^{t}(\tau ,\vec \sigma )\, B_{ u}(\tau
,\vec \sigma ) = - {[\vec \pi (\tau ,\vec \sigma )\times \vec
B(\tau ,\vec \sigma )]}_{s},\nonumber \\
 &&{}\nonumber \\
 &&{}\nonumber \\
 \lbrace {\cal H}_{\mu}(\tau ,\vec \sigma )&,&{\cal H}_{\nu}(\tau ,{\vec
\sigma}^{'} )\rbrace =
 \lbrace [l_{\mu}(\tau ,\vec
\sigma )\, z_{{r}\nu}(\tau ,\vec \sigma ) - l_{\nu}(\tau ,\vec
\sigma )\, z_{{r}\mu}(\tau ,\vec \sigma )]\, { {\pi^{ r}(\tau
,\vec \sigma )}\over {\sqrt{\gamma (\tau ,\vec \sigma )}} }
-\nonumber \\
 &-&z_{u\mu}(\tau ,\vec \sigma
)\, \gamma^{{u}{r}}(\tau ,\vec \sigma )\, F_{{r}{s}}(\tau ,\vec
\sigma )\, \gamma^{{s}{v}}(\tau ,\vec \sigma )\, z_{ v\nu}(\tau
,\vec \sigma )\rbrace \, \Gamma (\tau ,\vec \sigma )\,
\delta^3(\vec \sigma -{\vec \sigma}^{'})\approx 0.
 \label{VI74}
 \eea

\bigskip

The constraints $\pi^{\tau}(\tau ,\vec \sigma ) \approx 0$ and
$\Gamma (\tau ,\vec \sigma ) \approx 0$ are the canonical
generators of the electro-magnetic gauge transformations, while
${\cal H}_{\mu}(\tau ,\vec \sigma ) \approx 0$ generate the gauge
transformations from an admissible 3+1 splitting of Minkowski
space-time to another one with the associated change in the notion
of simultaneity.
\bigskip

Instead of the constraints ${\cal H}_{\mu}(\tau ,\vec \sigma )
\approx 0$, we can use their projections ${\cal H}_r(\tau ,\vec
\sigma )  = {\cal H}^{\mu}(\tau ,\vec \sigma )\, z_{r \mu}(\tau
,\vec \sigma ) \approx 0$, ${\cal H}_{\perp}(\tau ,\vec \sigma ) =
{\cal H}^{\mu}(\tau ,\vec \sigma )\, l_{\mu}(\tau ,\vec \sigma )
\approx 0$, normal and tangent to the simultaneity surfaces
$\Sigma_{\tau}$. Modulo the Gauss law constraint $\Gamma (\tau
,\vec \sigma ) \approx 0$, the new constraints satisfy the
universal Dirac algebra of the superhamiltonian and supermomentum
constraints of canonical metric gravity (see Ref.\cite{86a}). This
implies \cite{86a} that the gauge transformations generated by the
constraint ${\cal H}_{\perp}(\tau ,\vec \sigma )$ change the form
of the simultaneity surfaces $\Sigma_{\tau}$ (i.e. the 3+1
splitting), while those generated by the constraints ${\cal
H}_r(\tau ,\vec \sigma )$ change the 3-coordinates on such
surfaces.

\medskip

The Dirac Hamiltonian and the Hamilton-Dirac equations are
($\lambda^{\mu}(\tau ,\vec \sigma )$ and $\lambda_{\tau}(\tau
,\vec \sigma )$ are arbitrary Dirac multipliers; $\cir$ means
evaluated on the extremals of the action principle)

\bea
 H_D &=& \int d^3\sigma\, [ \lambda^{\mu}(\tau ,\vec \sigma )\, {\cal H}_{\mu}(\tau ,\vec
\sigma ) + \lambda_{\tau}(\tau ,\vec \sigma )\, \pi^{\tau}(\tau
,\vec \sigma ) - A_{\tau}(\tau ,\vec \sigma )\, \Gamma (\tau ,\vec
\sigma )] =\nonumber \\
 &&{}\nonumber \\
   &=& \int d^3\sigma\, [ {\tilde \lambda}_{\perp}(\tau ,\vec \sigma )\,
  {\cal H}_{\perp}(\tau ,\vec \sigma ) +\nonumber \\
  &+& {\tilde \lambda}^r(\tau ,\vec \sigma )\,
  {\cal H}_r(\tau ,\vec \sigma ) + \lambda_{\tau}(\tau ,\vec \sigma )\,
  \pi^{\tau}(\tau ,\vec \sigma ) - A_{\tau}(\tau ,\vec \sigma )\,
  \Gamma (\tau ,\vec \sigma )],\nonumber \\
  &&{}\nonumber \\
  &&{}\nonumber \\
 {{\partial A_{\tau}(\tau ,\vec \sigma )}\over {\partial \tau}} &\cir&
 \{ A_{\tau}(\tau ,\vec \sigma ), H_D\} = \lambda_{\tau}(\tau
 ,\vec \sigma ),\nonumber \\
 &&{}\nonumber \\
  {{\partial A_r(\tau ,\vec \sigma )}\over {\partial \tau}} &\cir&
  \{ A_r(\tau ,\vec \sigma ), H_D \} = - \int d^3\sigma^{'}\,
  \Big[ \lambda^{\mu}(\tau ,{\vec \sigma}^{'} )\, \{ A_r(\tau ,\vec
  \sigma ), \nonumber \\
  &&{\cal G}_{\mu}(z^{\alpha}(\tau ,{\vec \sigma}^{'}),
  A_u(\tau ,{\vec \sigma}^{'}), \pi^v(\tau ,{\vec \sigma }^{'}))\}
  +\nonumber \\
  &+& A_{\tau}(\tau ,{\vec \sigma}^{'})\, \{ A_r(\tau ,\vec \sigma
  ), \Gamma (\tau ,{\vec \sigma}^{'})\} \Big],\nonumber \\
 {{\partial \pi^r(\tau ,\vec \sigma )}\over {\partial \tau}} &\cir&
  \{ \pi^r(\tau ,\vec \sigma ), H_D \} =\nonumber \\
  &=& - \int d^3\sigma^{'}\,
  \lambda^{\mu}(\tau ,{\vec \sigma}^{'} )\, \{ \pi^r(\tau ,\vec
  \sigma ), {\cal G}_{\mu}(z^{\alpha}(\tau ,{\vec \sigma}^{'}),
  A_u(\tau ,{\vec \sigma}^{'}), \pi^v(\tau ,{\vec \sigma
  }^{'}))\},\nonumber \\
  &&{}
 \label{VI75}
 \eea

\bigskip

Due to the last two lines of Eqs.(\ref{VI74}), we see that two
successive gauge transformations, of generators $G_i(\tau ,\vec
\sigma ) = \lambda^{\mu}_i(\tau ,\vec \sigma )\, {\cal
H}_{\mu}(\tau ,\vec \sigma )$, $i=1,2$, do not commute but imply
an electro-magnetic gauge transformation. Since the effect of the
$i=1,2$ gauge transformations is to modify the notions of
simultaneity, also the definition of the Dirac observables of the
electro-magnetic field will change with the 3+1 splitting. In
general, given two different 3+1 splittings, the two sets of Dirac
observables associated with them will be connected by an
electro-magnetic gauge transformation.

\medskip

Since it is not clear whether it is possible to find a
quasi-Shanmugadhasan canonical transformation adapted to ${\cal
H}_r(\tau ,\vec \sigma ) = {\cal H}_{\mu}(\tau ,\vec \sigma )\,
z^{\mu}_r(\tau ,\vec \sigma ) \approx 0$, $\pi^{\tau}(\tau ,\vec
\sigma ) \approx 0$, $\Gamma (\tau ,\vec \sigma ) \approx 0$
\footnote{${\cal H}_{\perp}(\tau ,\vec \sigma ) = {\cal
H}^{\mu}(\tau ,\vec \sigma )\, l_{\mu}(\tau ,\vec \sigma ) \approx
0$, like an ordinary Hamiltonian, can be included in the adapted
Darboux-Shanmugadhasan basis only in case of integrability of the
equations of motion.}, the search of the electro-magnetic Dirac
observables must be done with the following strategy:

i) make the choice of an admissible 3+1 splitting by adding four
gauge-fixing constraints determining the embedding $z^{\mu}(\tau
,\vec \sigma )$, so that the induced 4-metric $g_{AB}(\tau ,\vec
\sigma )$ becomes a numerical quantity and is no more a
configuration variable;

ii) find the Dirac observables on the resulting completely fixed
simultaneity surfaces $\Sigma_{\tau}$ with a suitable
Shanmugadhasan canonical transformation adapted to the two
remaining electro-magnetic constraints.

\medskip

Let us remark that a similar scheme has to be followed also in the
canonical Einstein-Maxwell theory: only after having fixed a 3+1
splitting (a system of 4-coordinates on the solutions of
Einstein's equations) we can find the Dirac observables of the
electro-magnetic field.

\medskip

This strategy is induced by the fact that, while the Gauss law
constraint $\Gamma (\tau ,\vec \sigma ) = \partial_r\, \pi^r(\tau
,\vec \sigma ) \approx 0$ is a scalar under change of admissible
3+1 splittings \footnote{$\pi^r(\tau ,\vec \sigma )$ is a vector
density like in canonical metric gravity.},  the gauge vector
potential $A_r(\tau ,\vec \sigma )$ is the pull-back to the base
of a connection one-form and can be considered as a tensor only
with topologically trivial surfaces $\Sigma_{\tau}$ (like in the
case we are considering). Since a Shanmugadhasan canonical
transformation adapted to the Gauss law constraint transforms
$\Gamma (\tau ,\vec \sigma )$  in one of the new momenta, it is
not clear how to define a conjugate gauge variable $\eta_{em}(\tau
,\vec \sigma )$ such that  $\{ \eta_{em}(\tau ,\vec \sigma ),
\Gamma (\tau ,{\vec \sigma}_1) \} = \delta^3(\vec \sigma , {\vec
\sigma}_1)$ and two conjugate pairs of Dirac observables having
vanishing Poisson brackets with both  $\eta_{em}(\tau ,\vec \sigma
)$ and  $\Gamma (\tau ,\vec \sigma )$ when the 3-metric on
$\Sigma_{\tau}$ is not Euclidean ($g_{rs}(\tau ,\vec \sigma )
\not= -\sgn\, \delta_{rs}$).
\medskip

The only case studied till now \cite{87a} is the restriction to
the Wigner hyper-planes associated to the rest-frame instant form,
where both $A_r(\tau ,\vec \sigma )$ and $\pi^r(\tau ,\vec \sigma
)$ transform as Wigner spin-1 3-vectors under Lorentz boosts.
Since on Wigner hyper-planes we have $g_{rs}(\tau ,\vec \sigma ) =
-\sgn\, \delta_{rs}$, the Shanmugadhasan canonical transformation
leads to a {\it radiation gauge}

\bea
 &&\begin{minipage}[t]{1cm}
\begin{tabular}{|l|} \hline
$A_A$ \\  \hline
 $\pi^A$ \\ \hline
\end{tabular}
\end{minipage} \ {\longrightarrow \hspace{.2cm}} \
\begin{minipage}[t]{2 cm}
\begin{tabular}{|l|l|l|} \hline
$A_{\tau}$ & $\eta_{em}$   & $A_{\perp\, r}$   \\ \hline
$\pi^{\tau}\approx 0$& $\Gamma \approx 0$ &$\pi^r_{\perp}$ \\
\hline
\end{tabular}
\end{minipage},\nonumber \\
 &&{}\nonumber \\
 &&{}\nonumber \\
 A^r(\tau ,\vec \sigma )&=& -\sgn\, A_r(\tau ,\vec \sigma ) = {
{\partial}\over {\partial \sigma^r} }\eta_{em} (\tau ,\vec \sigma
)+A^r_{\perp}(\tau ,\vec \sigma ),\nonumber \\
 &&{}\nonumber \\
 \pi^r(\tau ,\vec \sigma )&=&\pi^r_{\perp}(\tau
,\vec \sigma )+{1\over {\triangle_{\sigma}} }{ {\partial}\over
{\partial \sigma^r}} \, \Gamma (\tau ,\vec \sigma ), \qquad
\triangle_{\sigma} = - {\vec \partial}^2_{\sigma}, \nonumber \\
 &&{}\nonumber \\
  \eta_{em}(\tau ,\vec \sigma )&=&-{1\over {\triangle_{\sigma}} }{ {\partial} \over
 {\partial \vec \sigma} }\cdot \vec A(\tau ,\vec \sigma ),\nonumber \\
 &&{}\nonumber \\
 A^r_{\perp}(\tau ,\vec \sigma ) &=& (\delta^{rs}+{{\partial^r_{\sigma}\partial^s_{\sigma}}
 \over {\triangle_{\sigma} }})\, A_s(\tau ,\vec \sigma ),\nonumber \\
  \pi^r_{\perp}(\tau ,\vec \sigma ) &=& (\delta^{rs}+{{\partial^r_{\sigma}\partial^s_{\sigma}}
 \over {\triangle_{\sigma} }})\, \pi_s(\tau ,\vec \sigma ),\qquad
 {\vec \pi}^2(\tau ,\vec \sigma ) \approx {\vec
 \pi}^2_{\perp}(\tau ,\vec \sigma ),\nonumber \\
 &&{}\nonumber \\
  &&\lbrace \eta_{em}(\tau ,\vec \sigma ),\Gamma
(\tau ,{\vec \sigma}^{'} ) \rbrace {}^{**}=- \sgn\, \delta^3(\vec
\sigma -{\vec \sigma}^{'}),\nonumber \\
 &&{}\nonumber \\
 &&\lbrace A^r_{\perp}(\tau ,\vec
\sigma ),\pi^s_{\perp}(\tau ,{\vec \sigma} ^{'})\rbrace {}^{**}=-
\sgn\,  (\delta^{rs}+{{\partial^r_{\sigma}\partial^s_{\sigma}}
\over {\triangle_{\sigma} }})\delta^3(\vec \sigma -{\vec
\sigma}^{'}).
 \label{VI76}
 \eea

\bigskip

With every fixed type of simultaneity surface $\Sigma_{\tau}$ with
non-trivial 3-metric, $g_{rs}(\tau ,\vec \sigma ) \not= -\sgn\,
\delta_{rs}$, we have to find suitable gauge variable
 $\eta_{em}(\tau ,\vec \sigma )$ and the Dirac observables
 replacing $A^r_{\perp}(\tau ,\vec \sigma )$ and $\pi^r_{\perp}(\tau ,\vec \sigma )$.
\medskip

To choose a 3+1 splitting with a foliation, whose simultaneity
surfaces are described by a given admissible embedding
$z^{\mu}_F(\tau ,\vec \sigma )$, we add the gauge-fixing
constraints

\begin{eqnarray*}
 \zeta^{\mu}(\tau ,\vec \sigma ) &=& z^{\mu}(\tau ,\vec
\sigma ) - z^{\mu}_F(\tau ,\vec \sigma ) \approx 0,\nonumber \\
 &&{}\nonumber \\
 &&z^{\mu}_F(\tau ,\vec \sigma ) = x^{\mu}(\tau ) + F^{\mu}(\tau
 ,\vec \sigma )\, \rightarrow_{|\vec \sigma | \rightarrow \infty}\,
 x^{\mu}_{(\infty )}(0) + l^{\mu}_F\, \tau + \epsilon^{\mu}_{F\,
 r}\, \sigma^r {\buildrel {def}\over =}\, x^{\mu}_{(\infty )}(\tau
 ) + \epsilon^{\mu}_{F\, r}\, \sigma^r,
 \end{eqnarray*}

 \bea
 &&x^{\mu}(\tau ) = l^{\mu}_F\, a(\tau ) - \epsilon^{\mu}_{F\,
 r}\, b^r(\tau ), \qquad F^{\mu}(\tau ,\vec 0) = 0,\nonumber \\
 &&{}\nonumber \\
 &&F^{\mu}(\tau ,\vec \sigma )\, \rightarrow_{|\vec \sigma | \rightarrow \infty}\,
 l^{\mu}_F\, [\tau - a(\tau )] + \epsilon^{\mu}_{F\, r}\,
 [\sigma^r + b^r(\tau )],\nonumber \\
 &&{}\nonumber \\
 &&z^{\mu}_{F\, \tau}(\tau ,\vec \sigma ) = {\dot x}^{\mu}(\tau )
 + {{\partial F^{\mu}(\tau ,\vec \sigma )}\over {\partial \tau}}\,
 \rightarrow_{|\vec \sigma | \rightarrow \infty}\, {\dot
 x}^{\mu}_{(\infty )}(\tau ) = l^{\mu}_F,\nonumber \\
 &&{}\nonumber \\
 &&det\, \Big( \{ \zeta^{\mu}(\tau ,\vec \sigma ), {\cal
 H}_{\nu}(\tau ,{\vec \sigma}_1) \} \Big) \not= 0,
 \label{VI77}
 \eea

\noindent where we have used the notation of Eqs.(\ref{III3}),
(\ref{III5}). Here $\epsilon^{\mu}_{F\, A} = \Big( l_F^{\mu} =
\epsilon^{\mu}_{F\, \tau}; \epsilon^{\mu}_{F\, r}\Big)$ is the
asymptotic tetrad at spatial infinity associated to the foliation,
with $l^{\mu}_F$ the normal to the asymptotic hyper-planes (see
Section I). If $l^{\mu}_{(F)}(\tau ,\vec \sigma )$ are the
components of the unit normal vector field to $\Sigma_{F\, \tau}$,
built in terms of the $z^{\mu}_{F\, r}(\tau ,\vec \sigma )$, we
have $l^{\mu}_{(F)}(\tau ,\vec \sigma )\, \rightarrow_{|\vec
\sigma | \rightarrow \infty}\, l^{\mu}_F$.

The functions $F^{\mu}(\tau ,\vec \sigma )$, assumed to satisfy
the admissibility conditions of Section III, describe the form of
the simultaneity surfaces $\Sigma_{F\, \tau}$, while the arbitrary
centroid $x^{\mu}(\tau )$ (a time-like, in general non-inertial,
observer chosen as origin of the 3-coordinates on $\Sigma_{F\,
\tau}$) describes how they are packed in the foliation. The
centroid corresponds to an observer of the rotating skew
congruence associated to the foliation, because ${\dot
x}^{\mu}(\tau ) = z^{\mu}_{F\, \tau}(\tau ,\vec 0)$. Instead
$x^{\mu}_{(\infty )}(\tau )$ is the world-line of an asymptotic
inertial observer at spatial infinity.

\bigskip

For instance, if we want to select foliations which arise from
those admitted in the rest-frame instant form of canonical metric
gravity \cite{86a} in the limit of vanishing gravitational field,
we must have that the hyper-surfaces $\Sigma_{\tau}$ tend in a
direction-independent way to Wigner hyper-planes at spatial
infinity, $z_W^{\mu}(\tau ,\vec \sigma ) = x^{\mu}(0) +
\epsilon^{\mu}_A(u(P_s))\, \sigma^A$. This can be obtained with
admissible embeddings of the type $z^{\mu}_F(\tau ,\vec \sigma ) =
x^{\mu}(0) + \epsilon^{\mu}_A(u(P))\, [\sigma^A + F^A(\tau ,\vec
\sigma )]$, $lim_{|\vec \sigma | \rightarrow \infty}\, F^A(\tau
,\vec \sigma )\, = 0$. Here $P^{\mu}$ is a time-like 4-vector and
$\epsilon^{\mu}_A(u(P)) = \Big( u^{\mu}(P) = P^{\mu}/\sqrt{\sgn\,
P^2 }; \epsilon^{\mu}_r(u(P))\Big)$ is the tetrad defined by the
standard Wigner boost sending $P^{\mu}$ at rest. To enforce the
requirement $P^{\mu} = P^{\mu}_s$ (asymptotism to Wigner
hyper-planes), we enlarge the phase space with four pairs
$X^{\mu}$, $P^{\mu}$ of conjugate variables and we add the four
first class constraints $P^{\mu} - P^{\mu}_s \approx 0$ to the
Dirac Hamiltonian. After having evaluated the Dirac brackets
associated to the gauge fixings (\ref{VI77}), the 4-metric
$g_{AB}(\tau ,\vec \sigma )$ becomes a function $g_{(F)AB}(\tau
,\vec \sigma | P)$ depending only on $P^{\mu}$.
\medskip

In special relativity a simpler set of admissible embeddings is
given by the foliations with parallel space-like hyper-planes of
Section IV. Before restricting to them let us delineate how the
method of canonical reduction to Wigner hyper-planes introduced in
Ref.\cite{87a} is modified when  there is a gauge fixing
(\ref{VI77}) restricting the description to specific admissible
simultaneity surfaces.
\bigskip

The preservation in time of the gauge fixing (\ref{VI77}) gives
the following determination of the Dirac multipliers
$\lambda^{\mu}(\tau ,\vec \sigma )$ appearing in the Dirac
Hamiltonian (\ref{VI75})

\bea
 {{\partial \zeta^{\mu}(\tau ,\vec \sigma )}\over {\partial \tau}}
 &\cir& \{ z^{\mu}(\tau ,\vec \sigma ), H_D \} - {{\partial
 z^{\mu}_F(\tau ,\vec \sigma )}\over {\partial \tau}} \approx
 0,\nonumber \\
 &&{}\nonumber \\
 &&\Downarrow \nonumber \\
 &&{}\nonumber \\
 \lambda^{\mu}(\tau ,\vec \sigma ) &=& \lambda^{\mu}(\tau ) -
 {{\partial F^{\mu}(\tau ,\vec \sigma )}\over {\partial
 \tau}},\qquad
 \lambda^{\mu}(\tau ) = - {\dot x}^{\mu}(\tau ) = -
 z^{\mu}_{F\, \tau}(\tau ,\vec 0),\nonumber \\
 &&{}\nonumber \\
 H_D &=& \lambda^{\mu}(\tau )\, H_{\mu}(\tau) + \int d^3\sigma\,
 \Big[ - {{\partial F^{\mu}}\over {\partial \tau}}\, {\cal
 H}_{\mu} + \lambda_{\tau}\, \pi^{\tau} - A_{\tau}\, \Gamma
 \Big](\tau ,\vec \sigma ),\nonumber \\
 &&{}\nonumber \\
 &&H_{\mu}(\tau ) = \int d^3\sigma\, {\cal H}_{\mu}(\tau ,\vec
 \sigma ),
 \label{VI78}
 \eea

\noindent with the arbitrariness reduced to $\lambda^{\mu}(\tau
)$.

\medskip

If we go to the reduced phase space by introducing the Dirac
brackets associated to the gauge fixing (\ref{VI77}), we see that
all the gauge degrees of freedom of the embedding are reduced to
the four variables $x^{\mu}(\tau )$. Therefore the first class
constraints ${\cal H}_{\mu}(\tau ,\vec \sigma )$ are reduced to
only four independent ones

\bea
 H^{\mu}(\tau ) &=& P^{\mu}_s - \int d^3\sigma\,
 [l_{(F)\mu}(\tau ,\vec \sigma )\, T_{(F)\tau\tau}(\tau ,\vec \sigma )
+ z_{F{r}\mu}(\tau ,\vec \sigma )\, \gamma_F^{{r}{s}} (\tau ,\vec
\sigma )\, T_{(F)\tau s}(\tau ,\vec \sigma )] =\nonumber \\
 &=& P^{\mu}_s - \int d^3\sigma
\,{\cal G}_\mu\,[z^\mu_{F\,r}(\tau,\vec{\sigma});
\,A_r(\tau,\vec{\sigma}),\pi^s(\tau,\vec{\sigma})] \approx 0,
 \label{VI79}
 \eea

 \noindent where $P^\mu_s(\tau)$ is the canonical momentum conjugate to $x^\mu(\tau)$

\beq
 \{ x^{\mu}(\tau ), P^{\nu}_s \}^{*} = - \eta^{\mu\nu}.
 \label{VI80}
 \eeq

Let us remark that the Dirac brackets for the electro-magnetic
field and their momenta remain equal to their Poisson brackets.
\bigskip

If we restrict Hamilton equations (\ref{VI75}) to the gauge
(\ref{VI77}) and we use Eqs.(\ref{VI78}), we get a form of these
equations, which can be reproduced as the Hamilton equations of
the reduced phase space only by using the new Dirac Hamiltonian
(it differs from $H_D$ of Eqs.(\ref{VI78}) by the term $-\int
d^3\sigma\, \rho_{\mu}(\tau ,\vec \sigma )\, {{\partial
F^{\mu}(\tau ,\vec \sigma )}\over {\partial \tau}}$, which is
ineffective in the reduced phase space)

\bea
 {\tilde H}_D &=& \lambda^{\mu}(\tau )\, H_{\mu}(\tau ) +
 \int d^3\sigma\,\frac{\partial F^\mu(\tau,\vec{\sigma})}{\partial\tau}\,
\,{\cal G}_\mu\,[z^\mu_{F\,r}(\tau,\vec{\sigma});
\,A_r(\tau,\vec{\sigma}),\pi^s(\tau,\vec{\sigma})] +\nonumber \\
  &+& \int d^3\sigma\,
 [\lambda_{\tau}(\tau ,\vec \sigma )\, \pi^{\tau}(\tau ,\vec
 \sigma ) - A_{\tau}(\tau ,\vec \sigma )\, \Gamma (\tau ,\vec
 \sigma )].
 \label{VI81}
 \eea
\medskip

The induced 4-metric $g_{(F)AB}(\tau ,\vec \sigma ) = z^{\mu}_{F\,
A}(\tau ,\vec \sigma )\, \eta_{\mu\nu}\, z^{\nu}_{F\, B}(\tau
,\vec \sigma )$ is completely determined except for its dependence
on the arbitrary velocity of the centroid, ${\dot x}^{\mu}(\tau )
= - \lambda^{\mu}(\tau )$. The constraints (\ref{VI79}) determine
the generator of translations $P^{\mu}_s$, given in
Eq.(\ref{VI73}), so that the {\it coordinates of the centroid are
gauge variables}, corresponding to the {\it arbitrariness in the
choice of the (non-inertial) observer}.
\medskip

From the second line of Eq.(\ref{VI73}) we get the following form
of the generator of Lorentz transformations

\bea
 J^{\mu\nu}_s &=& x^{\mu}\, P^{\nu}_s - x^{\nu}\, P^{\mu}_s +
S^{\mu\nu}_s,\nonumber \\
 &&{}\nonumber \\
 S^{\mu\nu}_s &=& \int d^3\sigma\,
\Big[F^{\mu} \,{\cal G}^\nu\,[z^\mu_{F\,r}; \,A_r,\pi^s]- F^\nu\,
{\cal G}^\nu\,[z^\mu_{F\,r}; \,A_r,\pi^s]\Big](\tau,\vec{\sigma}).
 \label{VI82}
 \eea

\bigskip

By using the asymptotic tetrad at spatial infinity the four
constraints $H^{\mu}(\tau ) \approx 0$ and the Dirac Hamiltonian
may be transformed in the following form
\medskip

\begin{eqnarray*}
 {\bar H}_l(\tau ) &=& l_{F\, \mu}\, P^{\mu}_s - l_{F\, \mu}\,
\int d^3\sigma\, [l_{(F)}^{\mu}\, T_{(F)\tau\tau} - z_{F\,
r}^{\mu}\, \gamma^{rs}_{F}\, T_{(F)\tau s}](\tau ,\vec
\sigma ) =\nonumber \\
 &=& l_{F\, \mu}\, P^{\mu}_s - l_{F\, \mu}\, \int d^3\sigma\,
 {\cal G}^\mu\,[z^\mu_{F\,r}(\tau,\vec{\sigma});
\,A_r(\tau,\vec{\sigma}),\pi^s(\tau,\vec{\sigma})] =
 l_F \cdot P_s - M_{(F)l} \approx 0,\nonumber \\
 &&{}\nonumber \\
 {\bar H}_r(\tau ) &=& \epsilon_{F\, r\mu}\, P_s^{\mu} -  \epsilon_{F\, r\mu}\,
\int d^3\sigma\,  {\cal G}^\mu\,[z^\mu_{F\,r}(\tau,\vec{\sigma});
\,A_r(\tau,\vec{\sigma}),\pi^s(\tau,\vec{\sigma})] (\tau ,\vec
\sigma ) \approx 0,
 \end{eqnarray*}

 \bea
 {\bar H}_D &=& {\bar \lambda}_l(\tau )\, {\bar H}_l(\tau ) - \sum_r\,
 {\bar \lambda}_r(\tau )\, {\bar H}_r(\tau ) +\nonumber \\
 &&{}\nonumber \\
 &+& \int d^3\sigma\,
\frac{\partial F^\mu(\tau,\vec{\sigma})}{\partial\tau}\, \,{\cal
G}_\mu\,[z^\mu_{F\,r}(\tau,\vec{\sigma});
\,A_r(\tau,\vec{\sigma}),\pi^s(\tau,\vec{\sigma})] +\nonumber \\
&&{}\nonumber \\
 &+& \int d^3\sigma\,
 [\lambda_{\tau}(\tau ,\vec \sigma )\, \pi^{\tau}(\tau ,\vec
 \sigma ) - A_{\tau}(\tau ,\vec \sigma )\, \Gamma (\tau ,\vec
 \sigma )].
 \label{VI83}
 \eea
\bigskip

Since $l_F \cdot x(\tau )$, $l_F \cdot P_s$ and $\epsilon_{F\, r}
\cdot x(\tau )$, $\epsilon_{F\, r} \cdot P_s$ are four pairs of
conjugate variables, we see that

i) the gauge fixing $l_F \cdot x(\tau ) - \tau \approx 0$ (so that
$a(\tau ) = \sgn\, \tau$) to ${\bar H}_l(\tau ) \approx 0$
identifies the mathematical time $\tau$ with the Lorentz-scalar
asymptotic time of the asymptotic inertial observer
$x^{\mu}_{(\infty )}(\tau )$and, through Dirac brackets, forces
the identity $l_F \cdot P_s \equiv M_{(F)l}$, with $M_{(F)l} =
l_{F\, \mu}\, \int d^3\sigma\, {\cal
G}^\mu\,[z^\mu_{F\,r}(\tau,\vec{\sigma});
\,A_r(\tau,\vec{\sigma}),\pi^s(\tau,\vec{\sigma})]$ being the {\it
internal mass} seen by the observer $x^{\mu}(\tau )$;
\medskip

ii) the gauge fixings $\epsilon_{F\, r} \cdot x(\tau ) -  g_r(\tau
) \approx 0$ (so that $b^r(\tau ) = \sgn\, g_r(\tau )$ and
$\lambda^{\mu}(\tau ) =- {\dot x}^{\mu}(\tau ) = -\sgn\,
[l^{\mu}_F - \epsilon^{\mu}_{F\, r}\, {\dot g}_r(\tau )]$) to
${\bar H}_r(\tau ) \approx 0$ identify the centroid $x^{\mu}(\tau
)$ with the world-line ${\tilde x}^{\mu}(\tau ) = \sgn\, [
l^{\mu}_F\, \tau - \epsilon^{\mu}_{F\, r}\, g_r(\tau )]$ of a
time-like (in general non-inertial) observer, whose {\it
3-momentum} is $\epsilon_{F\, r} \cdot P_s \equiv \epsilon_{F\,
r\, \mu}\, \int d^3\sigma\, {\cal
G}^\mu\,[z^\nu_{F\,s}(\tau,\vec{\sigma});
\,A_s(\tau,\vec{\sigma}),\pi^s(\tau,\vec{\sigma})]$. It plays the
same role of the {\it external 4-center of mass} of the rest-frame
instant form \cite{43a}. For $g_r(\tau ) = 0$, this observer
coincides with asymptotic inertial one: ${\tilde x}^{\mu}(\tau ) =
\sgn\, x^{\mu}_{(\infty )}(\tau )$.
\bigskip

The {\it internal angular momentum} is $S_{(F)s\, AB} =
\epsilon_{F\, A\, \mu}\, \epsilon_{F\, B\, \nu}\, S^{\mu\nu}_s$.
This quantity,  $M_{(F)l}$ and $\epsilon_{F\, r} \cdot P_s$,
replace the generators of the internal Poincare' group of the
rest-frame instant form on Wigner hyper-planes \cite{87a,43a}.

\bigskip

After the previous gauge fixings we arrive at a  phase space
containing only the electro-magnetic field restricted to the
simultaneity surfaces $\Sigma_{F\, \tau}$ of the completely fixed
embedding $z^{\mu}_F(\tau ,\vec \sigma ) = {\tilde x}^{\mu}(\tau )
+ F^{\mu}(\tau ,\vec \sigma )\, \rightarrow{|\vec \sigma |
\rightarrow \infty}\, x^{\mu}_{(\infty )}(\tau ) +
\epsilon^{\mu}_{F\, r}\, \sigma^r$ and with a non-vanishing Dirac
Hamiltonian given by the last two lines of ${\bar H}_D$ in
Eq.(\ref{VI83}). However, since the gauge fixings are explicitly
$\tau$-dependent, this restricted Hamiltonian does not reproduce
the same Hamilton equations for the electro-magnetic field that
would be obtained by using ${\bar H}_D$ of Eq.(\ref{VI83}) and
then restricting them with the gauge fixings. The same steps used
to get Eq.(\ref{VI81}) show that the true Hamiltonian ${\hat H}_D$
acting in the reduced phase space is obtained by adding the
projection ${\dot {\tilde x}}_{\mu}(\tau )\, \int d^3\sigma\,
{\cal G}^{\mu}(..)(\tau ,\vec \sigma )$ of the total 4-momentum
along the 4-velocity ${\dot {\tilde x}}^{\mu}(\tau ) = -
\lambda^{\mu}(\tau )$ of the observer ${\tilde x}^{\mu}(\tau )$ to
the reduced Dirac-Hamiltonian. The final Hamiltonian ${\hat H}_D$
is the sum of an {\it effective non-inertial Hamiltonian}
(containing the internal mass and the internal 3-momentum) and of
the generator of the electro-magnetic gauge transformations

\bea
 {\hat H}_D &=& \int d^3\sigma\, z^{\mu}_{F\tau}(\tau ,\vec \sigma
 )\, {\cal G}_\mu\,[z^\mu_{F\,r}(\tau,\vec{\sigma});
\,A_r(\tau,\vec{\sigma}),\pi^s(\tau,\vec{\sigma})] +\nonumber \\
&+&  \int d^3\sigma\,
 [\lambda_{\tau}(\tau ,\vec \sigma )\, \pi^{\tau}(\tau ,\vec
 \sigma ) - A_{\tau}(\tau ,\vec \sigma )\, \Gamma (\tau ,\vec
 \sigma )] =\nonumber \\
 &=& M_{(F)l} + \int d^3\sigma\, \Big[ \epsilon^{\mu}_{F\, r}\,
 {\dot g}_r(\tau ) + {{\partial F^{\mu}(\tau ,\vec \sigma )}\over
 {\partial \tau}}\Big]\, {\cal G}_\mu\,[z^\mu_{F\,r}(\tau,\vec{\sigma});
\,A_r(\tau,\vec{\sigma}),\pi^s(\tau,\vec{\sigma})] +\nonumber \\
&+&  \int d^3\sigma\,
 [\lambda_{\tau}(\tau ,\vec \sigma )\, \pi^{\tau}(\tau ,\vec
 \sigma ) - A_{\tau}(\tau ,\vec \sigma )\, \Gamma (\tau ,\vec
 \sigma )] =\nonumber \\
 &=& M_{(F)l} + {\dot g}_r(\tau )\, \epsilon_{F\, r} \cdot P_s +
 \int d^3\sigma\, {{\partial F^{\mu}(\tau ,\vec \sigma )}\over
 {\partial \tau}}\, {\cal G}_\mu\,[z^\mu_{F\,r}(\tau,\vec{\sigma});
\,A_r(\tau,\vec{\sigma}),\pi^s(\tau,\vec{\sigma})] +\nonumber \\
 &+& \int d^3\sigma\,
 [\lambda_{\tau}(\tau ,\vec \sigma )\, \pi^{\tau}(\tau ,\vec
 \sigma ) - A_{\tau}(\tau ,\vec \sigma )\, \Gamma (\tau ,\vec
 \sigma )].
 \label{VI84}
 \eea
 \bigskip

This is the generator of the evolution seen by the  non-inertial
observer ${\tilde x}^{\mu}(\tau ) = \sgn\, [l^{\mu}_F\, \tau -
\epsilon^{\mu}_{F\, r}\, g_r(\tau )]$ as a consequence of the
chosen notion of simultaneity with its asymptotic constant tetrad
$\epsilon^{\mu}_A$ at spatial infinity.

\medskip

Therefore the time-like non-inertial observer (not orthogonal to
the instantaneous 3-space $\Sigma_{F\, \tau}$) $x^{\mu}(\tau )
\equiv {\tilde x}^{\mu}(\tau )$ with ${\dot x}^{\mu}(\tau ) =
z^{\mu}_{F\tau}(\tau ,\vec 0)$ must

i) use the 3+1 point of view (instantaneous 3-space $\Sigma_{F\,
\tau}$) to describe the evolution in $\tau \equiv l_F \cdot \tilde
x$ with ${\hat H}_D$ as Hamiltonian: besides an  electro-magnetic
internal mass term, $M_{(F)l}$, like in the rest-frame instant
form, there are two extra terms interpretable as potentials of the
{\it inertial forces} associated to this notion of simultaneity;

ii) make the choice of a tetrad ${\cal V}^{\mu}_A = \Big( {\cal
V}^{\mu}_{\tau} = {\dot {\tilde x}}^{\mu}/\sqrt{\sgn\, {\dot
{\tilde x}}^2}; {\cal V}^{\mu}_r\Big)$ and use the 1+3 point of
view to measure the tetradic components ${\cal F}_{(F)AB} = {\cal
V}^{\mu}_A\, {\cal V}^{\nu}_B\, F_{\mu\nu} = {\cal V}^{\mu}_A\,
z^C_{F\, \mu}\,\, {\cal V}^{\nu}_B\, z^D_{F\, \nu}\,\, F_{CD}$ of
the electro-magnetic field.
\bigskip

The Hamilton equations for the vector potential of the
electro-magnetic field  on the simultaneity surfaces of the
foliation $z^{\mu}_F(\tau ,\vec \sigma ) = {\tilde x}^{\mu}(\tau )
+ F^{\mu}(\tau ,\vec \sigma )$ are

\bea
 && \frac {\partial\,A_r(\tau,\vec{\sigma})} {\partial\tau}
\,\cir\, \{A_r(\tau,\vec{\sigma}),{\hat H}_D\}=\left[
\frac{\sqrt{g_{F}}}{\gamma_{F}}\,g_{F\, rs}\,\pi^s+ g_{F\, \tau
u}\,\gamma_F^{us}\,F_{sr}+\partial_r A_\tau
\right](\tau,\vec{\sigma}),\nonumber\\
 &&\nonumber\\
  && \frac
{\partial\,\pi^r(\tau,\vec{\sigma})} {\partial\tau} \,\cir\,
\{\pi^r(\tau,\vec{\sigma}),{\hat H}_D
\}=\frac{\partial}{\partial\sigma^s}\,
\left[\sqrt{g_{F}}\,\gamma_{F}^{sv}\,\gamma_{F}^{ru}\, F_
{vu}-(g_{F\, \tau u}\,\gamma_{F}^{us}\,\pi^r - g_{F\, \tau
u}\,\gamma_{F}^{ur}\,\pi^s) \right](\tau,\vec{\sigma}).
\nonumber \\
 &&{}
 \label{VI85}
 \eea

\medskip

We can invert the first to obtain

 \bea
\pi^s(\tau,\vec{\sigma})&=&
-\left[\frac{\gamma_{F}}{\sqrt{g_{F}}}\, \gamma_{F}^{sr}\left(
F_{\tau r}-g_{F\, \tau u}\,\gamma_{F}^{uv}\,
F_{vr}\right)\right](\tau,\vec{\sigma})=\nonumber\\
 &&\nonumber\\
 &=& -\sqrt{g_{F}(\tau,\vec{\sigma})}\, g_{F}^{\tau
A}(\tau,\vec{\sigma})\, g_{F}^{s
B}(\tau,\vec{\sigma})\,F_{AB}(\tau,\vec{\sigma}).
 \label{VI86}
 \eea
\bigskip

It can be shown that Eqs.(\ref{VI85}) are equivalent to
\medskip

 \beq
\frac{\partial}{\partial \sigma^A}\left[
\sqrt{g_{F}(\tau,\vec{\sigma})}\, g_{F}^{A B}(\tau,\vec{\sigma})\,
g_{F}^{s D}(\tau,\vec{\sigma})\,F_{BD}(\tau,\vec{\sigma})
\right]\cir 0,
 \label{VI87}
  \eeq

\noindent and that by using Eq.(\ref{VI86}) the Gauss law
$\partial_r\, \pi^r(\tau ,\vec \sigma ) = 0$ becomes

 \beq
\frac{\partial}{\partial \sigma^A}\left[
\sqrt{g_{F}(\tau,\vec{\sigma})}\, g_{F}^{A B}(\tau,\vec{\sigma})\,
g_{F}^{\tau D}(\tau,\vec{\sigma})\,F_{BD}(\tau,\vec{\sigma})
\right]\cir 0.
 \label{VI88}
  \eeq

As a consequence Eqs.(\ref{VI87}) and (\ref{VI88}) imply

 \beq
\frac{1}{\sqrt{g_{F}(\tau,\vec{\sigma})}}\,
\frac{\partial}{\partial \sigma^A}\left[
\sqrt{g_{F}(\tau,\vec{\sigma})}\,g_{F}^{AB}(\tau,\vec{\sigma})\,
g_{F}^{CD}(\tau,\vec{\sigma})\,
F_{BD}(\tau,\vec{\sigma})\right]\cir 0.
 \label{VI89}
  \eeq

These are the expected equations for the field strengths  written
in a {\it manifestly covariant form}.

\medskip

If we add to the  Lagrangian (\ref{VI71}) a set of $N$ charged
point particles interacting with the electro-magnetic fields (see
Refs.\cite{36a}) and with 3-positions $\eta^r_i(\tau)$ on
$\Sigma_{F\, \tau}$, such that $x_i^\mu(\tau)=
z_F^\mu(\tau,\vec{\eta}_i(\tau)$, it can be shown \cite{36a} that
the Gauss law and the second half of the Hamilton equations
(\ref{VI85}) are modified to the form

 \bea
 \Gamma (\tau ,\vec \sigma ) &=&
 \frac{\partial}{\partial\sigma^r}\,\pi^r(\tau,\vec{\sigma}) -
\sum_{i=1}^N Q_i\,\delta(\vec{\sigma}-\vec{\eta}_i(\tau))\approx
0, \nonumber\\
 &&\nonumber\\
  \frac {\partial\,\pi^r(\tau,\vec{\sigma})}
{\partial\tau} &\,\cir\,& \frac{\partial}{\partial\sigma^s}\,
\left[\sqrt{g_{F}}\,\gamma_{F}^{sv}\,\gamma_{F}^{ru}\, F_
{vu}-(g_{F\, \tau u}\,\gamma_{F}^{us}\,\pi^r -g_{F\, \tau
u}\,\gamma_{F}^{ur}\,\pi^s)
\right](\tau,\vec{\sigma})-\nonumber\\
 &&\nonumber\\
 &-&\sum_{i=1}^N Q_i\,\dot{\eta}^r_i(\tau)\,
\delta(\vec{\sigma}-\vec{\eta}_i(\tau)).
 \label{VI90}
  \eea

If we introduce the charge density and the charge current density
on $\Sigma_{\tau}$

 \bea
  \overline{\rho}(\tau,\vec{\sigma})&=&
\frac{1}{\sqrt{\gamma_{F}(\tau,\vec{\sigma})}} \sum_{i=1}^N
Q_i\,\delta(\vec{\sigma}-\vec{\eta}_i(\tau)),\nonumber\\
 &&\nonumber\\
  \overline{J}^r(\tau,\vec{\sigma})&=&
\frac{1}{\sqrt{\gamma_{F}(\tau,\vec{\sigma})}} \sum_{i=1}^N
Q_i\,\dot{\eta}^r_i(\tau)\,\delta(\vec{\sigma}-\vec{\eta}_i(\tau)),
 \label{VI91}
  \eea

\noindent   so that the total charge is

 \beq
Q_{tot}=\sum_{i=1}^N=\int d^3\sigma\,
\sqrt{\gamma_{F}(\tau,\vec{\sigma})}\,
\overline{\rho}(\tau,\vec{\sigma}),
 \label{VI92}
  \eeq

\noindent then Eqs.(\ref{VI90}) can be rewritten in the more
general form

 \bea
\frac{\partial}{\partial\sigma^r}\,\pi^r(\tau,\vec{\sigma})&\approx&
\sqrt{\gamma_{F}(\tau,\vec{\sigma})}\,
\overline{\rho}(\tau,\vec{\sigma}),\nonumber\\
 &&\nonumber\\
  \frac
{\partial\,\pi^r(\tau,\vec{\sigma})} {\partial\tau} &\,\cir\,&
\frac{\partial}{\partial\sigma^s}\,
\left[\sqrt{g_{F}}\,\gamma_{F}^{sv}\,\gamma_{F}^{ru}\, F_
{vu}-(g_{F\, \tau u}\,\gamma_{F}^{us}\,\pi^r -g_{F\, \tau
u}\,\gamma_{F}^{ur}\,\pi^s)
\right](\tau,\vec{\sigma})-\nonumber\\
 &&\nonumber\\
&-&\sqrt{\gamma_{F}(\tau,\vec{\sigma})}\,
\overline{J}^r(\tau,\vec{\sigma}).
 \label{VI93}
  \eea
\medskip

If we introduce the current density  4-vector

 \bea
s^\tau(\tau,\vec{\sigma})&=&
\frac{1}{\sqrt{g_{F}(\tau,\vec{\sigma})}} \sum_{i=1}^N
Q_i\,\delta(\vec{\sigma}-\vec{\eta}_i(\tau)),\nonumber\\
 &&\nonumber\\
  s^r(\tau,\vec{\sigma})&=&
\frac{1}{\sqrt{g_{F}(\tau,\vec{\sigma})}} \sum_{i=1}^N
Q_i\,\dot{\eta}^r_i(\tau)\,
\delta(\vec{\sigma}-\vec{\eta}_i(\tau)),
 \label{VI94}
  \eea

\noindent Eqs.(\ref{VI89}) are replaced by

 \beq
\frac{1}{\sqrt{g_{F}(\tau,\vec{\sigma})}}\,
\frac{\partial}{\partial \sigma^A}\left[
\sqrt{g_{F}(\tau,\vec{\sigma})}\,g_{F}^{AB}(\tau,\vec{\sigma})\,
g_{F}^{CD}(\tau,\vec{\sigma})F_{BD}\,
(\tau,\vec{\sigma})\right]\,\cir\, s^C(\tau,\vec{\sigma}).
 \label{VI95}
  \eeq
\medskip

From these equations, using the skew-symmetry of  $F_{AB}$, we
obtain the continuity equation

 \beq
\frac{1}{\sqrt{g_{F}(\tau,\vec{\sigma})}}\,
\frac{\partial}{\partial\sigma^C}\,
\left[\sqrt{g_{F}(\tau,\vec{\sigma})}\,
s^C(\tau,\vec{\sigma})\right]\cir 0.
 \label{VI96}
  \eeq

This equation can be rewritten in the 3-dimensional form

 \beq
\frac{1}{\sqrt{\gamma_{F}(\tau,\vec{\sigma})}}\,
\frac{\partial}{\partial\tau}\,
\left[\sqrt{\gamma_{F}(\tau,\vec{\sigma})}\,
\overline{\rho}(\tau,\vec{\sigma})\right]+
\frac{1}{\sqrt{\gamma_{F}(\tau,\vec{\sigma})}}\,
\frac{\partial}{\partial\sigma^r}\,
\left[\sqrt{\gamma_{F}(\tau,\vec{\sigma})}\,
\overline{J}^r(\tau,\vec{\sigma})\right]\cir 0,
 \label{VI97}
  \eeq
\medskip

\noindent and implies

 \beq
\frac{d}{d\tau}\,Q_{tot}\,\cir 0.
 \label{VI98}
  \eeq

\bigskip

Let us define the following {\it generalized non-inertial electric
and magnetic fields}

 \bea
  {\cal E}^{s}_{(F)}(\tau,\vec{\sigma})&=&
-\left[\frac{\sqrt{\gamma_{F}}}{\sqrt{N_{F}}}\,
\gamma_{F}^{sr}\left( F_{\tau r} -N_{F}^v\,
F_{vr}\right)\right](\tau,\vec{\sigma})
\,\cir\,\pi^s(\tau,\vec{\sigma})),\nonumber\\
 &&\nonumber\\
  {\cal B}_{(F)}^{w}(\tau,\vec{\sigma})&=& -\frac{1}{2}\,\epsilon^{wsr}\,
\left[N_{F}\,\sqrt{\gamma_{F}}\,
\gamma_{F}^{sv}\,\gamma_{F}^{ru}\, F_ {vu}-
(N_{F}^s\,\pi^r-N_{F}^r\,\pi^s) \right](\tau,\vec{\sigma}),
 \label{VI99}
  \eea

\noindent where we have introduced the lapse and shift functions
$N_{F}=\sqrt{g_{F}/\gamma_{F}}$, $N_{F}^r=g_{F\, \tau
u}\gamma_{F}^{ur}$. With these new fields the Hamilton equations
(\ref{VI93}) can be written in the form (we use the vector
notation as in the 3-dimensional Euclidean case)

 \bea
  \nabla\cdot\vec{\cal E}{}_{(F)}(\tau,\vec{\sigma})&=&
\sqrt{\gamma_{F}(\tau,\vec{\sigma})}\,
\overline{\rho}(\tau,\vec{\sigma}),\nonumber\\
 &&\nonumber\\
 \frac{\partial{\cal E}_{(F)}^{r}(\tau,\vec{\sigma})} {\partial\tau}
- (\nabla\times\vec{\cal B}{}_{(F)}(\tau,\vec{\sigma}))^r&=&
\sqrt{\gamma_{F}(\tau,\vec{\sigma})}\,
\overline{J}^r(\tau,\vec{\sigma}).
 \label{VI100}
  \eea

{\it With these non-inertial electric and magnetic fields the
Hamilton equations look like the usual source dependent Maxwell
equations written in a inertial frame}.
\bigskip

However it can be useful to introduce the standard definition (see
Ref.\cite{106a}) of the {\it inertial electric and magnetic
fields}

 \beq
E^r(\tau,\vec{\sigma})=\eta^r_s\,F_{\tau s}(\tau,\vec{\sigma}),
\qquad
 B^r(\tau,\vec{\sigma})= - \frac{1}{2}\epsilon^{rsu}F_{rs}(\tau,\vec{\sigma}).
 \label{VI101}
  \eeq

These fields satisfy the source independent Maxwell equation
(existence of the gauge potential) by definition

 \beq
\nabla\times\vec{E}(\tau,\vec{\sigma}) = 0, \qquad
  \nabla\cdot\vec{B}(\tau,\vec{\sigma}) = 0.
\label{VI102}
 \eeq
\medskip

The source dependent equations for these fields can be found
observing that we have
\medskip

 \bea
  {\cal E}^{s}_{(F)}(\tau,\vec{\sigma})&=&
\left[-\frac{\sqrt{\gamma_{F}}}{\sqrt{N_{F}}}\, \gamma_{F}^{sr}
\,E^r +\frac{\sqrt{\gamma_{F}}}{\sqrt{N_{F}}}\, \gamma_{F}^{sr}
(\vec{N}{}_{F}\times\vec{B})^r \right](\tau,\vec{\sigma}),
\nonumber\\
 &&\nonumber\\
  {\cal B}_{(F)}^{w}(\tau,\vec{\sigma})&=& \left[\frac{1}{2}\,\epsilon^{wsr}\,
N_{F}\,\sqrt{\gamma_{F}}\, \gamma_{F}^{sv}\,\gamma_{F}^{ru}\,
\epsilon_{vu\ell}\,B^\ell+ (\vec{N}{}_{F}\times\vec{ E})^w
\right](\tau,\vec{\sigma}),
 \label{VI103}
  \eea
\medskip

\noindent so that we get

  \bea
\nabla\cdot\vec{E}(\tau,\vec{\sigma})&=&
\sqrt{\gamma_{F}(\tau,\vec{\sigma})}\,\Big[\,
\overline{\rho}(\tau,\vec{\sigma})-
\overline{\rho}_R(\tau,\vec{\sigma})\,\Big],\nonumber\\
 &&\nonumber\\
  \frac{\partial{E}^{r}(\tau,\vec{\sigma})}
{\partial\tau} - (\nabla\times\vec{B}(\tau,\vec{\sigma}))^r&=&
\sqrt{\gamma_{F}(\tau,\vec{\sigma})}\,
\Big[\,\overline{J}^r(\tau,\vec{\sigma})-
\overline{J}^r_R(\tau,\vec{\sigma})\,\Big],
 \label{VI104}
  \eea

\noindent where

\bea
 \overline{\rho}_R(\tau,\vec{\sigma})&=&
\frac{1}{\sqrt{\gamma_{F}(\tau,\vec{\sigma})}}\, \nabla\cdot
\left(\vec{\cal
E}_{(F)}(\tau,\vec{\sigma})-\vec{E}(\tau,\vec{\sigma})\right),
\nonumber\\
 &&\nonumber\\
  \overline{J}^r_R(\tau,\vec{\sigma})&=&
\frac{1}{\sqrt{\gamma_{F}(\tau,\vec{\sigma})}}\, \left[
\frac{\partial}{\partial \tau}\, \left({\cal
E}^r_{(F)}(\tau,\vec{\sigma})-{E}^r(\tau,\vec{\sigma})\right)
-\left(\nabla\times\vec{\cal
B}{}_{(F)}-\nabla\times\vec{B}\right)^r\right].
 \label{VI105}
  \eea

Due to Eqs(\ref{VI103}), these charge and current densities are
functions only of the metric tensor and of the fields $\vec{E}$, $
\vec{B}$.

\bigskip

Also when the gauge fixing constraints (\ref{VI77}) identify the
admissible embedding $z^{\mu}_F(\tau ,\vec \sigma ) = x^{\mu}(\tau
) + F^{\mu}(\tau ,\vec \sigma ) = x^{\mu}(\tau ) +
\epsilon^{\mu}_r\, \zeta^r(\tau ,\vec \sigma ) $ with
$\zeta^r(\tau ,\vec \sigma ) = R^r{}_s(\tau ,\sigma )\, \sigma^s$
of Eq.(\ref{IV1}), whose simultaneity surfaces $\Sigma_{\tau}$ are
space-like hyper-planes with normal $l^{\mu} =
\epsilon^{\mu}_{\tau}$, we must use Eqs.(\ref{VI105}),  because
the 3-metric $g_{rs}$ of Eqs.(\ref{IV5}) has a complicate inverse
3-metric.

\medskip

Instead in Ref.\cite{106a} Schiff  uses a {\it non-admissible}
[$F(\sigma)=1$, $\vec{\Omega}(\tau,\sigma) \equiv
\vec{\Omega}(\tau)$] coordinate system of the type (\ref{IV1})
with $x^\mu(\tau)=u^\mu\,\tau$. In this case, we have

\bea
 &&N_{F}(\tau,\vec{\sigma})=
\sqrt{\gamma_{F}(\tau,\vec{\sigma})}=1,\;\;\;
\gamma_{F}^{rs}(\tau,\vec{\sigma})=-\delta^{rs},\nonumber\\
 &&\nonumber\\
  &&N_{F}^r(\tau,\vec{\sigma})=
(\vec{\Omega}(\tau)\times\vec{\sigma})^r.
 \label{VI106}
  \eea

If we put these expressions in Eqs.(\ref{VI103}), we find the
results of the Appendix A of Ref.\cite{107a}

 \bea
  \vec{\cal
E}_{(F)}(\tau,\vec{\sigma})&=&\vec{E}(\tau,\vec{\sigma})
+(\vec{\Omega}(\tau)\times\vec{\sigma})\times\vec{B}(\tau,\vec{\sigma}),
\nonumber\\
 &&\nonumber\\
  \vec{\cal B}{}_{(F)}(\tau,\vec{\sigma})&=&\vec{B}
+(\vec{\Omega}(\tau)\times\vec{\sigma})\times\vec{E}(\tau,\vec{\sigma})+
(\vec{\Omega}(\tau)\times\vec{\sigma})\times[
(\vec{\Omega}(\tau)\times\vec{\sigma})\times\vec{B}(\tau,\vec{\sigma})],
 \label{VI107}
 \eea

\noindent but at the price of a coordinate singularity when
$g_{\tau\tau}(\tau ,\vec \sigma )$ vanishes (the horizon problem).

\bigskip

Let us make some remarks

a) Eqs.(\ref{VI89}) are  the generally covariant equations
$\nabla_{\nu}\, F^{\mu\nu} \cir 0$, suggested by the equivalence
principle, in the 3+1 point of view after having taken care of the
asymptotic properties at spatial infinity. Eqs. (\ref{VI104}) and
(\ref{VI105}), with the metric associated to the admissible notion
of simultaneity (\ref{IV1}), should be the starting point for the
calculations in the magnetosphere of pulsars, instead of Schiff's
equations \cite{109a} (\ref{VI104}) and (\ref{VI107}), used in
Ref.\cite{110a}, for the case of uniform rotations  or of the
variants of Refs.\cite{111a} (based on Refs.\cite{46a}) avoiding
the so-called {\it light cylinder} (the horizon problem) for
$\omega\, R = c$, like Eqs.(\ref{IV1}), but with a bad behavior at
spatial infinity.

b) These equations also show that the non-inertial electric and
magnetic fields ${\vec {\cal E}}_{(F)}$ and ${\vec {\cal
B}}_{(F)}$ are {\it not}, in general, {\it equal} to the fields
obtained from the inertial ones ${\vec E}$ and ${\vec B}$ with a
Lorentz transformations to the comoving inertial system like it is
usually assumed following Rohrlich \cite{112a} and the locality
hypothesis \footnote{See Ref.\cite{113a} for the study of
electro-magnetic waves in a standard uniformly rotating frame
using the geodesics coordinates of type B quoted in Subsection C
of Section I. This study relies on the locality hypothesis and is
used to elucidate the phenomenon of helicity-rotation coupling.} .
Elsewhere we shall study the system of $N$ charged particles with
Grassmann valued electric charges plus the electro-magnetic field
in a non-inertial system (till now it was studied only in inertial
systems \cite{36a}) to understand the energy balance in the case
of accelerated charges emitting radiation.

\bigskip

Regarding the {\it electro-magnetic Dirac observables} on the
surface $z^{\mu}_F(\tau ,\vec \sigma ) = {\tilde x}^{\mu}(\tau ) +
F^{\mu}(\tau ,\vec \sigma )$, let us observe that  $A_r$ and
$\pi^r$ admit both a {\it non-covariant decomposition} \cite{111a}
in a transverse and a longitudinal part

\bea
 \pi^r &=& \pi^r_{\perp} + \pi^r_L,\nonumber \\
 &&{}\nonumber \\
 &&\pi^r_{\perp} = \epsilon^{rsu}\, \partial_s\, V_u,\qquad
 \partial_r\, \pi^r_{\perp} \equiv 0,\nonumber \\
  &&\quad \vec \partial \times {\vec \pi}_{\perp} = \vec \partial \times
 (\vec \partial \times \vec V) = \vec \partial \times \vec
 \pi,\nonumber \\
 &&\pi^r_L = {\tilde \partial}^r\, V_L,\qquad {\tilde \partial}^r
 = \delta^{rs}\, \partial_r,\qquad \vec \partial \times {\vec
 \pi}_L \equiv 0,\nonumber \\
 &&\partial_r\, \pi^r_L \equiv \triangle V_L = \partial_r\, \pi^r
 = \Gamma ,\qquad \triangle = \partial_r\, {\tilde \partial}^r =
 {\vec \partial}^2,\nonumber \\
 &&V_L = {1\over {\triangle}}\, \Gamma,\qquad \pi^r_{\perp} =
 (\delta^r_s - {\tilde \partial}^r\, {1\over {\triangle}}\,
 \partial_s)\, \pi^s,\nonumber \\
 &&{}\nonumber \\
 &&{}\nonumber \\
 A_r &=& A_{\perp r} + A_{L\, r},\nonumber \\
 &&{}\nonumber \\
 && A_{\perp r} = \epsilon_{rsu}\, \partial_s\, W_u,\qquad {\tilde
 \partial}^r\, A_{\perp r} \equiv 0,\nonumber \\
 && \vec \partial \times {\vec A}_{\perp} = \vec \partial \times
 (\vec \partial \times \vec W) = \vec \partial \times \vec
 A,\nonumber \\
 &&A_{L r} = \partial_r\, \eta_{em},\qquad \vec \partial \times
 {\vec A}_L = 0,\nonumber \\
 &&{\tilde \partial}^r\, A_{\perp r} \equiv \triangle\, \eta_{em}
 = {\tilde \partial}^r\, A_r,\nonumber \\
 &&\eta_{em} = {1\over {\triangle}}\, {\tilde \partial}^r\,
 A_r,\qquad A_{\perp r} = (\delta^s_r - \partial_r\, {1\over
 {\triangle}}\, {\tilde \partial}^s)\, A_s,
 \label{VI108}
 \eea

\noindent and a {\it covariant decomposition}

\bea
 \pi^r &=& {\hat \pi}^r_{\perp} + {\hat \pi}^r_L,\qquad
  \pi_r = g_{F\, rs}\, \pi^s,\nonumber \\
 &&{}\nonumber \\
 &&{\hat \pi}^r_{\perp} = (\gamma^{rs}_{F} - \nabla^r_{F}\, {1\over
 {\triangle_{F}}}\, \nabla^s_{F})\, \pi_s,\qquad {\hat
 \pi}_{\perp r} = g_{F\, rs}\, {\hat \pi}^s_{\perp},\nonumber \\
 &&{\hat \pi}^r_L =  \nabla^r_{F}\, {1\over
 {\triangle_{F}}}\, \nabla^s_{F}\, \pi_s,\nonumber \\
 &&{}\nonumber \\
 &&{}\nonumber \\
 A_r &=& {\hat A}_{\perp r} + {\hat A}_{L\, r},\qquad A^r =
 \gamma^{rs}_{F}\, A_s,\nonumber \\
 &&{\hat A}_{\perp r} = (g_{F\, rs} - \nabla_{F\, r}\, {1\over {\triangle_{F}}}\,
 \nabla_{F\, s})\, A^r,\nonumber \\
 &&{\hat A}_{L\, r} = \nabla_{F\, r}\, {1\over {\triangle_{F}}}\,
 \nabla_{F\, s}\, A^r.
 \label{VI109}
  \eea

Here $\nabla^r_{F}$ and $\triangle_{F} = \nabla^r_{F}\,
\nabla_{F\, r} = {1\over { \sqrt{\gamma_{F}(\tau ,\vec \sigma
)}}}\, \partial_r\, \Big( \sqrt{ \gamma_{F}(\tau ,\vec \sigma )
}\, \gamma_{F}^{rs}(\tau ,\vec \sigma )\, \partial_s\Big)$ are the
covariant derivative and the Laplace-Beltrami operator associated
to $g_{F\, rs}(\tau ,\vec \sigma |P)$. Since $\pi^r$ is a vector
density, we have $\partial_r\, \pi^r = \nabla_{F\, r}\, \pi^r$.

While with the non-covariant decomposition we can easily find a
Shanmugadhasan canonical transformation adapted to the Gauss law
constraint with the standard canonically conjugate Dirac
observables ${\vec A}_{\perp}$ and ${\vec \pi}_{\perp}$ of the
radiation gauge, it is not clear whether the covariant
decomposition can produce such a canonical basis. In any case, as
shown in Ref.\cite{114a}, the radiation gauge formalism is well
defined in both cases if we add suitable gauge fixings.

\bigskip

See Appendix A for a sketch of the derivation of the Sagnac effect
from the non-inertial Maxwell equations, following a suggestion of
Ref. \cite{57a}.

\bigskip

With foliations with parallel hyper-planes  [$z^{\mu}_F(\tau ,\vec
\sigma ) = x^{\mu}(\tau ) + F^{\mu}(\tau ,\vec \sigma ) =
x^{\mu}(\tau ) + \epsilon^{\mu}_r\, \zeta^r(\tau ,\vec \sigma ) $
with $\zeta^r(\tau ,\vec \sigma ) = R^r{}_s(\tau ,\sigma )\,
\sigma^s$ of Eq.(\ref{IV1})] the constraints (\ref{VI77}) imply
$l^{\mu}(\tau ,\vec \sigma ) = l^{\mu} = const.$, i.e. an inertial
reference system. As a consequence, the action of the external
Lorentz boosts [with the generators (\ref{VI73})] on the reduced
phase space is broken, because the given conditions $l^{\mu} =
const.$ are compatible only with a subset of the inertial Lorentz
frames. Let us remark that the breaking of the canonical action of
Lorentz boosts happens also for simultaneity surfaces more general
than hyper-planes like those defined by the gauge fixing
(\ref{VI77}), since the form $z^{\mu}_F(\tau ,\vec \sigma )$ of
the embedding is defined in the given reference inertial system.
\medskip

To recover a good canonical action of the Lorentz group we have to
{\it select a family of admissible embeddings with parallel
hyper-planes containing the given embedding with $l^{\mu} =
const.$ and all all the embeddings obtained by it by means of
Lorentz transformations} (i.e. with $l^{\mu} = const.\, \mapsto
\Lambda^{\mu}{}_{\nu}\, l^{\nu} = const.$). A similar family of
embeddings has to be found also for every type of admissible
simultaneity hyper-surfaces identified by the $z^{\mu}_F$ of
Eq.(\ref{VI77}) with a fixed $F^{\mu}$.

If we define $x^{\mu}(\tau ) = x^{\mu}(0) + l^{\mu}\, x_l(\tau ) +
\epsilon^{\mu}_r\, x^r_{\epsilon}(\tau )$ with $x_l(\tau ) =
\sgn\, l_{\mu}\, [x^{\mu}(\tau ) - x^{\mu}(0)]$ and
$x^r_{\epsilon}(\tau ) = -\sgn\, \epsilon^r_{\mu}\, [x^{\mu}(\tau
) - x^{\mu}(0)]$ [$\epsilon^A_{\mu}$ are inverse tetrads], the
embedding (\ref{IV1}) is rewritten in the form

\bea
 z^{\mu}_F(\tau ,\vec \sigma ) &=& x^{\mu}(\tau ) + F^{\mu}(\tau
 ,\vec \sigma ) = x^{\mu}(0) + l^{\mu}\, x_l(\tau )
+ \epsilon^{\mu}_r\, \xi^r(\tau ,\vec \sigma ),\nonumber \\
 &&{}\nonumber \\
 && \xi^r(\tau ,\vec \sigma ) = x^r_{\epsilon}(\tau ) + \zeta^r(\tau
,\vec \sigma ),\qquad F^{\mu}(\tau ,\vec \sigma ) =
\epsilon^{\mu}_r\, \zeta^r(\tau ,\vec \sigma ),\quad F^{\mu}(\tau
,\vec 0) = 0.\nonumber \\
 &&{}
 \label{VI110}
 \eea

To describe the above more general family of embeddings, let us
modify the Lagrangian $L(\tau )$ of Eqs.(\ref{VI71}) to
$L^{'}(\tau ) = L(\tau ) - \sqrt{{\sgn\, \dot X}^2(\tau )}$, by
adding the degrees of freedom of a free relativistic particle of
unit mass. This amounts to enlarge the phase space by adding the
new pairs of canonical variables $X^{\mu}(\tau )$, $U^{\mu}(\tau )
= {\dot X}^{\mu}(\tau ) / \sqrt{\sgn\, {\dot X}^2(\tau )}$, $\{
X^{\mu}(\tau ), U^{\nu}(\tau ) \} = - \sgn \, \eta^{\mu\nu}$
restricted by the first class constraint $\chi (\tau ) = \sgn\,
U^2(\tau ) - 1 \approx 0$, which is added to the Dirac Hamiltonian
(\ref{VI75}), $H_D \mapsto H^{'}_D = H_D + \kappa (\tau )\, \chi
(\tau )$ ($\kappa (\tau )$ is a new Dirac multiplier).

Then we replace the embedding (\ref{IV1}) with the more general
embedding [$l^{\mu},\, \epsilon^{\mu}_r\, \mapsto {\hat
U}^{\mu}(\tau ) = U^{\mu}(\tau ) / \sqrt{\sgn\, U^2(\tau )}
\approx U^{\mu}(\tau )$, $\epsilon^{\mu}_r(\hat U(\tau ))$]

\bea
 z^{\mu}_{FU}(\tau ,\vec \sigma ) &=& x^{\mu}(0) +
 {\hat U}^{\mu}(\tau )\, x_U(\tau ) +
 \epsilon^{\mu}_r(\hat U(\tau ))\, \xi^r_U(\tau ,\vec \sigma ) =\nonumber \\
 &=& x^{\mu}_U(\tau ) + F^{\mu}_U(\tau ,\vec \sigma ),\nonumber \\
 &&{}\nonumber \\
 &&x^{\mu}_U(\tau ) = x^{\mu}(0) + {\hat U}^{\mu}(\tau )\, x_U(\tau ) +
 \epsilon^{\mu}_r(\hat U(\tau ))\, x^r_U(\tau ),\nonumber \\
 &&\xi^r_U(\tau ,\vec \sigma ) = x^r_U(\tau ) + \zeta^r(\tau ,\vec
 \sigma ),\qquad F^{\mu}_U(\tau ,\vec \sigma ) =
 \epsilon^{\mu}_r(\hat U(\tau ))\, \zeta^r(\tau ,\vec \sigma
 ),\quad F^{\mu}_U(\tau ,\vec 0) = 0,\nonumber \\
 &&{}
 \label{VI111}
 \eea

\noindent where $\epsilon^{\mu}_A(\hat U) = \Big( {\hat U}^{\mu};
\epsilon^{\mu}_r(\hat U)\Big)$,  are the column of the standard
Wigner boost sending ${\hat U}^{\mu}$ at rest. \bigskip

In this enlarged phase space, studied in detail in Appendix B, we
select a special family of 3+1 splittings by means of the gauge
fixing constraints [see Eq.(\ref{a7})]

\bea
 S(\tau ,\vec \sigma ) &=& {\hat U}_{\mu}(\tau )\, [z^{\mu}(\tau
,\vec \sigma ) - z^{\mu}(\tau ,\vec 0)] \approx 0,\nonumber \\
 &&{}\nonumber \\
 &&\Downarrow\nonumber \\
 &&{}\nonumber \\
 z^{\mu}(\tau ,\vec \sigma ) &\approx& \theta (\tau )\, {\hat U}^{\mu}(\tau
) + \epsilon^{\mu}_r(\hat U(\tau ))\,{\cal A}^r(\tau ,\vec \sigma
).
 \label{VI112}
 \eea

{\it The admissible foliations of this family have space-like
hyper-planes, orthogonal to the arbitrary unit vector ${\hat
U}^{\mu}(\tau )$, as simultaneity and Cauchy leaves}. The centroid
$z^{\mu}(\tau ,\vec 0) = \theta (\tau )\, {\hat U}^{\mu}(\tau ) +
\epsilon^{\mu}_r(\hat U(\tau ))\, {\cal A}^r(\tau ,\vec 0)$,
origin of the 3-coordinates,  describes an arbitrary non-inertial
time-like observer and on the hyper-planes there is an arbitrary
admissible rotating frame determined by the functions ${\cal
A}^r(\tau ,\vec \sigma )$. As shown by Eqs.
(\ref{a8})-(\ref{a10}), if we make the decomposition
$\rho^{\mu}(\tau ,\vec \sigma ) = \sgn\, \Big( [ M_U(\tau ) +
{\tilde \rho}_U(\tau ,\vec \sigma )]\, {\hat U}^{\mu}(\tau ) -
\epsilon^{\mu}_r(\hat U(\tau ))\, \rho_{Ur}(\tau ,\vec \sigma
)\Big)\,\,$ \footnote{With $\rho_U(\tau ,\vec \sigma ) = {\hat
U}^{\mu}(\tau )\, \rho_{\mu}(\tau ,\vec \sigma )$, $\rho_{Ur}(\tau
,\vec \sigma ) = \epsilon^{\mu}_r(\hat U(\tau ))\, \rho_{\mu}(\tau
,\vec \sigma )$, $M_U(\tau ) = \int d^3\sigma\, \rho_U(\tau ,\vec
\sigma )$, ${\tilde \rho}_U(\tau ,\vec \sigma ) = \rho_U(\tau
,\vec \sigma ) - M_U(\tau )$.}, of the momentum  of
Eq.(\ref{VI72}), then the gauge fixing (\ref{VI112}) together with
the constraint ${\hat U}^{\mu}(\tau )\, [{\cal H}_{\mu}(\tau ,\vec
\sigma ) - \int d^3\sigma_1\, {\cal H}_{\mu}(\tau ,{\vec \sigma}_1
)] \approx 0$ \footnote{Determining ${\tilde \rho}_U(\tau ,\vec
\sigma ) \approx {\hat U}^{\mu}(\tau )\,  [l_{\mu}(\tau ,\vec
\sigma )\, T_{\tau\tau}(\tau ,\vec \sigma ) + z_{r\mu}(\tau ,\vec
\sigma )\, \gamma^{rs}(\tau ,\vec \sigma )\, T_{\tau s}(\tau ,\vec
\sigma ) - \int d^3\sigma_1 (l_{\mu}\, T_{\tau\tau} + z_{r\mu}\,
\gamma^{rs}\, T_{\tau s})(\tau ,{\vec \sigma}_1)] $.} form a pair
of second class constraints, which can be eliminated by going to
Dirac brackets.

\medskip

As shown in Appendix B a canonical basis for this new reduced
phase space is $\theta (\tau )$, $M_U(\tau )$, ${\cal A}^r(\tau
,\vec \sigma )$, $\rho_{Ur}(\tau ,\vec \sigma )$, ${\tilde
X}^{\mu}(\tau )$, $U^{\mu}(\tau )$ plus the electro-magnetic
canonical variables. Now we have $l^{\mu}(\tau ,\vec \sigma )
\equiv {\hat U}^{\mu}(\tau )$, $z^{\mu}_r(\tau ,\vec \sigma )
\equiv \epsilon^{\mu}_s(\hat U(\tau ))\, {{\partial {\cal
A}^s(\tau ,\vec \sigma )}\over {\partial \sigma^r}}$, $g_{rs}(\tau
,\vec \sigma ) \equiv -\sgn\, \sum_u\, {{\partial {\cal A}^u(\tau
,\vec \sigma ) }\over {\partial \sigma^r}}\, {{\partial {\cal
A}^u(\tau ,\vec \sigma ) }\over {\partial \sigma^s}}$ and $\sgn\,
{\tilde \rho}_U(\tau ,\vec \sigma ) \equiv T_{\tau\tau}(\tau ,\vec
\sigma ) - \int d^3\sigma_1\, T_{\tau\tau}(\tau ,{\vec
\sigma}_1)$. The remaining first class constraints are $\chi (\tau
) \approx 0$, $H_{\perp} = M_U(\tau ) - \int d^3\sigma \,
T_{\tau\tau}(\tau ,\vec \sigma ) \approx 0$, ${\cal H}_r(\tau
,\vec \sigma ) = {{\partial {\cal A}^s(\tau ,\vec \sigma )}\over
{\partial \sigma^r}}\, \rho_{Us}(\tau ,\vec \sigma ) - \sgn\,
T_{\tau r}(\tau ,\vec \sigma ) \approx 0$ with the components of
the energy-momentum tensor given in Eqs.(\ref{VI74}). The
canonical 4-coordinate ${\tilde X}^{\mu}(\tau ) = X^{\mu}(\tau ) +
W^{\mu}(\tau )$ is not a 4-vector \footnote{Like it happens with
the decoupled external 4-center of mass ${\tilde x}^{\mu}(\tau )$
in the rest-frame instant form in inertial frames.}: ${\tilde
X}^{\mu}$ plays the role of the decoupled 4-center of mass of the
accelerated isolated system. See Appendix B for the form of the
Poincare' generators. {\it The resulting breaking of the canonical
action of the Lorentz boosts is restricted to the gauge variable
${\tilde X}^{\mu}$}. As in the rest-frame instant form we can make
a canonical transformation from the canonical basis ${\tilde
X}^{\mu}$, $U^{\mu}$ to one spanned by ${\hat U}_{\mu}\, {\tilde
X}^{\mu} = {\hat U}_{\mu}\, X^{\mu}$ [since ${\hat U}_{\mu}\,
W^{\mu} = 0$], $\sqrt{\sgn\, U^2} \approx 1$, $\vec z =
\sqrt{\sgn\, U^2}\, [{\vec {\tilde X}} - {\tilde X}^o\, {\vec U} /
U^o] \approx {\vec {\tilde X}} - {\tilde X}^o\, {\vec U} / U^o$,
$\vec k = {\vec U} / \sqrt{\sgn\, U^2}\, \vec U \approx {\hat
{\vec U}}$, with $\vec z$ and $\vec k$ non-evolving Jacobi initial
data.

\bigskip

If we want to recover the embedding (\ref{VI111}), we have to add
the gauge fixings $\theta (\tau ) - x_U(\tau ) - {\hat
U}_{\mu}(\tau )\, x^{\mu}(0) \approx 0$, ${\cal A}^r(\tau ,\vec
\sigma ) - \xi^r_U(\tau ,\vec \sigma ) - \epsilon^r_{\mu}(\hat
U(\tau ))\, x^{\mu}(0) \approx 0$ with $x_U(\tau )$, $\xi^r_U(\tau
,\vec \sigma ) = x^r_U(\tau ) + \zeta^r(\tau ,\vec \sigma )$
[$\zeta^r(\tau ,\vec 0) = 0$] given ($U$-independent) functions.
This implies $z^{\mu}(\tau ,\vec \sigma ) \approx
z^{\mu}_{FU}(\tau ,\vec \sigma )$ and $z^{\mu}(\tau ,\vec 0) =
x^{\mu}_U(\tau ) = x^{\mu}(0) + {\hat U}^{\mu}(\tau )\, x_U(\tau )
+ \epsilon^{\mu}_r(\hat U(\tau ))\, x^r_U(\tau )$, i.e. a family
of admissible 3+1 splittings whose whose simultaneity leaves are
hyper-planes orthogonal to ${\hat U}^{\mu}(\tau )$ and with
rotating 3-coordinates determined by the functions $\zeta^r(\tau
,\vec \sigma )$ [for instance the admissible ones of
Eqs.(\ref{IV1})]. By going to new Dirac brackets we get a new
reduced phase space  spanned by ${\tilde X}^{\mu}(\tau )$,
$U^{\mu}(\tau )$ and the electro-magnetic canonical variables.

\bigskip

The natural gauge fixing to the constraint $\chi (\tau ) = \sgn\,
U^2(\tau ) - 1 \approx 0$ is ${\hat U}_{\mu}(\tau)\, {\tilde
X}^{\mu}(\tau ) - \sgn\, \theta (\tau ) \approx 0$: it replaces
the gauge fixing $\tau - u \cdot \tilde x \approx 0$ of the
rest-frame instant form. After this gauge fixing we have ${\tilde
X}^{\mu}(\tau ) = X^{\mu}(\tau ) + W^{\mu}(\tau ) = z^{\mu}(\tau ,
{\vec \sigma}_{\tilde X}(\tau ))$ and $X^{\mu}(\tau ) =
z^{\mu}(\tau ,{\vec \sigma}_X(\tau ))$ for some ${\vec
\sigma}_{\tilde X}(\tau )$ and ${\vec \sigma}_X(\tau )$.
\bigskip

If finally we want  to recover the embedding (\ref{IV1}), we must
add by hand three more first class constraints, the independent
ones in ${\hat U}^{\mu}(\tau ) \approx l^{\mu} = {\hat
U}^{\mu}(\vec k) = const.$, which determine $\vec k$. As gauge
fixings to these three extra constraints it is natural to choose
$\vec z \approx 0$. In this way ${\vec \sigma}_{\tilde X}(\tau )$
is determined.

\medskip

Therefore the description of non-inertial isolated systems follows
a pattern similar to that needed for their description in the
inertial system of the rest-frame instant form. There is a
decoupled non-covariant canonical variable, needed for the
canonical implementation of the external Lorentz transformations.
However now it does not carry the conserved 4-momentum of the
isolated system, which is associated to the centroid describing a
non-inertial observer.

\bigskip

The formalism developed in this Subsection and in Appendix B will
be needed to implement the program of quantization of the
electro-magnetic field in non-inertial systems along the lines
under study for relativistic particles in Ref.\cite{115a}.

\bigskip

Finally see Refs.\cite{57a,69a,113a,116a} for what is known on the
open problem of the constitutive equations for electrodynamics in
material media in non-inertial systems.

\vfill\eject

\section{Conclusions.}

In the Introduction we have reviewed  many old and new physical
problems in special relativity, which are naturally formulated in
accelerated (in particular rotating) frames. We have stressed that
they present pathologies (coordinate singularities) originating
from the absence of an admissible notion of simultaneity (i.e. of
a frame-dependent rule for the synchronization of distant clocks
to the reference clock of a given observer). Then we have analyzed
in detail which are the conditions to be imposed on coordinate
transformations, starting from the standard Cartesian coordinates
of Minkowski space-time, so that the new {\it equal time surfaces}
(instantaneous 3-spaces) are the space-like leaves of the
foliation associated to an admissible 3+1 splitting of Minkowski
space-time. Einstein's convention is a very special case and
corresponds to the space-like hyper-planes orthogonal to the
world-line of an inertial observer. More in general the leaves of
a foliation  will have both a linear acceleration, describing how
they are packed, and a parametrization with differentially
rotating 3-coordinate systems. It turns out that, while there is
no restriction on linear accelerations, on the contrary angular
velocities and rotational accelerations cannot be given
arbitrarily, but must be suitably restricted. In particular rigid
rotations are not allowed.
\bigskip

In this paper it is pointed out that it is convenient to
characterize the admissible 3+1 splittings of Minkowski space-time
with {\it intrinsic Lorentz-invariant radar 4-coordinates }
$\sigma^A = (\tau , \vec \sigma )$ [$\tau$ labels the leaves and
$\vec \sigma = (\sigma^r )$ are curvilinear 3-coordinates on the
leaf $\Sigma_{\tau}$ with respect to an arbitrary centroid
$x^{\mu}(\tau )$], which parametrize the {\it embedding} $x^{\mu}
= z^{\mu}(\tau ,\vec \sigma )$ of the leaves in Minkowski
space-time. We have explicitly built a family of such admissible
radar coordinates implementing the {\it locality hypothesis}. A
sub-family of these 4-coordinates corresponds to foliations of
Minkowski space-time with parallel (but in general not
equi-spaced) hyper-planes endowed with suitable differentially
rotating 3-coordinates. In particular all these admissible
foliations of Minkowski space-time are associated to arbitrary
accelerated time-like observers, whose world-line $x^{\mu}(\tau )$
is the centroid origin of the 3-coordinates. The {\it equal time
($\tau = const.$) surfaces} (instantaneous 3-spaces) are {\it not
orthogonal} to the world-line of the observer and their associated
notion of simultaneity corresponds to a modification of Einstein's
convention for the synchronization of clocks. Moreover we have
shown that given an admissible foliation with simultaneity
surfaces, there are two associated congruences of time-like (in
general non-inertial) observers. One non-rotating determined by
the normals to the simultaneity surfaces and one rotating
(non-surface-forming) determined by the $\tau$-time derivative of
the embedding $z^{\mu}(\tau ,\vec \sigma )$.

\bigskip

All the admissible notions of simultaneity are {\it gauge
equivalent} for the description of an isolated system, when it
admits a formulation in terms of a parametrized Minkowski theory.
In this case the Lagrangian density depends on the embedding
$z^{\mu}(\tau ,\vec \sigma )$ besides on the variables of the
system and there is a special type of general covariance peculiar
to special relativity. As a consequence the embeddings are {\it
gauge variables} and the transition from a 3+1 splitting to
another one (i.e. the change of the notion of simultaneity) is a
gauge transformation like the change of the 3-coordinates $\vec
\sigma$. Therefore parametrized Minkowski theories, which have
local invariance (second Noether theorem) under the sub-group of
frame preserving diffeomorphisms $\tau \mapsto f(\tau ,\vec \sigma
)$, $\vec \sigma  \mapsto \vec g(\vec \sigma )$, are generally
covariant theories like general relativity, where the embedding
$z^{\mu}(\tau ,\vec \sigma )$ are replaced by the 4-metric and the
Hilbert action has local invariance under the full group of
diffeomorphisms. In canonical metric gravity the change of the
foliation (i.e. of the notion of simultaneity) is a Hamiltonian
gauge transformation generated by the super-Hamiltonian constraint
\cite{86a}, while the super-momentum constraints are the
generators of the change of the 3-coordinates adapted to the
foliation. Parametrized Minkowski theories are a non-trivial
genuine (i.e. not artificial) example which validates
Kretschmann's rejection of Einstein's argument that only general
relativity has a general covariance group \cite{117a}.

\bigskip

Let us emphasize that the real novelty of canonical gravity is
that each solution of Einstein's equations, i.e. of the Hamilton
equations in a completely fixed gauge, with given boundary
conditions and allowed initial data also determines the extrinsic
curvature of the Cauchy surfaces, which are then found solving an
inverse problem. As a consequence, {\it the dynamics of the
gravitational field determines the admissible notions of
simultaneity in general relativity} \cite{8}: they are much less
than the admissible ones of special relativity, because in absence
of matter, as said in the second paper of Ref.\cite{94a}, they
must have the leaves 3-conformally flat.

\bigskip

Let us remark that there is a distinction between notions and/or
results independent from the choice of the notion of simultaneity
(i.e. they are observer-independent, like the statement that two
events $A$ and $B$ have a space-like separation) and those which
are frame-dependent (like the solutions of the equations of motion
after a well defined choice of the Cauchy-simultaneity surface).
However, like in general relativity \cite{8a}, the definition of
an extended laboratory, with its instruments and its standard
units, corresponds to a well defined choice of the notion of
simultaneity, of an associated notion of spatial distance and of
an associated set of adapted 4-coordinates (or of their intrinsic
variant, the radar coordinates). Therefore a laboratory will
always give a frame-dependent description of physical systems. A
change of the notion of simultaneity and of the 4-coordinates will
only produce {\it inertial} effects in the description of the
motion of particles and fields.

\bigskip

The 3+1 point of view, in which the simultaneity surfaces are also
Cauchy surfaces for the equation of motion of isolated systems,
has been contrasted in this paper with the 1+3 point of view of
either an accelerated observer or  a rotating congruence of
observers like the one determined by the $\tau$-time derivative of
an embedding $z^{\mu}(\tau ,\vec \sigma )$. While, after having
endowed the observer with a tetrad (whose space axes are
arbitrarily chosen), the 1+3 point of view is the only one
allowing to define the tetradic (coordinate-independent but
tetrad-dependent) components of the fields (like the
electro-magnetic field) to be measured locally by the observer,
only in the 3+1 point of view we have a well posed Cauchy problem
and a control on the predictability of the theory.

\bigskip

We have analyzed the gauge nature (frame-dependence) of the
notions of one-way velocity of light and spatial distance and
compared the results of the 3+1 point of view (admissible global
notions of simultaneity and instantaneous 3-space) with those of
the 1+3 one (only approximate non-global synchronizability of
clocks and non-existence of an instantaneous  3-space, locally
replaced with the 3-space of the vectors orthogonal to the
observer world-line).

\bigskip

Then we have applied the formalism of the admissible 4-coordinate
transformations to various problems.
\medskip

A) We have delineated a method for building a grid of radar
4-coordinates after having assigned an admissible modification of
Einstein convention plus a convention on how to build fixed-time
3-coordinates. This method could be used by a set of spacecrafts
or satellites like in the GPS setting.
\medskip

B) We have given the 3+1 point of view on the rotating disk and
the Sagnac effect.
\medskip

C) We have evaluated the correction of order $c^{-3}$ to the
one-way time transfer between an Earth station and a satellite due
to the rotation of the Earth, after having established a grid of
radar coordinates like in A) and we have given a re-interpretation
of the ACES mission as the determination of the deviation from
Einstein's convention of the chosen notion of simultaneity.
Similar calculations could be done for LISA and VLBI.
\medskip

D) We have studied the description of the electro-magnetic field
with a parametrized Minkowski theory and analyzed in detail the
restriction to an arbitrary notion of simultaneity. As a
consequence we have determined the most general form of Maxwell
equations in a non-inertial system and studied in detail the case
of foliations with parallel hyper-planes and rotating
3-coordinates. This will be useful in the study of the
magneto-sphere of pulsars and of the energy balance for the
radiation emitted by accelerated charges.

\bigskip

The technology developed in D) will be needed for the study of a
new method of quantization of relativistic particles and of the
electro-magnetic field in non-inertial frames \cite{115a}.

\vfill\eject

\appendix

\section{The Sagnac Effect from Non-Inertial Maxwell Equations.}

Let us now sketch how it is possible to derive the Sagnac effect
from Maxwell equations in non-inertial system as suggested in
Ref.\cite{57a}.

\medskip

To find the bridge between the geometric derivation of the Sagnac
effect and the non-inertial equations of motion of the
electro-magnetic field, we can use the {\em eikonal
approximation}. To do this we specify a embedding $z^{\mu}_F(\tau
,\vec \sigma )$ of the form (\ref{VI110}), that is such that the
hyper-surfaces are parallel hyper-planes with constant normal
$l^\mu$.

It is convenient to use gauge fixed vector potentials
$A_B(\tau,\vec{\sigma})$ satisfying the conditions

 \bea
  &&A_N(\tau,\vec{\sigma})=
\left[\frac{1}{N_{F}}\left(
A_\tau-N^r_{F}\,A_r\right)\right](\tau,\vec{\sigma})=0,
\nonumber\\
 &&\nonumber\\
 &&\frac{1}{\sqrt{\gamma_{F}(\tau,\vec{\sigma})}}
\frac{\partial}{\partial\sigma^r}\,\left(
{\sqrt{\gamma_{F}(\tau,\vec{\sigma})}}\,
\gamma^{rs}_{F}(\tau,\vec{\sigma})\,
A_s(\tau,\vec{\sigma})\right)=0.
 \label{b1}
  \eea

These conditions correspond to a radiation gauge for the inertial
observers with coordinates $\tau$, $\xi^r(\tau ,\vec \sigma ) =
R^r{}_s(\tau ,\sigma )\, \sigma^s$. Using the notations of
Eq.(\ref{VI110}), they imply

\beq
 \frac{1}{\sqrt{g_{F}(\tau,\vec{\sigma})}}
\frac{\partial}{\partial\sigma^A}\,\left(
{\sqrt{g_{F}(\tau,\vec{\sigma})}}\, g_F^{AB}(\tau,\vec{\sigma})\,
A_B(\tau,\vec{\sigma})\right)=0.
 \label{b2}
  \eeq

Then we make the following {\it ansatz} for the potential

 \beq
A_B(\tau,\vec{\sigma})=\frac{1}{\omega^2}\, {\cal
A}_B(\tau,\vec{\sigma})\,\exp[\,i\,\omega\,
\Phi(\tau,\vec{\sigma})],
 \label{b3}
  \eeq

\noindent where $\omega$ is a frequency. We assume the validity of
the following conditions ({\em eikonal approximation})

 \beq
  \omega>>1,\qquad \left|\frac{\partial
{\cal A}_B(\tau,\vec{\sigma})} {\partial\sigma^C}\right|<<1.
 \label{b4}
  \eeq

At order $1/\omega$ the equations of motion (\ref{VI89}) give

\beq
 \left[{\cal A}_D\; g_{F}^{AB}\;
\frac{\partial\Phi}{\partial\sigma^A}\,
\frac{\partial\Phi}{\partial\sigma^B}\,-
\frac{\partial\Phi}{\partial\sigma^D}\, {\cal A}_B\; g_{F}^{AB}\;
\frac{\partial\Phi}{\partial\sigma^A}\,
\right](\tau,\vec{\sigma})+{\cal O}(1/\omega)=0.
 \label{b5}
  \eeq

The  condition (\ref{b2}) gives at the same order

 \beq
  \left[{\cal
A}_B\; g_{F}^{BA}\;
\frac{\partial\Phi}{\partial\sigma^A}\right](\tau,\vec{\sigma})+{\cal
O}(1/\omega)=0.
 \label{b6}
  \eeq

\noindent Therefore we get the {\em eikonal equation}

 \beq
g_{F}^{AB}(\tau,\vec{\sigma})\;
\frac{\partial\Phi}{\partial\sigma^A}(\tau,\vec{\sigma})\;
\frac{\partial\Phi}{\partial\sigma^B}(\tau,\vec{\sigma})=0.
 \label{b7}
  \eeq

\bigskip

Let us make the {\it ansatz} that there is a solution of the type
(this is the weak point of the derivation, because strictly
speaking this solution requires a static metric; one should show
that the deviations from the static case are negligible!)

 \beq
\Phi(\tau,\vec{\sigma})=\tau+\Psi(\vec{\sigma}).
 \label{b8}
  \eeq

We want to evaluate the infinitesimal variation of the $\Psi$
along a infinitesimal 3-dimensional displacement tangent to a
curve $\sigma^r(\lambda)$. Namely we want to evaluate

 \beq
d\Psi(\vec{\sigma}(\lambda))=
\frac{\partial\Psi}{\partial\sigma^r} (\vec{\sigma}(\lambda))\cdot
\frac{d\sigma^r(\lambda)}{d\lambda}\,d\lambda .
 \label{b9}
  \eeq

To do this, we transform the 3-dimensional curve
$\sigma^r(\lambda)$ in Minkowski space-time in a world-line by
introducing a $\tau(\lambda)$ such that

 \beq
\Phi(\tau(\lambda),\vec{\sigma}(\lambda))=const.,
 \label{b10}
  \eeq

\noindent so that we get

\beq
\frac{\partial\Phi}{\partial\sigma^B}(\tau(\lambda),\vec{\sigma}(\lambda))\cdot
\frac{d\sigma^B(\lambda)}{d\lambda}=0.
 \label{b11}
  \eeq

From the eikonal equation (\ref{b7}) we obtain

 \beq
\alpha(\lambda)\,\frac{d\sigma^B(\lambda)}{d\lambda}=
\left[\,g^{BA}_{F}\,
\frac{\partial\Phi}{\partial\sigma^A}\,\right](\tau(\lambda),\vec{\sigma}(\lambda)),
 \label{b12}
  \eeq

\noindent where $\alpha(\lambda)$ is a multiplier depending of the
choice of the affine parameter $\lambda$. Then we also have

\beq
 \frac{d\sigma^A(\lambda)}{d\lambda}\cdot
g^{AB}_{F}(\tau(\lambda),\vec{\sigma}(\lambda))\cdot
\frac{d\sigma^B(\lambda)}{d\lambda}=0.
 \label{b13}
  \eeq

\noindent Therefore $\sigma^A(\lambda)$ is a null curve. From
Eq.(\ref{b12}) and using the (\ref{b8}), we obtain

 \bea
  \alpha(\lambda)\,
\frac{d\tau(\lambda)}{d\lambda}&=&\left[\, g^{\tau r}_{F}\,
\frac{\partial\Psi}{\partial\sigma^r}+g^{\tau\tau}_{F}\,\right]
(\tau(\lambda),\vec{\sigma}(\lambda)),\nonumber\\
 &&\nonumber\\
 \alpha(\lambda)\, \frac{d\sigma^r(\lambda)}{d\lambda}&=&
\left[\,g^{rs}_{F}\, \frac{\partial\Psi}{\partial\sigma^s}
+g^{\tau\,r}_{F}\,\right] (\tau(\lambda),\vec{\sigma}(\lambda)).
 \label{b14}
  \eea

Using the second of these equations we obtain

 \beq
\alpha(\lambda)\,\frac{\partial\Psi}{\partial\sigma^r}
(\vec{\sigma}(\lambda))\cdot \frac{d\sigma^r(\lambda)}{d\lambda}
=\left[\, \frac{\partial\Psi}{\partial\sigma^r}\, g^{rs}_{F}\,
\frac{\partial\Psi}{\partial\sigma^s}
+\frac{\partial\Psi}{\partial\sigma^r} g^{\tau r}_{F}\,\right]
(\tau(\lambda),\vec{\sigma}(\lambda)).
 \label{b15}
  \eeq

Summing the first of Eqs.(\ref{b14}) with Eq. (\ref{b15}) and
using Eqs.(\ref{b7}) and (\ref{b8}), we obtain

 \beq
 \frac{\partial\Psi}{\partial\sigma^r}
(\vec{\sigma}(\lambda))\cdot\frac{d\sigma^r(\lambda)}{d\lambda}
=-\frac{d\tau(\lambda)}{d\lambda}.
 \label{b16}
  \eeq

We can think a ray of light constrained to follow a curve
$\vec{\sigma}(\lambda)$ as a sequence of wave plane solutions of
the type (\ref{b3}) covering infinitesimal distances
$(d\sigma^r(\lambda)/d\lambda)d\lambda$. Then the phase shift
accumulate by this ray of light along a finite length on the curve
is

 \beq
  \Delta\Psi=-\int d\lambda\,\frac{d\tau(\lambda)}{d\lambda},
 \label{b17}
  \eeq

\noindent where $\tau(\lambda),\vec{\sigma}(\lambda)$ is a null
curve.

This justifies the geometrical calculus of Subsection C of Section
VI. In that case the 3-dimensional curve $\vec{\sigma}(\varphi)$
was the circle on the hyperplane and the $\tau(\varphi)$ was built
imposing Eqs.(\ref{b13}) [see Eq.(\ref{VI26}) for $\varphi (\tau
)$] obtaining so the two solutions corresponding to the two
directions.

\vfill\eject

\section{A Family of Foliations with Hyper-planes Closed under
Lorentz Transformations.}

In this Appendix we study a family of admissible embeddings with
parallel hyper-planes closed under the action of the Lorentz
transformations of the inertial system, namely such that a Lorentz
transformation maps one member of the family onto another of its
members.\medskip

Let us consider the $U$-dependent family of embeddings of
Eq.(\ref{VI111})

\bea
 z^{\mu}_{FU}(\tau ,\vec \sigma ) &=& x^{\mu}(0) + {\hat U}^{\mu}(\tau )\, x_U(\tau ) +
 \epsilon^{\mu}_r({\hat U}(\tau ))\, \xi^r_U(\tau ,\vec \sigma ) =\nonumber \\
 &=& x^{\mu}_U(\tau ) + F^{\mu}_U(\tau ,\vec \sigma ),\qquad
 F^{\mu}_U(\tau ,\vec 0) = 0,\nonumber \\
 &&{}\nonumber \\
 &&x^{\mu}_U(\tau ) = x^{\mu}(0) + {\hat U}^{\mu}(\tau )\, x_U(\tau ) +
 \epsilon^{\mu}_r({\hat U}(\tau ))\, x^r_U(\tau ),\nonumber \\
 &&\xi^r_U(\tau ,\vec \sigma ) = x^r_U(\tau ) + \zeta^r(\tau ,\vec
 \sigma ),\qquad F^{\mu}_U(\tau ,\vec \sigma ) =
 \epsilon^{\mu}_r({\hat U}(\tau ))\, \zeta^r(\tau ,\vec \sigma ),
 \label{a1}
 \eea

\noindent where ${\hat U}^\mu(\tau )$ is the unit normal to the
hyper-surface and ${\hat U}^{\mu}$, $\epsilon^{\mu}_r({\hat U})$
are the columns of the standard Wigner boost sending ${\hat
U}^{\mu}$ at rest. As a consequence the hyper-sufaces of this
foliations are parallel hyper-planes. The $\epsilon^\mu_a({\hat
U})$ are a triad of space-like four-vector such that \cite{87a}

\begin{eqnarray}
&&{\hat U}_\mu\,\epsilon^\mu_a({\hat U})=0,\quad
\epsilon^\mu_a({\hat U})\epsilon_{\mu\,b}({\hat
U})=\eta_{ab},\quad {\hat U}_\mu\,
\frac{\partial\epsilon^\lambda_a({\hat U})}{\partial {\hat
U}_\mu}=0, \nonumber\\
 &&\nonumber\\
  &&\epsilon^\mu_a(\Lambda\, {\hat U})=\Lambda^\mu{}_\nu\,
\epsilon^\nu_b({\hat U})R_{ba}(\Lambda,{\hat U}),
 \label{a2}
\end{eqnarray}

\noindent where $R_{ba}(\Lambda,{\hat U})$ is the Wigner rotation
associed to the Lorentz transformation $\Lambda$ by the standard
Wigner boost $L({\hat U},{\hat U}_o)$, such that ${\hat
U}^\mu=L^\mu_\nu({\hat U},{\hat U}_o)\, {\hat U}^\nu_o$ (${\hat
U}_o^\nu=(1,0,0,0)$). By definition we have $[L({\hat U},{\hat
U}_o)\, \Lambda\, L^{-1}({\hat U}, {\hat U}_o)]^i{}_j =
R_{a=i,b=j}(\Lambda,{\hat U})$, $[L({\hat U}, {\hat U}_o)\,
\Lambda\, L^{-1}({\hat U},{\hat U}_o)]^o{}_o=1$, $[L({\hat U},
{\hat U}_o)\, \Lambda\, L^{-1}({\hat U},{\hat U}_o)]^i{}_o=
[L({\hat U},{\hat U}_o)\, \Lambda\, L^{-1}({\hat U},{\hat
U}_o)]^o{}_j =0$.

\bigskip

When we add the free relativistic particle $X^{\mu}(\tau )$ of
unit mass to the Lagrangian (\ref{VI71}), its conjugate momentum
$U^{\mu}(\tau )$ ($\{X^\mu(\tau),U_\nu(\tau)\}=-\eta^\mu_\nu$)
realizes the  parameter of the family as a canonical variable,
which satisfies the extra first class constraint

\begin{equation}
\chi(\tau)= \sgn\, U^2(\tau)-1\approx 0,\quad \Rightarrow\quad
{\hat U}^{\mu}(\tau ) = {{U^{\mu}(\tau )}\over {\sqrt{}\sgn \,
U^2(\tau )}} \approx U^{\mu}(\tau ).
 \label{a3}
\end{equation}

\noindent The new Dirac hamiltonian (see Eq.(\ref{VI75}); we
momentarily ignore the electro-magnetic constraints) is

 \begin{equation}
H_D(\tau)=\int d^3\sigma \left[{\tilde \lambda}_\perp
(\tau,\vec{\sigma}) {\cal H}_\perp (\tau,\vec{\sigma})+ {\tilde
\lambda}^r(\tau,\vec{\sigma}){\cal H}_r(\tau,\vec{\sigma})
\right]+\kappa(\tau)\,\chi(\tau).
 \label{a4}
\end{equation}

The canonical generators (\ref{VI73}) of the Poincar\'e group are
replaced by

\begin{eqnarray}
P_s^\mu(\tau)&=& U^\mu(\tau)+\int
d^3\sigma\,\rho^\mu(\tau,\vec{\sigma}), \nonumber\\
 &&\nonumber\\
 J_s^{\mu\nu}(\tau) &=& X^\mu(\tau)U^\nu(\tau)-
X^\nu(\tau)U^\mu(\tau)+ \int d^3\sigma\,[
z^\mu(\tau,\vec{\sigma})\rho^\nu(\tau,\vec{\sigma})-
z^\nu(\tau,\vec{\sigma})\rho^\mu(\tau,\vec{\sigma})].
 \label{a5}
\end{eqnarray}

To identify the embeddings (\ref{a1}) we cannot use the gauge
fixings (\ref{VI77}) implying $z^{\mu}(\tau ,\vec \sigma ) \approx
z^{\mu}_F(\tau ,\vec \sigma ) = x^{\mu}(\tau ) + F^{\mu}(\tau
,\vec \sigma )$. Instead we have to introduce the gauge fixing

\begin{equation}
S(\tau,\vec{\sigma})={\hat U}^\mu(\tau)\,\left[
z_\mu(\tau,\vec{\sigma})-z_\mu(\tau,0)\right]\approx 0,
 \label{a6}
\end{equation}

\noindent implying

\bea
 z^{\mu}(\tau ,\vec \sigma ) &\approx& \theta (\tau )\, {\hat
 U}^{\mu}(\tau ) + \epsilon^{\mu}_r(\hat U(\tau ))\, {\cal
 A}^r(\tau ,\vec \sigma ) =\nonumber \\
 &=& z^{\mu}(\tau ,\vec 0) +
\epsilon^{\mu}_r(U(\tau ))\, \Big[{\cal A}^r(\tau ,\vec \sigma ) -
{\cal A}^r(\tau ,\vec 0)\Big] ,\nonumber \\
 &&{}\nonumber \\
 &&\theta (\tau ) = \sgn\, {\hat U}^{\mu}(\tau )\, z_{\mu}(\tau ,\vec
 0),\qquad {\cal A}^r(\tau ,\vec \sigma ) = -\sgn\,
 \epsilon^r_{\mu}(\hat U(\tau ))\,
 z^{\mu}(\tau ,\vec \sigma ),\nonumber \\
 &&{}\nonumber \\
 &&z^{\mu}(\tau ,\vec 0) = \theta (\tau )\, {\hat U}^{\mu}(\tau )
 + \epsilon^{\mu}_r(\hat U(\tau ))\, {\cal A}^r(\tau ,\vec
 0),\nonumber \\
 &&{}\nonumber \\
 &&z^{\mu}_r(\tau ,\vec \sigma ) \approx \epsilon^{\mu}_s(\hat U(\tau
 ))\, {{\partial {\cal A}^s(\tau ,\vec \sigma )}\over {\partial
 \sigma^r}},\quad \Rightarrow \quad l^{\mu}(\tau ,\vec \sigma )
 \approx {\hat U}^{\mu}(\tau ),\nonumber \\
 &&{}\nonumber \\
 \Rightarrow&& g_{rs}(\tau ,\vec \sigma ) \approx -\sgn\, \sum_u\,
 {{\partial  {\cal A}^u(\tau ,\vec \sigma )}\over {\partial \sigma^s}}\,
 {{\partial {\cal A}^u(\tau ,\vec \sigma )}\over {\partial \sigma^s}}.
 \label{a7}
 \eea

Therefore the gauge fixing (\ref{a6}) implies that the
simultaneity surfaces $\Sigma_{\tau}$ are hyper-planes orthogonal
to the arbitrary time-like unit vector ${\hat U}^{\mu}(\tau )$,
which is a constant of the motion since Eq.(\ref{a4}) implies ${{d
{\hat U}^{\mu}(\tau )}\over {d \tau}} \cir 0$.

\bigskip

The time preservation of the gauge fixing (\ref{a6}) implies

\begin{equation}
\frac{d}{d\tau}\,S(\tau,\vec{\sigma})\approx 0\,
\Rightarrow\,{\tilde \lambda}_\perp(\tau,\vec{\sigma})- {\tilde
\lambda}_\perp(\tau,0)\approx 0\, \Rightarrow\,{\tilde
\lambda}_\perp(\tau,\vec{\sigma})\approx\mu(\tau),
 \label{a8}
\end{equation}

\noindent and then in the reduced theory we have the Dirac
Hamiltonian (still ignoring the electro-magnetic constraints)

 \begin{equation}
H_D(\tau)=\mu (\tau) {\cal H}_\perp (\tau)+\int d^3\sigma\,
{\tilde \lambda}^r(\tau,\vec{\sigma}){\cal H}_r(\tau,\vec{\sigma})
+\kappa(\tau)\,\chi(\tau),
 \label{a9}
\end{equation}

\noindent where

\begin{equation}
H_\perp (\tau)=\int d^3\sigma\, {\cal H}_\perp
(\tau,\vec{\sigma})\approx 0,
 \label{a10}
\end{equation}

\noindent and [see Eqs.(\ref{VI74}) ]

\begin{eqnarray}
{\cal H}_\perp (\tau,\vec{\sigma}) &=& l^{\mu}(\tau ,\vec \sigma
)\, {\cal H}_{\mu}(\tau ,\vec \sigma ) \approx {\cal H}_U(\tau
,\vec \sigma ) = {\hat U}^{\mu}(\tau )\, {\cal H}_{\mu}(\tau ,\vec
\sigma ) \approx \nonumber \\
 &\approx& \rho_U(\tau,\vec{\sigma})- \sgn\,
T_{\tau\tau}(z^\mu(\tau,\vec{\sigma}),{\cal I})\approx
0,\nonumber\\
 &&\nonumber\\
  {\cal H}_r(\tau,\vec{\sigma}) &=& z^{\mu}_r(\tau ,\vec \sigma
  )\, {\cal H}_{\mu}(\tau ,\vec \sigma ) =
z^\mu_r(\tau,\vec{\sigma})\rho_\mu(\tau,\vec{\sigma})- \sgn\,
T_{\tau r}(z^\mu(\tau,\vec{\sigma}),{\cal I})\approx \nonumber \\
 &\approx& \epsilon^{\mu}_s(\hat U(\tau ))\, {{\partial {\cal
 A}^s(\tau ,\vec \sigma )}\over {\partial \sigma^r}}\, {\cal
 H}_{\mu}(\tau ,\vec \sigma ) = {{\partial {\cal A}^s(\tau ,\vec
 \sigma )}\over {\partial \sigma^r}}\, {\cal H}_{Us}(\tau ,\vec
 \sigma ) =\nonumber \\
 &=& {{\partial {\cal A}^s(\tau ,\vec \sigma )}\over
 {\partial \sigma^r}}\, \rho_{Us}(\tau ,\vec \sigma ) - \sgn\, T_{\tau
 r}(\tau ,\vec \sigma ) \approx 0,\nonumber \\
 &&{}\nonumber \\
 &&{}\nonumber \\
 \rho_U(\tau ,\vec \sigma ) &=& {\hat U}^{\mu}(\tau )\,
 \rho_{\mu}(\tau ,\vec \sigma ),\qquad
 \rho_{Ur}(\tau ,\vec \sigma ) = \epsilon^{\mu}_r(\hat U(\tau
 ))\, \rho_{\mu}(\tau ,\vec \sigma ).
 \label{a11}
\end{eqnarray}

\noindent Here we introduced the notation ${\cal I} = \Big(
A_A(\tau ,\vec \sigma ), \pi^A(\tau ,\vec \sigma )\Big) $
[$\{A_A(\tau,\vec{\sigma}),
\pi^B(\tau,\vec{\sigma}')\}=\delta^\beta_\alpha\,
\delta^3(\vec{\sigma}-\vec{\sigma}')$] to denote the
electro-magnetic canonical variables.
\medskip

Introducing the variable (the {\it internal mass} of the
electro-magnetic field on the simultaneity and Cauchy surface
$\Sigma_{\tau}$)

\begin{equation}
M_U(\tau)=\int d^3\sigma\, \rho_U(\tau,\vec{\sigma}), \qquad \{
\theta (\tau ), M_U(\tau ) \} = \sgn,
 \label{a12}
\end{equation}

\noindent the constraint (\ref{a9}) can be written in the form

\begin{eqnarray}
H_\perp (\tau)&=&M_U(\tau)- {\cal E}[{\cal
A}^r(\tau,\vec{\sigma}),{\cal I}]\approx 0,\nonumber\\
 &&\nonumber\\
  {\cal E}[{\cal A}^r(\tau,\vec{\sigma}),{\cal I}]&=&\int d^3\sigma\,
[T_{\tau\tau}(z^\mu(\tau,\vec{\sigma}),{\cal I})]
_{\mbox{\scriptsize evaluated on the gauge fixing}}.
 \label{a13}
\end{eqnarray}

We also have

\bea
 && \{ {\cal A}^r(\tau ,\vec \sigma ), \rho_{Us}(\tau ,{\vec
\sigma}^{'}) \} = -\sgn\,
 \delta^r_s\, \delta^3(\vec \sigma - {\vec \sigma}^{'}),\nonumber \\
 &&{}\nonumber \\
  \rho^{\mu}(\tau ,\vec \sigma ) &\approx& \sgn\, \Big[
 \rho_U(\tau ,\vec \sigma)\, {\hat U}^{\mu}(\tau ) - \epsilon^{\mu}_r(\hat U(\tau ))\,
 \rho_{Ur}(\tau ,\vec \sigma )\Big],\nonumber \\
 &&{}\nonumber \\
 \rho_U(\tau ,\vec \sigma ) &\approx& \sgn\, {\tilde T}_{\tau\tau}({\cal
 A}^s(\tau ,\vec \sigma ), {\cal I})
 = \sgn\, [T_{\tau\tau}(z^\mu(\tau,\vec{\sigma}),{\cal I})]
_{\mbox{\scriptsize evaluated on the gauge fixing}}.
 \label{a14}
\eea

Then we can rewrite the constraints ${\cal H}_r(\tau,\vec{\sigma})
\approx 0 $ in the form

\begin{eqnarray}
{\cal H}_r(\tau,\vec{\sigma})&=& \frac{\partial {\cal
A}^s(\tau,\vec{\sigma})}{\partial \sigma^r}\,
\rho_{Us}(\tau,\vec{\sigma})- \sgn\, {\tilde T}_{\tau r}({\cal
A}^s(\tau,\vec{\sigma}),{\cal I})\approx 0,\nonumber\\
 &&\nonumber\\
 {\tilde T}_{\tau r}({\cal A}^s(\tau,\vec{\sigma}),{\cal I})&=&
[T_{\tau r}(z^\mu(\tau,\vec{\sigma}),{\cal I})]
_{\mbox{\scriptsize evaluated on the gauge fixing}}.
 \label{a15}
\end{eqnarray}
\medskip

Eqs.(\ref{a13}) show that the gauge fixing (\ref{a6}) and the
constraints ${\cal H}_U(\tau ,\vec \sigma ) - \delta^3(\vec \sigma
)\, {\cal H}_U(\tau ) = {\hat U}^{\mu}(\tau )\, [{\cal
H}_{\mu}(\tau ,\vec \sigma ) - \delta^3(\vec \sigma )\, \int
d^3\sigma_1\, {\cal H}_{\mu}(\tau ,{\vec \sigma}_1] = {\hat
U}(\tau )\, {\cal H}_{\mu}(\tau ,\vec \sigma ) - H_{\perp}(\tau
)\, \delta^3(\vec \sigma ) \approx {\tilde \rho}_U(\tau ,\vec
\sigma ) - \sgn\, [T_{\tau\tau}(\tau ,\vec \sigma ) -
\delta^3(\vec \sigma )\, \int d^3\sigma_1\, T_{\tau\tau}(\tau
,{\vec \sigma}_1)] \approx 0$, with ${\tilde \rho}_U(\tau ,\vec
\sigma ) = \rho_U(\tau ,\vec \sigma ) - \sgn\, M_U(\tau )$, form a
pair of second class constraints and the surviving first class
constraints are $H_{\perp}(\tau ) \approx 0$ and ${\cal H}_r(\tau
,\vec \sigma ) \approx 0$.

\medskip

After the gauge fixing (\ref{a6}), a set of canonical variables
for the reduced embedding are $\theta (\tau )$, $M_U(\tau )$,
${\cal A}^r(\tau ,\vec \sigma )$, $\rho_{Ur}(\tau ,\vec \sigma )$.
Note that they have non zero Poisson brackets with $X^{\mu}(\tau
)$, which therefore has to be replaced with a new ${\tilde
X}^{\mu}(\tau )$ to complete the canonical basis with ${\tilde
X}^{\mu}(\tau )$ and $U^{\mu}(\tau )$.

\bigskip

It can be shown \footnote{To show the validity of Eq.(\ref{a16}),
let us consider the constraints ${\cal H}_U(\tau ) = \int
d^3\sigma\, {\cal H}_U(\tau ,\vec \sigma ) = {\hat U}^{\mu}(\tau
)\, \int d^3\sigma\, {\cal H}_{\mu}(\tau ,\vec \sigma ) \approx 0$
and ${\cal H}_{Ur}(\tau ,\vec \sigma ) = \epsilon^{\mu}_r(\hat
U(\tau ))\, {\cal H}_{\mu}(\tau ,\vec \sigma ) \approx 0$, which
are weakly equal to $H_{\perp}(\tau ) \approx 0$ and ${\cal
H}_r(\tau ,\vec \sigma ) \approx 0$ when we add the gauge fixing
$S(\tau ,\vec \sigma ) \approx 0$. These constraints have weakly
vanishing Poisson brackets among themselves when $S(\tau ,\vec
\sigma ) \approx 0$. We have $\{ S(\tau ,\vec \sigma ), {\cal
H}_U(\tau ,{\vec \sigma}^{'}\} = \delta^3(\vec \sigma ) -
\delta^3(\vec \sigma - {\vec \sigma}^{'})$ [compatible with
$S(\tau ,\vec 0) = 0$] and this implies $\{ S(\tau ,\vec \sigma ),
{\cal H}_U(0) \} = \int d^3\sigma^{'}\, \{ S(\tau ,\vec \sigma ),
{\cal H}_U(\tau ,{\vec \sigma}^{'}) \} = 0$ and $\{ S(\tau ,\vec
\sigma ), {\cal H}_U(\tau ,{\vec \sigma}^{'} - \delta^3({\vec
\sigma}^{'})\, {\cal H}_U(\tau )\} = 0$.
 \medskip
To find the Dirac brackets associated to the second class
constraints $S(\tau ,\vec \sigma ) \approx 0$, ${\cal H}_U(\tau
,\vec \sigma ) - \delta^3(\vec \sigma )\, {\cal H}_U(\tau )
\approx 0$, we make their expansion around $\vec \sigma = 0$. Then
the multipoles
 \medskip
${\cal H}_{m_1m_2m_3}(\tau ) = {1\over {\sqrt{m_1!\, m_2!\,
m_3!}}}\, \int d^3\sigma\, (\sigma^1)^{m_1}\, (\sigma^2)^{m_2}\,
(\sigma^3)^{m_3}\, {\cal H}_U(\tau ,\vec \sigma )$,
 \medskip
$S_{m_1m_2m_3}(\tau ) = {1\over {\sqrt{m_1!\, m_2!\, m_3!}}}\,
{{\partial^{m_1+m_2+m_3}S(\tau ,\vec \sigma )}\over {\partial
\sigma^{1\, m_1} \partial \sigma^{2\, m_2}
\partial \sigma^{3\, m_3}}} {|}_{\vec \sigma =0}$,
\noindent  satisfy the algebra
\medskip
$\{ S_{m_1m_2m_3}(\tau ), {\cal H}_{n_1n_2n_3}(\tau )\} =
\delta_{m_1n_1}\, \delta_{m_2n_2}\, \delta_{m_3n_3}$, $\{
S_{m_1m_2m_3}(\tau ), S_{n_1n_2n_3}\} =0$, $\{{\cal
H}_{m_1m_2m_2}(\tau ), {\cal H}_{n_1n_2n_3}(\tau )\} \approx 0$.
 \medskip
This allows to get Eq.(\ref{a16}) with ${\cal H}_U(\tau ,\vec
\sigma )$ in place of ${\cal H}_{\perp}(\tau ,\vec \sigma )$. Then
this result weakly implies Eq.(\ref{a16}).} that the Dirac
brackets associated to the gauge fixing (\ref{a6}) are

\begin{eqnarray}
&&\{A(\tau),B(\tau)\}^* \approx \{A(\tau),B(\tau)\}+\nonumber\\
 &&\nonumber\\
 &+&\int d^3\sigma[ \{A(\tau),S(\tau,\vec{\sigma})\}
\{{\cal H}_U (\tau,\vec{\sigma}),B(\tau)\}-
\{B(\tau),S(\tau,\vec{\sigma})\} \{{\cal H}_U
(\tau,\vec{\sigma}),A(\tau)\}],\nonumber \\
 &&{}\nonumber \\
 \Rightarrow&& \rho^{\mu}(\tau ,\vec \sigma ) \approx
 {\tilde T}_{\tau\tau}({\cal A}^s(\tau ,\vec \sigma ), {\cal
 I})\, {\hat U}^{\mu}(\tau ) - \sgn\, \epsilon^{\mu}_r(\hat U(\tau ))\,
 \rho_{Ur}(\tau ,\vec \sigma ),\nonumber \\
 &&{}\nonumber \\
 P_s^{\mu}(\tau ) &=& \Big[ \sqrt{\sgn\, U^2(\tau )} +  \int
 d^3\sigma\, {\tilde T}_{\tau\tau}({\cal A}^s(\tau ,\vec \sigma ),
 {\cal I})\Big]\, {\hat U}^{\mu}(\tau ) - \sgn\,
 \epsilon^{\mu}_r(\hat U(\tau ))\, \int d^3\sigma\, \rho_{Ur}(\tau
 ,\vec \sigma ) \approx\nonumber \\
 &\approx& [1 +  M_U(\tau )]\, {\hat U}^{\mu}(\tau ) - \sgn\,
 \epsilon^{\mu}_r(\hat U(\tau ))\, \int d^3\sigma\, \rho_{Ur}(\tau
 ,\vec \sigma ).
 \label{a16}
\end{eqnarray}
\bigskip

It is easy to verify the following brackets [here  $F({\cal I})$
is a function of the canonical variables ${\cal I}$ only]

\begin{eqnarray*}
&&\{F_1({\cal I}),F_2({\cal I})\}^*= \{F_1({\cal I}),F_2({\cal
I})\} ,\nonumber \\
 &&\nonumber\\
 &&\{{\cal A}^r(\tau,\vec{\sigma}),{\cal A}^s(\tau,\vec{\sigma}')\}^*=
\{\rho_{Ur}(\tau,\vec{\sigma}),\rho_{Us}(\tau,\vec{\sigma}')\}^*=0,
\nonumber\\
 &&\nonumber\\
 &&\{{\cal A}^r(\tau,\vec{\sigma}),\rho_{Us}(\tau,\vec{\sigma}')\}^*=
 - \sgn\, \delta^r_s\,\delta(\vec{\sigma}-\vec{\sigma}'), \nonumber \\
 &&\nonumber\\
  &&\{{\cal A}^r(\tau,\vec{\sigma}),F({\cal I})\}^*=
\{\rho_{Ur}(\tau,\vec{\sigma}),F({\cal I})\}^*=0,
 \end{eqnarray*}

\bea
 &&\{M_U(\tau),\theta(\tau)\}^*=\{M_U(\tau),\theta(\tau)\}= \sgn,\nonumber\\
 &&\nonumber\\
 &&\{M_U(\tau),M_U(\tau)\}^*=\{\theta(\tau),\theta(\tau)\}=0,\nonumber \\
 &&\nonumber\\
  &&\{M_U(\tau),F({\cal I})\}^*=
\{M_U(\tau),{\cal A}^r(\tau,\vec{\sigma})\}^*=
\{M_U(\tau),\rho_{Ur}(\tau,\vec{\sigma})\}^*=0,\nonumber\\
 &&\nonumber\\
  &&\{\theta(\tau),F({\cal I})\}^*=
\{\theta(\tau),{\cal A}^r(\tau,\vec{\sigma})\}^*=
\{\theta(\tau),\rho_{Ur}(\tau,\vec{\sigma})\}^*=0.
 \label{a17}
 \end{eqnarray}

Moreover we have

\begin{eqnarray}
&&\{U^\mu(\tau),F({\cal I})\}^*= \{U^\mu(\tau),{\cal A}
^r(\tau,\vec{\sigma})\}^*=
\{U^\mu(\tau),\rho_{Ur}(\tau,\vec{\sigma})\}^*=\nonumber\\
&&\nonumber\\ &=& \{U^\mu(\tau),M_U(\tau)\}^*=
\{U^\mu(\tau),\theta(\tau)\}^* =0.
 \label{a18}
\end{eqnarray}

Since the Dirac brackets of $X^\mu(\tau)$ with the other canonical
variables are very complicated, we do not give them.
\bigskip

All these brackets show us that the pairs $\theta(\tau)$, $
M_U(\tau)$, ${\cal A} ^r(\tau,\vec{\sigma})$, $
\rho_{Ur}(\tau,\vec{\sigma})$, ${\tilde X}^{\mu}(\tau )$,
$U^{\mu}(\tau )$ together with the original variables ${\cal I}$
are a canonical basis for the reduced phase space, if ${\tilde
X}^{\mu}(\tau )$ is a suitable replacement of $X^{\mu}(\tau )$. To
find ${\tilde X}^\mu(\tau)$ we have first to study the Lorentz
covariance of the new variables on the reduced phase space.
\medskip

Let us first observe that  Eqs.(\ref{a6}) imply

\begin{equation}
\{J_s^{\mu\nu}(\tau),S(\tau,\vec{\sigma})\}=
\{J_s^{\mu\nu}(\tau),{\cal H}_\perp(\tau,\vec{\sigma})\}=0.
 \label{a19}
\end{equation}

\noindent so that the Dirac brackets  (\ref{a16}) change neither
the Poisson algebra of the generators $J^{\mu\nu}$

\begin{eqnarray}
\{J_s^{\mu\nu}(\tau),
J_s^{\sigma\rho}(\tau)\}^*&=&\{J_s^{\mu\nu}(\tau),
J_s^{\sigma\rho}(\tau)\}=
C^{\mu\nu\sigma\rho}_{\alpha\beta}J_s^{\alpha\beta}(\tau),\nonumber\\
 &&\nonumber\\
  C^{\mu\nu\rho\sigma}_{\alpha\beta}&=&
\eta^\nu_\alpha\eta^\rho_\beta\eta^{\mu\sigma}+
\eta^\mu_\alpha\eta^\sigma_\beta\eta^{\nu\rho}-
\eta^\nu_\alpha\eta^\sigma_\beta\eta^{\mu\rho}-
\eta^\mu_\alpha\eta^\rho_\beta\eta^{\nu\sigma},
 \label{a20}
\end{eqnarray}

\noindent nor the transformations properties of the canonical
variables on the reduced phase space. In particular we have the
Lorentz scalar variables

\begin{equation}
\{J_s^{\mu\nu}(\tau),F({\cal
I})\}^*=\{J_s^{\mu\nu}(\tau),M_U(\tau)\}^*=
\{J_s^{\mu\nu}(\tau),\theta (\tau )\}^*=0.
 \label{a21}
\end{equation}
\medskip

On the contrary the variables ${\cal A}^r(\tau,\vec{\sigma})$,
$\rho^r_U(\tau,\vec{\sigma})= \eta^{rs}\,
\rho_{Us}(\tau,\vec{\sigma})$ are not scalar, but they transform
as {\em Wigner spin-1 3-vectors} since the tetrad fields
$\epsilon^{\mu}_A(\hat U)$ are the columns of the standard Wigner
boost $L(\hat U, {\hat U}_o)$ for time-like Poincare' orbits.
\medskip

In fact by using the infinitesimal transformations

\begin{eqnarray}
\Lambda_{\mu\nu}&=&\eta_{\mu\nu}+\delta\omega_{\mu\nu},\qquad
\delta\omega_{\mu\nu}=-\delta\omega_{\nu\mu},\nonumber\\
 &&\nonumber\\
  R_{sr}(\Lambda,\hat U)&=&\delta_{sr} +D_{sr}{}^{
\mu\nu}(\hat U)\,\delta\omega_{\mu\nu}, \qquad D_{sr}{}^{
\mu\nu}(\hat U)= -D_{rs}{}^{ \mu\nu}(\hat U)= -D_{sr}{}^{
\nu\mu}(\hat U),
 \label{a22}
\end{eqnarray}

\noindent in the last of Eqs.(\ref{a2}), we obtain

\begin{equation}
\left(\eta_{\mu\nu}+\delta\omega_{\mu\nu}\right)\,\epsilon^\nu_s(\hat
U)\, \left(\delta_{sr}+D_{sr}{}^{ \sigma\rho}(\hat
U)\,\delta\omega_{\sigma\rho}\right)= \epsilon^\mu_r(\hat
U)+\frac{1}{2}\delta\omega_{\sigma\rho}\{ \epsilon^\mu_r(\hat
U),J_s^{\sigma\rho}(\tau)\}.
 \label{a23}
\end{equation}
\medskip

Then we get

\begin{equation}
\{\epsilon^\mu_r(\hat U),J_s^{\sigma\rho}(\tau)\}=
-\eta^{\rho\mu}\epsilon^\sigma_r(\hat U) +
\eta^{\sigma\mu}\epsilon^\rho_r(\hat U)+ 2\,D_{sr}{}^{
\sigma\rho}(\hat U)\epsilon^\mu_s(\hat U),
 \label{a24}
\end{equation}

\noindent and finally

\begin{eqnarray}
\{{\cal A}^r(\tau,\vec{\sigma}),J_s^{\sigma\rho}(\tau)\}^*&=&
\{\epsilon^r_\mu(\hat
U)\,z^\mu(\tau,\vec{\sigma}),J_s^{\sigma\rho}(\tau)\}=
-2\,D_{rs}{}^{ \sigma\rho}(\hat U)\,{\cal A}
^s(\tau,\vec{\sigma}),\nonumber\\
 &&\nonumber\\
 \{\rho^r_U(\tau,\vec{\sigma}),J_s^{\sigma\rho}(\tau)\}^*&=&
\{\epsilon^r_\mu(\hat
U)\,z^\mu(\tau,\vec{\sigma}),J_s^{\sigma\rho}(\tau)\}=
-2\,D_{rs}{}^{ \sigma\rho}(\hat U)\,\rho^s_U(\tau,\vec{\sigma}).
 \label{a25}
\end{eqnarray}
\medskip

Since we have

\begin{equation}
\{\epsilon^\mu_r(\hat U),J_s^{\sigma\rho}(\tau)\}= \frac{\partial
\epsilon^\mu_r(\hat U)}{\partial {\hat U}_\gamma}\, \{{\hat
U}_\gamma(\tau),J_s^{\sigma\rho}(\tau)\}= -\frac{\partial
\epsilon^\mu_r(\hat U)}{\partial {\hat U}_\gamma}
\left[\eta^\rho_\gamma\,{\hat U}^\sigma(\tau)
-\eta^\sigma_\gamma\,{\hat U}^\rho(\tau) \right],
 \label{a26}
\end{equation}

\noindent we  obtain the following expression for the matrix $D$

\begin{equation}
D^{\alpha\beta}_{rs}(\hat U)= \frac{1}{2}\left[
\epsilon^\alpha_r(\hat U)\epsilon^\beta_s(\hat U)-
\epsilon^\alpha_s(\hat U)\epsilon^\beta_r(\hat U)-\left( {\hat
U}^\alpha\frac{\partial \epsilon^\mu_r(\hat U)}{\partial {\hat
U}_\beta}- {\hat U}^\beta\frac{\partial \epsilon^\mu_r(\hat
U)}{\partial {\hat U}_\alpha} \right)\,\epsilon_{s\mu}(\hat
U)\right].
 \label{a27}
\end{equation}
\bigskip

To find the last canonical variables ${\tilde X}^\mu(\tau)$, let
us define

\begin{equation}
L^{\mu\nu}(\tau)=J_s^{\mu\nu}(\tau)-D_{rs}{}^{ \mu\nu}(\hat
U)\,\int d^3\sigma\, \left[ {\cal A}
^r(\tau,\vec{\sigma})\,\rho^s_U(\tau,\vec{\sigma})- {\cal A}
^s(\tau,\vec{\sigma})\,\rho^r_U(\tau,\vec{\sigma}) \right].
 \label{a28}
\end{equation}

Then from Eq.(\ref{a5}) we get

\begin{equation}
L^{\mu\nu}(\tau)=X^\mu(\tau)U^\nu(\tau)-X^\nu(\tau)U^\mu(\tau)
+\widetilde{I}^{\mu\nu}(\tau),
 \label{a29}
\end{equation}

\noindent with

\begin{eqnarray}
\widetilde{I}^{\mu\nu}(\tau)&=&\int d^3\sigma\,\left[
z^\mu(\tau,\vec{\sigma})\rho^\nu(\tau,\vec{\sigma})-
z^\nu(\tau,\vec{\sigma})\rho^\mu(\tau,\vec{\sigma})\right]
_{\mbox{\scriptsize evaluated on the gauge fixing}} -\nonumber\\
 &&\nonumber\\
 &-&D_{rs}{}^{ \mu\nu}(\hat U)\,\int d^3\sigma\,
\left[ {\cal A}^r(\tau,\vec{\sigma})\,\rho^s_U(\tau,\vec{\sigma})-
{\cal A}^s(\tau,\vec{\sigma})\,\rho^r_U(\tau,\vec{\sigma})
\right]=\nonumber\\
 &&\nonumber\\
  &=& U^\mu(\tau)\,\left[
\frac{1}{\sqrt{\sgn\, U^2(\tau)}}\,\int
d^3\sigma\,\left(\theta(\tau)\, \epsilon^\nu_r(\hat U(\tau )
)\rho^r_U(\tau,\vec{\sigma})- \epsilon^\nu_r(\hat U(\tau ))\,{\cal
A}^r(\tau,\vec{\sigma})\,
\rho_U(\tau,\vec{\sigma})\right)\right.+\nonumber\\
 &&\nonumber\\
 &+&\left. \frac{\partial \epsilon^\alpha_r(\hat U(\tau )}{\partial
{\hat U}_\nu} \,\epsilon_{s\alpha}(\hat U(\tau )) \int
d^3\sigma\,{\cal A}
^r(\tau,\vec{\sigma})\,\rho^s_U(\tau,\vec{\sigma})\right]-(\mu
\leftrightarrow\nu) =\nonumber\\
 &&\nonumber\\
 &{\buildrel {def}\over =}&
 U^\mu(\tau)\,\widetilde{W}^\nu(\tau)-(\mu\leftrightarrow\nu).
 \label{a30}
\end{eqnarray}

Therefore we get

\bea
 L^{\mu\nu}(\tau) & =&
(X^\mu(\tau)-\widetilde{W}^\mu(\tau))U^\nu(\tau)
-(\mu\leftrightarrow\nu) ={\tilde X}^\mu(\tau)U^\nu(\tau)-{\tilde
X}^\nu(\tau)U^\mu(\tau),\nonumber \\
 &&{}\nonumber \\
 J_s^{\mu\nu} &=& {\tilde X}^{\mu}(\tau )\, U^{\nu}(\tau ) - {\tilde
 X}^{\nu}(\tau )\, U^{\mu}(\tau ) + D_{rs}{}^{\mu\nu}(\hat U)\,
 \int d^3\sigma\, [{\cal A}^r\, \rho^s_U - {\cal A}^s\,
 \rho^r_U](\tau ,\vec \sigma ) =\nonumber \\
 &{\buildrel {def}\over =}& {\tilde X}^{\mu}(\tau )\, U^{\nu}(\tau ) - {\tilde
 X}^{\nu}(\tau )\, U^{\mu}(\tau ) + {\tilde S}^{\mu\nu},\nonumber \\
 &&{}\nonumber \\
 &&\{ {\tilde S}^{\mu\nu}, {\tilde S}^{\alpha\beta} \} =
 C^{\mu\nu\alpha\beta}_{\rho\sigma}\, {\tilde S}^{\rho\sigma} +
 \Big({{\partial D_{rs}{}^{\mu\nu}(\hat U)}\over {\partial {\hat U}_{\beta}}}\,
 U^{\alpha} - {{\partial D_{rs}{}^{\mu\nu}(\hat U)}\over {\partial {\hat U}_{\alpha}}}\,
 U^{\beta} -\nonumber \\
 &&\quad - {{\partial D_{rs}{}^{\alpha\beta}(\hat U)}\over {\partial {\hat U}_{\nu}}}\,
 U^{\mu} + {{\partial D_{rs}{}^{\alpha\beta}(\hat U)}\over {\partial {\hat U}_{\mu}}}\,
 U^{\nu}\Big)\, S^{rs},\nonumber \\
 &&S^{rs} = \int d^3\sigma\, ({\cal A}^r\, \rho^s_U - {\cal A}^s\,
 \rho^r_U)(\tau ,\vec \sigma ),
 \label{a31}
 \eea

\noindent where

\bea
 {\tilde X}^\mu(\tau) &=& X^\mu(\tau)-\widetilde{W}^\mu(\tau)
 =\nonumber \\
 &=& X^{\mu}(\tau ) + \frac{1}{\sqrt{\sgn\, U^2(\tau)}}\,\int
d^3\sigma\,\left(\theta(\tau)\, \epsilon^\mu_r(\hat U(\tau )
)\rho^r_U(\tau,\vec{\sigma})- \epsilon^\mu_r(\hat U(\tau ))\,{\cal
A}^r(\tau,\vec{\sigma})\, \rho_U(\tau,\vec{\sigma}) \right)
+\nonumber \\
 &+& \frac{\partial \epsilon^\alpha_r(\hat U(\tau )}{\partial
{\hat U}_\mu} \,\epsilon_{s\alpha}(\hat U(\tau )) \int
d^3\sigma\,{\cal A}
^r(\tau,\vec{\sigma})\,\rho^s_U(\tau,\vec{\sigma})
 \label{a32}
\eea

\medskip

We see that ${\tilde S}^{\mu\nu}$ does not satisfy the right
algebra for a spin tensor: this suggests that a further
modification of ${\tilde X}^{\mu}(\tau )$ should be possible so to
obtain a real spin tensor.

\medskip

From Eqs.(\ref{a2}) we get

\begin{equation}
{U}_\mu(\tau)\,\widetilde{W}^\mu(\tau)=0\,\Rightarrow\,
{U}_\mu(\tau)\,{\tilde X}^\mu(\tau)={U}_\mu(\tau)\,X^\mu(\tau),
 \label{a33}
\end{equation}

\noindent so that we can write

\begin{eqnarray}
{\tilde X}
^\mu(\tau)&=&(\hat{U}^\sigma(\tau)\,X_\sigma(\tau))\,\hat{U}^\sigma(\tau)+
L^{\mu\rho}(\tau)\hat{U}_\rho(\tau)\frac{1}{\sqrt{\sgn\,
U^2(\tau)}}= \nonumber\\
 &&\nonumber\\
 &=&(\hat{U}^\sigma(\tau)\,X_\sigma(\tau))\,\hat{U}^\sigma(\tau)+
J^{\mu\rho}(\tau)\hat{U}_\rho(\tau)\frac{1}{\sqrt{\sgn\,
U^2(\tau)}}- \nonumber\\
 &&\nonumber\\
  &-&\frac{\partial
\epsilon^\alpha_r(\hat U(\tau )}{\partial {\hat U}_\nu}
\,\epsilon_{s\alpha}(\hat U)\,\int d^3\sigma\, \left[ {\cal A}
^r(\tau,\vec{\sigma})\, \rho^s_U(\tau,\vec{\sigma})- {\cal A}
^s(\tau,\vec{\sigma})\, \rho^r_U(\tau,\vec{\sigma}) \right].
 \label{a34}
\end{eqnarray}

By construction we have

\begin{eqnarray}
&&\{L^{\mu\nu}(\tau),F({\cal I})\}^*= \{L^{\mu\nu}(\tau),{\cal A}
^r(\tau,\vec{\sigma})\}^*=
\{L^{\mu\nu}(\tau),\rho_{Ur}(\tau,\vec{\sigma})\}^*=\nonumber\\
&&\nonumber\\ &=& \{L^{\mu\nu}(\tau),M_U(\tau)\}^*=
\{L^{\mu\nu}(\tau),\theta(\tau)\}^* =0,\nonumber\\
 &&\nonumber\\
 &&\{L^{\mu\nu}(\tau),U^\sigma(\tau)\}^*
=\eta^{\nu\sigma}U^\mu(\tau)-\eta^{\mu\sigma}U^\nu(\tau),
 \label{a35}
\end{eqnarray}

\noindent and then we can get

\begin{eqnarray}
&&\{{\tilde X}^\mu(\tau),F({\cal I})\}^*= \{{\tilde X}
^\mu(\tau),{\cal A}^r(\tau,\vec{\sigma})\}^*= \{{\tilde X}
^\mu(\tau),\rho_{Ur}(\tau,\vec{\sigma})\}^*=\nonumber\\
&&\nonumber\\
 &=& \{{\tilde X}^\mu(\tau),M_U(\tau)\}^*=
\{{\tilde X}^\mu(\tau),\theta(\tau)\}^* =0.
 \label{a36}
\end{eqnarray}
\medskip

A long and tedious calculation allows to get

\begin{eqnarray}
\{{\tilde X}^\mu(\tau),{\tilde X}^\nu(\tau)\}^*&=&0,\nonumber\\
&&\nonumber\\
 \{{\tilde X}^\mu(\tau),U^\nu(\tau)\}^*&=&-\eta^{\mu\nu},\nonumber\\
&&\nonumber\\
 \{L^{\mu\nu}(\tau), L^{\sigma\rho}(\tau)\}^*&=&
C^{\mu\nu\sigma\rho}_{\alpha\beta}L^{\alpha\beta}(\tau),\nonumber\\
&&\nonumber\\
 \{L^{\mu\nu}(\tau),{\tilde X}^\sigma(\tau)\}^*
&=&\eta^{\nu\sigma}{\tilde X}^\mu(\tau)-\eta^{\mu\sigma}{\tilde
X}^\nu(\tau).
 \label{a37}
\end{eqnarray}

\bigskip

The looked for final pairs of canonical variables are given by
${\tilde X}^\mu(\tau)$, $U^{\mu}(\tau )$. Let us remark that
${\tilde X}^{\mu}$ is {\it not a Lorentz four-vector} since we
have

\begin{eqnarray}
\{J^{\mu\nu}(\tau),{\tilde X}^\sigma(\tau)\}^*
&=&\eta^{\nu\sigma}{\tilde X}^\mu(\tau)-\eta^{\mu\sigma}{\tilde X}
^\nu(\tau)+ \nonumber\\
 &&\nonumber\\
  &+&\frac{\partial D_{rs}{}^{
\mu\nu}(\hat U)} {\partial {\hat U}_\sigma} \,\int d^3\sigma\,
\left[ {\cal A}
^r(\tau,\vec{\sigma})\,\rho^s_U(\tau,\vec{\sigma})- {\cal A}
^s(\tau,\vec{\sigma})\,\rho^r_U(\tau,\vec{\sigma}) \right].
 \label{a38}
\end{eqnarray}

\medskip

Following Ref.\cite{43a} we can make the canonical transformation:
$({\tilde X}^{\mu}, U^{\mu})\, \mapsto \, ({\hat U}_{\mu}\,
{\tilde X}^{\mu} = {\hat U}_{\mu}\, X^{\mu}, \sqrt{\sgn\, U^2}
\approx 1),\, (\vec z, \vec k = \vec U / \sqrt{\sgn\, U^2} \approx
\vec U)$ with the Newton-Wigner-like non-covariant 3-vector $\vec
z = \sqrt{\sgn\, U^2}\, ({\vec {\tilde X}} - {\tilde X}_o\, \vec U
/ U^o) \approx {\vec {\tilde X}} - {\tilde X}^o \, \vec U /
\sqrt{1 - {\vec U}^2}$.

\bigskip

In conclusion with these Dirac brackets the Poincare' algebra is
still satisfied [$\{ P^{\mu}_s, P^{\nu}_s\}^* = 0$, $\{ P^{\mu}_s,
J_s^{\rho\sigma}\}^* = \eta^{\mu\rho}\, P^{\sigma}_s -
\eta^{\mu\sigma}\, P^{\rho}_s$, $\{ J^{\mu\nu}_s,
J^{\sigma\rho}_s\}^* = C^{\mu\nu\sigma\rho}_{\alpha\beta}\,
J_s^{\alpha\beta}$] and the final algebra of the surviving first
class constraints is

\begin{eqnarray}
&& \{{\cal H}_r(\tau,\vec{\sigma}) ,{\cal
H}_s(\tau,\vec{\sigma}')\}^*= {\cal H}_r(\tau,\vec{\sigma}')
\frac{\partial}{\partial\sigma^{\prime\,s}}
\delta^3(\vec{\sigma}-\vec{\sigma}')- {\cal
H}_s(\tau,\vec{\sigma}) \frac{\partial}{\partial\sigma^{s}}
\delta^3(\vec{\sigma}-\vec{\sigma}'),\nonumber\\
 &&\nonumber\\
  &&\{H_\perp(\tau),{\cal H}_r(\tau,\vec{\sigma})\}^*=0.
   \label{a39}
\end{eqnarray}

\bigskip

If we want to recover the embedding (\ref{a1}), i.e.
Eq.(\ref{VI110}), we must add the following gauge fixings to the
first class constraints $H_{\perp}(\tau ) \approx 0$ and ${\cal
H}_r(\tau ,\vec \sigma ) \approx 0$

\begin{eqnarray*}
 &&\theta (\tau ) - x_U(\tau ) - {\hat U}_{\mu}(\tau )\,
 x^{\mu}(0) \approx 0,\nonumber \\
 &&{}\nonumber \\
 &&{\cal A}^r(\tau ,\vec \sigma ) - \xi^r_U(\tau ,\vec \sigma ) -
 \epsilon^r_{\mu}(\hat U(\tau ))\, x^{\mu}(0) \approx 0,\nonumber \\
 &&{}\nonumber \\
 &&\Downarrow\nonumber \\
 &&{}\nonumber \\
 &&z^{\mu}(\tau ,\vec 0) = x^{\mu}_U(\tau ),\nonumber \\
 &&{}\nonumber \\
 &&\rho_{Ur}(\tau ,\vec \sigma ) \approx \sgn\, A^s_r(\tau ,\vec
 \sigma )\, {\tilde T}_{\tau s}(\xi_U^u(\tau ,\vec \sigma ) +
 \epsilon^u_{\mu}(\hat U(\tau ))\, x^{\mu}(0), {\cal I}),\nonumber \\
 &&{}\nonumber \\
 &&{}\nonumber \\
 &&\qquad \Big[A^s_r(\tau ,\vec \sigma )\, inverse\, of\,\,
 {{\partial \xi^r_U(\tau ,\vec \sigma )}\over {\partial \sigma^s}}\Big],
 \end{eqnarray*}

\bea
 &&S^{rs} = \int d^3\sigma\, \Big[ (\xi^r_U\, A^{sv} - \xi^s_U\,
 A^{rv})\, {\tilde T}_{\tau v}\Big](\tau ,\vec \sigma ) +\nonumber \\
 &&\qquad + x^{\mu}(0)\, \Big[\epsilon^r_{\mu}(\hat U(\tau ))\,
 \int d^3\sigma\, A^{sv}(\tau ,\vec \sigma ) - \epsilon^s_{\mu}(\hat U(\tau ))\,
 \int d^3\sigma\, A^{rv}(\tau ,\vec \sigma )\Big]\nonumber \\
 &&\qquad {\tilde T}_{\tau v}(\xi_U^u(\tau ,\vec \sigma ) +
 \epsilon^u_{\mu}(\hat U(\tau ))\, x^{\mu}(0), {\cal I}),\nonumber \\
 &&{}\nonumber \\
 &&P_s^{\mu}(\tau ) \approx \Big[1 + {\cal E}[\xi^u_U(\tau ,\vec \sigma )
 + \epsilon^u_{\alpha}(\hat U(\tau ))\, x^{\alpha}(0), {\cal I}]\Big]\, {\hat U}^{\mu}(\tau )
 -\nonumber \\
 &&\qquad - \epsilon^{\mu}_r(\hat U(\tau ))\, \int d^3\sigma\,
 A^s_r(\tau ,\vec \sigma )\, {\tilde T}_{\tau s}(\xi^u_U(\tau
 ,\vec \sigma ) + \epsilon^u_{\alpha}(\hat U(\tau ))\,
 x^{\alpha}(0), {\cal I}).
 \label{a40}
 \eea

\medskip

The stability of the gauge fixings (\ref{a40}) and ${{d {\hat
U}^{\mu}(\tau )}\over {d \tau}} \cir 0$ imply $\mu (\tau ) = -
\dot \theta (\tau ) = - {\dot x}_U(\tau ) = - {\dot
x}^{\mu}_U(\tau )\, {\hat U}_{\mu}(\tau )$, $\lambda^r(\tau ,\vec
\sigma ) = -\sgn\, {\cal A}^r_s(\tau ,\vec \sigma )\, {{\partial
{\cal A}^s(\tau ,\vec \sigma )}\over {\partial \tau}} = -\sgn\,
{\cal A}^r_s(\tau ,\vec \sigma )\, \Big({\dot x}^{\mu}_U(\tau )\,
\epsilon_{s\mu}(\hat U(\tau )) + {{\partial \zeta^s(\tau ,\vec
\sigma )}\over {\partial \tau}}\Big)$ for the Dirac multipliers
appearing in the Dirac Hamiltonian (\ref{a9}) and in the
associated Hamilton equations. If we go to new Dirac brackets, in
the new reduced phase space we get $H_D = \kappa (\tau )\, \chi
(\tau ) + (electro-magnetic\, constraints)$ and this Dirac
Hamiltonian does not reproduce the just mentioned Hamilton
equations after their restriction to Eqs.(\ref{a40}) due to the
explicit $\tau$-dependence of the gauge fixings. As a consequence,
in analogy to what was done to get Eqs.(\ref{VI81}) and
(\ref{VI84}), we have to find the correct Hamiltonian ruling the
evolution in the reduced phase space. A look at the Hamilton
equations shows that this Hamiltonian is

\begin{eqnarray*}
 H &=& - \mu (\tau )\, {\cal E}[\xi^r_U(\tau ,\vec \sigma ) +
 \epsilon^r_{\mu}(\hat U(\tau ))\, x^{\mu}(0), {\cal I}] -\nonumber \\
 &-& \int
 d^3\sigma\, \lambda^r(\tau ,\vec \sigma )\, {\tilde T}_{\tau
 r}(\xi^u_U(\tau ,\vec \sigma ) + \epsilon^u_{\mu}(\hat U(\tau
 ))\, x^{\mu}(0), {\cal I}) +\nonumber \\
&+&  \int d^3\sigma\,
 [\lambda_{\tau}(\tau ,\vec \sigma )\, \pi^{\tau}(\tau ,\vec
 \sigma ) - A_{\tau}(\tau ,\vec \sigma )\, \Gamma (\tau ,\vec
 \sigma )] =\nonumber \\
 &&{}\nonumber \\
 &=& {\dot x}^{\mu}_U(\tau )\, \Big[ {\hat U}_{\mu}(\tau )\,
{\cal E}[\xi^r_U(\tau ,\vec \sigma ) +
 \epsilon^r_{\mu}(\hat U(\tau ))\, x^{\mu}(0), {\cal I}]
 -\nonumber \\
 &-& \epsilon_{r\mu}(\hat U(\tau ))\, \int d^3\sigma\,
 {\tilde T}_{\tau
 r}(\xi^u_U(\tau ,\vec \sigma ) + \epsilon^u_{\mu}(\hat U(\tau
 ))\, x^{\mu}(0), {\cal I}) \Big] +\nonumber \\
 &+& \int d^3\sigma\, {\cal A}^r_s(\tau ,\vec \sigma )\,
 {{\partial \zeta^s(\tau ,\vec \sigma )}\over
 {\partial \tau}}\, {\tilde T}_{\tau
 r}(\xi^u_U(\tau ,\vec \sigma ) + \epsilon^u_{\mu}(\hat U(\tau
 ))\, x^{\mu}(0), {\cal I}) +\nonumber \\
&+&  \int d^3\sigma\,
 [\lambda_{\tau}(\tau ,\vec \sigma )\, \pi^{\tau}(\tau ,\vec
 \sigma ) - A_{\tau}(\tau ,\vec \sigma )\, \Gamma (\tau ,\vec
 \sigma )] =
 \end{eqnarray*}

\bea
 &=& {\dot x}^{\mu}_U(\tau )\, \Big[P_{s\mu} - {\hat U}_{\mu}(\tau
 )\Big] +\nonumber \\
 &+& \int d^3\sigma\, {\cal A}^r_s(\tau ,\vec \sigma )\,
 {{\partial \zeta^s(\tau ,\vec \sigma )}\over
 {\partial \tau}}\, {\tilde T}_{\tau
 r}(\xi^u_U(\tau ,\vec \sigma ) + \epsilon^u_{\mu}(\hat U(\tau
 ))\, x^{\mu}(0), {\cal I}) +\nonumber \\
  &+&  \int d^3\sigma\,
 [\lambda_{\tau}(\tau ,\vec \sigma )\, \pi^{\tau}(\tau ,\vec
 \sigma ) - A_{\tau}(\tau ,\vec \sigma )\, \Gamma (\tau ,\vec
 \sigma )] .
 \label{a41}
  \eea

\medskip

We find that, apart from the contribution of the remaining first
class constraints, the {\it effective non-inertial Hamiltonian}
ruling the $\tau$-evolution seen by the (in general non-inertial)
observer $x^{\mu}_U(\tau )$ (the centroid origin of the
3-coordinates) is the sum of the projection of the total
4-momentum along the 4-velocity of the observer (without the term
pertaining to the decoupled unit mass particle it is the {\it
effective internal mass}) plus a term induced by the differential
rotation of the 3-coordinate system around the world-line of the
observer (the potential of {\it inertial forces}).

\bigskip

 To eliminate the constraint $\chi (\tau ) = \sgn\, U^2(\tau ) - 1
 \approx 0$ we add the gauge fixing

\bea
 &&{\hat U}_{\mu}(\tau )\, {\tilde X}^{\mu}(\tau ) - \sgn\, \theta
 (\tau ) = {\hat U}_{\mu}(\tau )\, X^{\mu}(\tau ) - \sgn\, \theta
 (\tau ) \approx 0,\quad \Rightarrow\quad \kappa (\tau ) = - {{\sgn}\over 2}\,
 \dot \theta (\tau ),\nonumber \\
 &&{}\nonumber \\
 &&\Downarrow\nonumber \\
 &&{}\nonumber \\
 &&{\tilde X}^{\mu}(\tau ) = z^{\mu}(\tau , {\vec \sigma}_{\tilde
 X}(\tau )),\quad for\, some\quad {\vec \sigma}_{\tilde X}(\tau
 ),\nonumber \\
 &&X^{\mu}(\tau ) = z^{\mu}(\tau ,{\vec \sigma}_X(\tau )),\quad
 for\, some\quad {\vec \sigma}_X(\tau ),\nonumber \\
 &&{}\nonumber \\
 U^{\mu}(\tau ) &=& \Big( \sqrt{1 + {\vec k}^2}; k^i(\tau )
 \Big) = {\hat U}^{\mu}(\vec k),\nonumber \\
 {\tilde X}^{\mu}(\tau ) &=& \Big( \sqrt{1 + {\vec k}^2}\, [\sgn\,
 \theta (\tau ) + \vec k(\tau ) \cdot \vec z(\tau )];\nonumber \\
 &&z^i(\tau ) + k^i(\tau )\, [\sgn\, \theta (\tau ) + \vec k(\tau
 ) \cdot \vec z(\tau )] \Big) = z^{\mu}(\tau ,{\vec
 \sigma}_{\tilde X}(\tau )),\nonumber \\
 &&{}\nonumber \\
 &&L^{ij} = z^i\, k^j - z^j\, k^i,\qquad L^{oi} = - L^{io} = -
 z^i\, \sqrt{1 + {\vec k}^2}.
 \label{a42}
 \eea
 \bigskip

After having introduced new Dirac brackets, the extra added point
particle of unit mass is reduced to the decoupled non-evolving
variables $\vec z$, $\vec k$ and the not yet determined ${\vec
\sigma}_{\tilde X}(\tau )$ and ${\vec \sigma}_X(\tau )$ give the
3-location of ${\tilde X}^{\mu}(\tau )$ and $X^{\mu}(\tau )$,
respectively, which do not coincide with the world-line
$x^{\mu}_U(\tau )$ of the non-inertial observer. Now we get ${\dot
{\tilde X}}^{\mu}(\tau ) = \dot \theta (\tau )\, {\hat
U}^{\mu}(\tau )$ and this determines ${\vec \sigma}_{\tilde
X}(\tau )$ as solution of the equation ${{\partial {\cal A}^r(\tau
,{\vec \sigma}_{\tilde X}(\tau ))}\over {\partial \tau}} +
{{\partial {\cal A}^r(\tau ,\vec \sigma )}\over {\partial
\sigma^s}}{|}_{\vec \sigma = {\vec \sigma}_{\tilde X}(\tau ) }\,
{\dot \sigma}^s_{\tilde X}(\tau ) = 0$.
\medskip

Since Eq.(\ref{a35}) remains true, we still have that under a
Lorentz transformation $\Lambda$ we get $U^{\mu} \mapsto
\Lambda^{\mu}{}_{\nu}\, U^{\nu}$. Moreover, we still have ${\dot
x}^{\mu}_U(\tau ) = \dot \theta (\tau )\, {\hat U}^{\mu}(\tau ) +
\epsilon^{\mu}_r(\hat U(\tau ))\, {\dot {\cal A}}^r(\tau ,\vec
0)$, namely the 4-velocity of the non-inertial observer is {\it
not orthogonal} to the hyper-planes $\Sigma_{\tau}$.

\bigskip

Finally the embedding (\ref{IV1}) with a fixed unit normal
$l^{\mu}$, implying the breaking of the action of Lorentz boosts,
is obtained by adding by hand the first class constraints

\beq
 {\hat U}^{\mu}(\vec k ) - l^{\mu} \approx 0,
 \label{a43}
 \eeq

\noindent which determine the non-evolving constant $\vec k$. The
conjugate constant $\vec z$ can be eliminated with the
non-covariant gauge fixings

\beq
 \vec z \approx \vec 0,\,\, \Rightarrow \,\,{\tilde X}^{\mu}(\tau )
 \approx \sgn\, \theta (\tau )\, {\hat U}^{\mu}(\tau ).
 \label{a44}
 \eeq

\medskip

The constraints (\ref{a43}) and (\ref{a44}) eliminate the extra
non-evolving degrees of freedom $\vec k$ and $\vec z$ of the added
decoupled point particle, respectively.

\bigskip

At this stage only the electro-magnetic canonical variables  are
left and Eq.(\ref{a40}) determine the Poincare' generators. It is
not clear if in this case there is a non-inertial analogue of the
internal Poincare' group of the rest-frame instant form.

\medskip

To recover the rest-frame instant form, having the Wigner
hyper-planes orthogonal to the total 4-momentum as simultaneity
surfaces, we must require ${\hat U}^{\mu}(\tau ) - p^{\mu} /
\sqrt{\sgn\, p^2} \approx 0$ instead of Eq.(\ref{a43}). Then from
Eq.(\ref{a40}) we get the rest-frame conditions
$\epsilon^r_{\mu}(\hat U)\, p^{\mu} \approx 0$ (whose gauge fixing
is the vanishing (\ref{a44}) of the internal center of mass ${\vec
\sigma}_X(\tau ) \approx 0$, see Ref.\cite{43a}) and the invariant
mass ${\cal E} + 1$, which is the correct one if we neglect the
constant extra mass $1$.
\bigskip

The technology of this Appendix could be used to study a family of
admissible embeddings, whose leaves are general space-like
hyper-surfaces, closed under the action of the Lorentz group.

\vfill\eject

\end{document}